\documentclass[journal]{IEEEtran}
\usepackage{balance}
\usepackage{multirow}

\usepackage{cite}

\ifCLASSINFOpdf
   \usepackage[pdftex]{graphicx}
\else
   \usepackage[dvips]{graphicx}
\fi
\usepackage{amsmath}
\ifCLASSOPTIONcompsoc
  \usepackage[caption=false,font=normalsize,labelfont=sf,textfont=sf]{subfig}
\else
  \usepackage[caption=true,font=footnotesize]{subfig}
\fi

\usepackage{overpic}
\usepackage{color,xcolor}

\begin{document}
%
\title{A three-dimensional dual-domain deep network for high-pitch and sparse helical CT reconstruction}
%
%
%

\author{Wei Wang, Xiang-Gen Xia, Chuanjiang He, Zemin Ren and Jian Lu
	\thanks{This work was supported partly by National Natural Science Foundation of China (Nos.12001381 and 61871274),  Peacock Plan (No. KQTD2016053112051497), Shenzhen Key Basic Research Project (Nos. JCYJ20180507184647636, JCYJ20170818094109846, and JCYJ20190808155618806).}
	\thanks{Wei Wang is with the School of Biomedical Engineering, Shenzhen University, Shenzhen, China. (e-mail: wangwei@szu.edu.cn). }
	\thanks{Xiang-Gen Xia is with the Department of Electrical and Computer Engineering, University of Delaware, Newark, DE 19716, USA.  (e-mail: xxia@ee.udel.edu).}
	\thanks{Chuanjiang He is with the
		College of Mathematics and Statistics, Chongqing University, Chongqing, China (e-mail: cjhe@cqu.edu.cn).}
	\thanks{Zemin Ren is with the
		College of Mathematics and Physics, Chongqing University of Science and Technology, Chongqing, China (e-mail: zeminren@cqu.edu.cn).}
	\thanks{Jian Lu is with the
		Shenzhen Key Laboratory of Advanced Machine Learning and Applications, Shenzhen University, Shenzhen, China (e-mail: jianlu@szu.edu.cn).}
}

%
%

\markboth{Journal of \LaTeX\ Class Files,~Vol.~14, No.~8, August~2015}%
{Shell \MakeLowercase{\textit{et al.}}: Bare Demo of IEEEtran.cls for IEEE Journals}
%



\maketitle

\begin{abstract}
In this paper, we propose a new GPU implementation of the Katsevich algorithm for helical CT reconstruction. Our implementation divides the sinograms and reconstructs the CT images pitch by pitch. By utilizing the periodic properties of the parameters of the Katsevich algorithm, our method only needs to calculate these parameters once for all the pitches and so has lower GPU-memory burdens and is very suitable for deep learning. By embedding our implementation into the network, we propose an end-to-end deep network for the high pitch helical CT reconstruction with sparse detectors. Since our network utilizes the features extracted from both sinograms and CT images, it can simultaneously reduce the streak artifacts caused by the sparsity of sinograms and preserve fine details in the CT images. Experiments show that our network outperforms the related methods both in subjective and objective evaluations.
\end{abstract}


%
\IEEEpeerreviewmaketitle

\section{Introduction}
\label{sec:introduction}
\IEEEPARstart{C}{omputed} tomography (CT)  has been an important tool in medical diagnosis and the helical scan with multi-row detector is the most used scanning modality in hospitals. When scanning, the X-ray source and detector rotate around the central axis (i.e., the z-direction) while the table with a patient remaining stationary on it is translated along the z-direction at a constant velocity. Therefore, the trajectory of the X-ray source relative to the patient forms a helical curve, where the pitch of the helical curve is related to the moving velocity of the table. 

The patient needs to stay stationary during the scanning process, otherwise severe artifacts will be caused in the reconstructed CT images. Thus, the high pitch scanning that can complete the data acquiring in small time has many merits and is very desirable in clinical diagnosis. For example, by utilizing the high pitch helical CT, the whole lung can be scanned within a very small interval and so the patient needs not to hold the breath. This is very useful for patients that can’t perform breath-holding for a long time, such as infant, old patients and unconscious patients. For cardiac CT, scanning the whole heart in one heart beat is a reasonable request. 

In \cite{Noo2003a}, Noo et al. used the Tam-Danielsson window \cite{Tam_1998}\cite{Danielsson97:F3D} and Katsevich’s inverse formula \cite{Katsevich2004a} to derive the numbers of required detector rows for perfectly reconstructing the helical CT images with a curved detector or a flat detector. When other parameters are fixed, the higher the pitch is, the greater number of the detector rows is needed. However, as the number of the detector rows increasing, the price of the CT scanner will also rise significantly. One way to reduce the price is to use sparse detector \cite{Zheng2020} with fixed detector row numbers. However,  CT images reconstructed from sparse sinograms  usually have  streak artifacts and lose fine details especially when the scanning pitch is high.

In the literature, many algorithms were proposed to reconstruct the helical CT images. These algorithms can be mainly categorized into four classes: (a) rebinning algorithms, (b) FDK-like algorithms, (c) model-based iterative methods, (d) exact algorithms. 

The rebinning algorithms \cite{Noo1999}\cite{Stierstorfer2004a} first converted the measured helical sinograms to 2D fan-beam or parallel-beam sinograms and then reconstructed the CT images by using any 2D reconstruction methods. To reduce the interpolation error and accurately estimate the 2D sinogram, the sampling rate of the helical scan is usually high. 

The FDK-like algorithms \cite{WangSept}\cite{Guo2011}\cite{Zhao2007} are extended from the circle cone-beam reconstruction algorithm proposed by Feldkamp, Davis and Kress \cite{Feldkamp1984}. The FDK-like algorithms usually involve three steps: 1) filtering the sinograms, 2) weighting the filtered data, 3) Backprojecting the weighted data. By utilizing the fast Fourier transform (FFT), the FDK-like algorithms can be implemented efficiently. However, since the FDK-like algorithms were proposed heuristically and lack of theory supports, the artifacts in the reconstruction is usually hard to avoided especially when the cone angles are large.

Model-based methods usually first propose the minimizing functionals by combining the statistical property of the sinogram with different priori hypotheses, and then convert the functionals to iterative reconstruction algorithms by using optimizing methods. In the literature, total variation (TV) \cite{Sidky2008}\cite{Liu2012}, nonlocal means (NLM) \cite{Chen2009}, dictionary learning \cite{XuSept}\cite{BaoNov.} are the commonly used priori hypotheses. In \cite{Stierstorfer2004a}, Stierstorfer et al. proposed an iterative weighted filtered backprojection (WFBP) method to reconstruct helical CT images. In \cite{Sunnegaardh2021}, Sunneg$\mathring{a}$rdh proposed an improved regularized iterative weighted filtered backprojection (RIWFBP) method for the helical cone-beam CT. In \cite{Yu2021}, Yu and Zeng developed a fast TV-based iterative reconstruction algorithm for the limited-angle helical cone-beam CT reconstruction. Model-based iterative methods are very suitable for ill-posed problems, such as limited angle or sparse angle CT reconstructions. But the computational burdens of the model-based iterative methods are too high to reconstruct the images in real time especially for 3D helical CT.

Exact helical CT reconstruction can be obtained by using Katsevich’s algorithms \cite{Katsevich2021}\cite{Katsevich2004a}, Zou and Pan’s PI-line algorithms \cite{Zou2004}\cite{Zou2004a}. These algorithms have mathematical inverse formulas for helical CT reconstructions and so are theoretically exact. In \cite{Ye2021}\cite{Zou2021}, by extending Katsevich’s algorithms and the PI-line algorithms, the exact methods for cone-beam CT reconstructions with general scanning curves were proposed. In \cite{YanJuly}, Yan et al. accelerated the implementation of the Katsevich algorithm \cite{Katsevich2004a} by utilizing the graphics processing unit (GPU). In their implementation, the Tam–Danielsson window rather than the PI line is used to determine whether a sinogram corresponding to a scanning angle  needs to be backprojected to reconstruct a slice of CT image. Their method doesn't utilize the periodic properties of the PI-line and so needs to calculate the backprojection positions on the detector for pixels in all the CT images.
 Exact helical algorithms can reconstruct good quality CT images even if the helical pitch is high. But when the sinograms are sparse, streak artifacts  in the CT images reconstructed by the exact algorithms can’t be avoided. 

Recently, convolutional neural network (CNN) based methods have also been used for CT reconstructions. These methods can be roughly divided into three categories. The first category uses CNNs only to pre-process the sinograms \cite{LeeMarc}\cite{Fu2020} or post-process the CT images \cite{ChenDec.}\cite{JinSept}\cite{ZhangJune}. This type of methods needs an extra step to convert the sinograms to CT images.
The second category utilizes end-to-end CNNs to reconstruct CT images directly from measured sinograms \cite{Ge2020}\cite{Lin15-2}\cite{Zhang2020}\cite{Zheng2020}\cite{Wang2020}. Since the reconstruction algorithm is modelled as network layers embedded in CNNS, no extra step is needed when testing or deploying these trained CNNs. However, embedding reconstruction algorithm in CNNs is GPU-memory expensive and may result in GPU-memory exhausting error when training this type of CNNs especially for 3D CT. The third category uses CNNs to approximate the solution of the unroll iterative algorithms \cite{ChenJune}\cite{HeFeb.}, where each iteration of the algorithms was approximated by a subnetwork. This type of algorithms exhausts more GPU-memory than those in the second category since in the networks the operators of  projection and backprojection need to be performed  $n$ times, where $n$ is the iteration numbers.

In this paper, we focus on embedding the Katsevich algorithm \cite{Katsevich2004a} into CNNs to propose an end-to-end deep network to reconstruct helical CT images with high pitch and sparsely spaced detectors. Due to the limited GPU-memory, to the best of our knowledge, in the literature no exact algorithm (Katsevich’s algorithms or PI-line algorithms) was embedded in CNNs to reconstruct helical CT images. In \cite{Zheng2020}, an end-to-end deep network was proposed for helical CT, but the sinograms measured from helical scans were converted to 2D fan-beam sinograms and the embedded reconstruction algorithm was also for fan-beam geometry. As the helical pitch increases, the image volume involved to generate the rebinned fan-beam sinogram and the error of the estimated fan-beam sinogram will both increase. Therefore, for helical CT with high pitch, the CNN using rebining methodology may not work well. The Katsevich algorithm \cite{Katsevich2004a} can reconstruct good CT images even if the helical pitch is high. Our network uses Katsevich algorithm as the domain transfer layer and so can also reconstruct good CT images directly from high pitch helical sinograms. Moreover, because our network utilizes the features extracted from both sinograms and CT images, it can simultaneously reduce the streak artifacts caused by the sparsity of sinograms and reconstruct fine details in the CT images.

The sinogram measured from the helical scan usually has large size in the z-direction. If directly input the whole sinogram into CNN to reconstruct all the CT images, huge GPU-memory is needed. Besides, most of the parameters of Katsevich algorithm \cite{Katsevich2004a} depend on the position of the CT images and computing them is also time consuming and GPU-memory consuming. By observing that the parameters of Katsevich algorithm \cite{Katsevich2004a} are periodic, we divide the sinograms into parts and reconstruct the CT images cycle-by-cycle. By this way, the parameters of Katsevich algorithm for all cycles can be pre-calculated and the GPU-memory loading can be reduced. 

The rest of the paper is organized as follows. Section II gives our new GPU implementation of the Katsevich  algorithm \cite{Katsevich2004a}. Section III describes the detailed structure of our proposed network. Section IV presents the experimental results. Discussion and conclusion are given in Section V.

\section{GPU implementation of Katsevich Algorithm}
\label{secGPU}
In this section, we  give a new GPU implementation of the Katsevich  algorithm \cite{Katsevich2004a}, which reconstructs the helical CT images pitch by pitch and has low computational and GPU-memory burdens.
Our implementation of Katsevich  algorithm \cite{Katsevich2004a} is very suitable for deep learning since its paramerters are the same for all pitches  and so need only to  calculate once.

 Katsevich \cite{Katsevich2004a} proposed an exact reconstruction formula for helical cone-beam CT and Noo et al. \cite{Noo2003a} researched how to implement it efficiently and accurately. Let 
 \begin{equation}\label{key}
 \underline{a}(\lambda)=(R\cos(\lambda+\lambda_0), R\sin(\lambda+\lambda_0), z_0+p\lambda/2\pi)	
 \end{equation}
 be the helical trajectory formed by the X-ray source position with radius $R$ and pitch $p$, where 
 \begin{equation}
 	\underline{a}(0)=(R\cos(\lambda_0), R\sin(\lambda_0), z_0)
 \end{equation}
 is the start point of the helix,
 \begin{equation}\label{key}
 U=\{\underline{x}: x^2+y^2\le r^2\}	
 \end{equation}
is a cylinder inside the helix,  $0<r<R$ and $\underline{x}=(x,y,z)$.

 Then, for any $\underline{x}\in U$, the reconstruction formula can be written as 
 \begin{equation}\label{e1}
 	f(\underline{x})=-\frac{1}{2 \pi} \int_{\lambda_{i}(\underline{x})}^{\lambda_{o}(\underline{x})} \mathrm{d} \lambda \frac{1}{\|\underline{x}-\underline{a}(\lambda)\|} g^{F}\left(\lambda, \frac{x-\underline{a}(\lambda)}{\|\underline{x}-\underline{a}(\lambda)\|}\right)
 \end{equation}
where $\lambda_{i}(\underline{x})$ and $\lambda_{o}(\underline{x})$ are the extremities of the  $\pi$-line passing through $\underline{x}$ with $\lambda_{i}(\underline{x})<\lambda_{o}(\underline{x})$.
\begin{equation}\label{e2}
	g^{F}(\lambda, \underline{\theta})=\int_{0}^{2 \pi} \mathrm{d} \gamma h_{H}(\sin \gamma) g^{\prime}(\lambda, \cos \gamma \underline{\theta}+\sin \gamma(\underline{\theta} \times \underline{m}(\lambda, \underline{\theta})))
\end{equation}
\begin{equation}\label{e3}
	g^{\prime}(\lambda, \underline{\theta})=\lim _{\varepsilon \rightarrow 0} \frac{g(\lambda+\varepsilon, \underline{\theta})-g(\lambda, \underline{\theta})}{\varepsilon},
\end{equation}

\begin{equation}\label{e4}
	h_{H}(s)=-\int_{-\infty}^{+\infty} \mathrm{d} \sigma \mathrm{i} \mathop{sign}(\sigma) \mathrm{e}^{\mathrm{i} 2 \pi \sigma s}=\frac{1}{\pi s}
\end{equation}
is the Hilbert filter, $g(\lambda, \underline{\theta})$ is the measured projection data at X-ray source position $\underline{a}(\lambda)$ in the direction $\underline{\theta}$,  vector $\underline{m}(\lambda, \underline{\theta})$ is  normal to the $\kappa$-plane $\mathcal{K}(\lambda, \psi)$ of the smallest $|\psi|$ value that contains the line of direction $\underline{\theta}$ through $\underline{a}(\lambda)$ and the $\kappa$-plane $\mathcal{K}(\lambda, \psi)$ is any 2D plane that has three intersections $\underline{a}(\lambda)$, $\underline{a}(\lambda+\psi)$ and $\underline{a}(\lambda+2\psi)$ with the helix, where $\psi\in (-\pi,\pi)$. In \cite{Noo2003a}, to numerically calculate $g^{F}(\lambda, \underline{\theta})$, the integral  over $\gamma$ on the $\kappa$-plane with the minimum value $\left|\psi\right|$ is converted to the line integral along the curve that the $\kappa$-plane intersects with the detector plane. 

Let $g(\lambda,\alpha,w)$ be the sinogram measured from the helical scan by curved detector, then the  implementation of (\ref{e1})  in \cite{Noo2003a} is described as follows:
\begin{enumerate}
	\item Taking the derivative at constant direction via chain rule:
	\begin{equation}\label{Deriv}
		g_1(\lambda,\alpha,w)=\left.\left(\frac{\partial g(q, \alpha, w)}{\partial q}+\frac{\partial g(q, \alpha, w)}{\partial \alpha}\right)\right|_{q=\lambda}
	\end{equation}
   \item Length-correction weighting:
   \begin{equation}\label{LW}
   	g_{2}(\lambda, \alpha, w)=\frac{D}{\sqrt{D^{2}+w^{2}}} g_{1}(\lambda, \alpha, w)
   \end{equation}
   \item Forward height rebinning:
   computing 
   \begin{equation}\label{FR}
   	g_{3}(\lambda, \alpha, \psi)=g_{2}\left(\lambda, \alpha, w_{\kappa}(\alpha, \psi)\right)
   \end{equation}
 by interpolation for all $\psi \in\left[-\pi / 2-\alpha_{m}, \pi / 2+\alpha_{m}\right]$, where 
  $\alpha_m=\arcsin(r/R)$ is  the half fan angle, 
  \begin{equation}\label{w_k}
  	w_{\kappa}(\alpha, \psi)=\frac{D P}{2 \pi R_{o}}\left(\psi \cos \alpha+\frac{\psi}{\tan \psi} \sin \alpha\right)
  \end{equation}
\item 1-D Hilbert transform in $\alpha$:
at constant $\psi$, compute
\begin{equation}\label{HF}
	g_{4}(\lambda, \alpha, \psi)=\int_{-\pi / 2}^{+\pi / 2} \mathrm{~d} \alpha^{\prime} h_{H}\left(\sin \left(\alpha-\alpha^{\prime}\right)\right) g_{3}\left(\lambda, \alpha^{\prime}, \psi\right)
\end{equation}
where $h_H$ is the  Hilbert filter defined in (\ref{e4}).
\item Backward height rebinning: compute
\begin{equation}\label{BR}
	g_{5}(\lambda, \alpha, w)=g_{4}(\lambda, \alpha, \hat{\psi}(\alpha, w))
\end{equation}
where $\hat{\psi}(\alpha, w)$ is angle $\psi$ of the smallest absolute value that satisfies the equation
\begin{equation}
	w=\frac{D P}{2 \pi R_{o}}\left(\psi \cos \alpha+\frac{\psi}{\tan \psi} \sin \alpha\right)
\end{equation}
\item Post-cosine weighting:
\begin{equation}\label{PW}
	g^{F}(\lambda, \alpha, w)=\cos \alpha g_{5}(\lambda, \alpha, w)
\end{equation}
\item Backprojection:
\begin{equation}
	f(\underline{x})=\frac{1}{2 \pi} \int_{\lambda_{i}(\underline{x})}^{\lambda_{o}(\underline{x})} \mathrm{d} \lambda \frac{1}{v^{*}(\lambda, \underline{x})} g^{F}\left(\lambda, \alpha^{*}(\lambda, \underline{x}), w^{*}(\lambda, \underline{x})\right)
\end{equation}
where 
\begin{equation}
	v^{*}(\lambda, \underline{x})=R_{o}-x \cos \left(\lambda+\lambda_{0}\right)-y \sin \left(\lambda+\lambda_{0}\right)
\end{equation}
\begin{equation}
	\begin{aligned}
		\alpha^{*}(\lambda, \underline{x})=&\\
		\arctan& \left(\frac{\left(-x \sin \left(\lambda+\lambda_{0}\right)+y \cos \left(\lambda+\lambda_{0}\right)\right)}{v^{*}(\lambda, \underline{x})}\right)
	\end{aligned}
\end{equation}
\begin{equation}
	w^{*}(\lambda, \underline{x})=\frac{D \cos \alpha^{*}(\lambda, \underline{x})}{v^{*}(\lambda, \underline{x})}\left(z-z_{0}-\frac{P}{2 \pi} \lambda\right)
\end{equation}
\end{enumerate}

In the above steps, the parameters $\lambda_{i}(\underline{x})$, $\lambda_{o}(\underline{x})$,	$v^{*}(\lambda, \underline{x})$, $\alpha^{*}(\lambda, \underline{x})$, $w^{*}(\lambda, \underline{x})$, $	w_{\kappa}(\alpha, \psi)$ and $\hat{\psi}(\alpha, w)$  need to calculate. The parameters $w_{\kappa}(\alpha, \psi)$ and $\hat{\psi}(\alpha, w)$ can be pre-calculated only once but the others $\lambda_{i}(\underline{x})$, $\lambda_{o}(\underline{x})$,	$v^{*}(\lambda, \underline{x})$, $\alpha^{*}(\lambda, \underline{x})$, $w^{*}(\lambda, \underline{x})$  need  to be calculated for every $\underline{x}$, which is time consuming. Luckily, these parameters have the following periodic properties:
\begin{equation}\label{key}
\begin{aligned}
	&\lambda_{i}(\underline{x}^{k})= \lambda_{i}(\underline{x})+2k\pi\\
	&\lambda_{o}(\underline{x}^{k})= \lambda_{o}(\underline{x})+2k\pi\\
	&v^{*}(\lambda+2k\pi, \underline{x}^{k})=v^{*}(\lambda, \underline{x})\\
	&\alpha^{*}(\lambda+2k\pi, \underline{x}^{k})=\alpha^{*}(\lambda, \underline{x})\\
	&w^{*}(\lambda+2k\pi, \underline{x}^{k})=w^{*}(\lambda, \underline{x})
\end{aligned}	
\end{equation}
where $\underline{x}^{k}=(x,y,z+kp)$, $k$ is an integer and  $p$ is the pitch.
Thus, if we reconstruct the CT images in the z-direction pitch by pitch, the parameters $\lambda_{o}(\underline{x})$, $\lambda_{p}(\underline{x})$,	$v^{*}(\lambda, \underline{x})$, $\alpha^{*}(\lambda, \underline{x})$, $w^{*}(\lambda, \underline{x})$ can also be pre-calculated only once, which can reduce the computation and GPU-memory burdens a lot. 

In the following,  we describe the detailed GPU implementation scheme of  the reconstruction formula (\ref{e1}), which reconstruct the CT images pitch by pitch. In our scheme, to reduce the GPU-memory consuming, the backprojection step is performed  slice by slice in the z-direction while the other steps are performed on the whole sinogram  needed to reconstruct the  CT images of one pitch.

 Let $f(\underline{x}_j^{k})$ be the $j$-th sliced CT image  of the $k$-th pitch in the z-direction, where $j=0,1,2,...,N-1$, $k=1,...M$, $N$ is the total number of CT images to be reconstruct in one pitch, $M$ is the number of divided pitches of the scanning helix, $\underline{x}_j^{k}=(x,y,z_j+kp)$, $(x,y)$ is a point of the mesh grids formed by the x-coordinate and y-coordinate and $z_j$ is the  relative position on the z-coordinate in one pitch. Then our scheme can be divided into parameters pre-calculating step and reconstruction step.
 
The  pre-calculating step is described as follows:
\begin{enumerate}
	\item Pre-calculate  $w_{\kappa}(\alpha, \psi)$ and $\hat{\psi}(\alpha, w)$ by using equation (\ref{w_k}) and  the method in \cite{Noo2003a}.
	\item Define  $\underline{x}_j:=\underline{x}_j^{1}$, For $j=0,1,2,...,N-1$, pre-calculate the extremities $\lambda_{i}(\underline{x}_j)$ and $\lambda_{o}(\underline{x}_j)$ of the  $\pi$-line passing through $\underline{x}_j$ using the algorithm described in \cite{PI}. Note that an extra projection step that constrains $\theta$ on the interval $[-\pi,\pi]$ needs to perform at each iteration of the algorithm, which was not explicitly stated  in \cite{PI}.
    Also note that here  $\lambda_{i}(\underline{x}_j)$ and $\lambda_{o}(\underline{x}_j)$  are two matrices corresponding to  the   mesh grid points $(x,y)$ of  $\underline{x}_j$. 
	 After obtaining $\lambda_{i}(\underline{x}_j)$ and $\lambda_{o}(\underline{x}_j)$, we can calculate the minimal and maximal values of $\lambda$ needed to reconstruct $f(\underline{x}_j)$: 
	 \begin{equation}\label{key}
	 	\begin{aligned}
	 		\lambda_{min}(\underline{x}_j)&=\min_{(x,y)~ \text{of}~ x_j}(\lambda_{i}(\underline{x}_j)),\\
	 		\lambda_{max}(\underline{x}_j)&=\max_{(x,y) ~\text{of}~ x_j}(\lambda_{o}(\underline{x}_j)).
	 	\end{aligned}
	 \end{equation}

	\item  For $j=0,1,2,...,N-1$, pre-calculate $v^*(\lambda, \underline{x}_j)$, $\alpha^*(\lambda, \underline{x}_j)$ and $w^*(\lambda, \underline{x}_j)$:
	\begin{equation}
		v^{*}(\lambda, \underline{x}_j)=R_{o}-x \cos \left(\lambda+\lambda_{0}\right)-y \sin \left(\lambda+\lambda_{0}\right)
	\end{equation}
\begin{equation}
\begin{aligned}
\alpha^{*}(\lambda, \underline{x}_j)&=\\
\arctan& \left(\frac{\left(-x \sin \left(\lambda+\lambda_{0}\right)+y \cos \left(\lambda+\lambda_{0}\right)\right)}{v^{*}(\lambda, \underline{x}_j)}\right)	
\end{aligned}
\end{equation}
\begin{equation}
	w^{*}(\lambda, \underline{x}_j)=\frac{D \cos \alpha^{*}(\lambda, \underline{x}_j)}{v^{*}(\lambda, \underline{x}_j)}\left(z_i-z_{0}-\frac{P}{2 \pi} \lambda\right),
\end{equation}
where $\underline{x}_j=(x,y,z_j)$ and $\lambda = [\lambda_{min}(\underline{x}_j),\lambda_{max}(\underline{x}_j)]$.
Here $u^*(\lambda, \underline{x}_j)$, $\alpha^*(\lambda, \underline{x}_j)$ and $w^*(\lambda, \underline{x}_j)$ are three tensors of dimension three, where the first dimension corresponds to $\lambda = [\lambda_{min}(\underline{x}_j),\lambda_{max}(\underline{x}_j)]$. 
\end{enumerate}

Note that in the above step 3), since 
\begin{equation}\label{key}
	\begin{aligned}
&v^*(\lambda, \underline{x}_j^{1})=v^*(\lambda, \underline{x}_j^{k})\\
		&\alpha^*(\lambda, \underline{x}_j^{1})=\alpha^*(\lambda, \underline{x}_j^{k})\\
		&w^*(\lambda, \underline{x}_j^{1})=w^*(\lambda, \underline{x}_j^{k})
	\end{aligned}
\end{equation}
for any positive integer $k$, $v^*(\lambda, \underline{x}_j)$, $\alpha^*(\lambda, \underline{x}_j)$ and $w^*(\lambda, \underline{x}_j)$
can be pre-calculated only once.

 After the pre-calculating step, we can compute the minimal and maximal values of $\lambda$, $\alpha$ and $w$ needed to reconstruct the CT images of one pitch.
 These values can be calculated by 
 \begin{equation}\label{maxvalue}
 	\begin{aligned}
 &\lambda_{min}=\min\limits_{j=0,1,...N-1}\lambda_{min}(\underline{x}_j)\\
 &\lambda_{max}=\max\limits_{j=0,1,...N-1}\lambda_{max}(\underline{x}_j)\\
 &\alpha_{min}=\min\limits_{j=0,1,...N-1}(\min\limits_{\lambda \in [\lambda_{min}(\underline{x}_j),\lambda_{max}(\underline{x}_j)]}\alpha^*(\lambda, \underline{x}_j))\\
 &\alpha_{max}=\max\limits_{j=0,1,...N-1}(\max\limits_{\lambda \in [\lambda_{min}(\underline{x}_j),\lambda_{max}(\underline{x}_j)]}\alpha^*(\lambda, \underline{x}_j))\\
  &w_{min}=\min\limits_{j=0,1,...N-1}(\min\limits_{\lambda \in [\lambda_{min}(\underline{x}_j),\lambda_{max}(\underline{x}_j)]}w^*(\lambda, \underline{x}_j))\\
 &w_{max}=\max\limits_{j=0,1,...N-1}(\max\limits_{\lambda \in [\lambda_{min}(\underline{x}_j),\lambda_{max}(\underline{x}_j)]}w^*(\lambda, \underline{x}_j))
\end{aligned}
 \end{equation}
The values $\alpha_{min}$, $\alpha_{max}$, $w_{min}$ and $w_{max}$ can be used to determine the numbers of rows and columns of the detector, and $\lambda_{min}$ and $\lambda_{max}$ are used to determine the required sinogram to reconstruct the CT images of one pitch. To reconstruct the $k$-th pitch CT images, the sinogram corresponding to $\lambda=[\lambda_{min}+2k\pi,\lambda_{max}+2k\pi]$ is needed.

Let $g(\lambda^k,\alpha,w)$ be the $\lambda$-sliced sinogram from $g(\lambda,\alpha,w)$, where $\lambda^k=[\lambda_{min}+2k\pi,\lambda_{max}+2k\pi]$ for $k=1,...M$. Then the reconstruction step can be described as follows:
\begin{enumerate}
	\item Perform Derivative (\ref{Deriv}), Length-correction weighting (\ref{LW}), Forward height rebinning (\ref{FR}), Hilbert transform (\ref{HF}), Backward height rebinning (\ref{BR}) and  Post-cosine weighting (\ref{PW}) on $g(\lambda^k,\alpha,w)$ and obtain  $g^{F}(\lambda^k, \alpha, w)$.
	\item For $j=0,1,2,...,N-1$, backproject $g^{F}(\lambda^k, \alpha, w)$ slice by slice:
	\begin{equation}
		f(\underline{x}_j^{kp})=\frac{1}{2 \pi} \int_{\lambda_{min}(\underline{x}_j)+2k\pi}^{\lambda_{max}(\underline{x}_j)+2k\pi} \mathrm{d} \lambda \frac{1}{v^{*}_j} g^{F}\left(\lambda^k, \alpha^{*}_j, w^{*}_j\right)
	\end{equation}
where 
\begin{equation}\label{key}
	\begin{aligned}
		v^{*}_j&=v^{*}(\lambda, \underline{x}_j)\\
		\alpha^{*}_j&=\alpha^{*}(\lambda, \underline{x}_j)\\
		w^{*}_j&=w^{*}(\lambda, \underline{x}_j)
	\end{aligned}
\end{equation}
\end{enumerate}

To reconstruct the helical CT images pitch by pitch, we need to perform the pre-calculating step  once and store its outputs in memory, and the reconstruction step $M$ times for $M$ pitches. In this paper, the pre-calculating step was implemented by the central processing unit (CPU) while  the reconstruction step by GPU.

\section{Proposed Network}
In this section, we propose a 3D deep learning network to reconstruct CT images directly from sparse helical sinograms with high pitch, where our GPU implementation of the  Katsevich  algorithm \cite{Katsevich2004a} is used as the domain transfer layer in the network.
\subsection{Architecture of Proposed Network}
\begin{figure*}[htbp]	
	\centering{	
			\includegraphics[scale=0.9]{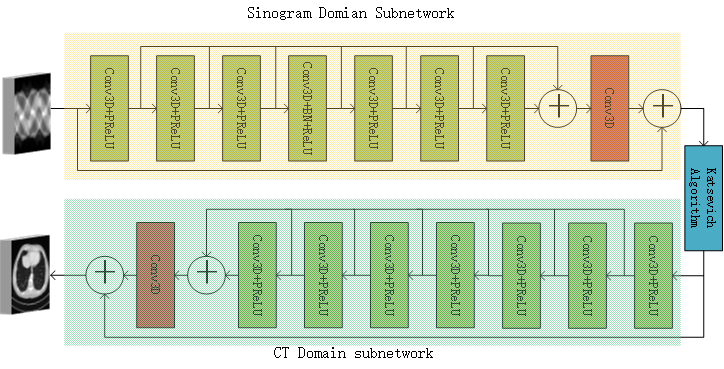}}
	\caption{The detailed structure of our network.}	
	\label{Fig1}
\end{figure*}

Our network (shown in Fig. \ref{Fig1}) is composed of three parts, the sinogram domain subnetwork, the domain  transfer layer and the CT  image domain subnetwork, where the sinogram domain subnetwork is used to denoise and upsample the sparse sinograms, the  domain  transfer layer is used to reconstruct the CT images from the sinograms output by the sinogram domain subnetwork, the CT image domain subnetwork is used to  denoise CT images reconstructed by the domain  transfer layer. 

The architecture of the  sinogram domain subnetwork is  based on the Resnet \cite{7780459}. 
In detail, the  sinogram domain subnetwork is composed of 7 cascaded 'Conv3D+PReLU' blocks and followed by  a 'Conv3D' block, where 'Conv3D' is the 3D convolutional layer and 'PReLu' is the activate function, parametric rectified linear unit (PReLU). The feature channels of 3D convolutional filters in 'Conv3D+PReLU' blocks are all 16 and 1 in the 'Conv3D' block, and the widths of the filters in 'Conv3D+PReLU' blocks and 'Conv3D' block are all 3.
 Each output of the 'Conv3D+PReLU' block is skipped-connected with the output of the last 'Conv3D+PReLU' block and the input of the sinogram domain subnetwork is skipped-connected with the output of the 'Conv3D' block.
The input to the sinogram domain subnetwork is the sliced 3D sinogram ($g(\lambda^k,\alpha,w)$) that can exactly reconstruct one pitch helical CT images. The output of the sinogram domain subnetwork is the denoised sinogram which has the same size of the input sinogram.

The domain  transfer layer implements the Katsevich  algorithm \cite{Katsevich2004a} as described in Section \ref{secGPU}. The input to the domain  transfer layer is the denoised sinograms output by the  sinogram domain subnetwork and the output is the reconstructed helical CT images of one pitch by the Katsevich  algorithm \cite{Katsevich2004a}.

The CT image domain subnetwork has the same architecture of the sinogram domain subnetwork, but the inputs and outputs are different. The input to the CT image domain subnetwork is the CT images of one pitch output by the domain  transfer layer and the output is the denosied CT images which has the same size as the input.
\subsection{Loss Function}
Let $(g^k, f^k)$ be the outputs of our network and  $(g_{GT}^k, f_{GT}^k)$ be the  corresponding labels,
where $g^k$ and  $g_{GT}^k$ are, respectively,  the denoised sinogram and label sinogram, and 
$f^k$ and $f_{GT}^k$ are, respectively,  the denoised CT image and label CT image.
Then the loss function used to train our network is 
\begin{equation}\label{loss}
\mathcal{L}_\Theta =\frac{1}{N_D}\sum\nolimits_{k = 1}^{N_D} {{{\left\| {{g_{GT}^{k}} - g^k} \right\|}^2}}+
 {{{\left\| {{f_{GT}^k} -f^k} \right\|}^2}},	
\end{equation}
where $N_D$ is the total number of training samples and $\Theta $ is the set of parameters of our network to be learned.

Our network is trained by optimizing the loss function (\ref{loss}) using the backpropagation algorithm (BP) \cite{INSPEC:3589863}. In this study, the loss function is optimized by the Adam algorithm \cite{kingma2014adam}, where  the learning rate is set as ${{1}}{{{0}}^{ - 3}}$.

The training code of our network is implemented by using the software Tensorflow 2.5  on a Dell PowerEdge T640 server with a Ubuntu 20.04 operating system, 377GB random access memory (RAM), a Intel(R) Xeon(R) Silver 4210R CPU with 2.4GHz frequency and a NVIDIA RTX A6000 GPU card with 45631MB memory.
The initial values of the training parameters of our network  are randomly set by Tensorflow using the default setup. The gradients of the loss function with respect to the training parameters are calculated by Tensorflow automatically using the automatic differentiation technique. 
The runtime of our network to complete one batch learning (a forward step and a backward step) with $batch=1$ is about $5\sim6$ seconds.
\subsection{Data Preparation} 
\label{Data_preparation}
\subsubsection{CT image labels} 
The CT images from  “the 2016 NIH-AAPM-Mayo Clinic Low Dose CT Grand Challenge” \cite{data} that was established and authorized by Mayo Clinics are used as the ground truth CT images $f_{GT}$. The AAPM challenge has also released the 3D helical sinograms corresponding to the ground truth CT images. But the pitches of the helical scan used to obtain the sinograms are low and not equal, so we need to generate the simulated sinograms with constant high pitch by performing the helical projection. 

In this paper, we perform our methods on two types of helical sinograms, which correspond to $p=7\pi$ and $p=14\pi$, respectively.
\subsubsection{Parameters of the scanning geometry} 
If not specified, all the length unit in our paper is millimetre ($mm$).
The radius of scanning helix is  $R=595$ and  the distance of the X-ray source to the detector is   $D=1085.6$, where the detector is a curved panel. The x-coordinate and y-coordinate of the ground truth CT images are, resectively   
\begin{equation}\label{key}
	\begin{aligned}
		x_{cor}&=[-256:1:255]\times 1\\
		y_{cor}&=[-256:1:255]\times 1
	\end{aligned}
\end{equation}
For simplify, here we assume the size of one pixel  of the CT images is  $1\times 1$. The half fan-angle is 
\begin{equation}
	\arcsin(\sqrt {2}\times256/R)\approx 19.4808 ^{\circ}
\end{equation}
and we set
\begin{equation}\label{key}
	\begin{aligned}
		\alpha_{cor}=&[-19.5625:1:19.5625]\times0.0625^{\circ}+\frac{1}{4}\times0.0625^{\circ}\\
		=&([-313:1:313]+0.25)\times\pi/180/16
	\end{aligned}
\end{equation}
for the $\alpha$-coordinate of the detector.

We constraint the number of detector rows to be 16 and so the positions of the detector units distributed on the w-coordinate are different for different  $p$.  
For $p=7\pi$, the max $w$ in equation (\ref{maxvalue}) is 
\begin{equation}\label{key}
	w_{L}=\max(w_{max},-w_{min}) =3.8819,
\end{equation}
and so $\nabla w=2w_{max}/15=0.5176$. Therefore, we set 
\begin{equation}\label{key}
	w_{cor}=[-7.5:1:7.5]\times0.5176
\end{equation}
 for $p=7\pi$ on the $w$-coordinate. Similarly, for $p=14\pi$, we have
\begin{equation}\label{key}
	w_{L}=\max(w_{max},-w_{min})=7.7499,
\end{equation}
 and so $\nabla w=2w_{max}/15=1.0333$ and set
 \begin{equation}\label{key}
 	w_{cor}=[-7.5:1:7.5]\times1.0333.
 \end{equation}

In summary, the scanning parameters of the helical CT are listed in Table \ref{T1}.

\subsubsection{Data generation}
We use the above described geometry to simulate scanning the CT image labels to generate the sinogram labels $y_{GT}$, where the slices of the ground truth CT images are equally spaced on the z-coordinate. For $p=7\pi$, the z-coordinate of the ground truth CT images on one pitch is 
\begin{equation}\label{key}
	\begin{aligned}
	z_{cor}=&linspace(0,2\pi h,ceil(3h))\\
	 =&[0:1:10]\times2.1991
	\end{aligned} 
\end{equation}
and for $p=14\pi$ is 
\begin{equation}\label{key}
	\begin{aligned}
		z_{cor}=&linspace(0,2\pi h,ceil(2h))\\
		=&[0:1:13]\times3.3833
	\end{aligned}
\end{equation}
Therefore, in one pitch, we need to reconstruct $11$ slices of  CT images for $p=7\pi$ and $14$ slices for $p=14\pi$.  

The sinogram $g_{GT}$ is obtained by scanning a whole patient, so we need to divide it into pitches to obtain $g_{GT}^k$, where 
\begin{equation}\label{key}
	\begin{aligned}
	g_{GT}^k&= g_{GT}[\lambda_{1}^k:1:\lambda_{2}^k,:,:],\\
	\lambda_{1}^k&=(\lambda_{min}-\lambda_0+2k\pi)/\nabla\lambda,\\
	\lambda_{2}^k&=(\lambda_{max}-\lambda_0+2k\pi)/\nabla\lambda
	\end{aligned}
\end{equation}
$k=2,..,M-1$, $\lambda_{min}$ and $\lambda_{max}$ are defined in equation (\ref{maxvalue}), $\nabla\lambda=\pi/180$ and $\lambda_0$ is the start angle of the scanning helix. 
For $p=7\pi$, $\lambda_{min}=-2.1720$, $\lambda_{max}=7.9625$, and for $p=14\pi$, $\lambda_{min}=-2.1875$, $\lambda_{max}=8.1025$. 
To avoid the  zero values, the data of the first and last pitches are discarded. 

 Correspondingly, the ground truth CT images $f_{GT}$ are also needed to divide into pitches to get $f_{GT}^k$, where
\begin{equation}\label{key}
	\begin{aligned}
		f_{GT}^k&= f_{GT}[:,:,z_1^k:1:z_2^k],\\
		z_1^k&=k\times z_L+1,\\
		z_2^k&=(k+1)\times z_L,\\
	\end{aligned}
\end{equation}
$k=2,..,M-1$, $z_L=11$ for $p=7\pi$ and $z_L=14$ for $p=14\pi$.

After obtaining the sinogram labels $g^k_{GT}$, we downsample it on the $\alpha$-coordinate at
\begin{equation}\label{key}
	\alpha_{sp}=\alpha_{cor}[0:4:end]
\end{equation}
to obtain the sparse sinogram $g^k_{sp}$. We then upsample $g^k_{sp}$ by interpolation to make it have the same size as $g^k_{GT}$ and add the 'Gaussian+Poisson' noise \cite{ISI:000311351400019}  to the upsampled sinogram $g^k_{up}$ to obtain the input sinograms $g^k_{in}$ of our network, where
\begin{equation}
	\begin{aligned}
		&g=I_{0}\times\exp (-g^k_{up} / M) \\
		&g=g+poission(g)+Gaussin\left(m, \text { var }\right) \\
		&g^k_{in}=\log \left(I_{0} / g\right) \times M,
	\end{aligned}
\end{equation}
$M$ is the maximal value of $g^k_{up}$, $I_0=10^5$ is the photon count, $m=0$ and $\text { var }=0.5$ are the mean and variance of the Gaussian noise, respectively.

 The CT image $f^k_{in}$ reconstructed from $g^k_{in}$ is used as the input of the compared post-processing methods.

\begin{table}[!t]
	\renewcommand{\arraystretch}{1.3}
	\caption{The Parameters to Generate the Training Data}
	\label{T1}
	\centering
\begin{tabular}{lc}
	\hline
	Radius of  helix                                         &              595               \\
	Distance of source to detector                           &             1085.6             \\
	$x$-coordinate of CT image                               &      $[-256:1:255]\times1$       \\
	$y$-coordinate of CT image                               &      $[-256:1:255]\times1$       \\
	\multirow{2}{*}{$z$-coordinate of CT image in one pitch} &      $[0:1:10]\times2.1991$      \\
	                                                         &      $[0:1:13]\times3.3833$      \\
	\multirow{2}{*}{$\lambda$-coordinate in one pitch}       &    $[-125:1:455]\times0.0175$    \\
	                                                         &    $[-125:1:463]\times0.0175$    \\
	$\alpha$-coordinate of detector                          & $[-312.75:1:313.25]\times0.0011$ \\
	\multirow{2}{*}{$w$-coordinate of detector}              &    $[-7.5:1:7.5]\times0.5176$    \\
	                                                         &    $[-7.5:1:7.5]\times1.0333$    \\ \hline
\end{tabular}
\end{table}%

\newcommand\size{0.41}
\begin{figure*}[htbp]	
	\centering{	
		\subfloat{
			\begin{overpic}[width=\size\columnwidth,percent]{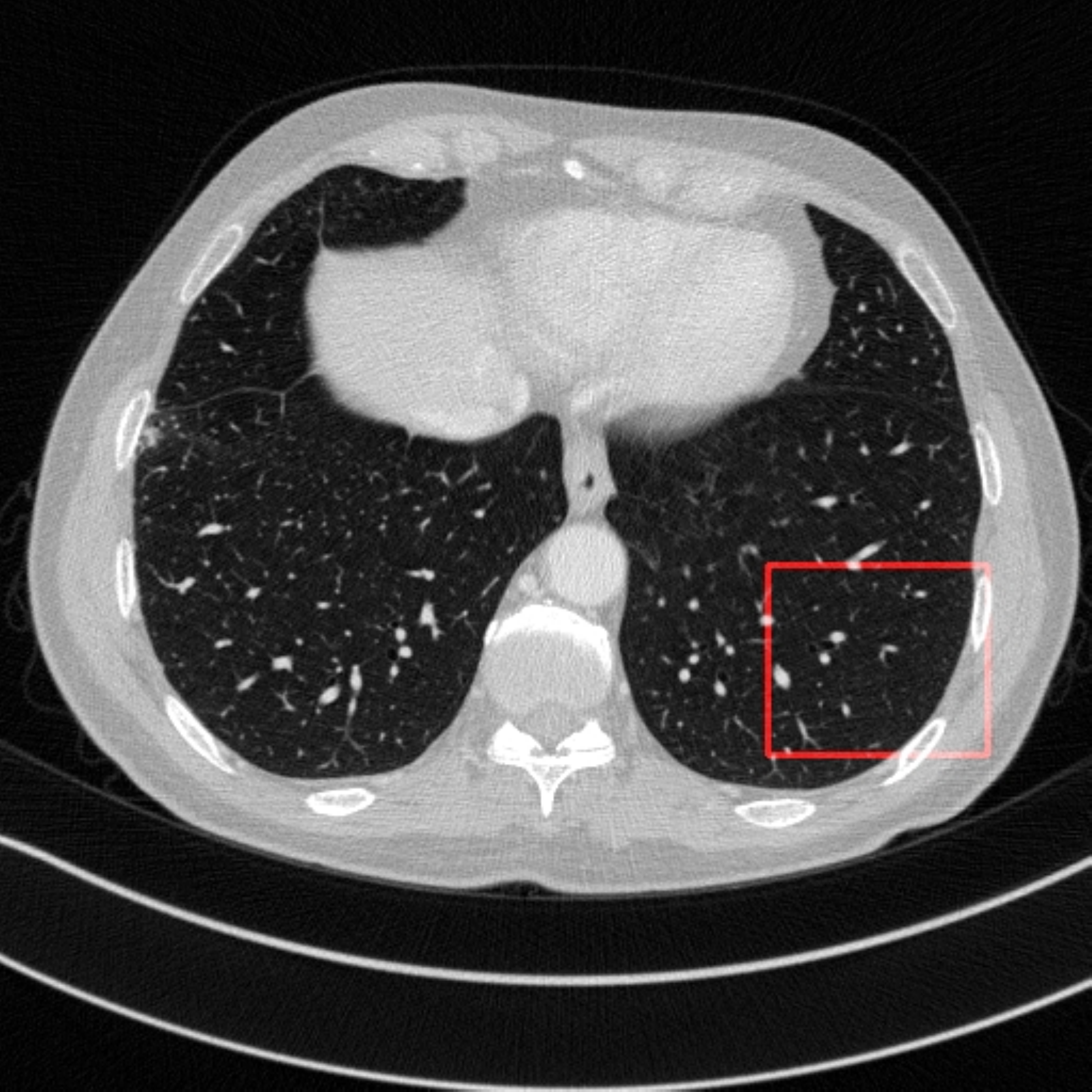}	
				\put(0,0){\color{red}%
					\frame{\includegraphics[scale=0.08]{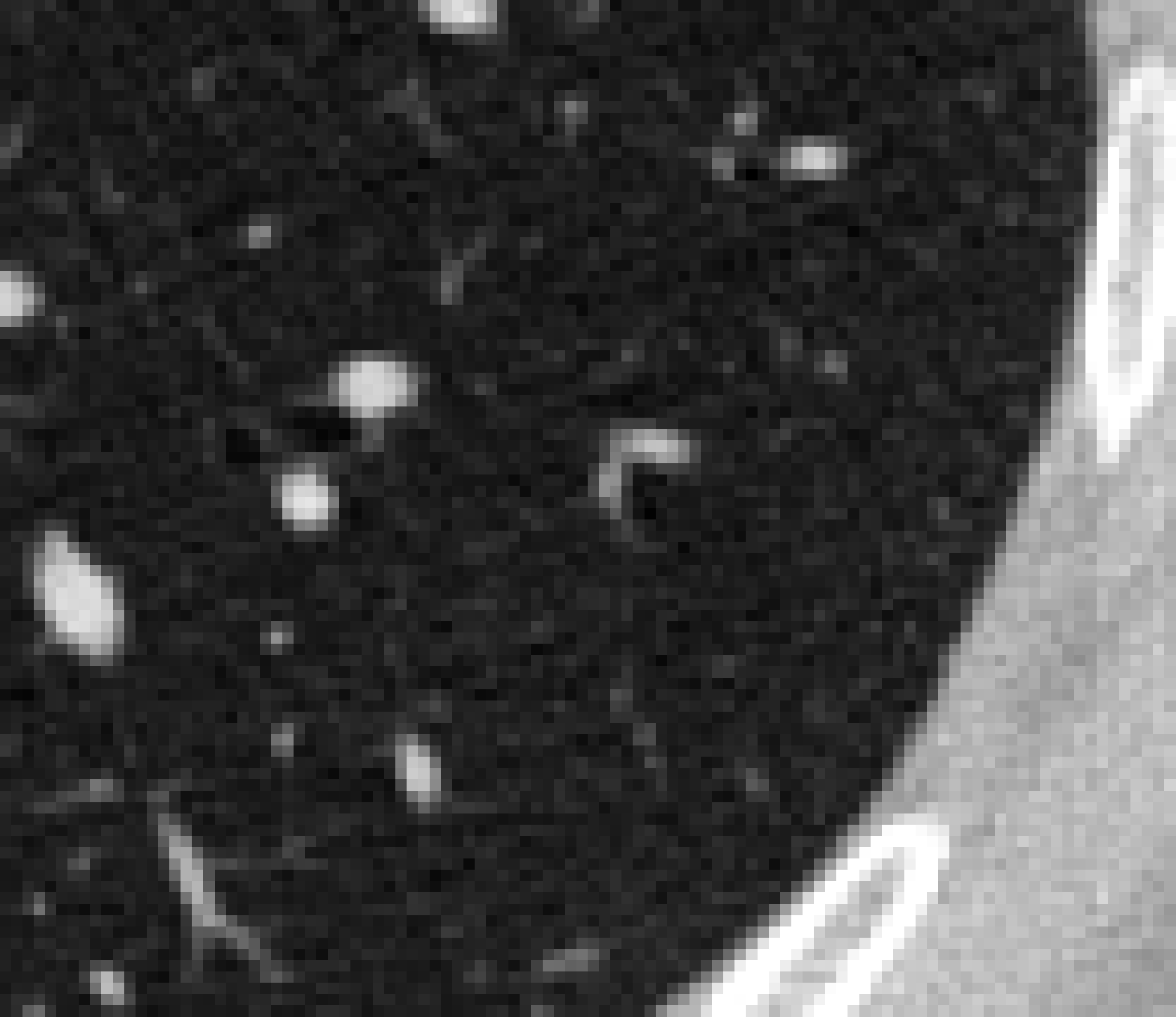}}
				}		
		\end{overpic}}
		\subfloat{
			\begin{overpic}[width=\size\columnwidth]{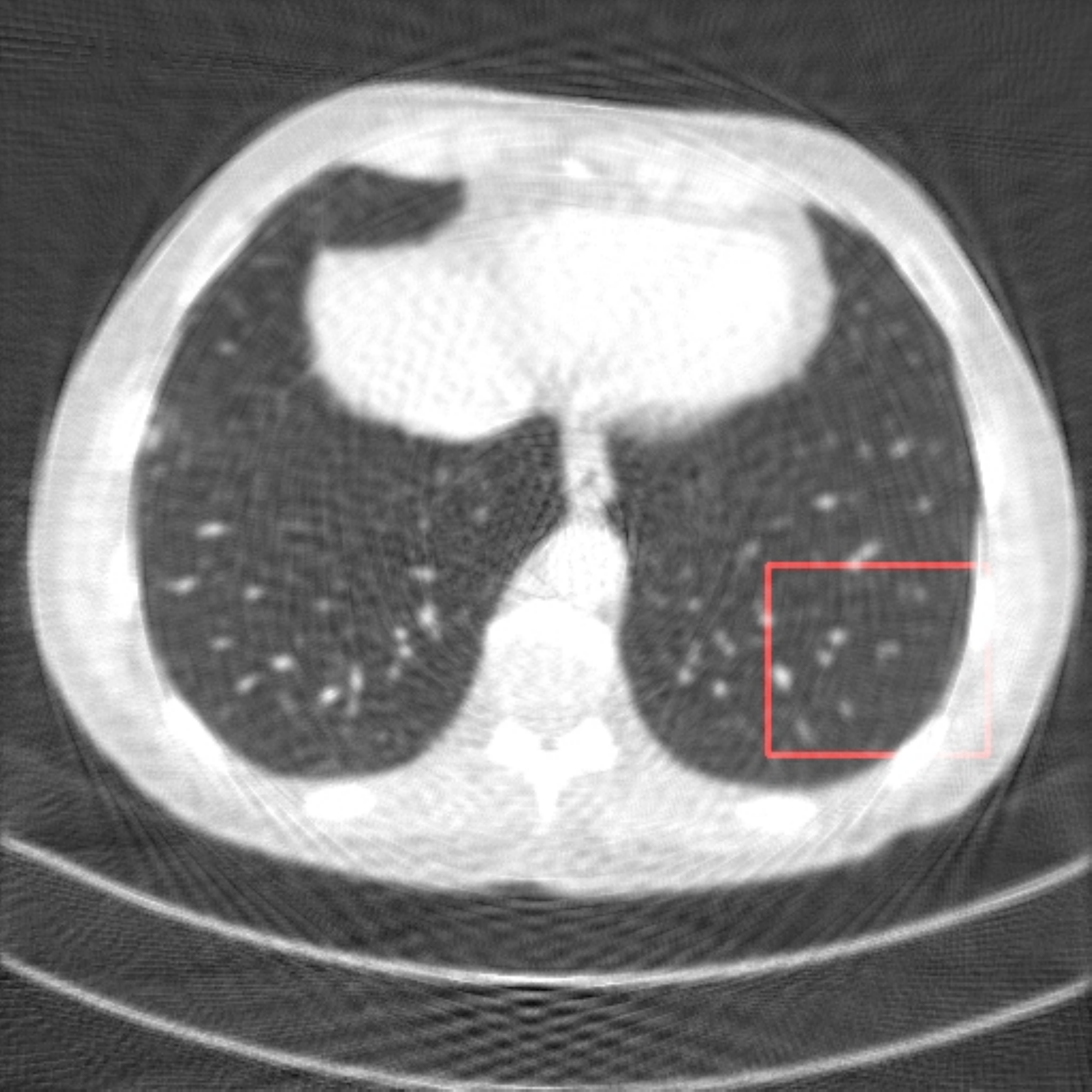}	
				\put(0,0){\color{red}%
					\frame{\includegraphics[scale=0.08]{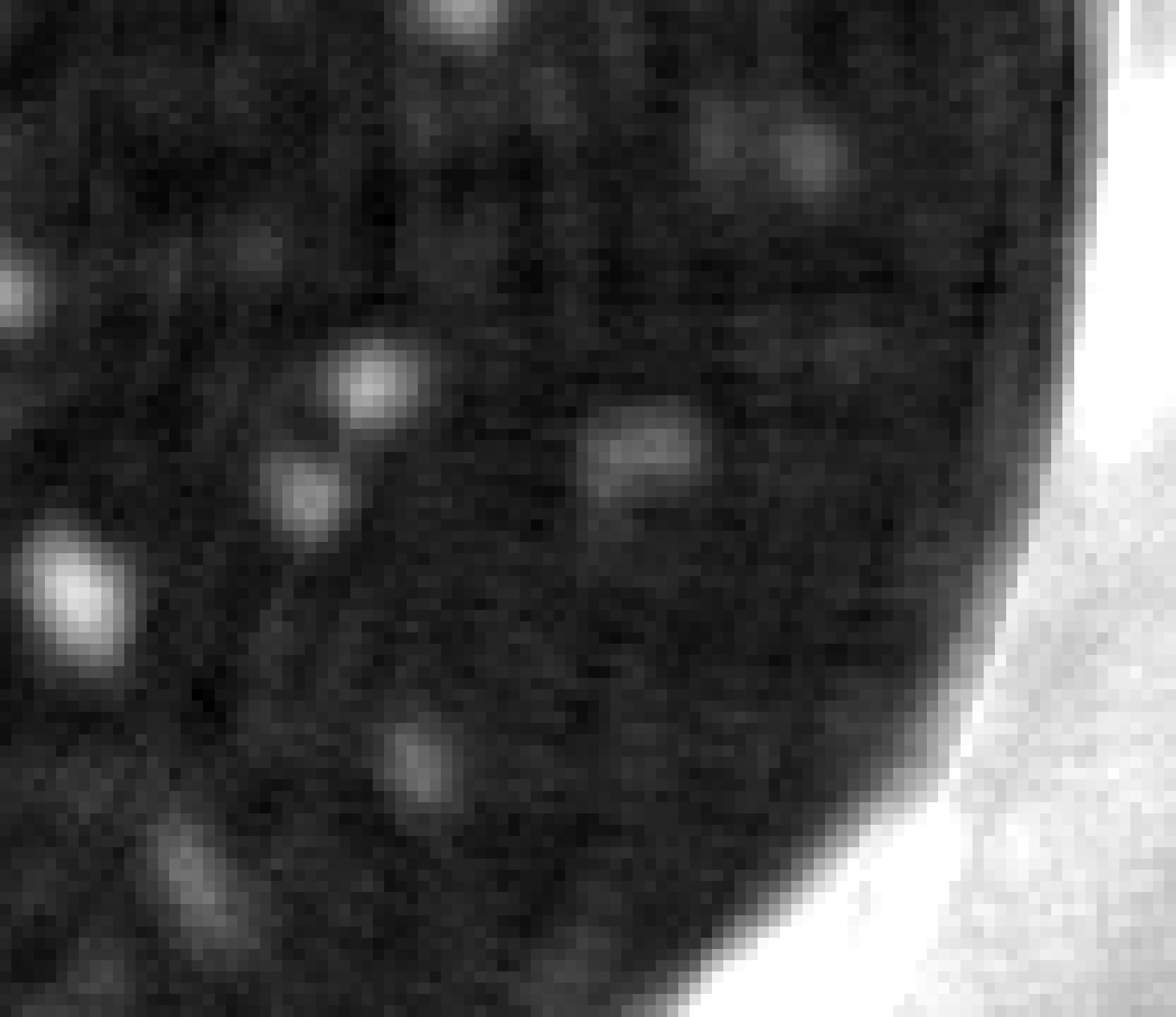}}
				}		
		\end{overpic}}
		\subfloat{
			\begin{overpic}[width=\size\columnwidth]{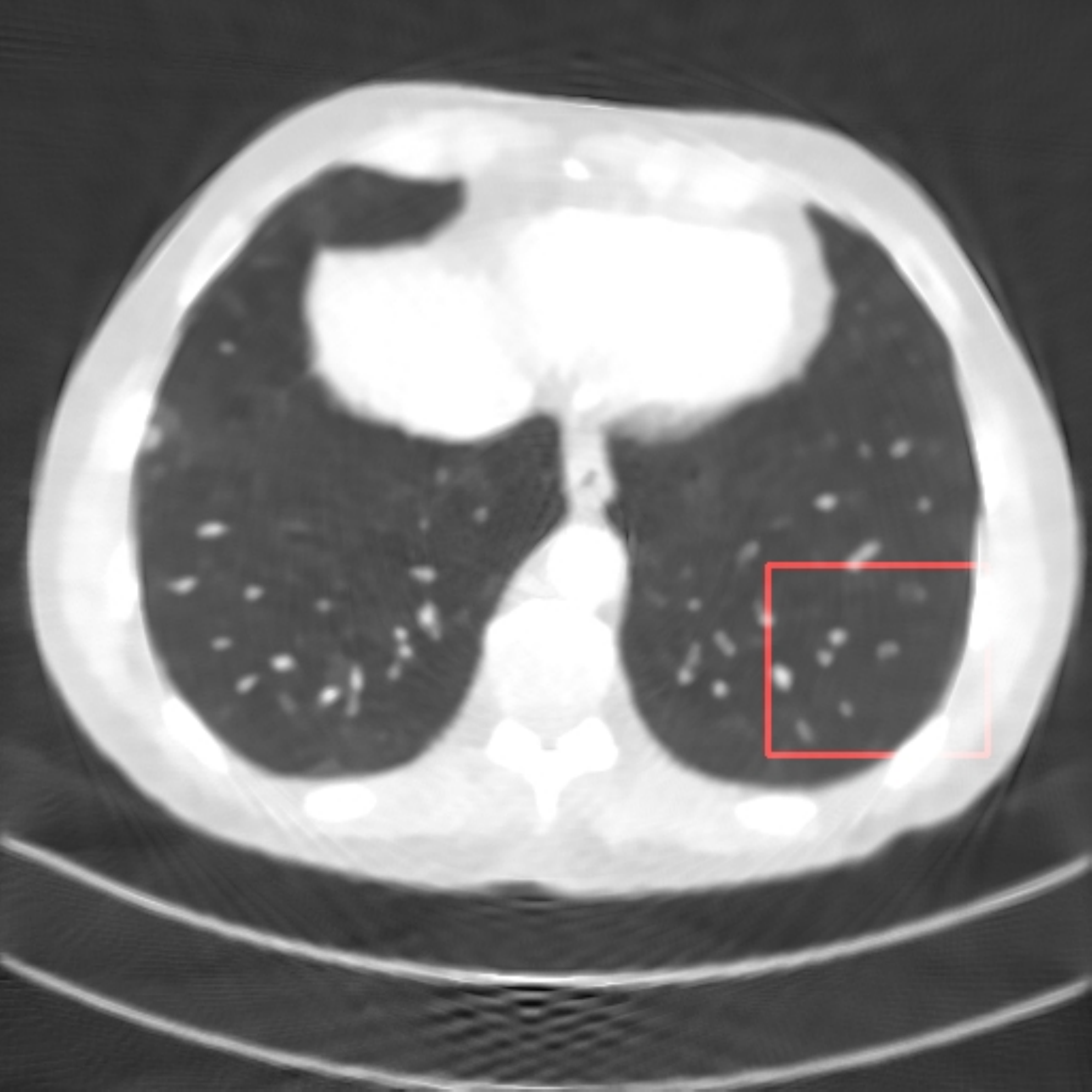}	
				\put(0,0){\color{red}%
					\frame{\includegraphics[scale=0.08]{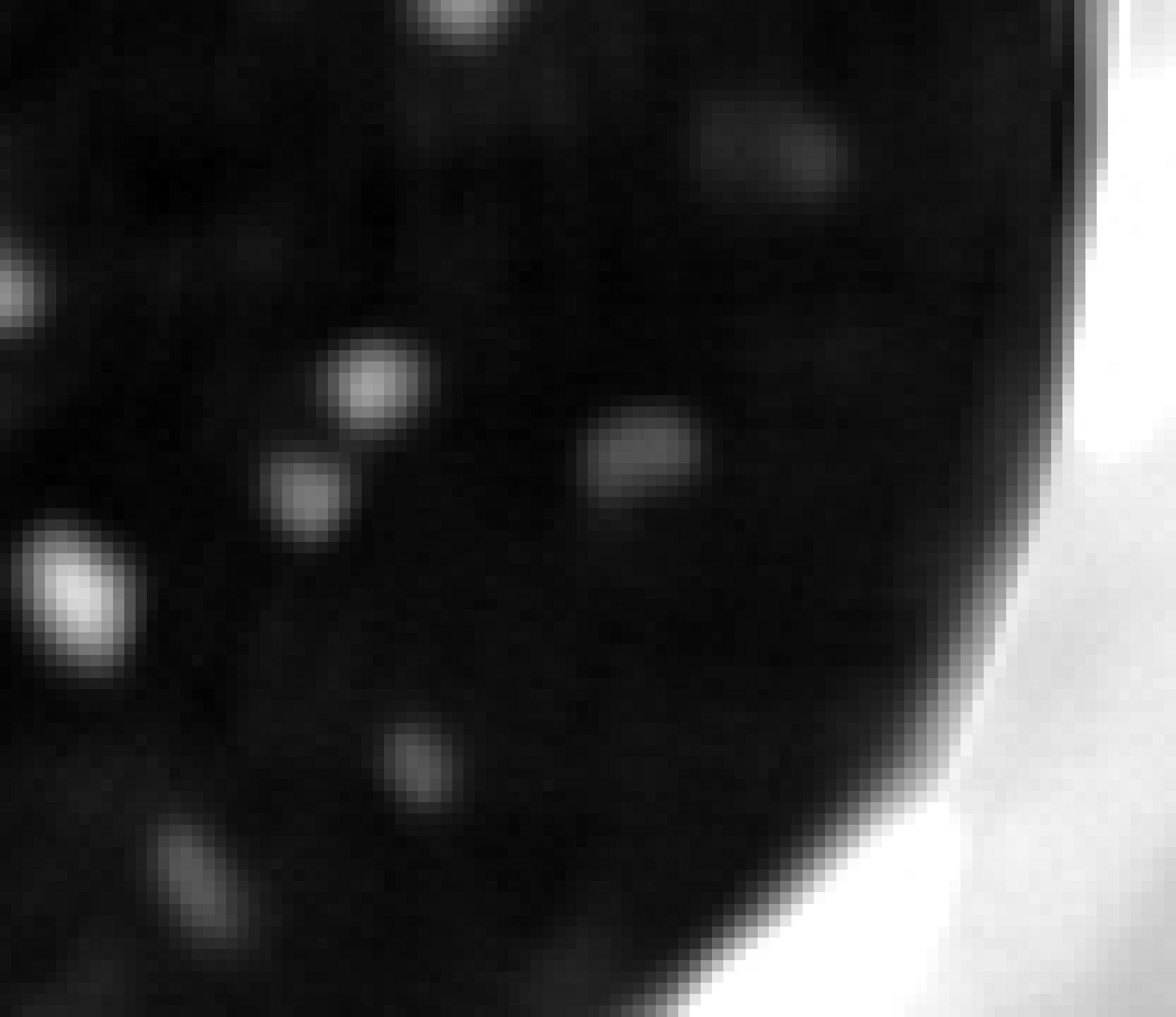}}
				}		
		\end{overpic}}
		\subfloat{
			\begin{overpic}[width=\size\columnwidth]{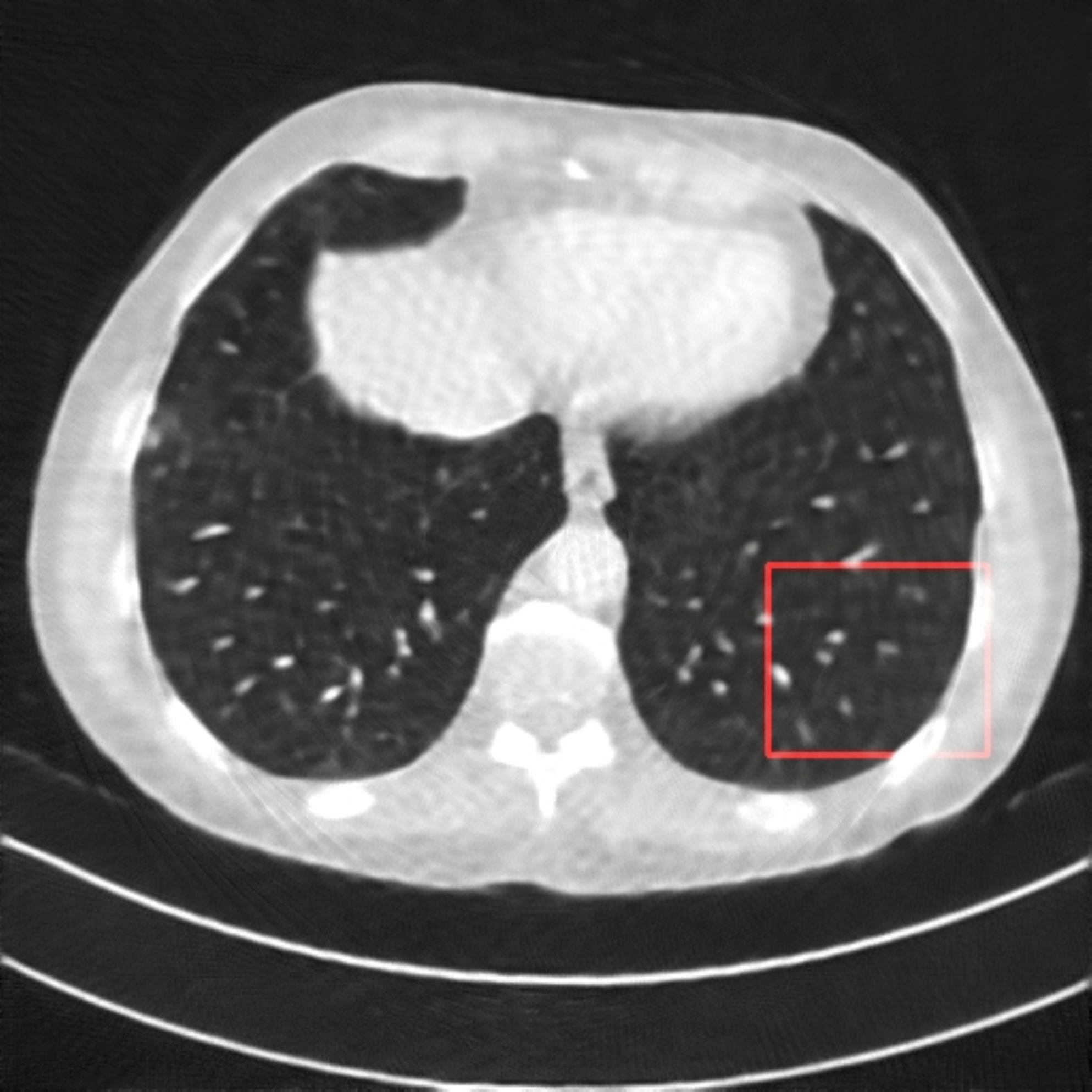}	
				\put(0,0){\color{red}%
					\frame{\includegraphics[scale=0.08]{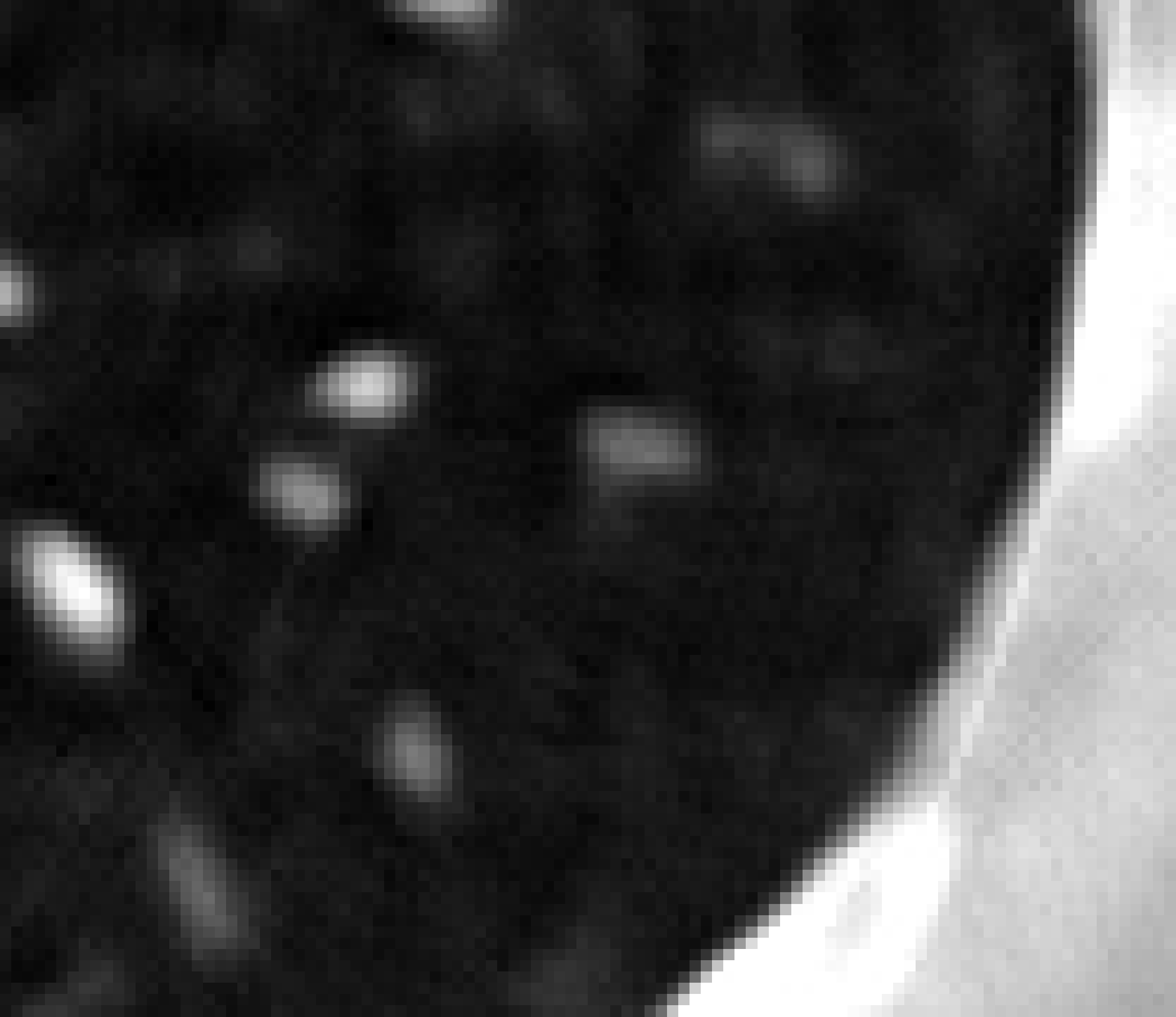}}
				}		
		\end{overpic}}
		\subfloat{
			\begin{overpic}[width=\size\columnwidth]{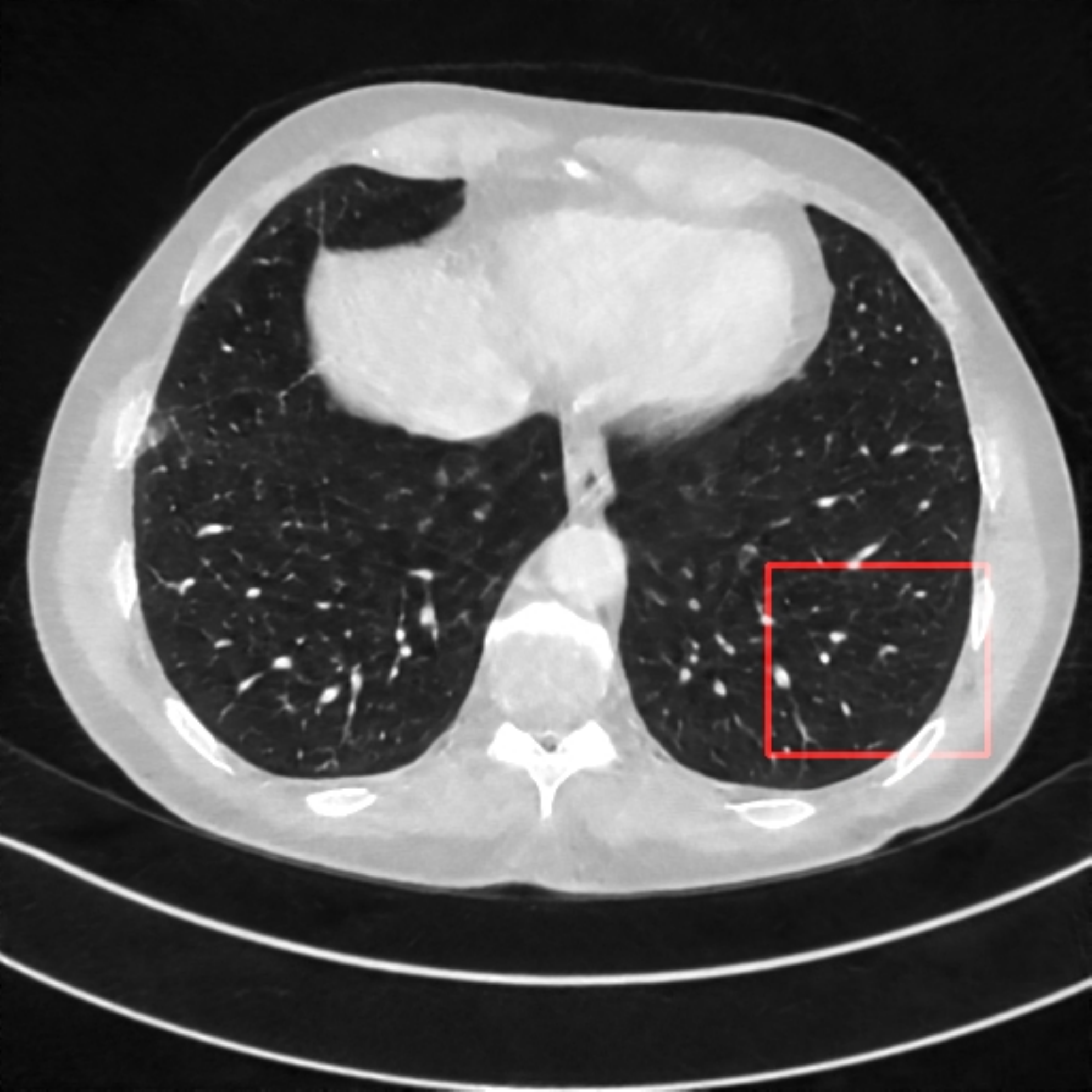}	
				\put(0,0){\color{red}%
					\frame{\includegraphics[scale=0.08]{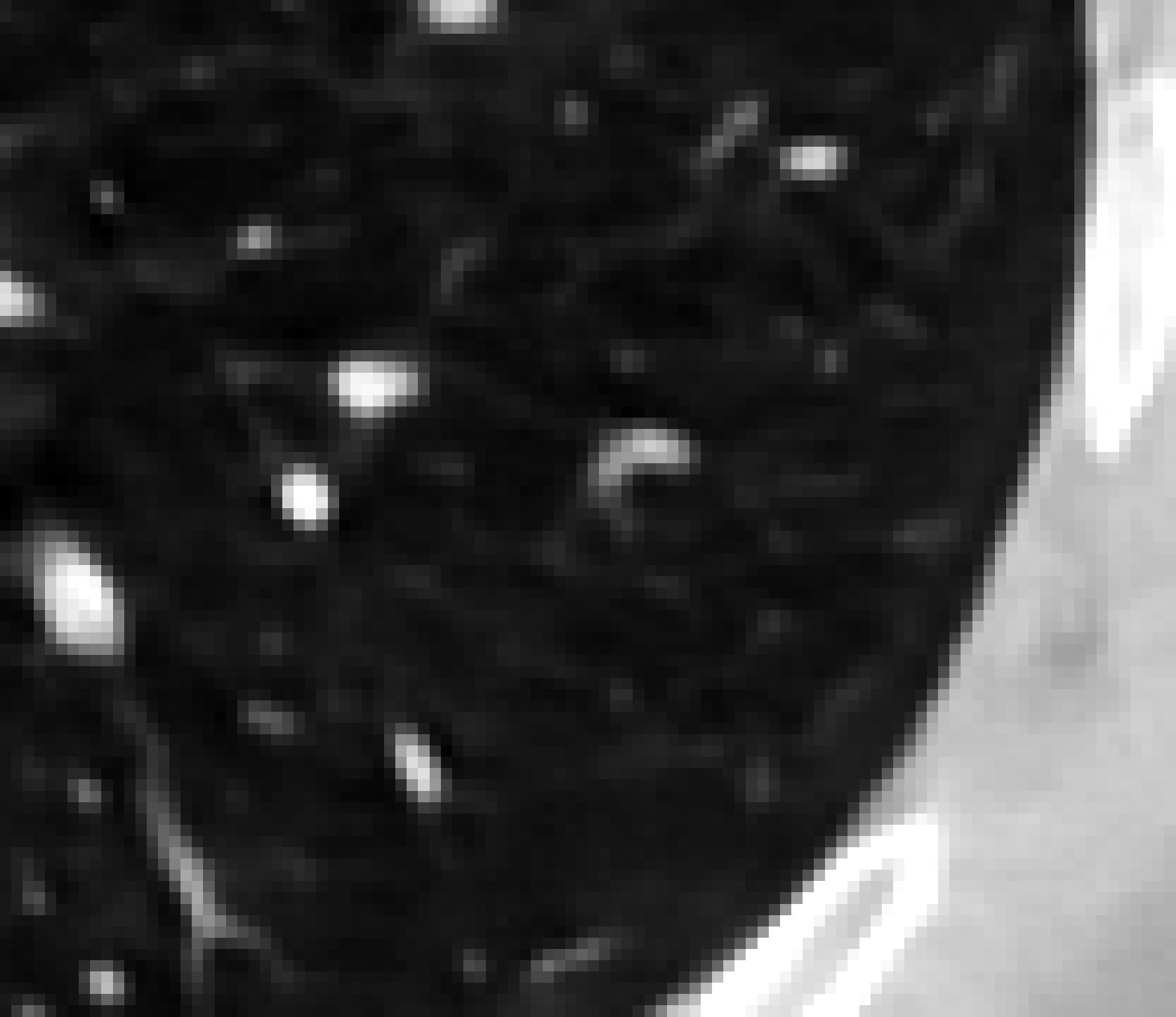}}
				}		
		\end{overpic}}
		
		\vspace{-3mm}
		\subfloat{
			\begin{overpic}[width=\size\columnwidth,percent]{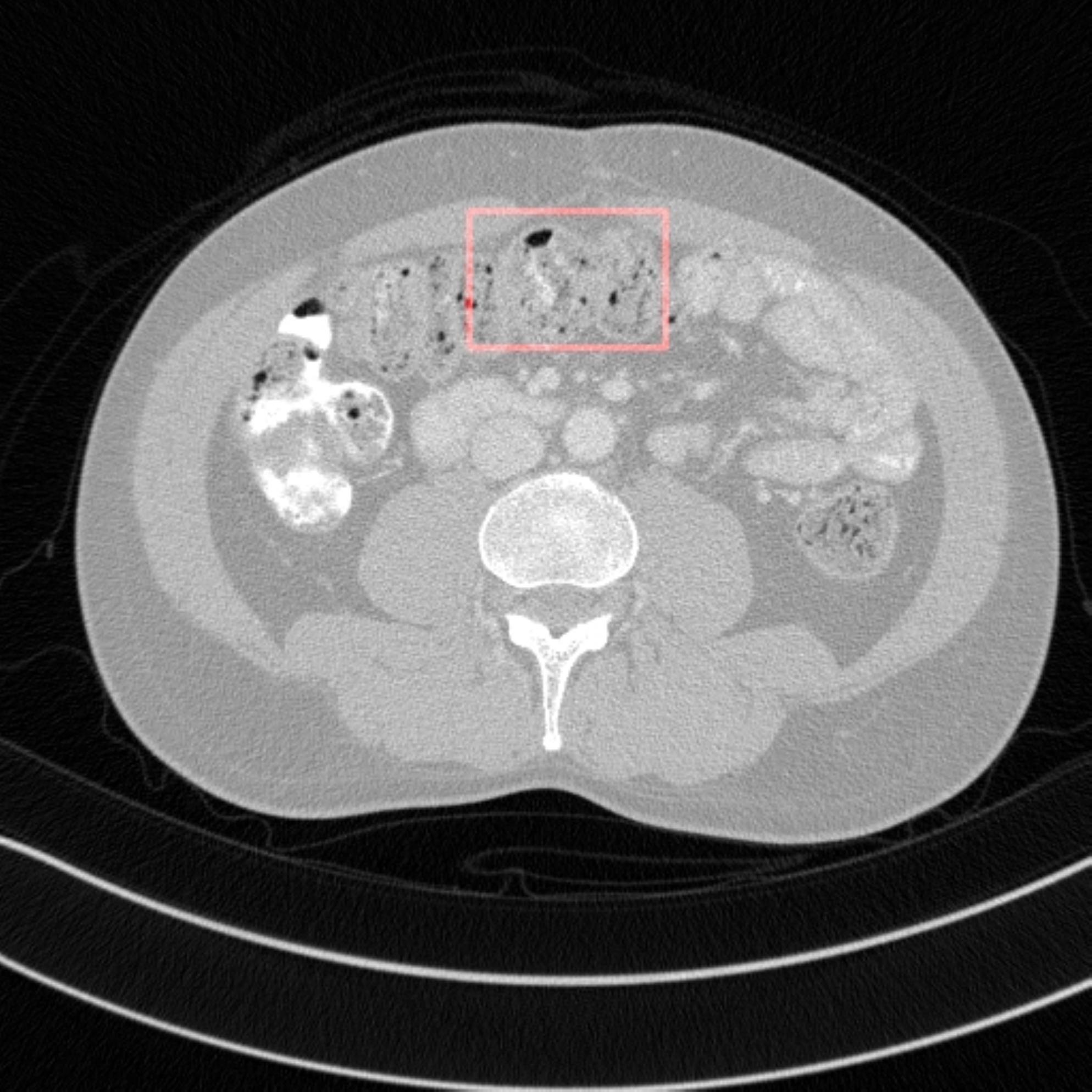}	
				\put(0,0){\color{red}%
					\frame{\includegraphics[scale=0.08]{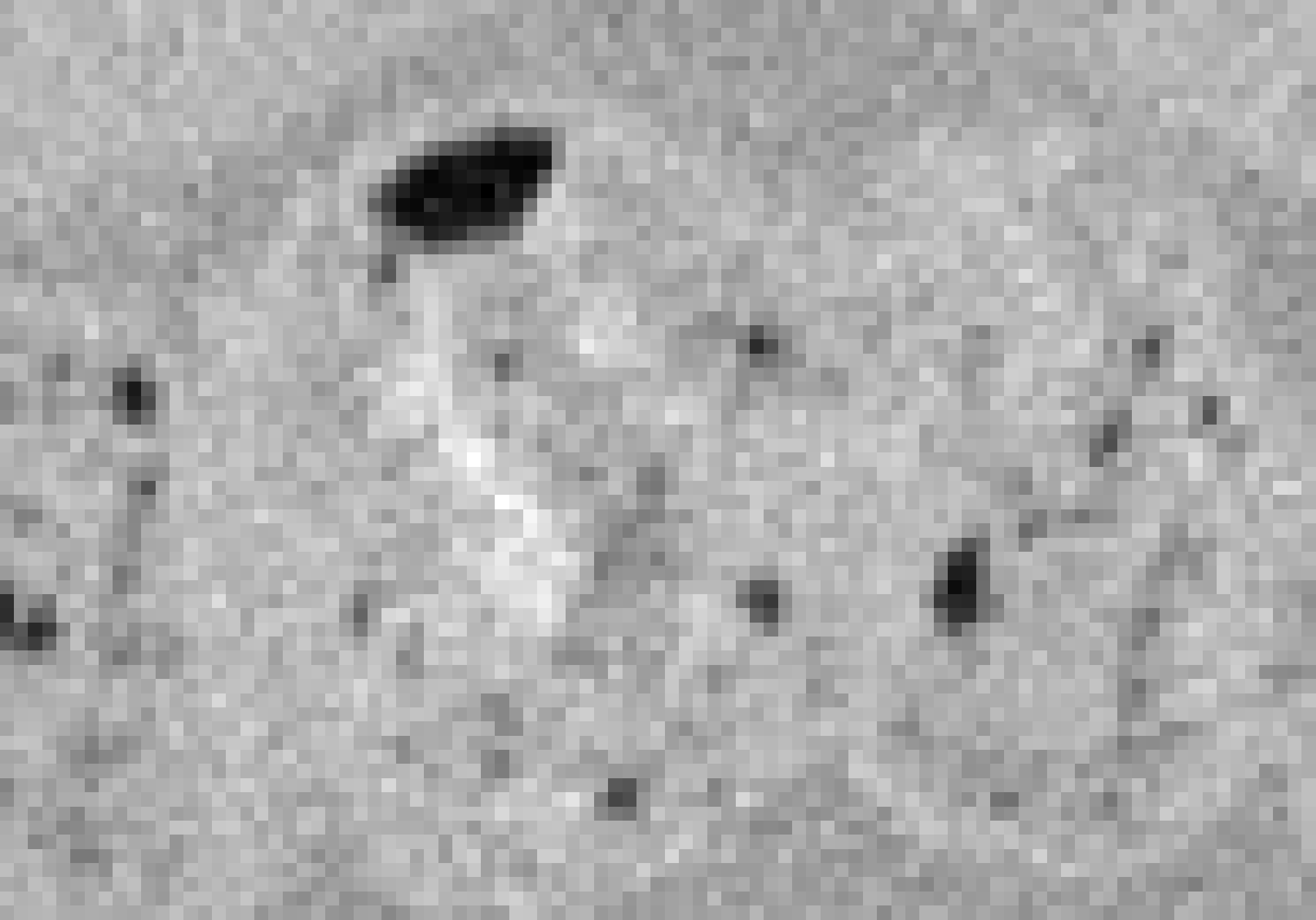}}
				}		
		\end{overpic}}
		\subfloat{
			\begin{overpic}[width=\size\columnwidth]{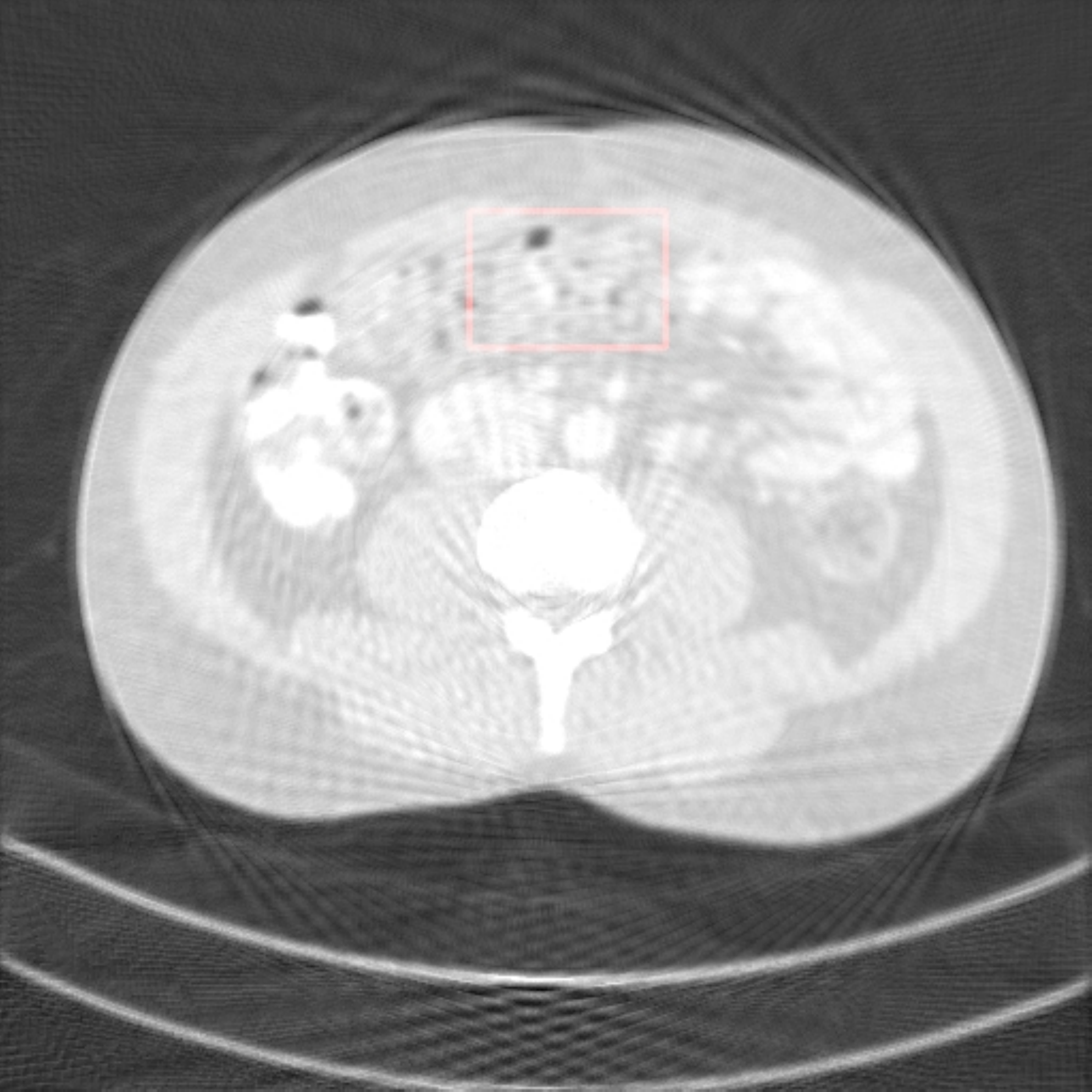}	
				\put(0,0){\color{red}%
					\frame{\includegraphics[scale=0.08]{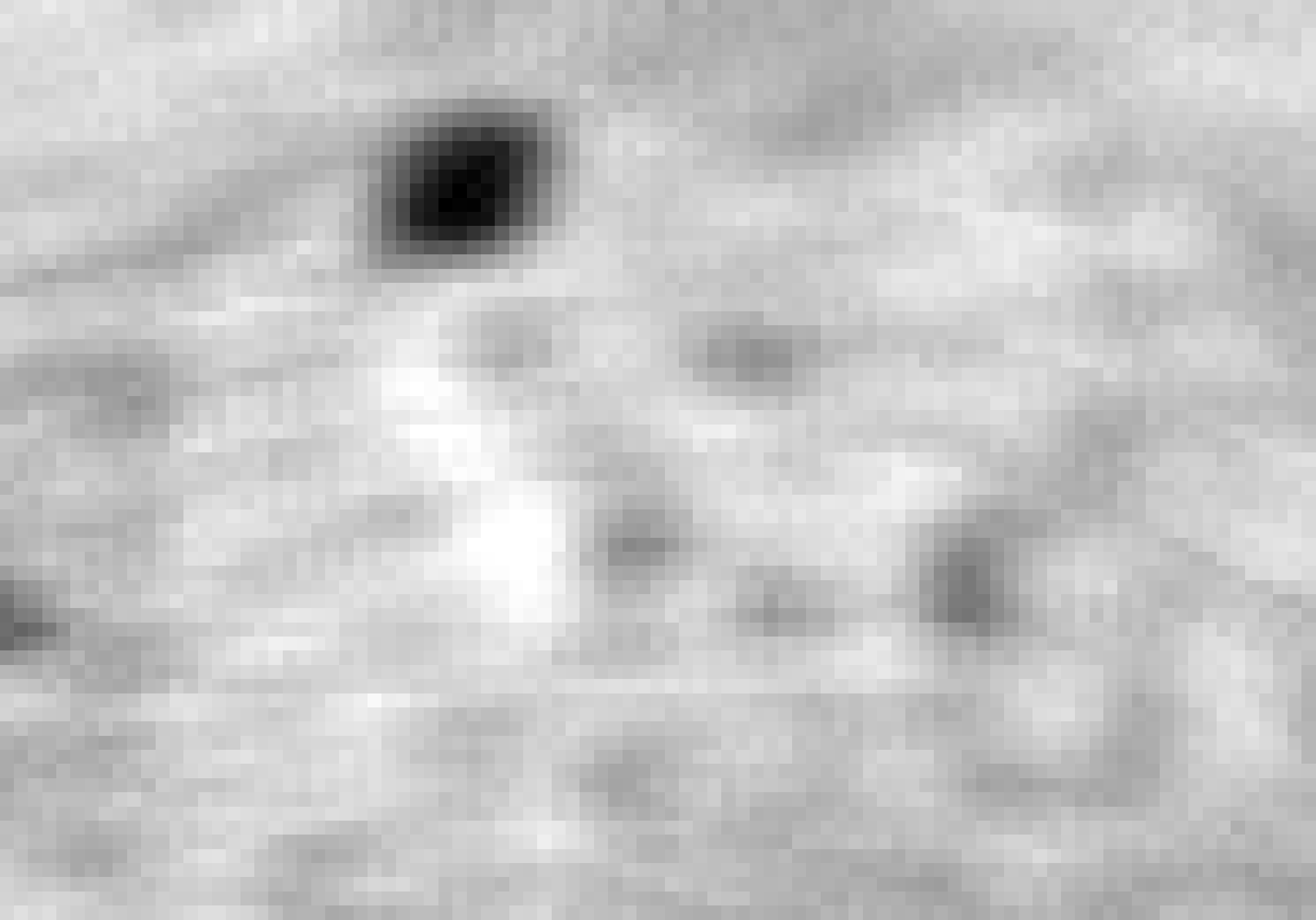}}
				}		
		\end{overpic}}
		\subfloat{
			\begin{overpic}[width=\size\columnwidth]{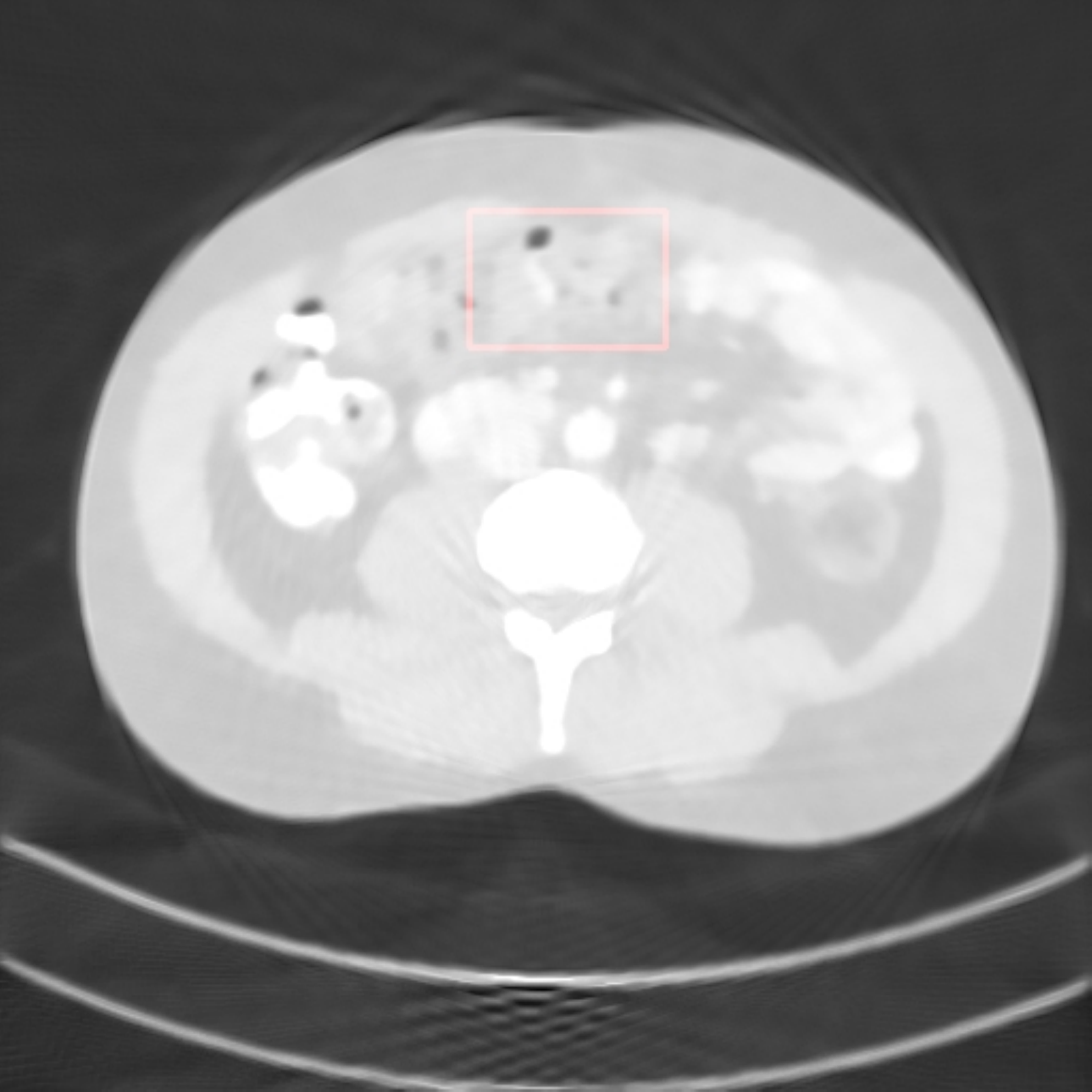}	
				\put(0,0){\color{red}%
					\frame{\includegraphics[scale=0.08]{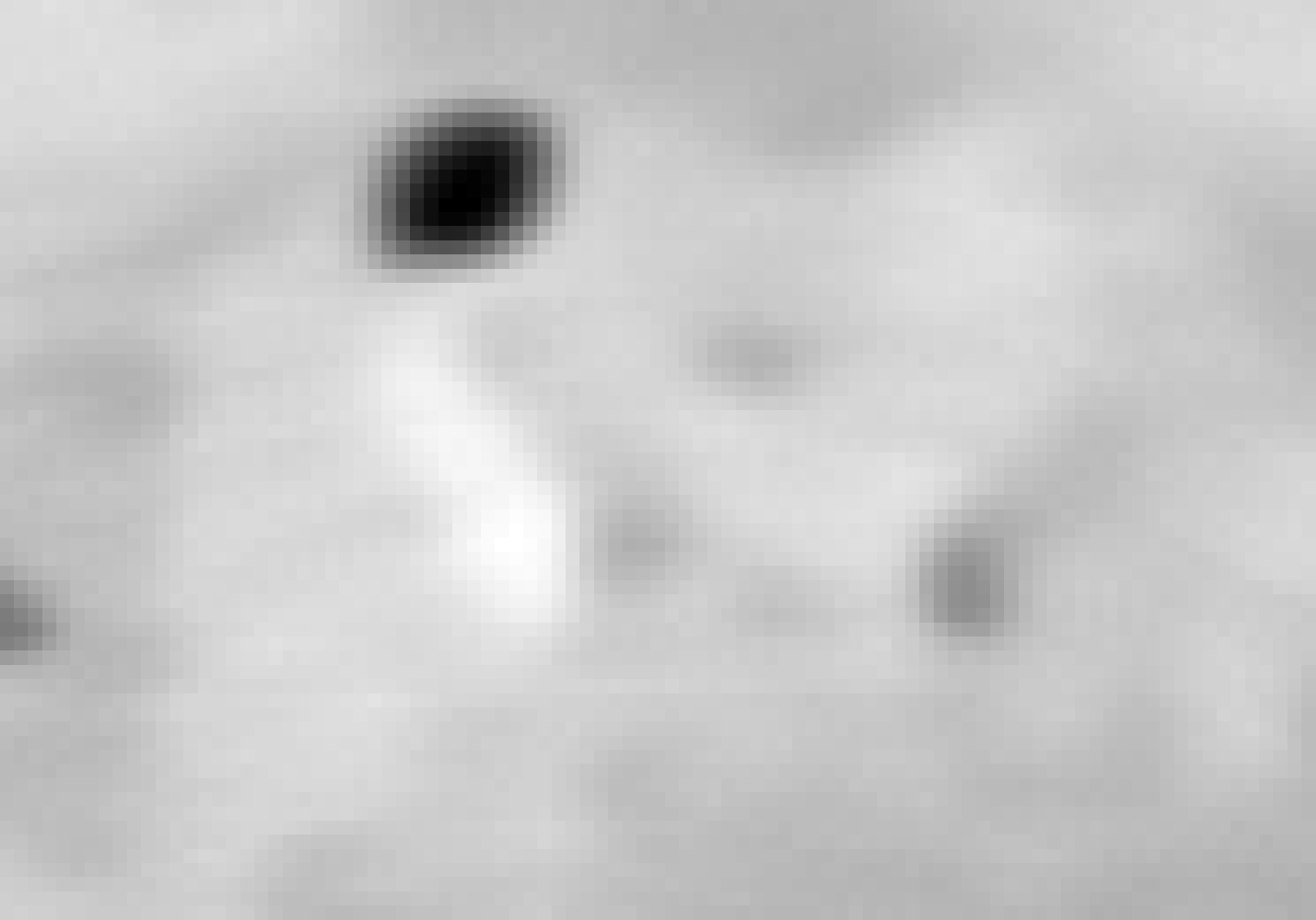}}
				}		
		\end{overpic}}
		\subfloat{
			\begin{overpic}[width=\size\columnwidth]{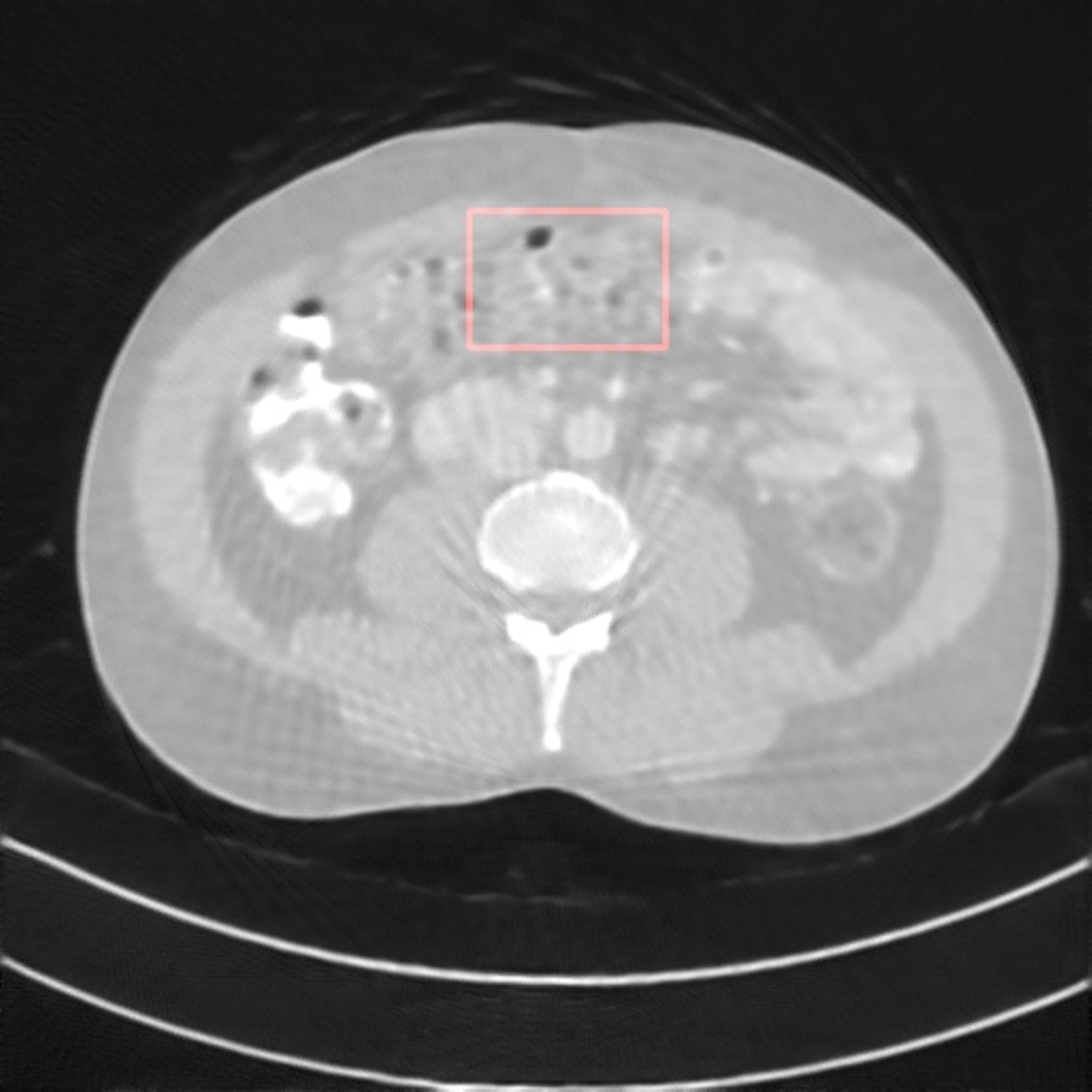}	
				\put(0,0){\color{red}%
					\frame{\includegraphics[scale=0.08]{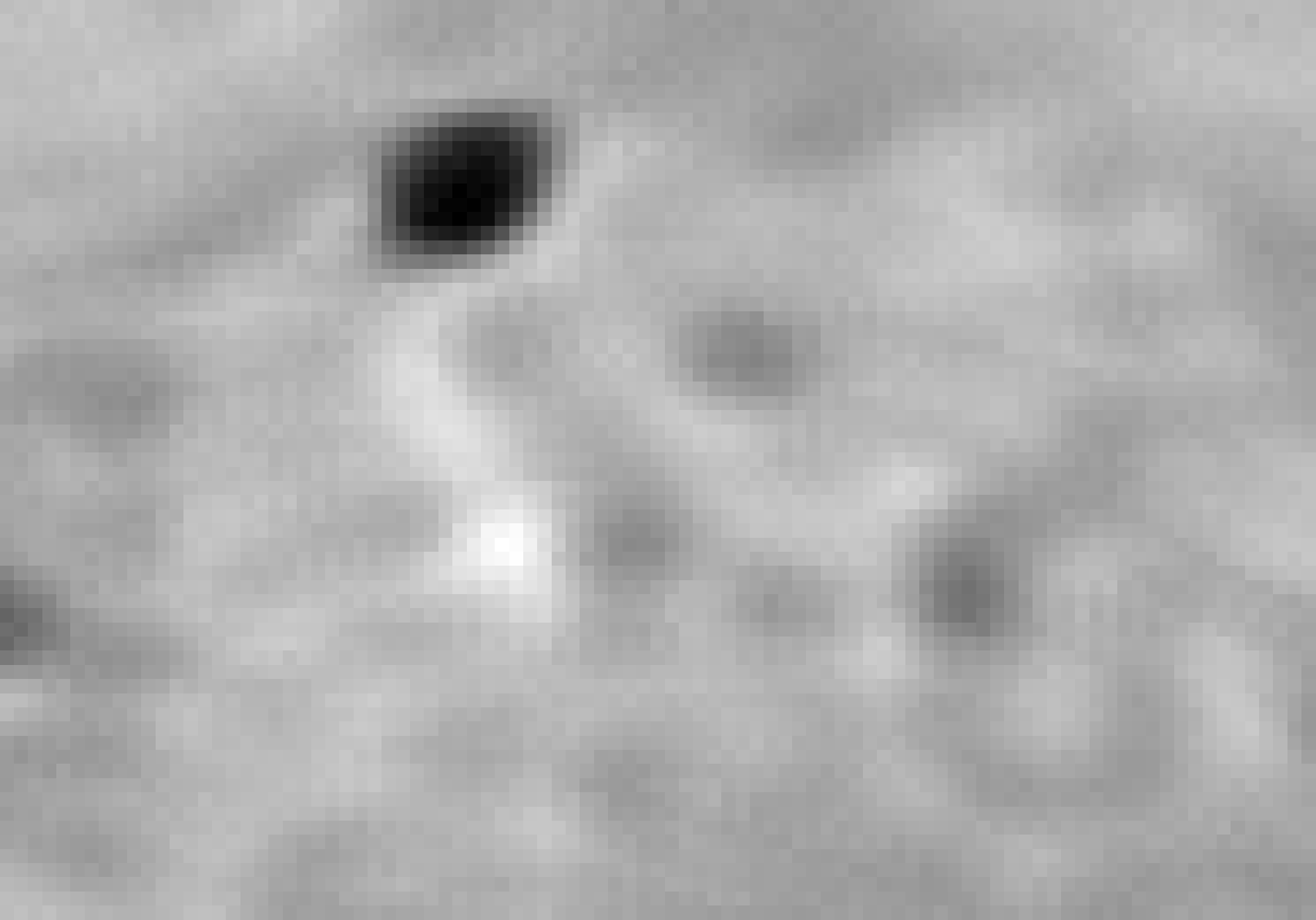}}
				}		
		\end{overpic}}
		\subfloat{
			\begin{overpic}[width=\size\columnwidth]{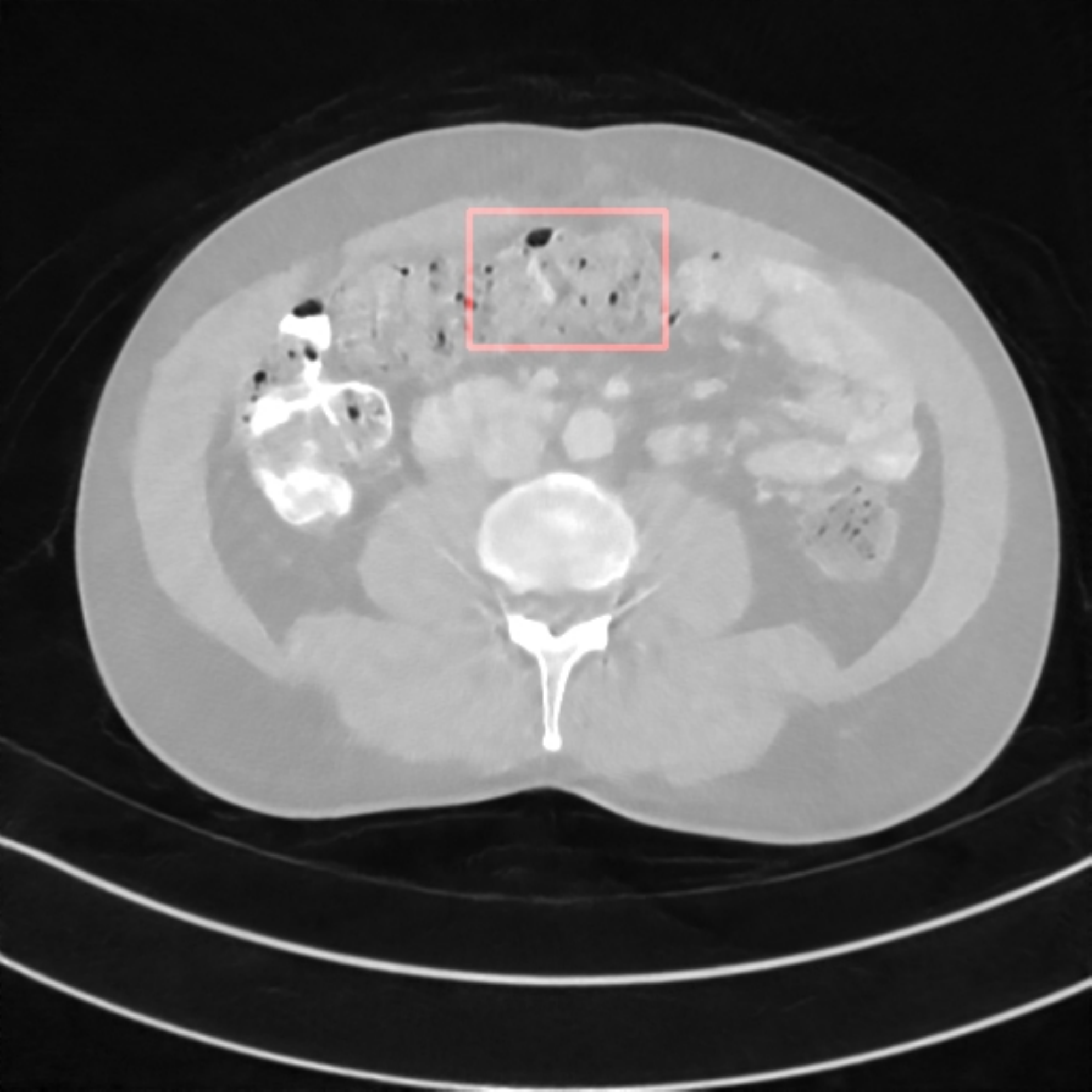}	
				\put(0,0){\color{red}%
					\frame{\includegraphics[scale=0.08]{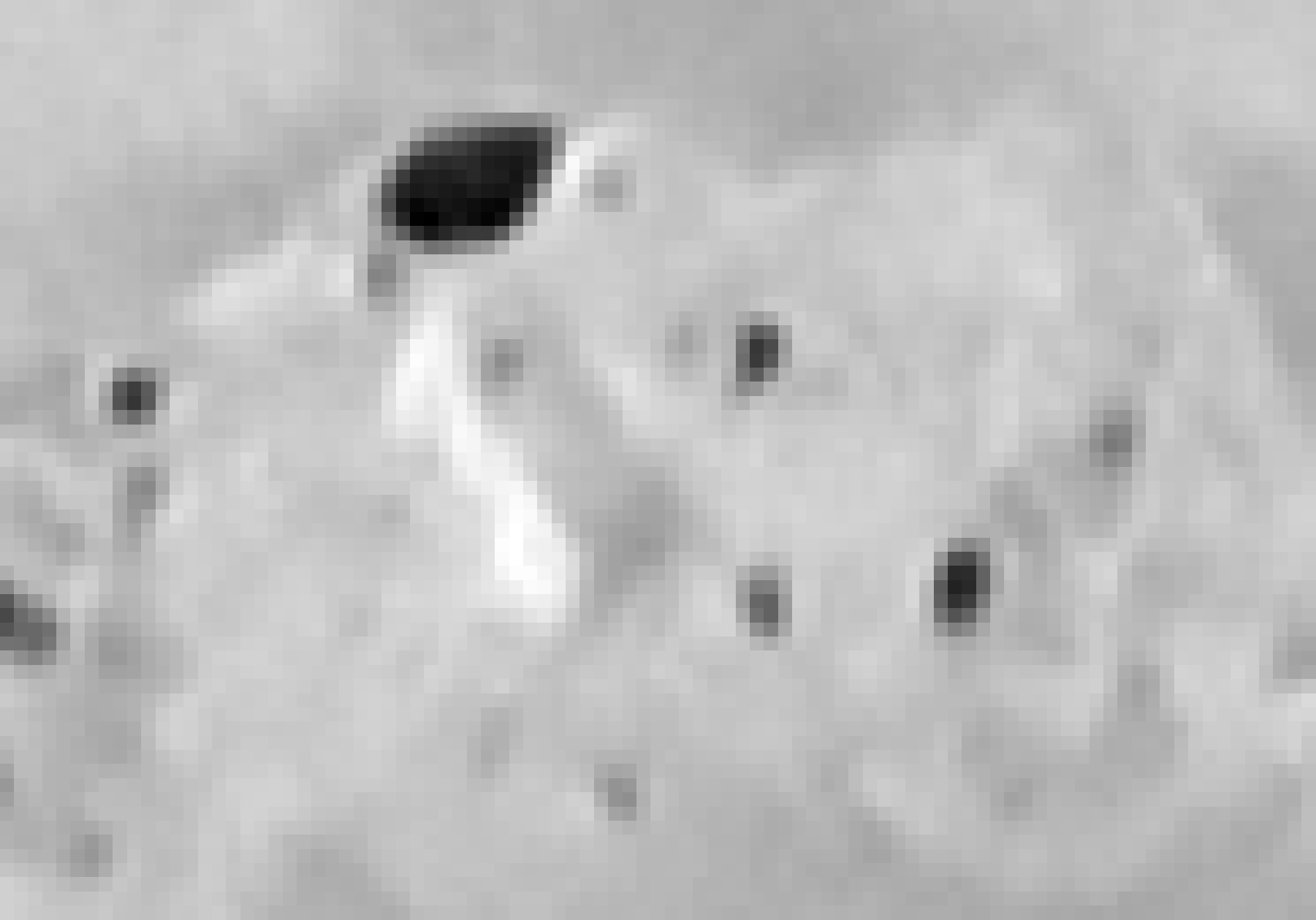}}
				}		
		\end{overpic}}
		\vspace{-3mm}
		
		\subfloat{
			\begin{overpic}[width=\size\columnwidth,percent]{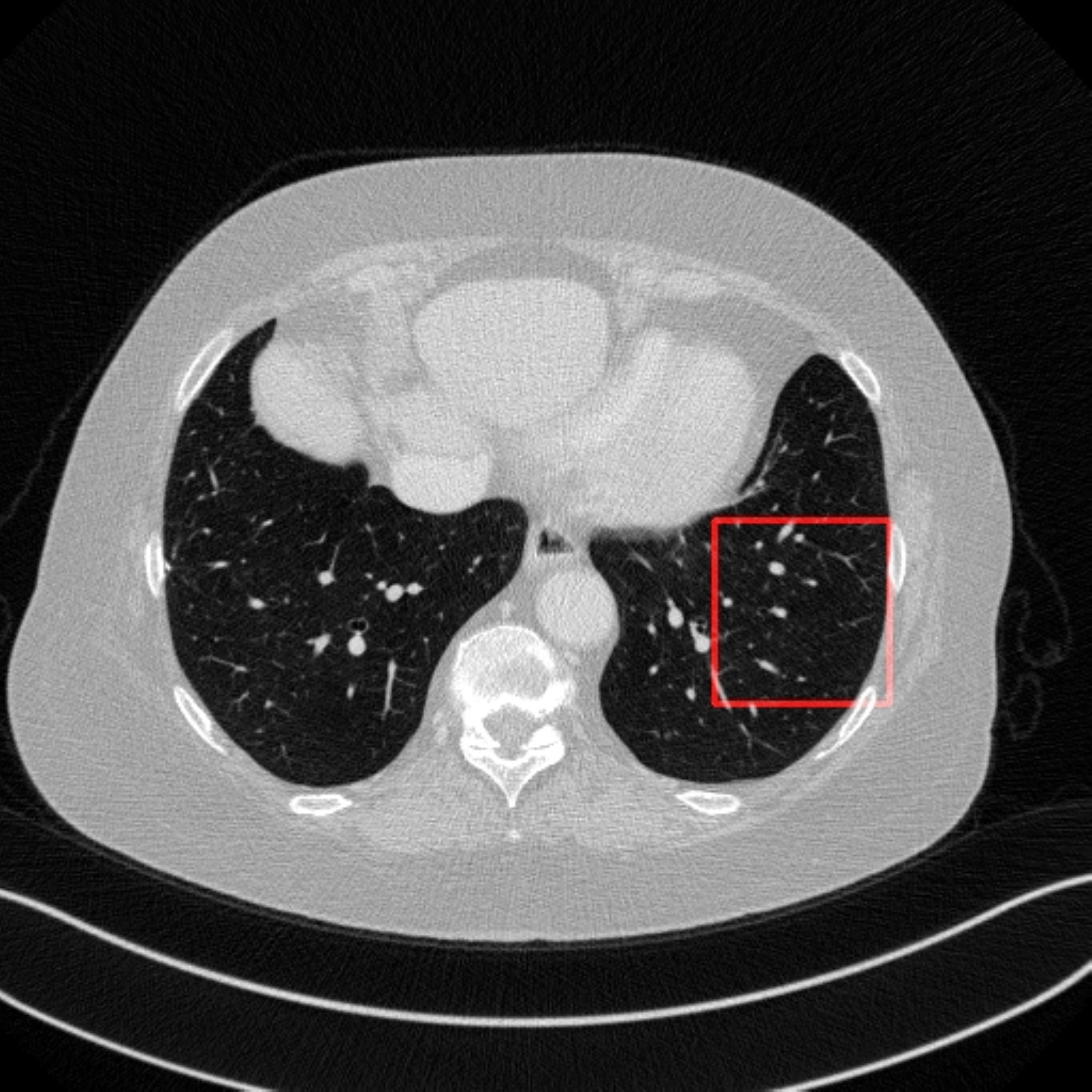}	
				\put(0,0){\color{red}%
					\frame{\includegraphics[scale=0.08]{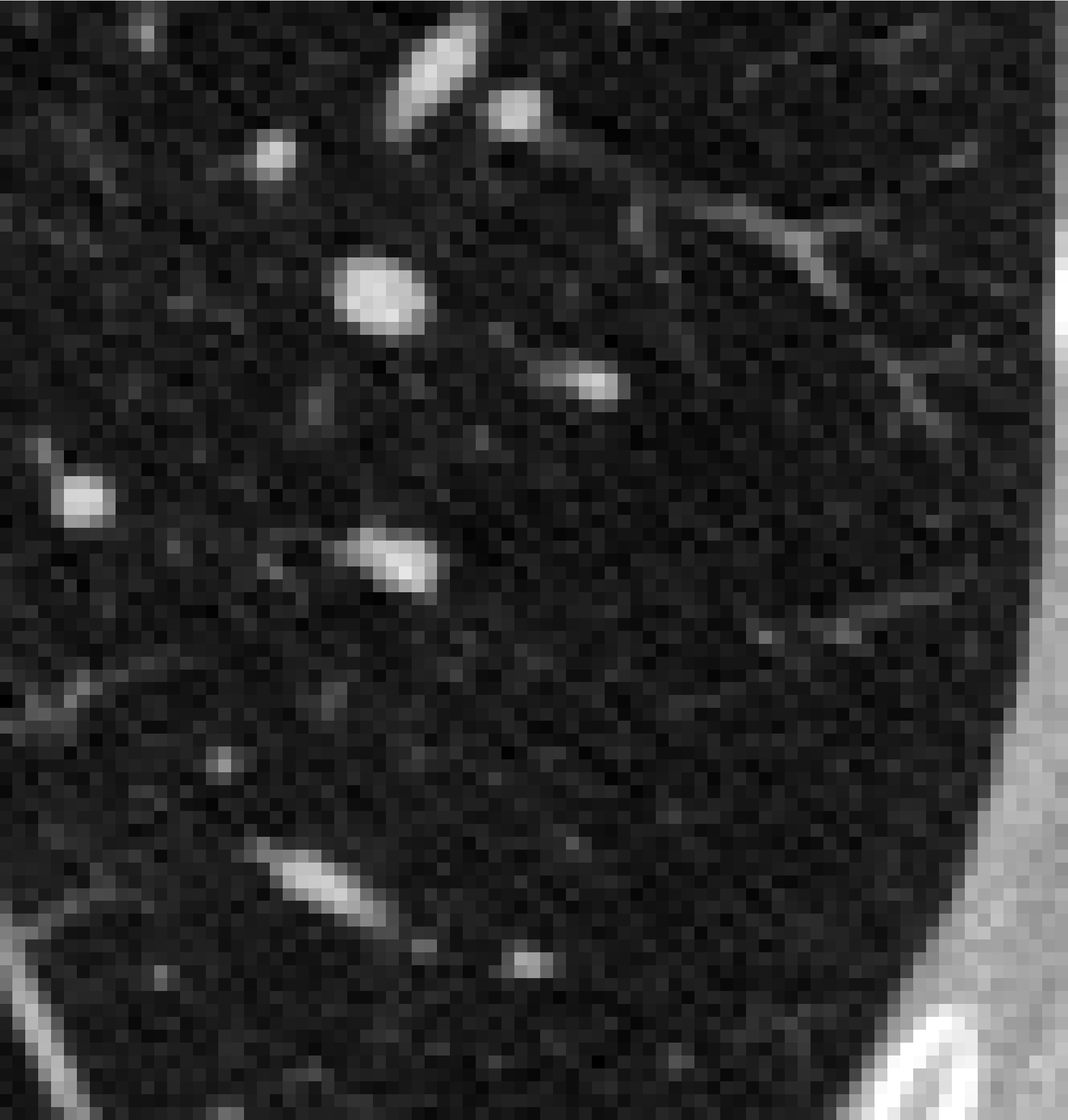}}
				}		
		\end{overpic}}
		\subfloat{
			\begin{overpic}[width=\size\columnwidth]{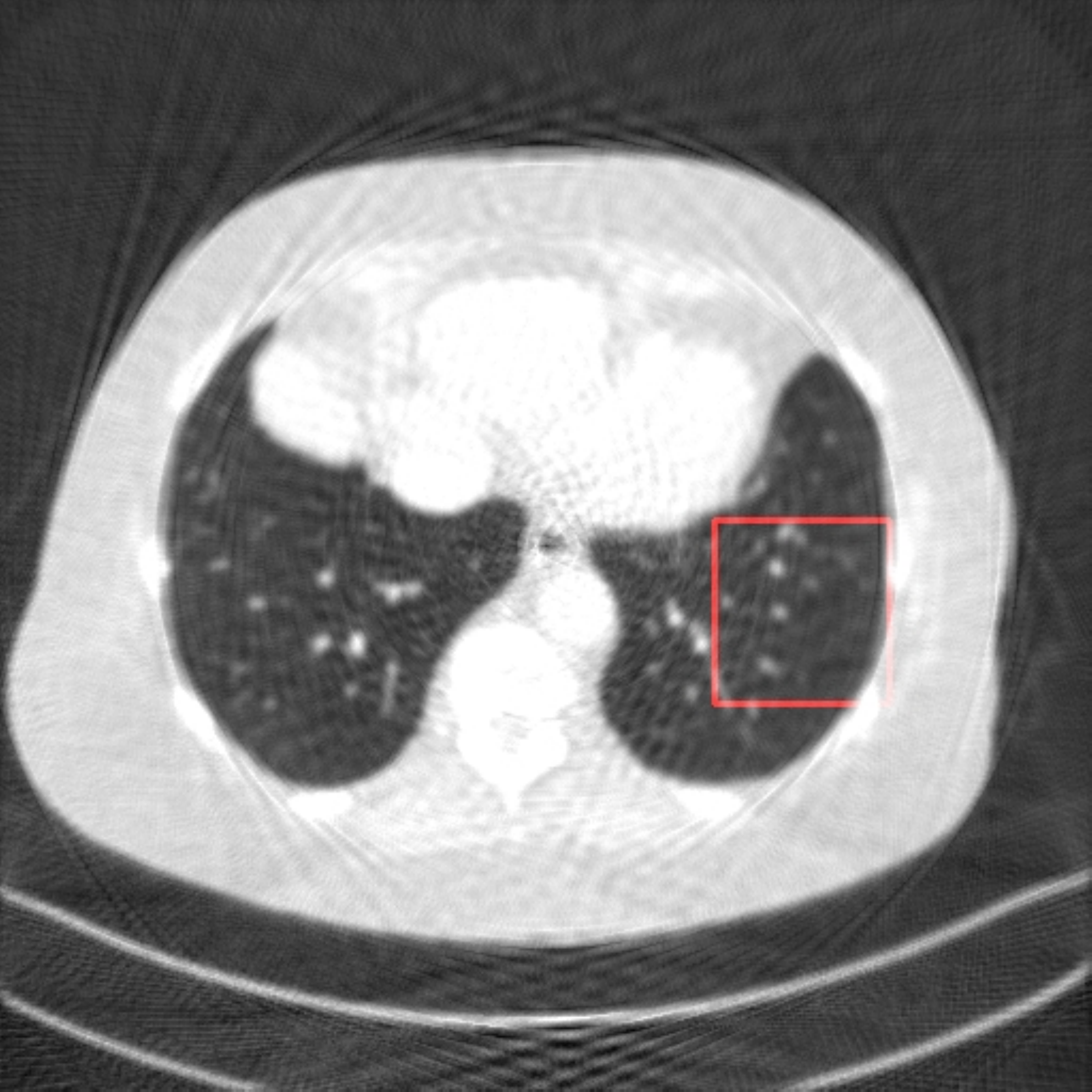}	
				\put(0,0){\color{red}%
					\frame{\includegraphics[scale=0.08]{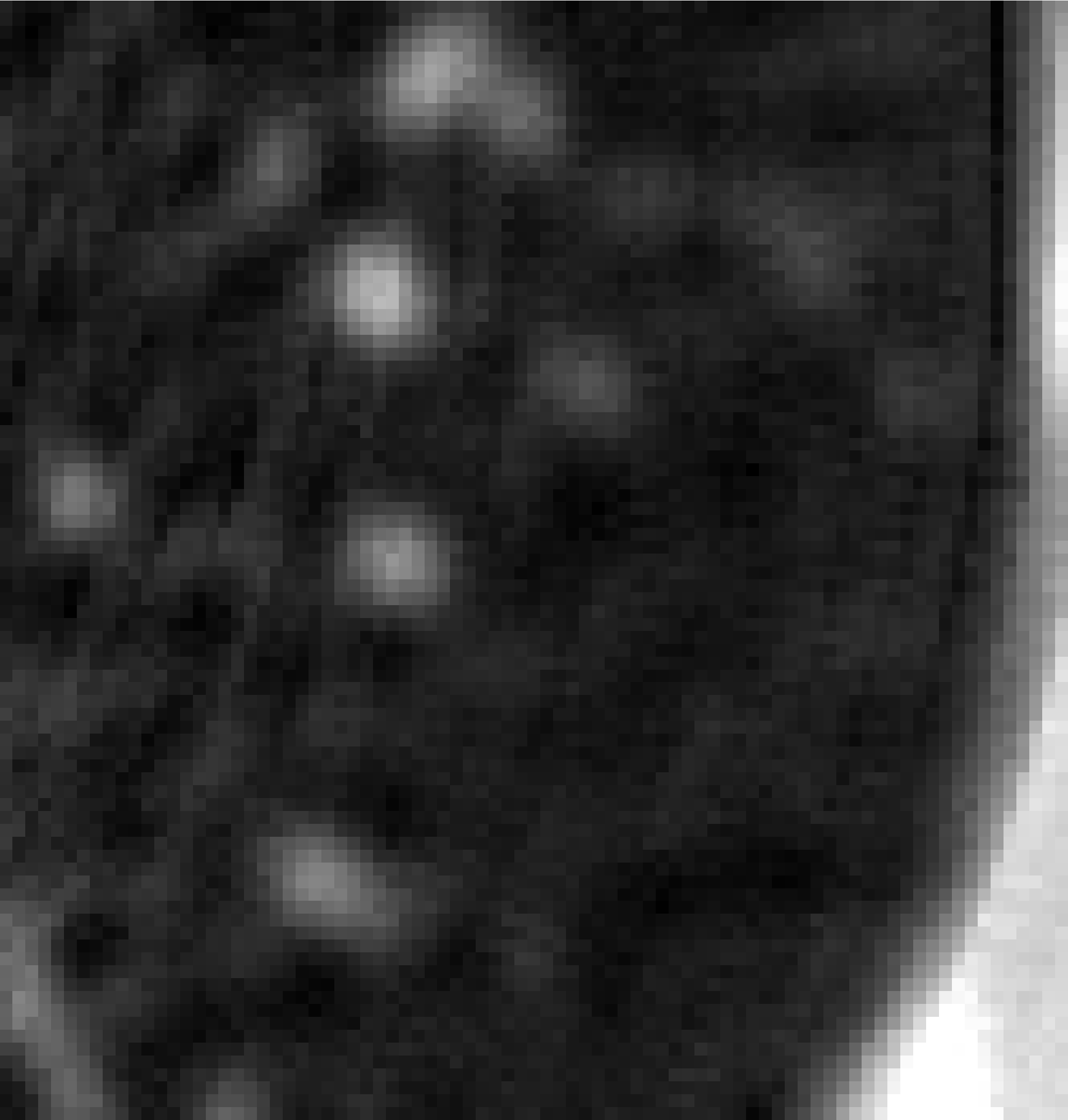}}
				}		
		\end{overpic}}
		\subfloat{
			\begin{overpic}[width=\size\columnwidth]{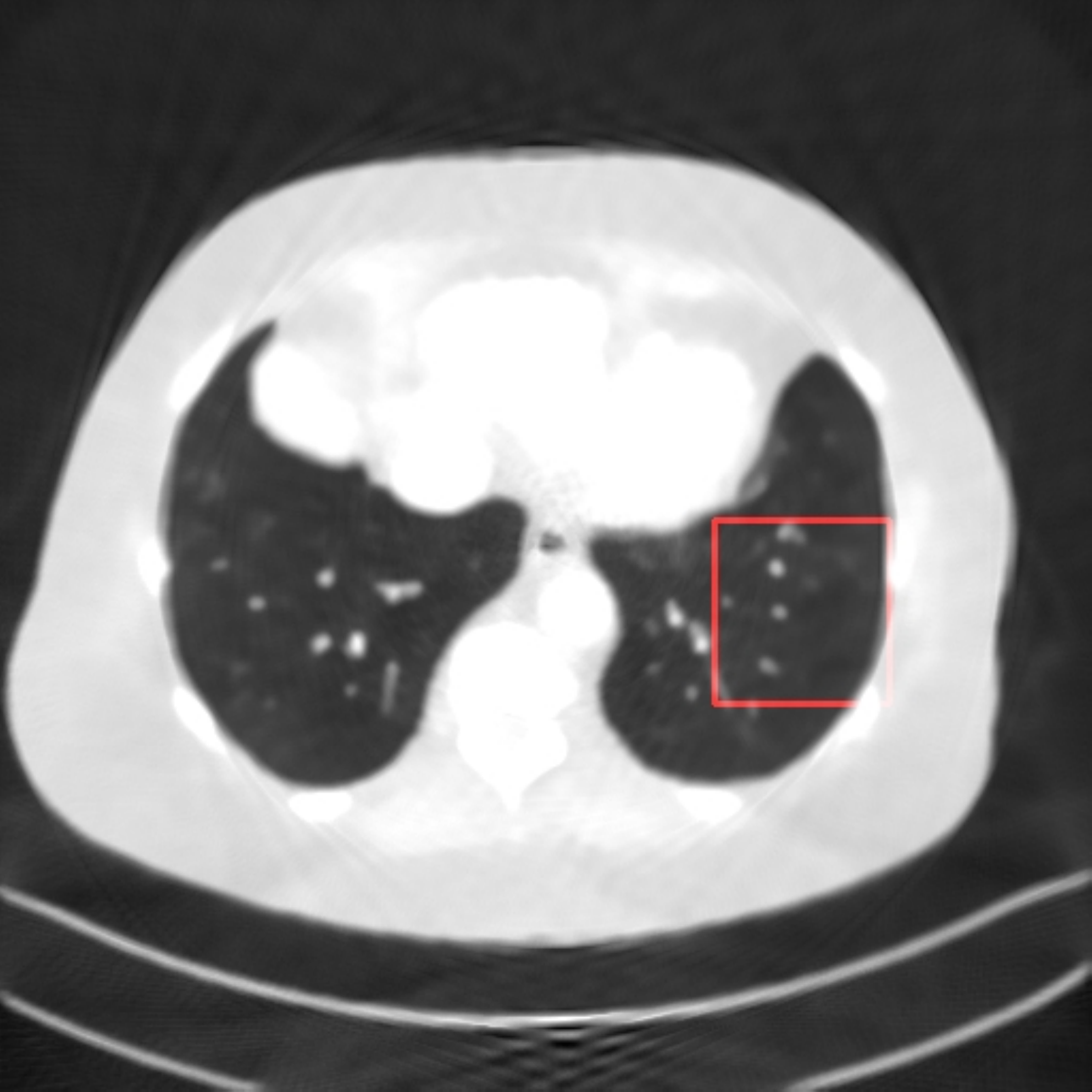}	
				\put(0,0){\color{red}%
					\frame{\includegraphics[scale=0.08]{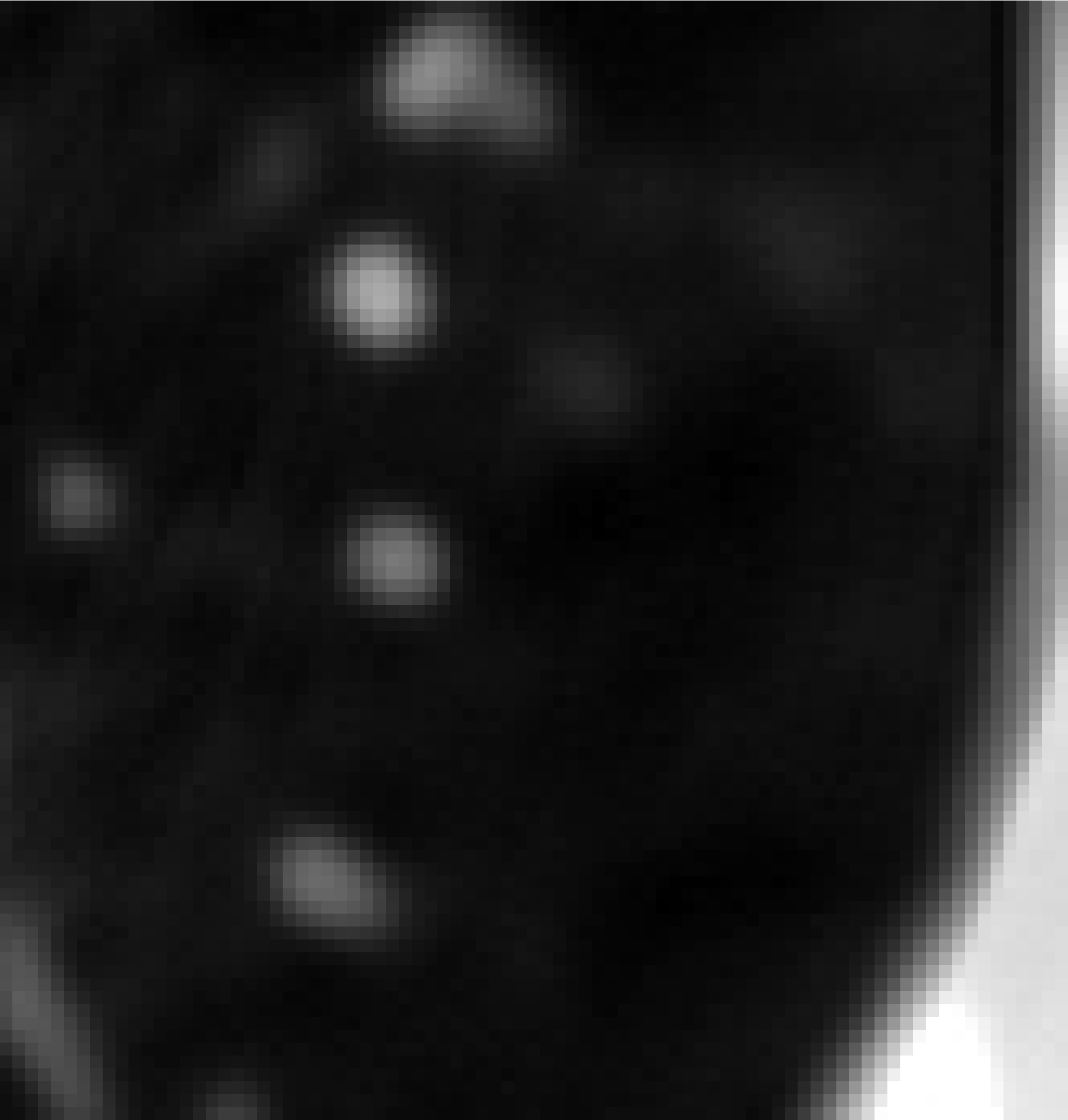}}
				}		
		\end{overpic}}
		\subfloat{
			\begin{overpic}[width=\size\columnwidth]{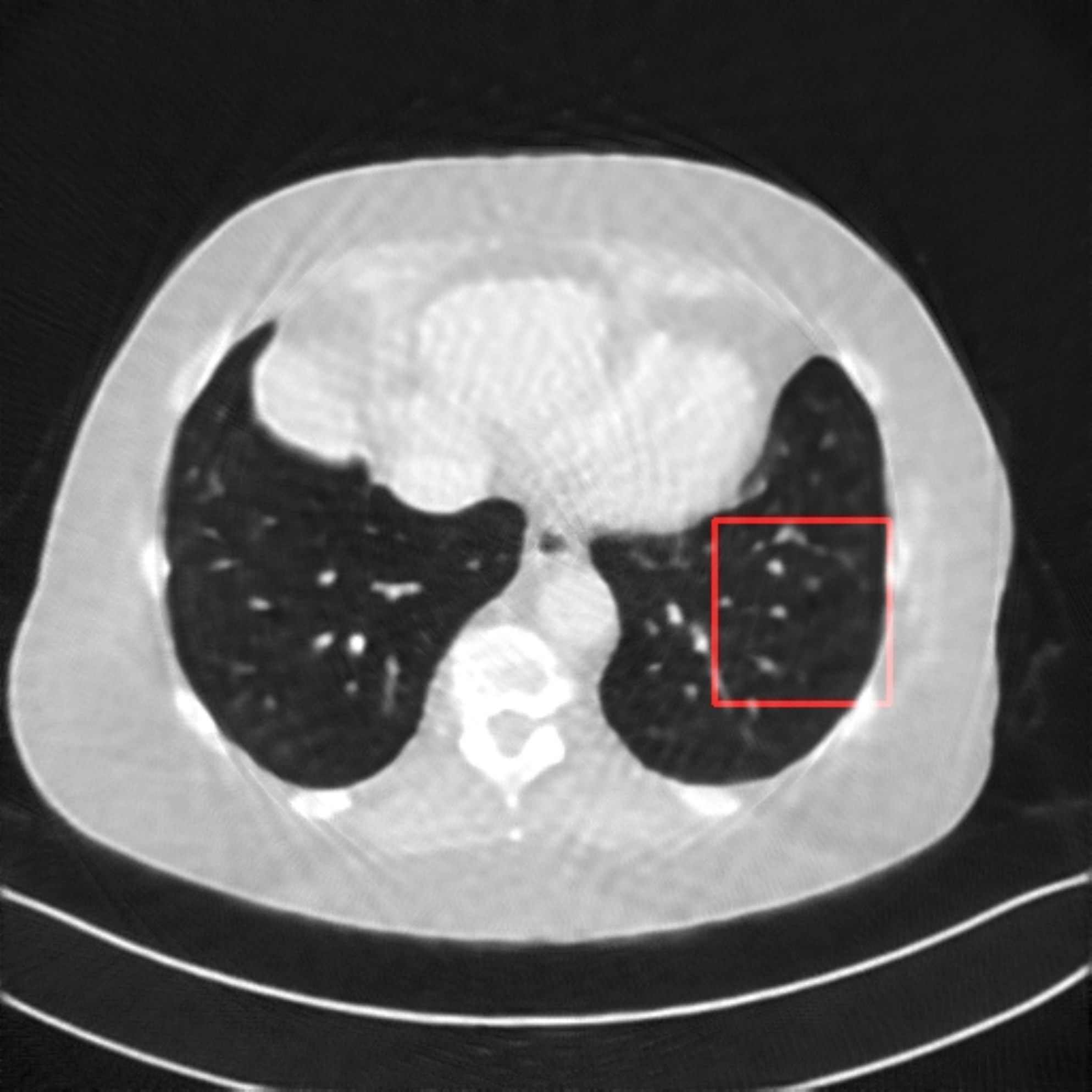}	
				\put(0,0){\color{red}%
					\frame{\includegraphics[scale=0.08]{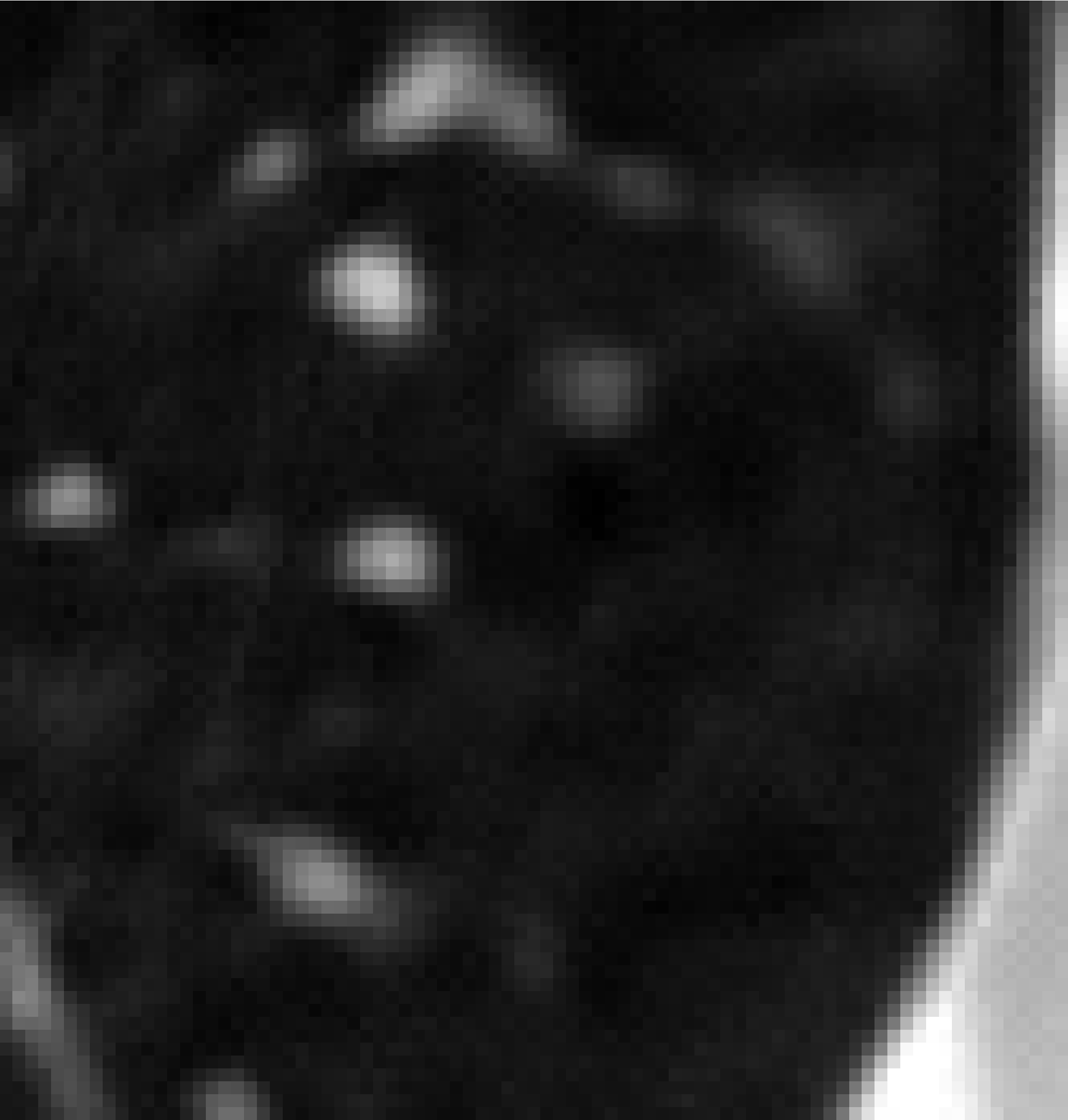}}
				}		
		\end{overpic}}
		\subfloat{
			\begin{overpic}[width=\size\columnwidth]{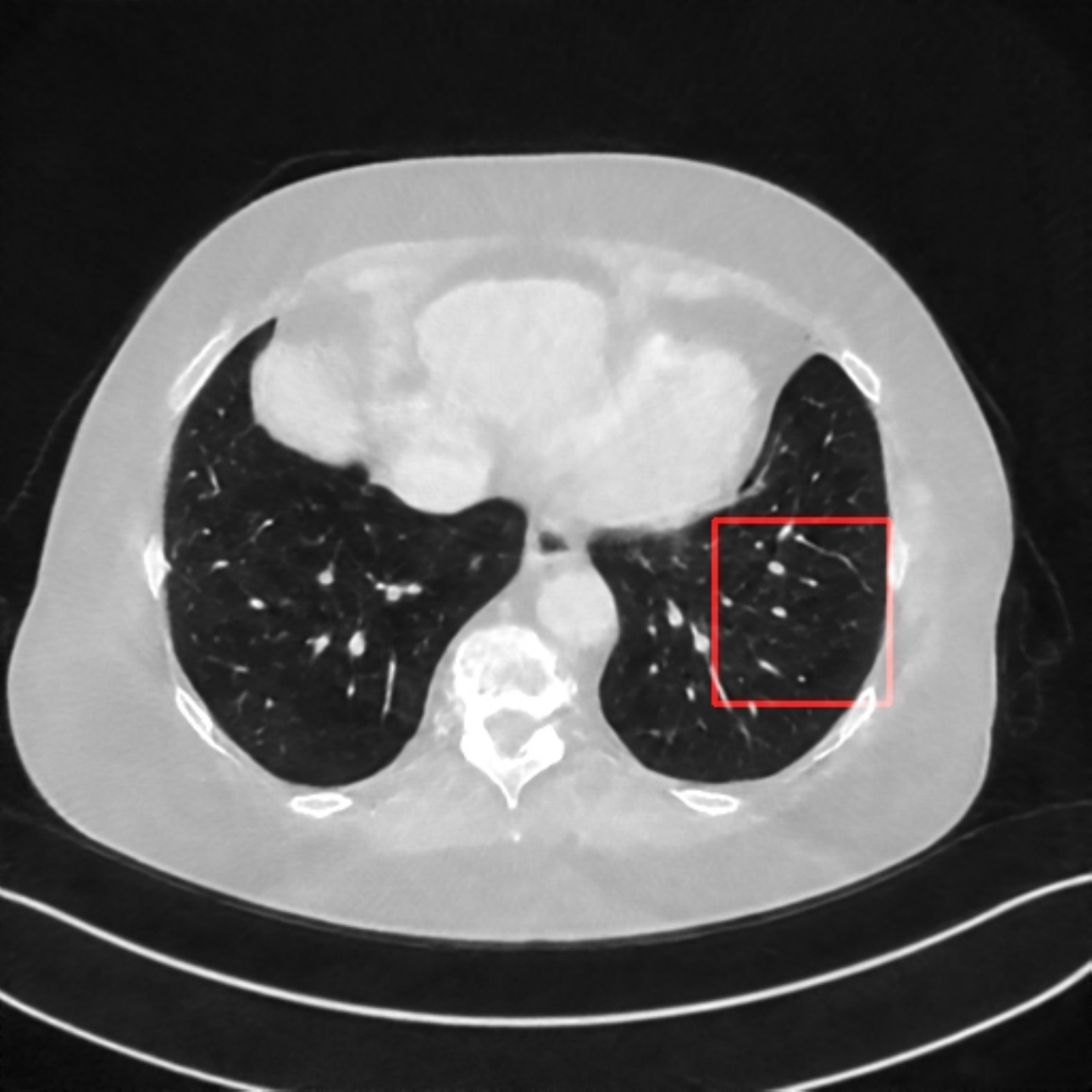}	
				\put(0,0){\color{red}%
					\frame{\includegraphics[scale=0.08]{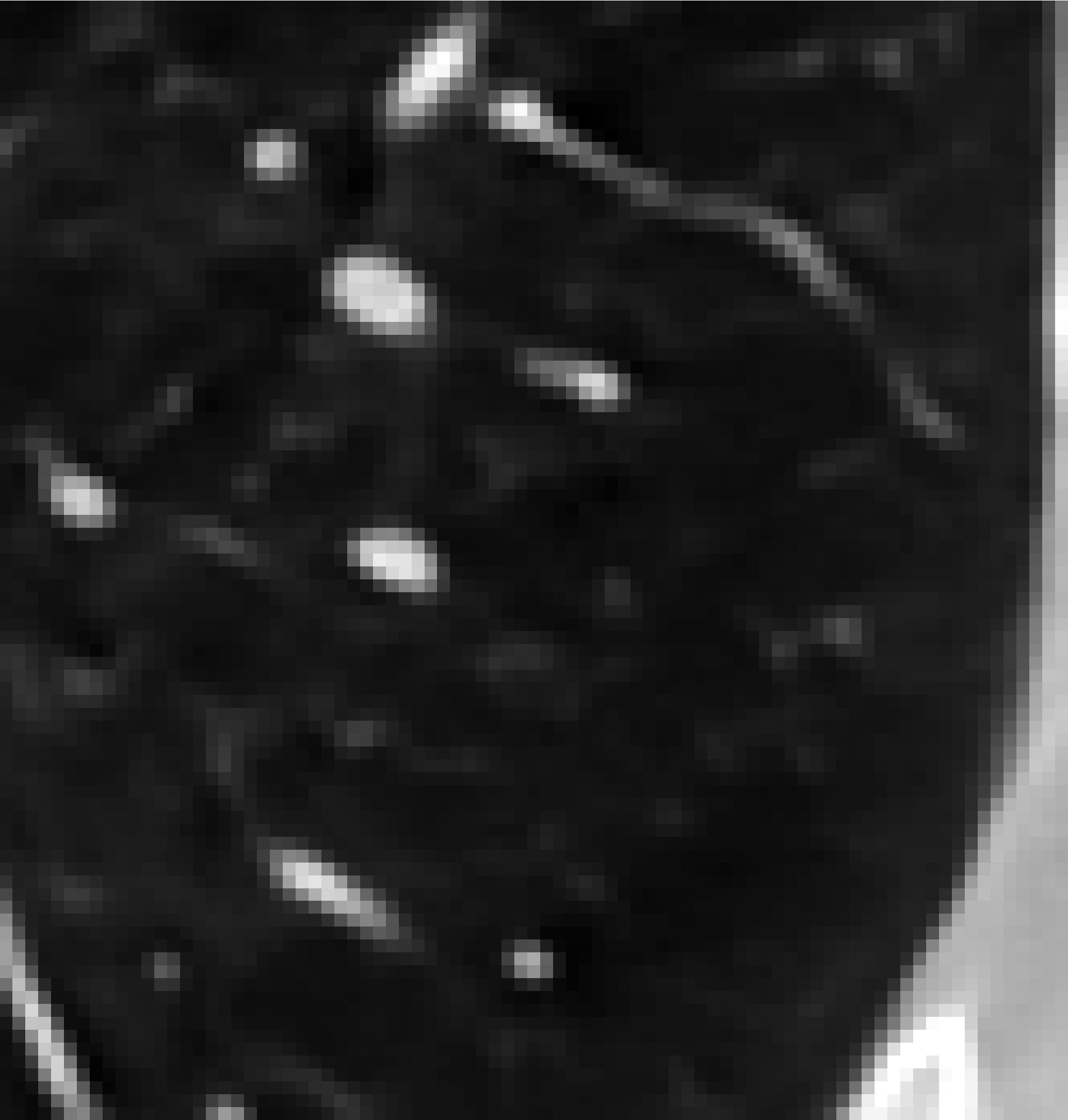}}
				}		
		\end{overpic}}
		\vspace{-3mm}
		\subfloat{
			\begin{overpic}[width=\size\columnwidth,percent]{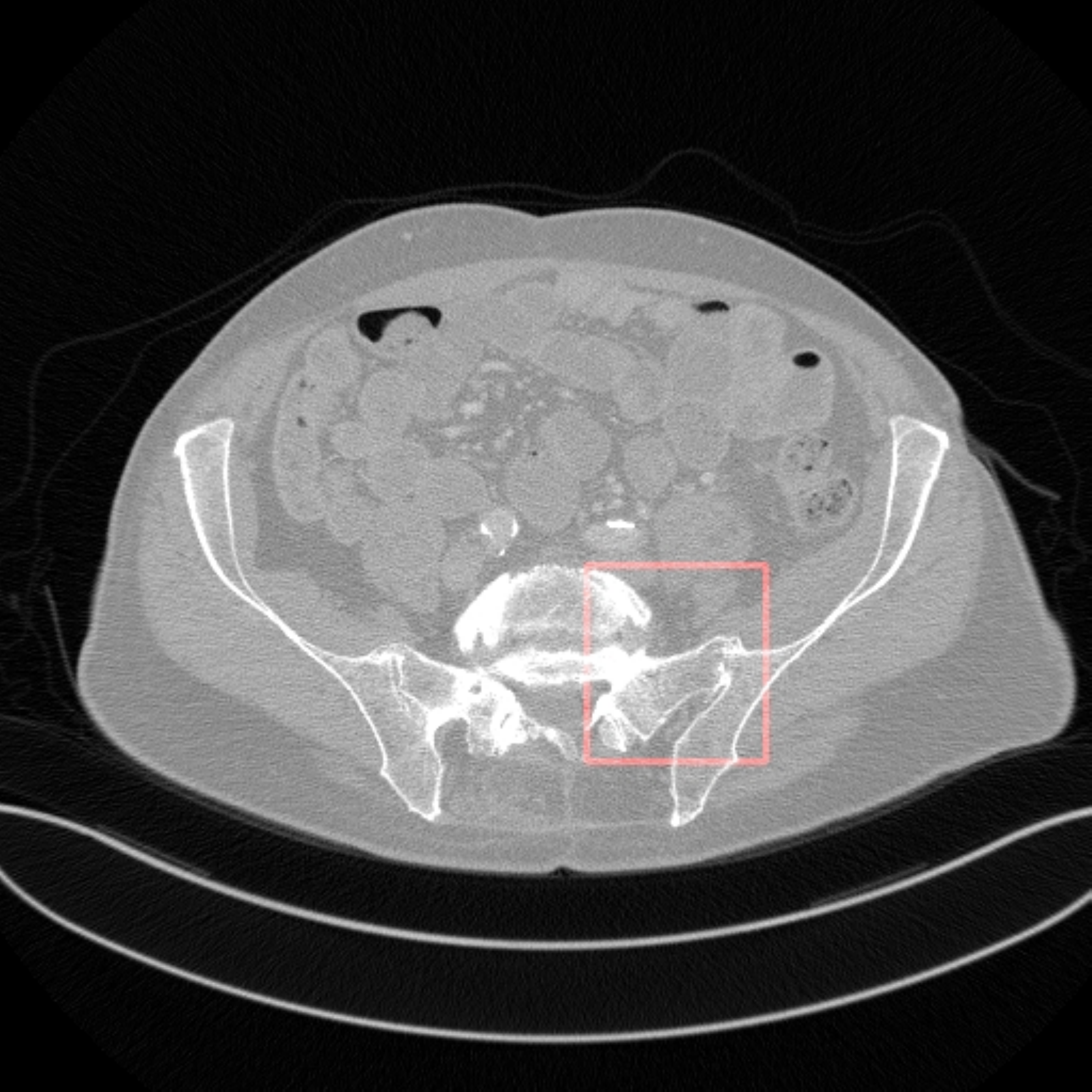}	
				\put(0,0){\color{red}%
					\frame{\includegraphics[scale=0.08]{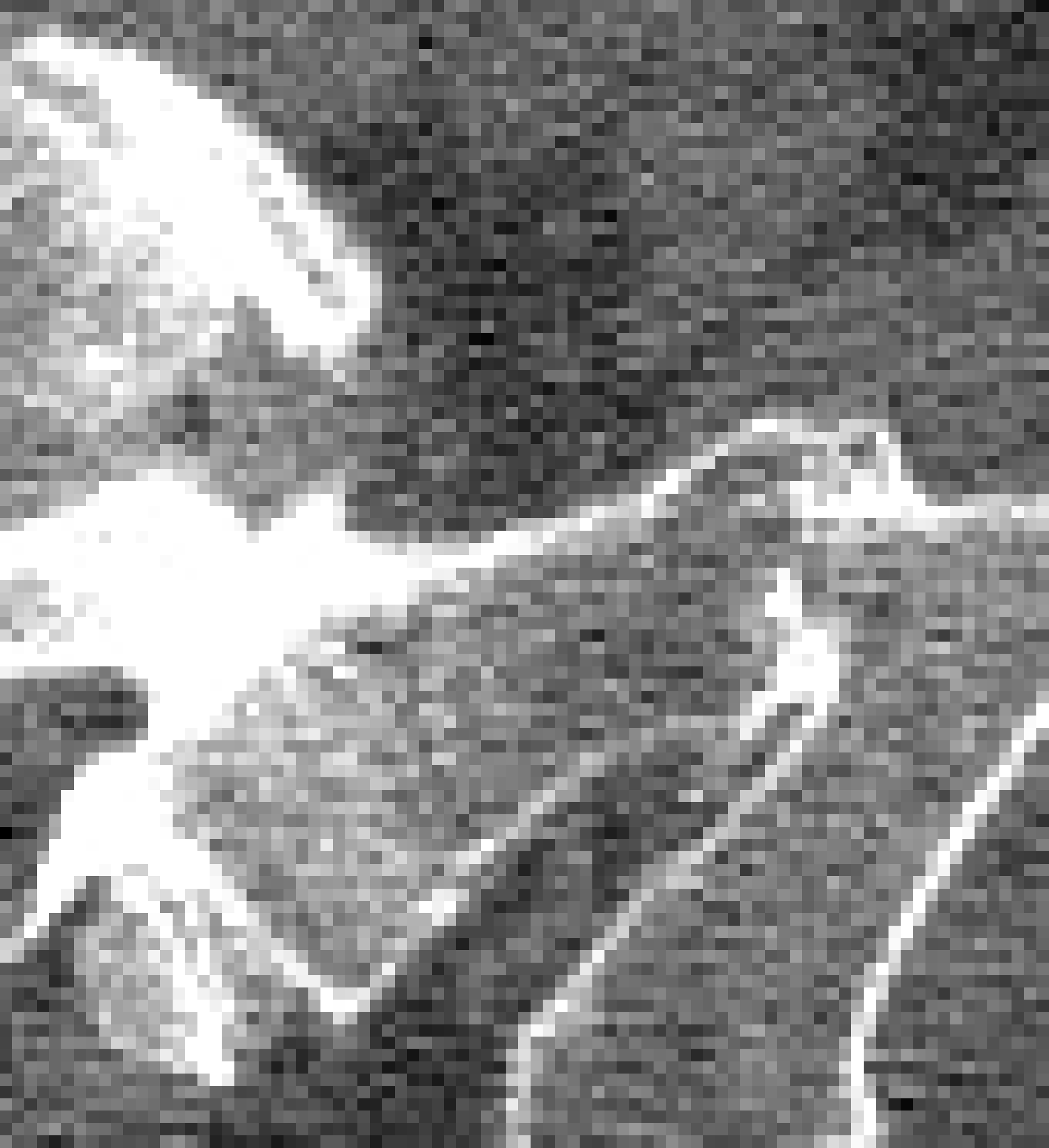}}
				}		
		\end{overpic}}
		\subfloat{
			\begin{overpic}[width=\size\columnwidth]{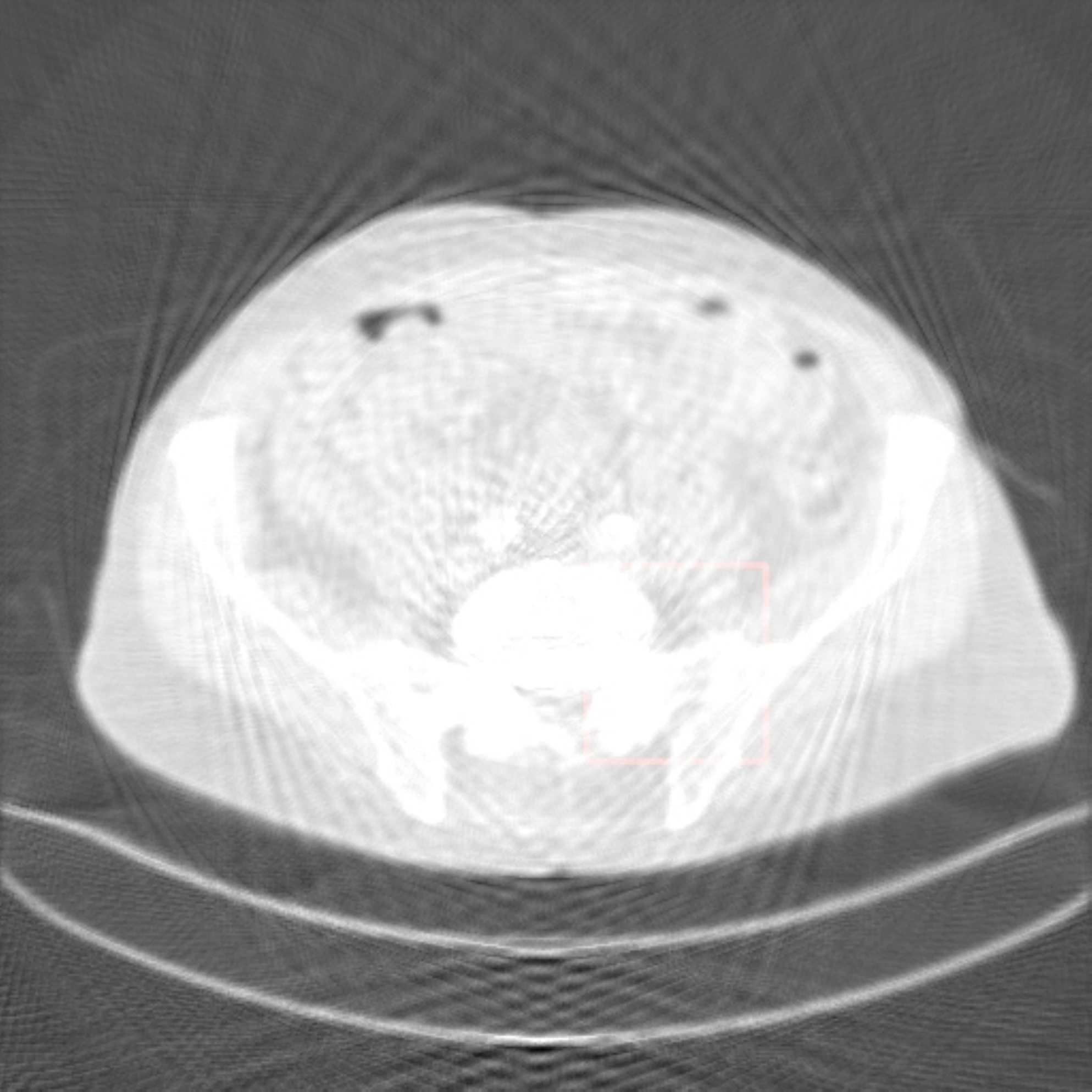}	
				\put(0,0){\color{red}%
					\frame{\includegraphics[scale=0.08]{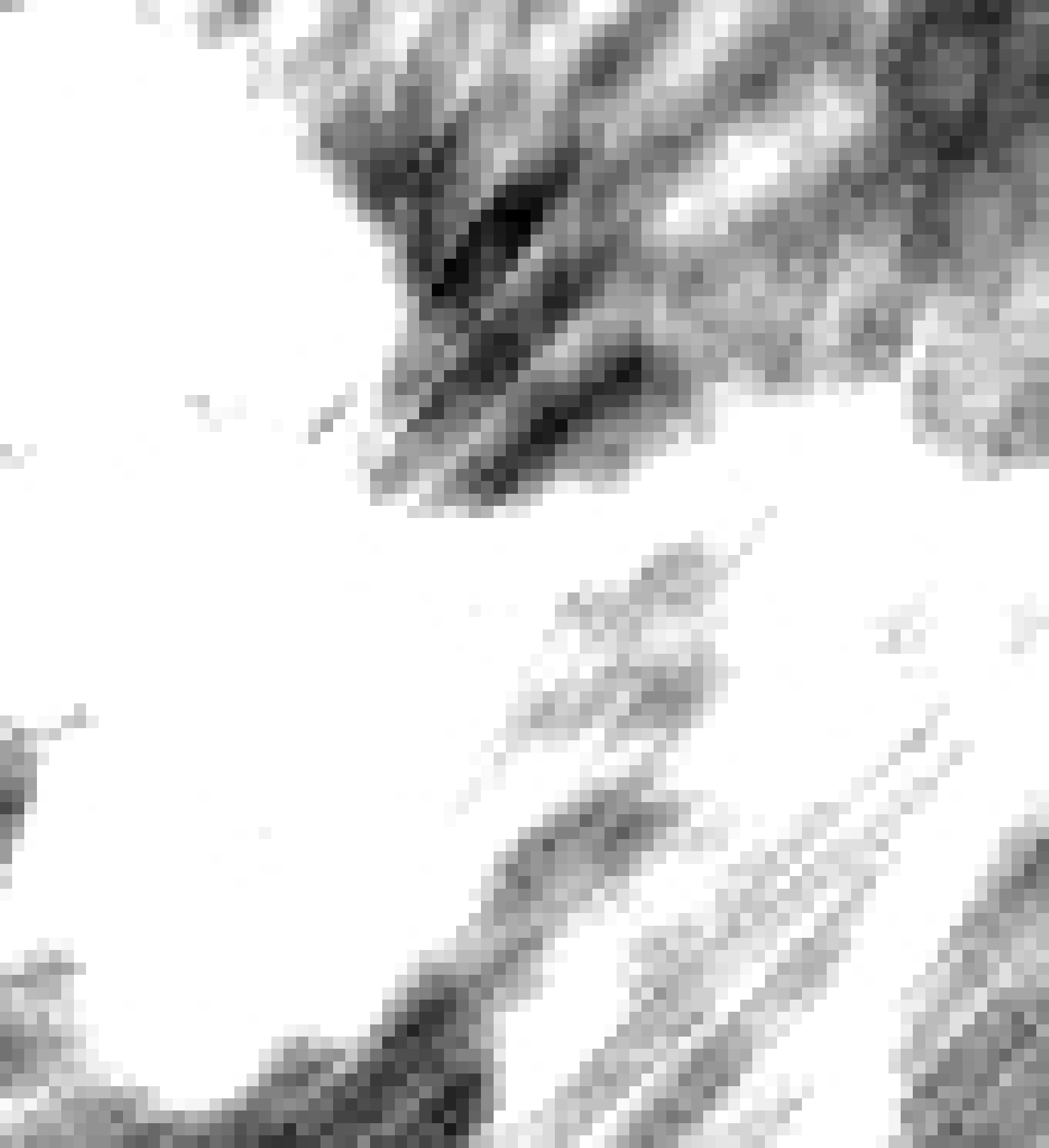}}
				}		
		\end{overpic}}
		\subfloat{
			\begin{overpic}[width=\size\columnwidth]{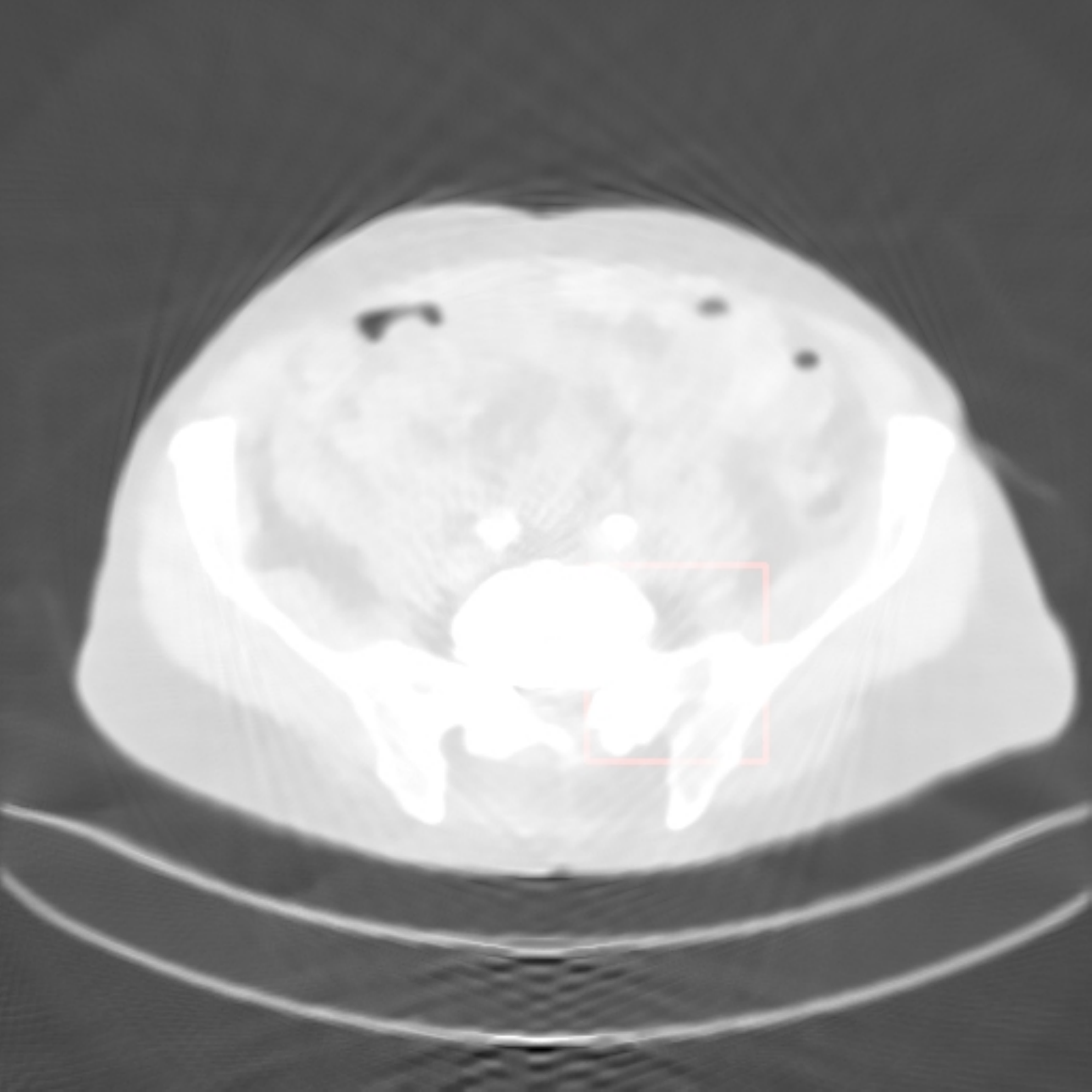}	
				\put(0,0){\color{red}%
					\frame{\includegraphics[scale=0.08]{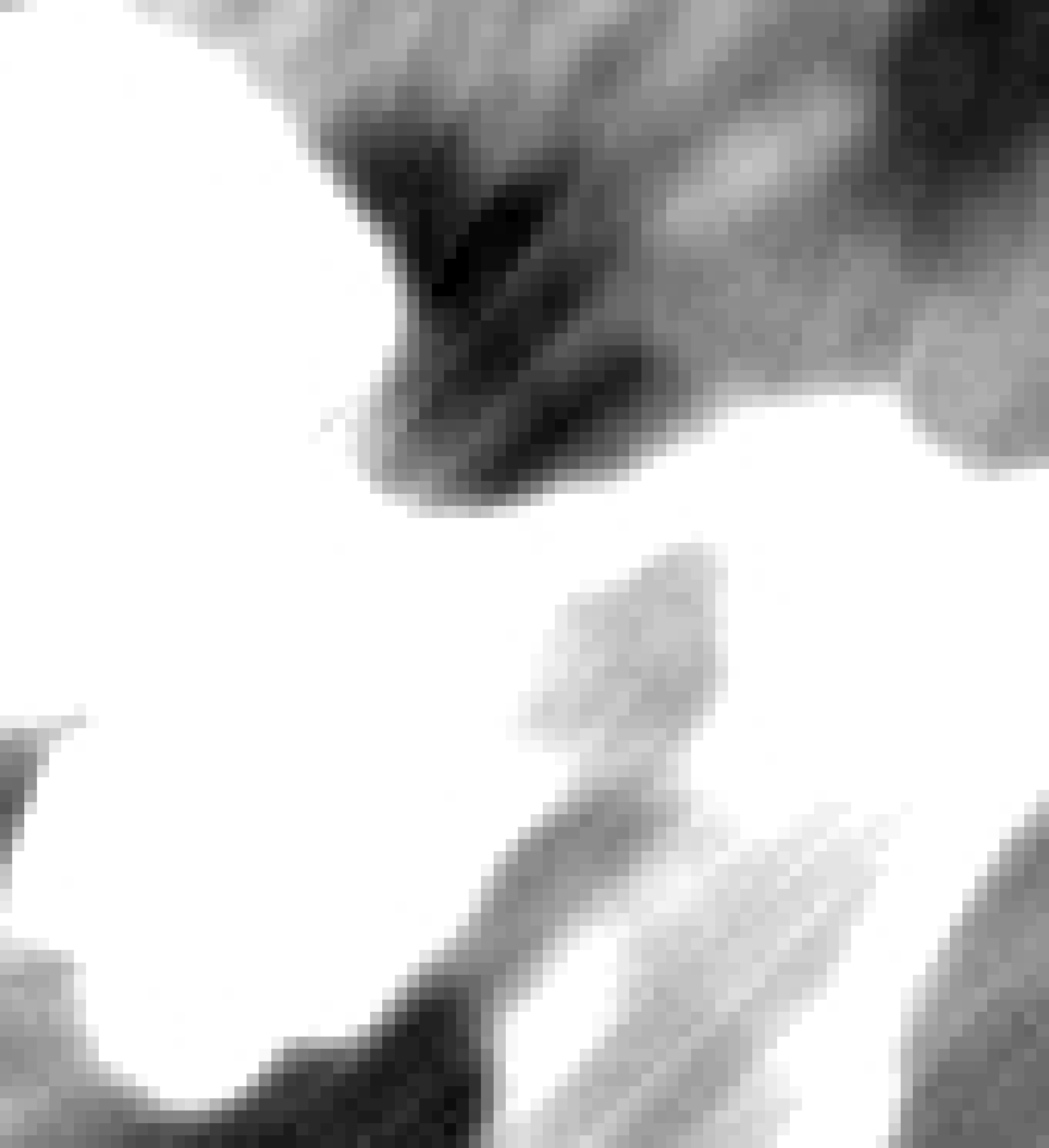}}
				}		
		\end{overpic}}
		\subfloat{
			\begin{overpic}[width=\size\columnwidth]{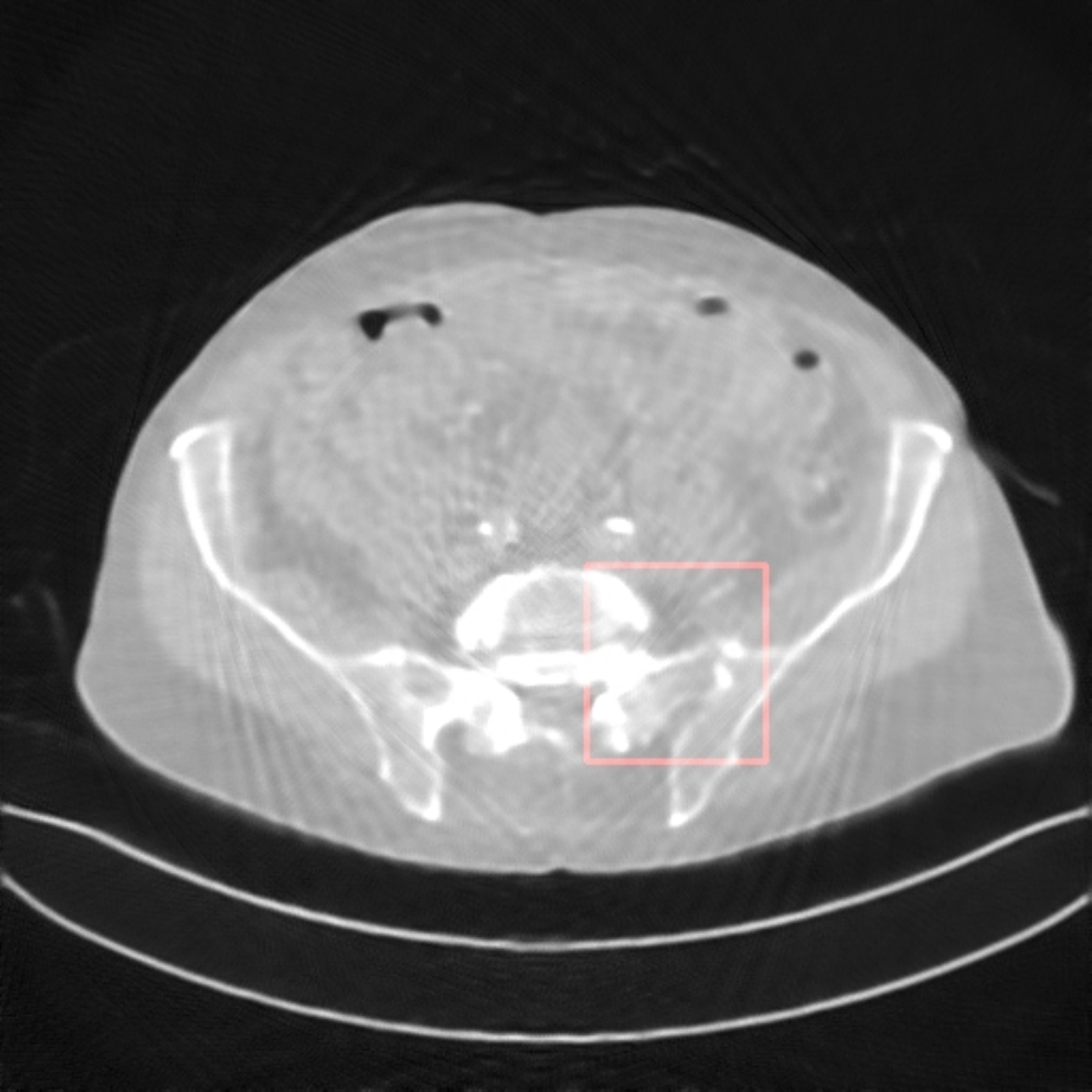}	
				\put(0,0){\color{red}%
					\frame{\includegraphics[scale=0.08]{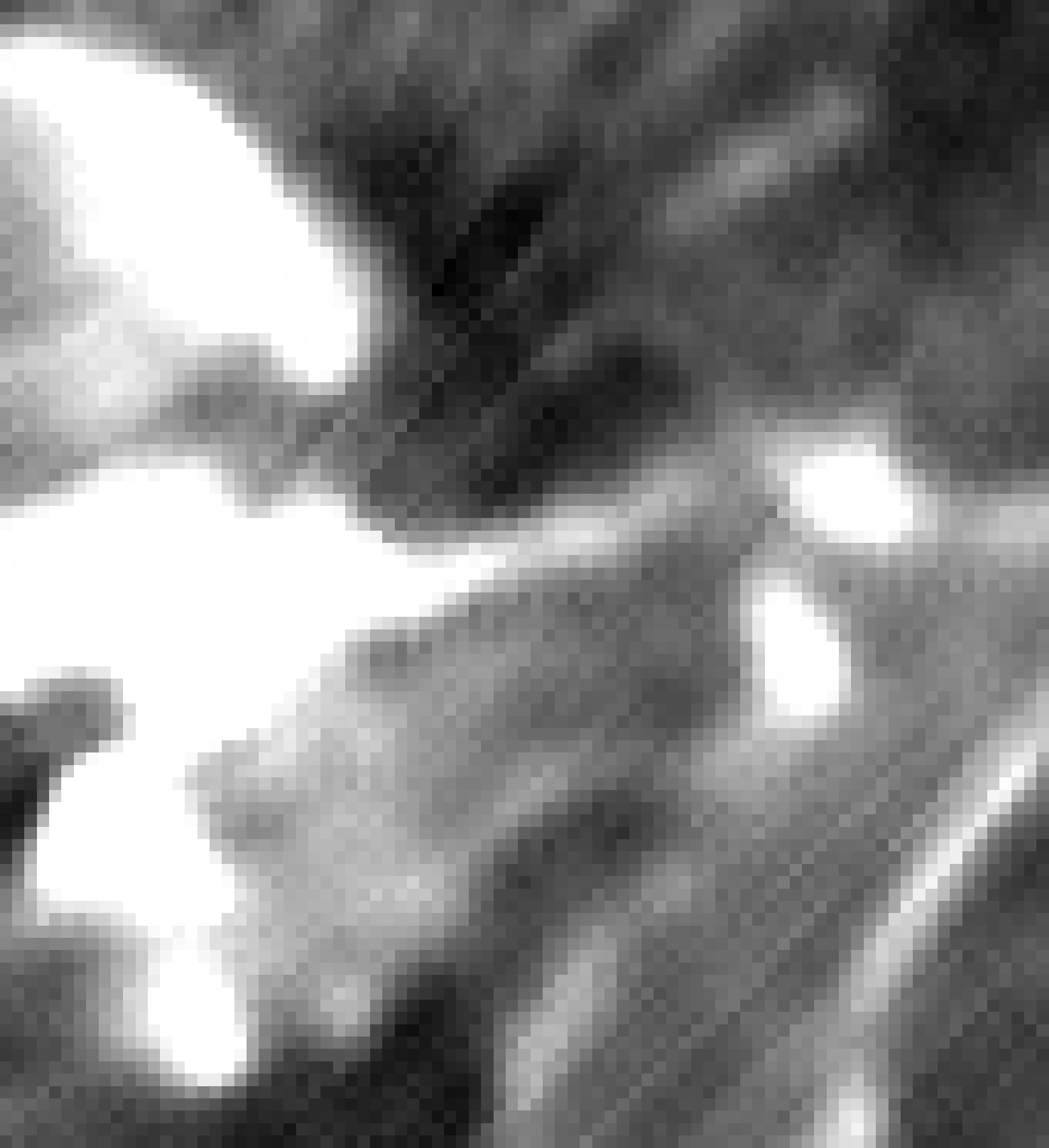}}
				}		
		\end{overpic}}
		\subfloat{
			\begin{overpic}[width=\size\columnwidth]{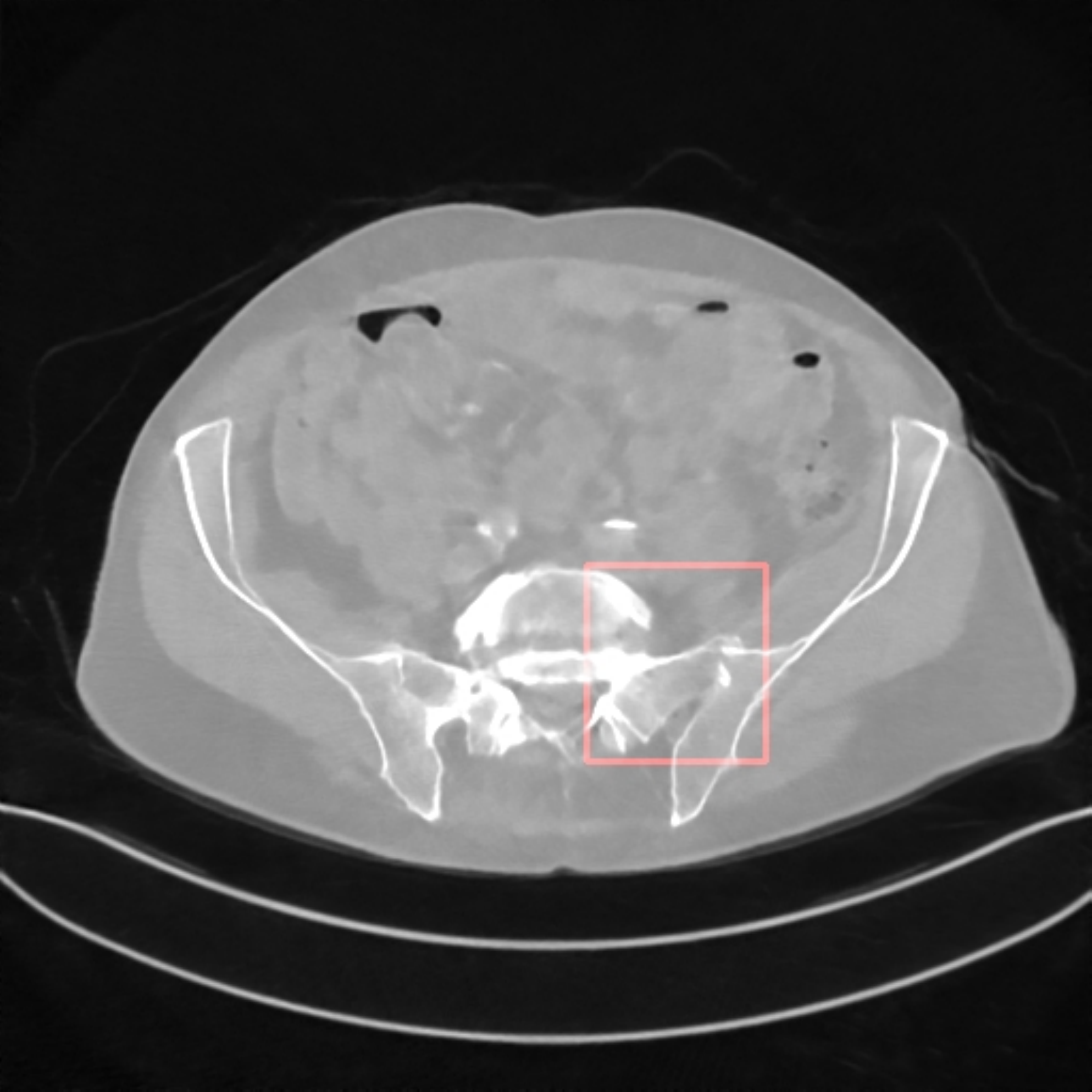}	
				\put(0,0){\color{red}%
					\frame{\includegraphics[scale=0.08]{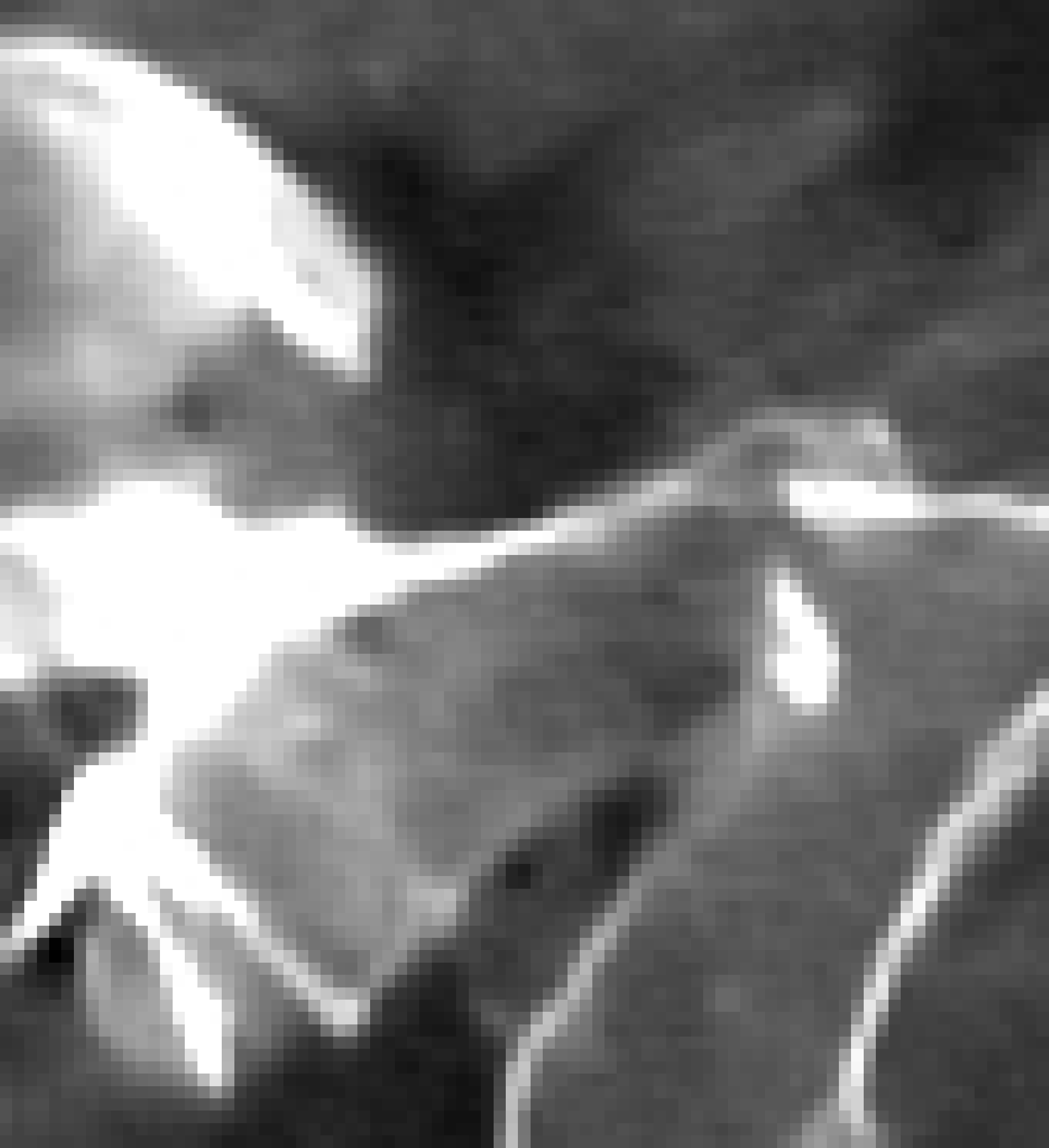}}
				}		
		\end{overpic}}
		\vspace{-3mm}
		\setcounter{subfigure}{0}
		\subfloat[Label]{
			\begin{overpic}[width=\size\columnwidth,percent]{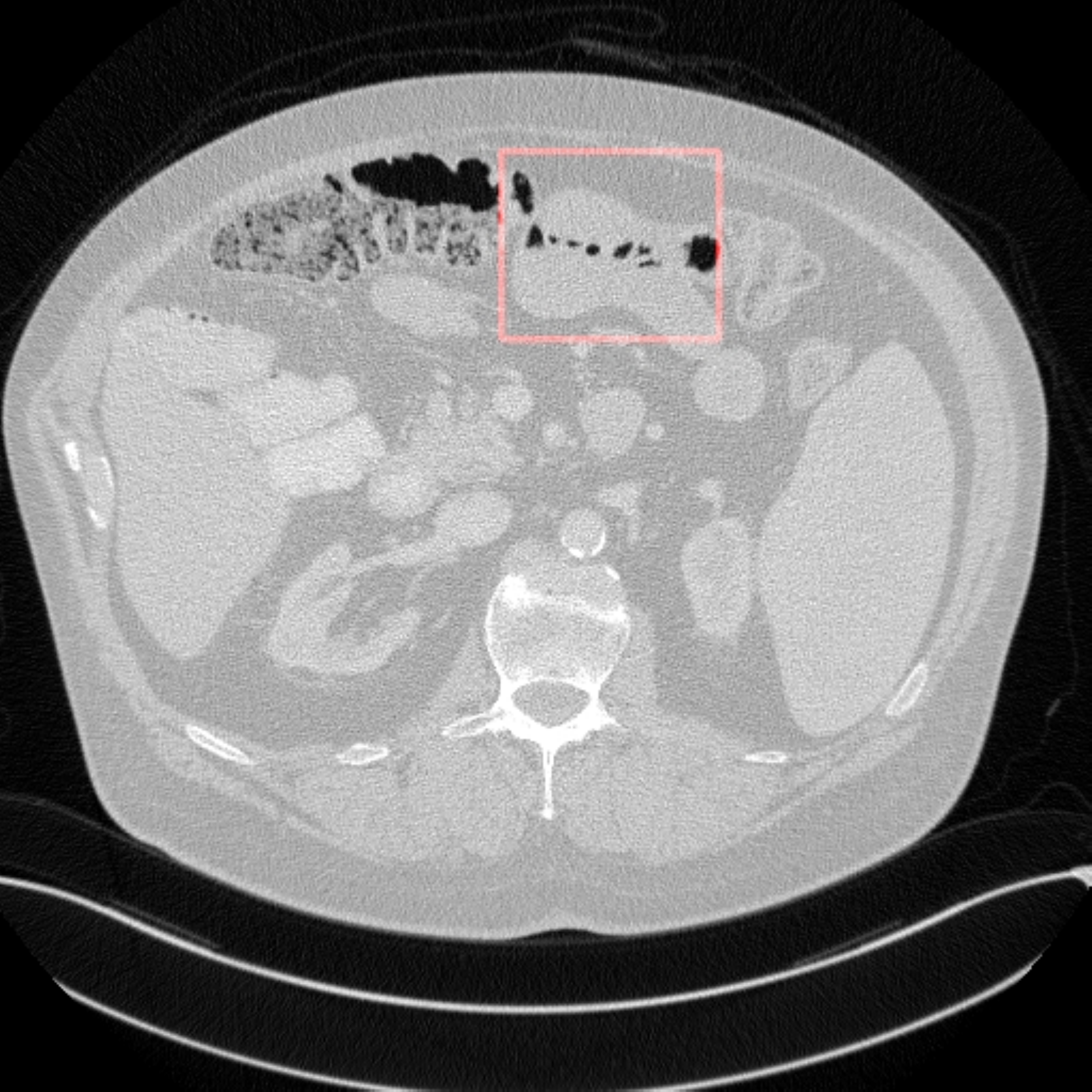}	
				\put(0,0){\color{red}%
					\frame{\includegraphics[scale=0.08]{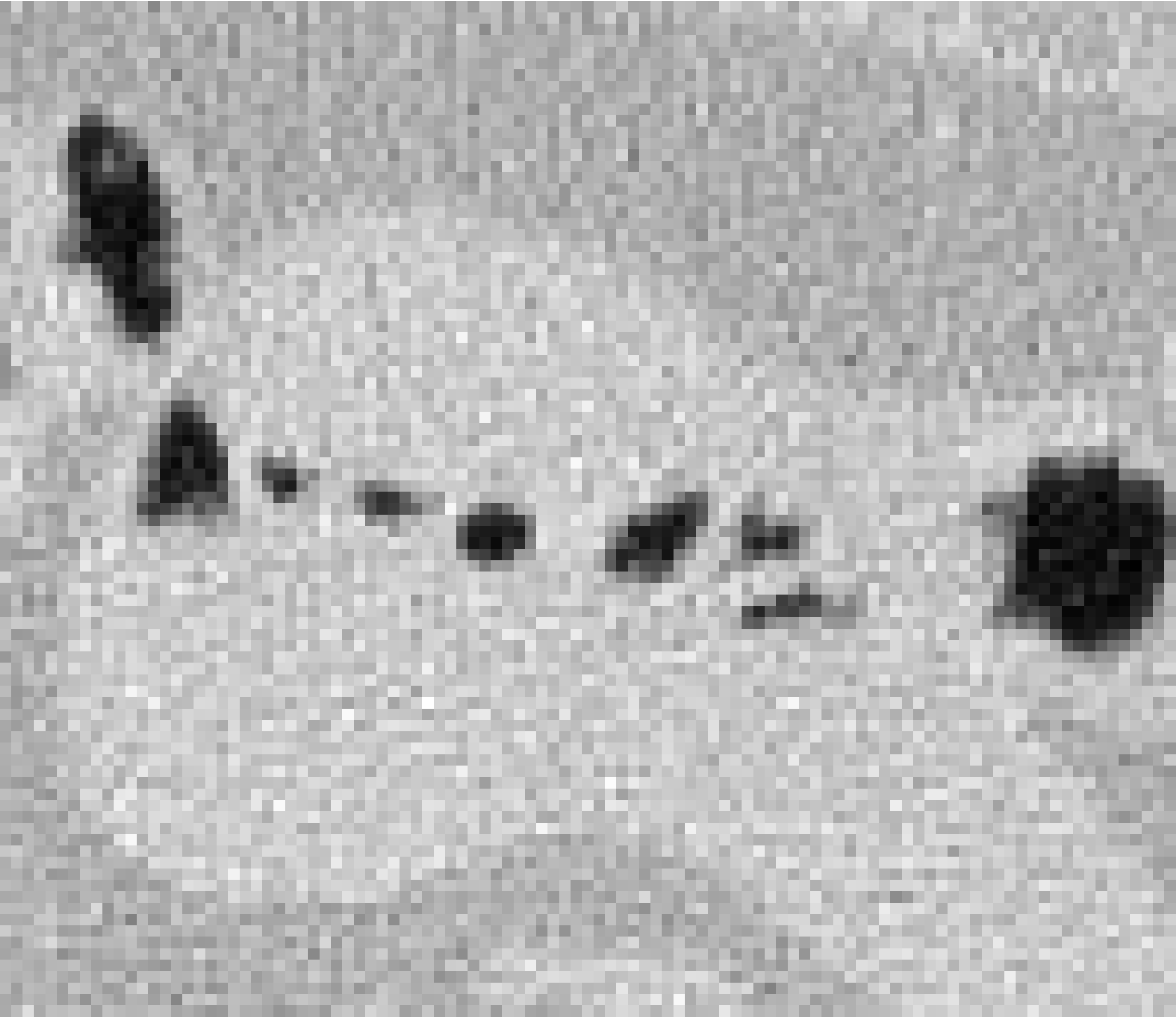}}
				}		
		\end{overpic}}
		\subfloat[Katsevich algorithm]{
			\begin{overpic}[width=\size\columnwidth]{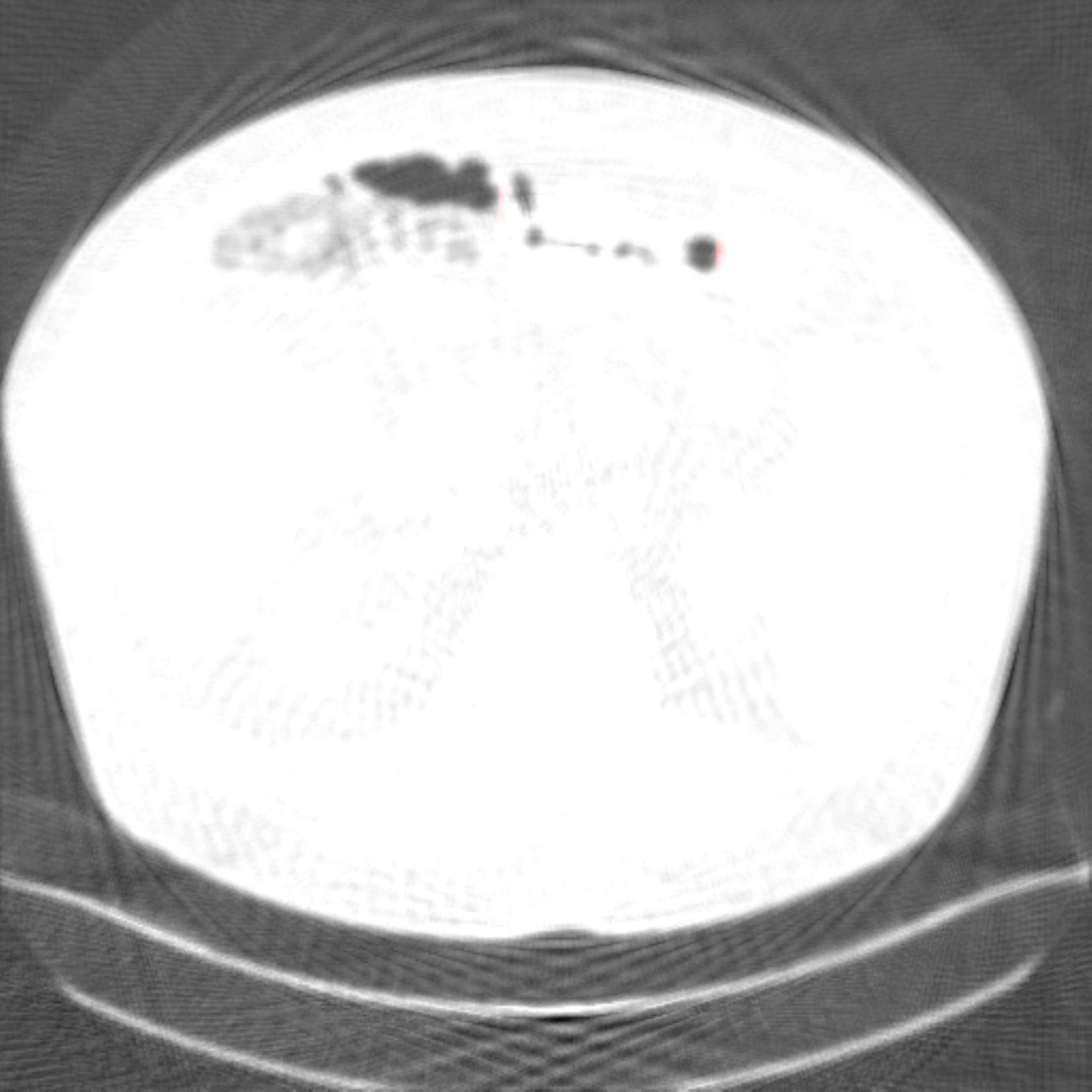}	
				\put(0,0){\color{red}%
					\frame{\includegraphics[scale=0.08]{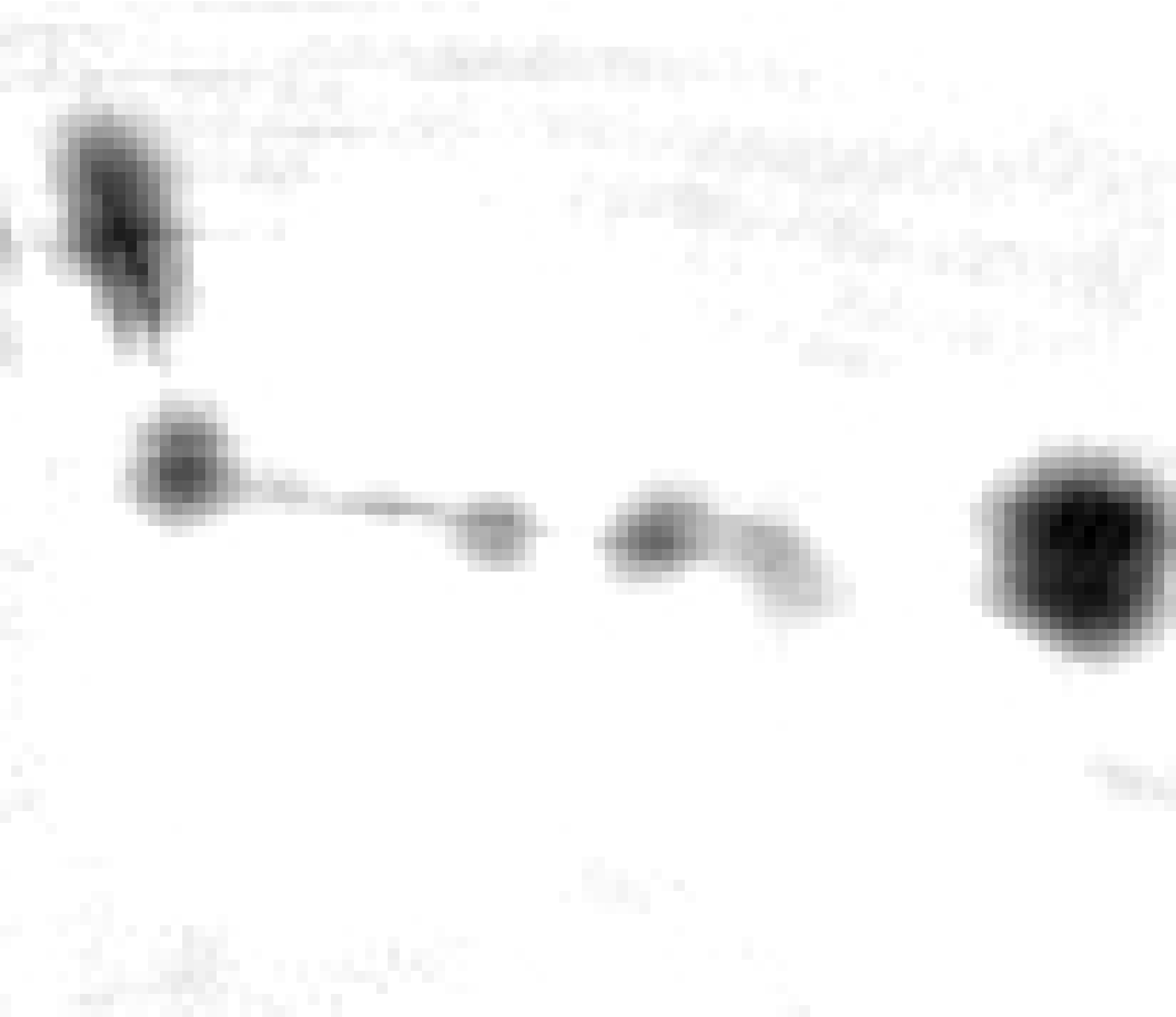}}
				}		
		\end{overpic}}
		\subfloat[Nonlocal-TV]{
			\begin{overpic}[width=\size\columnwidth]{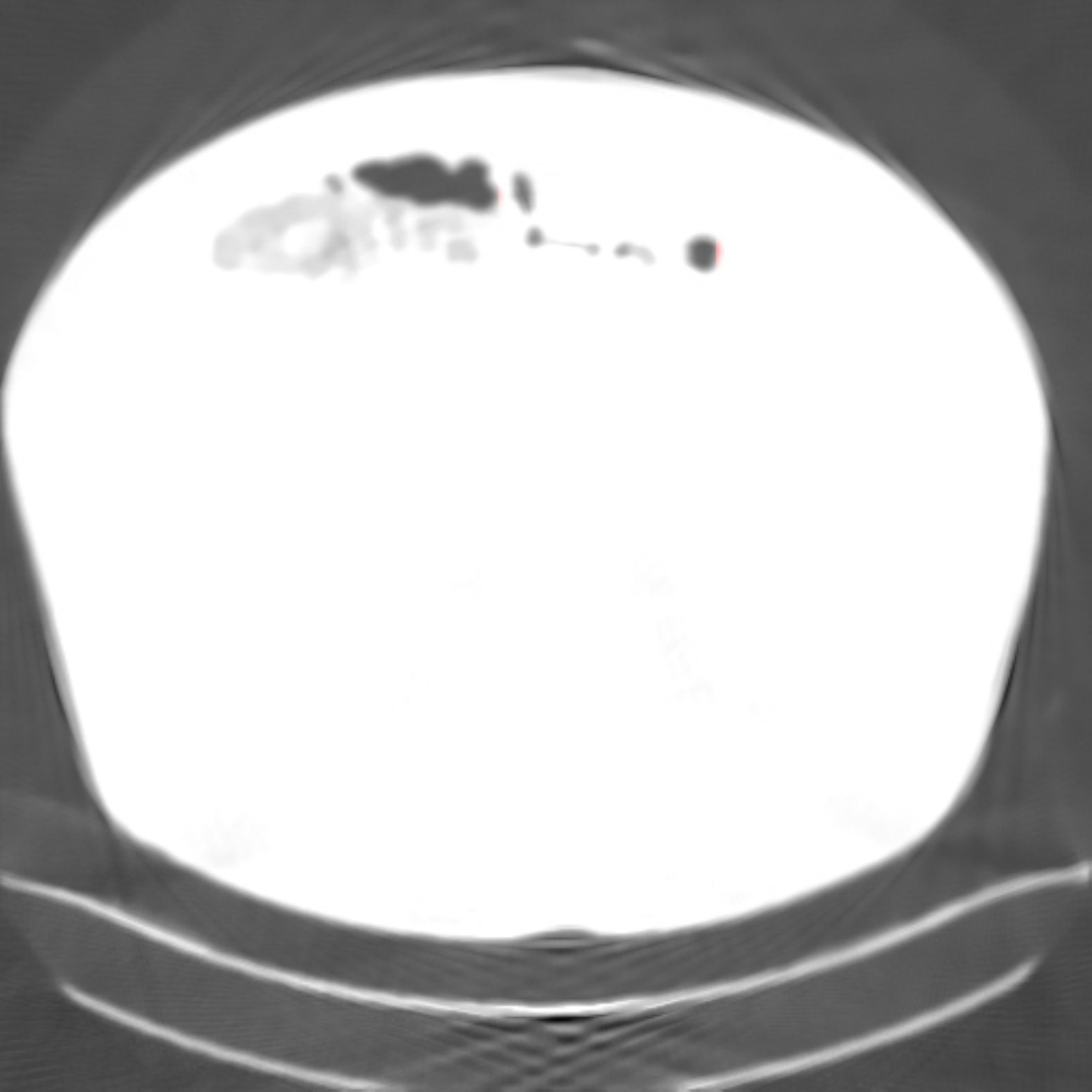}	
				\put(0,0){\color{red}%
					\frame{\includegraphics[scale=0.08]{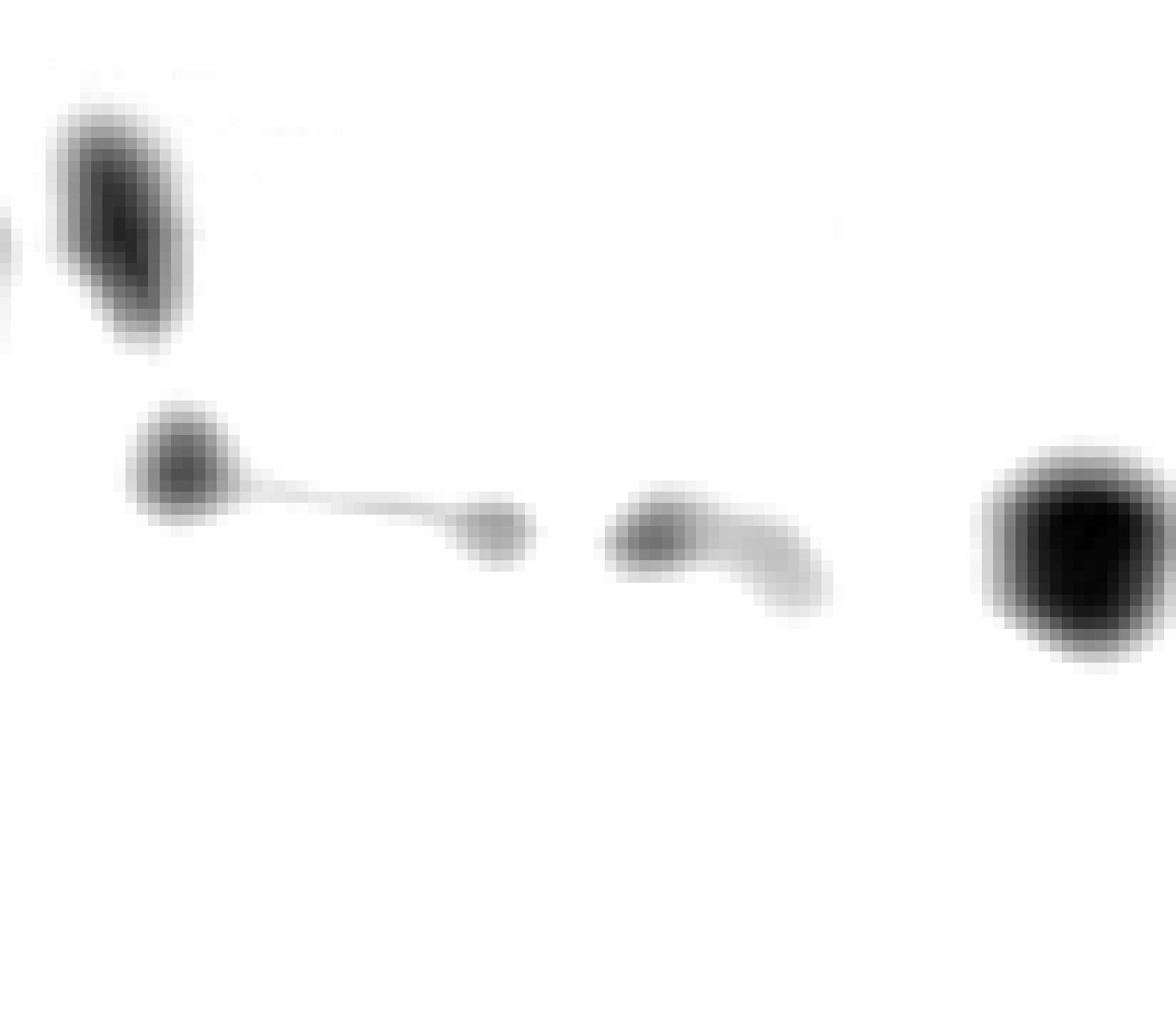}}
				}		
		\end{overpic}}
		\subfloat[FBPConvNet]{
			\begin{overpic}[width=\size\columnwidth]{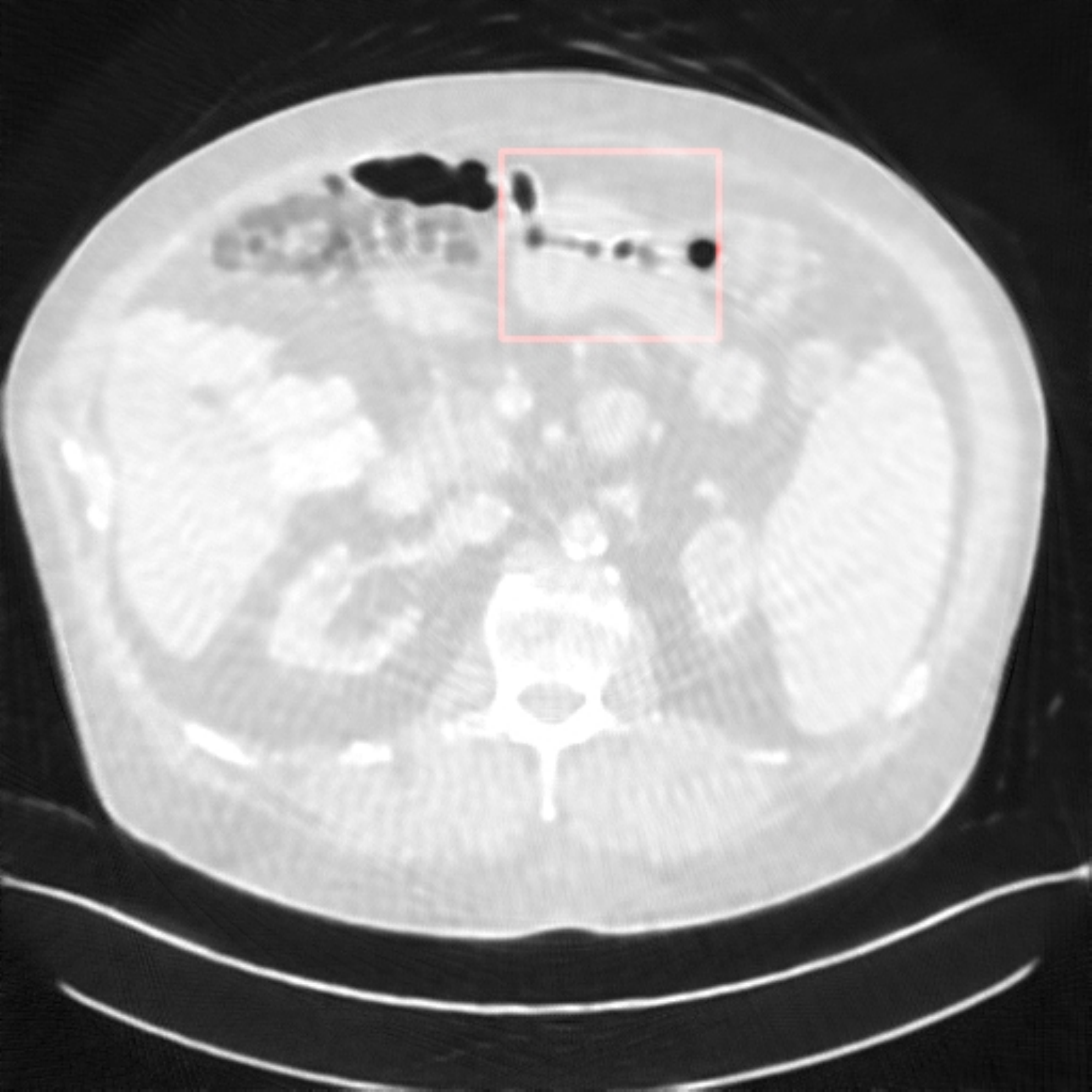}	
				\put(0,0){\color{red}%
					\frame{\includegraphics[scale=0.08]{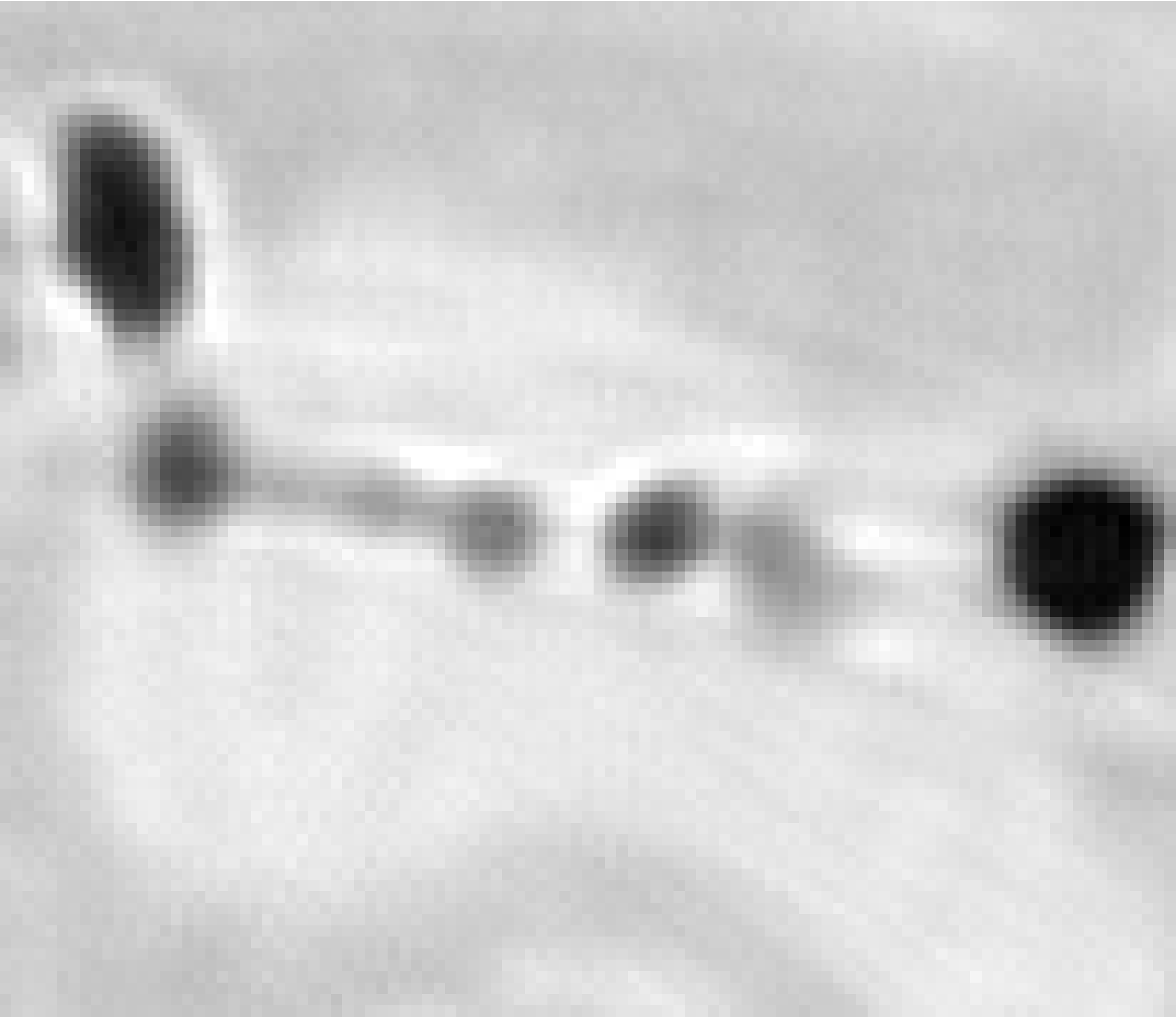}}
				}		
		\end{overpic}}
		\subfloat[Ours]{
			\begin{overpic}[width=\size\columnwidth]{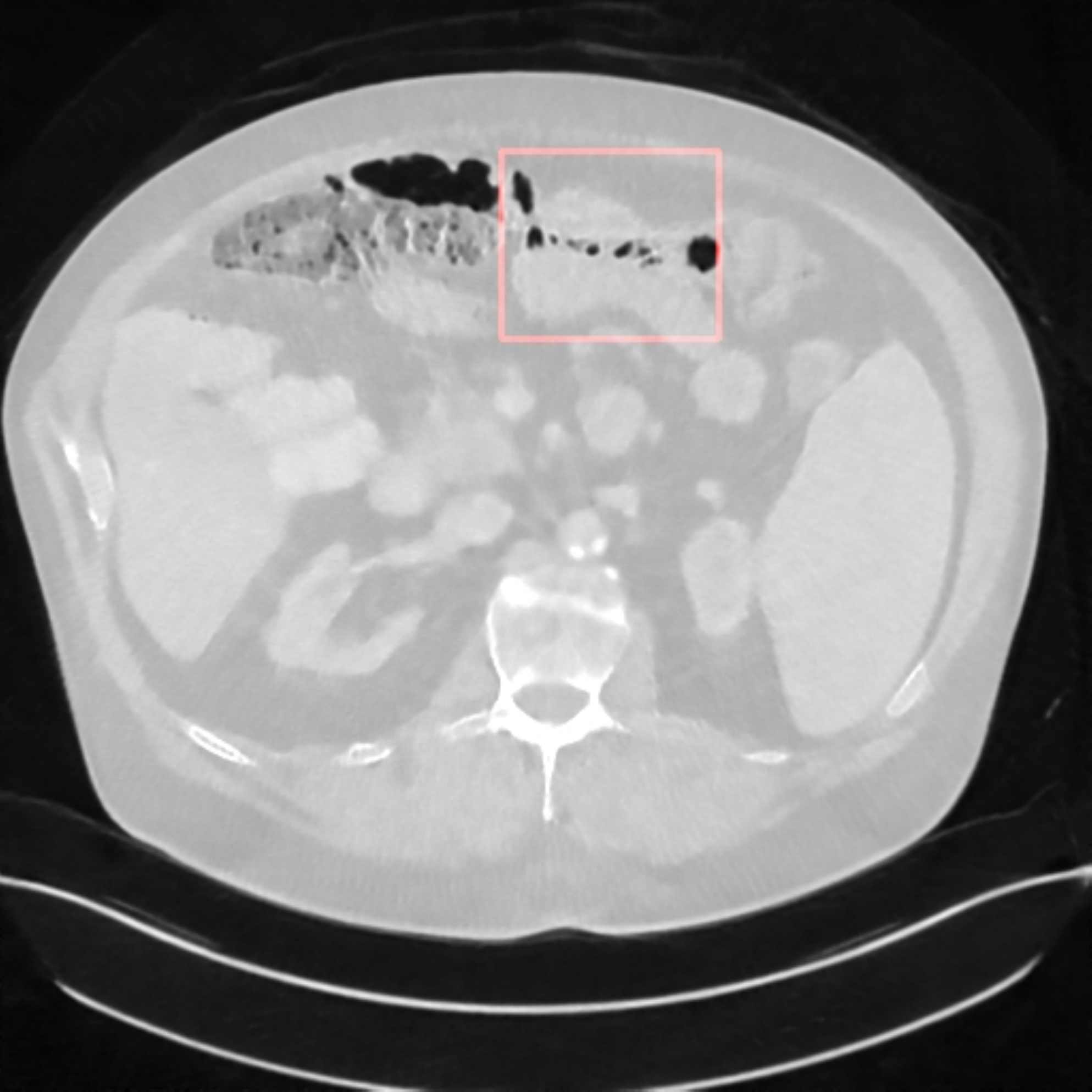}	
				\put(0,0){\color{red}%
					\frame{\includegraphics[scale=0.08]{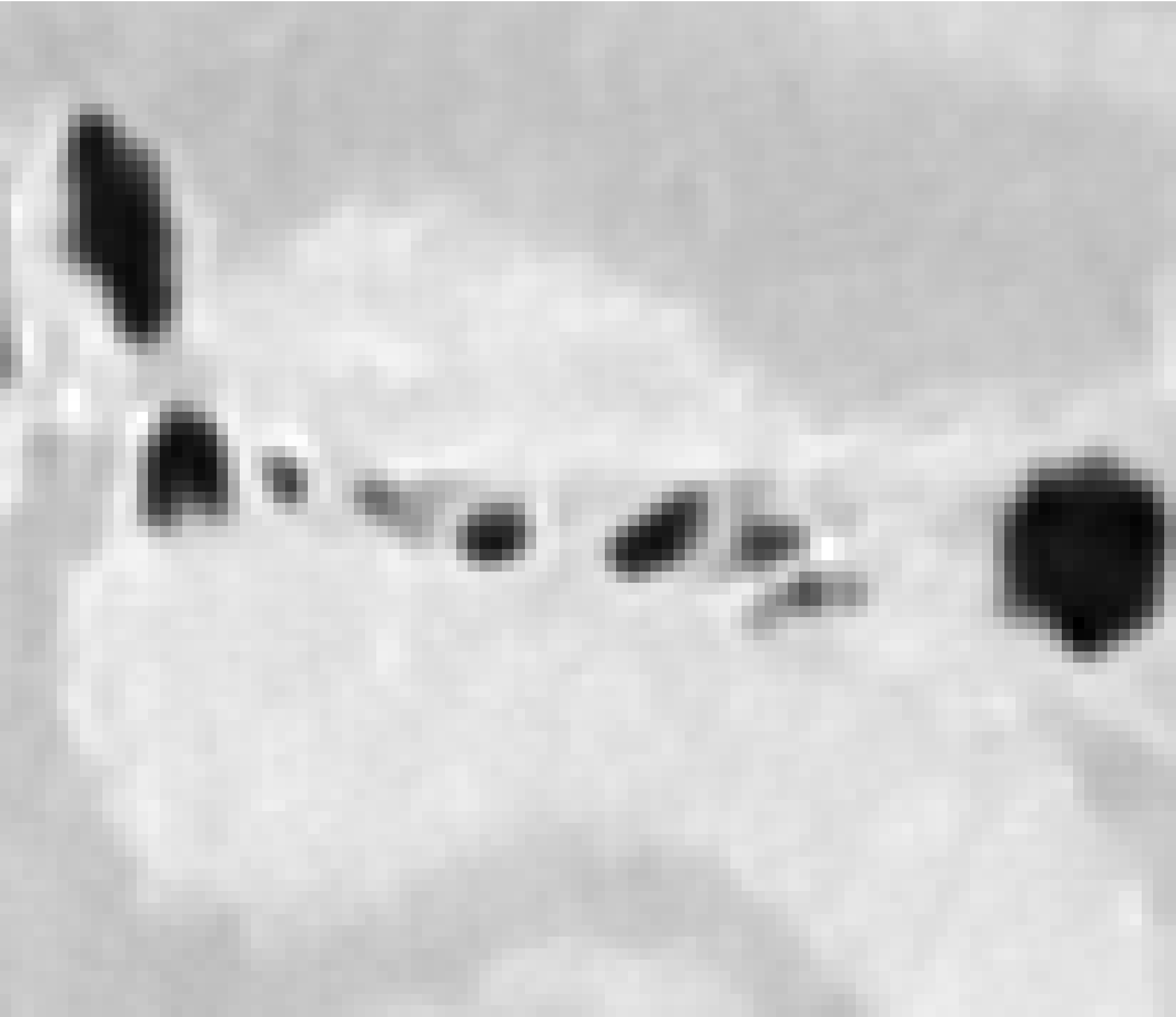}}
				}		
			\end{overpic}
			\label{Fig2_our}}
		
	}
	\caption{The reconstructed results of NIH-AAPM-Mayo data for $p=7\pi$. All images are linearly stretched to [0, 1] and the display window is [0, 0.6].}
	\label{Fig2}
\end{figure*}

\begin{figure*}[htbp]	
	\centering{	
		\newcommand{\id}{60}
		\subfloat{
			\begin{overpic}[width=\size\columnwidth,percent]{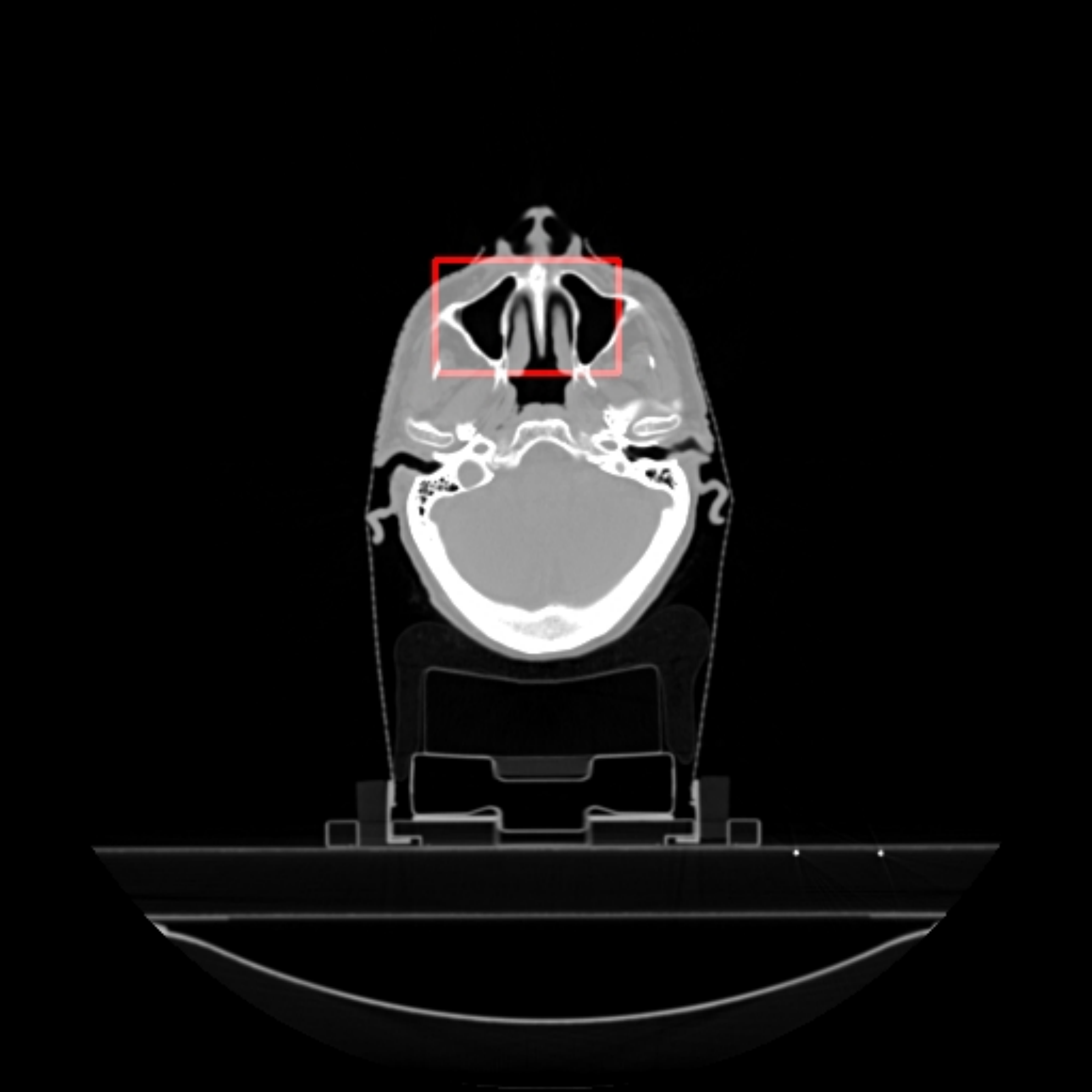}	
				\put(0,0){\color{red}%
					\frame{\includegraphics[scale=0.08]{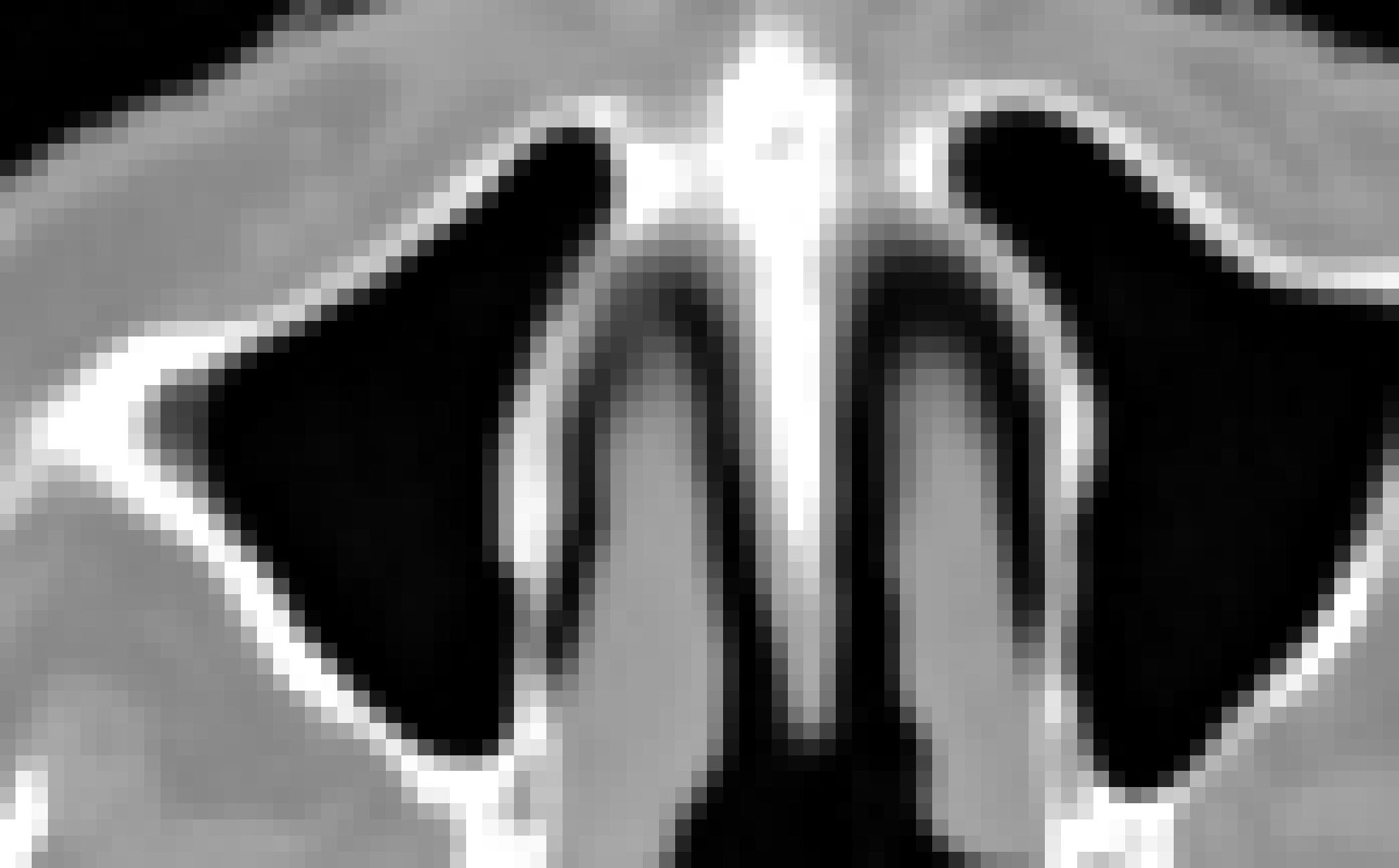}}
				}		
		\end{overpic}}
		\subfloat{
			\begin{overpic}[width=\size\columnwidth]{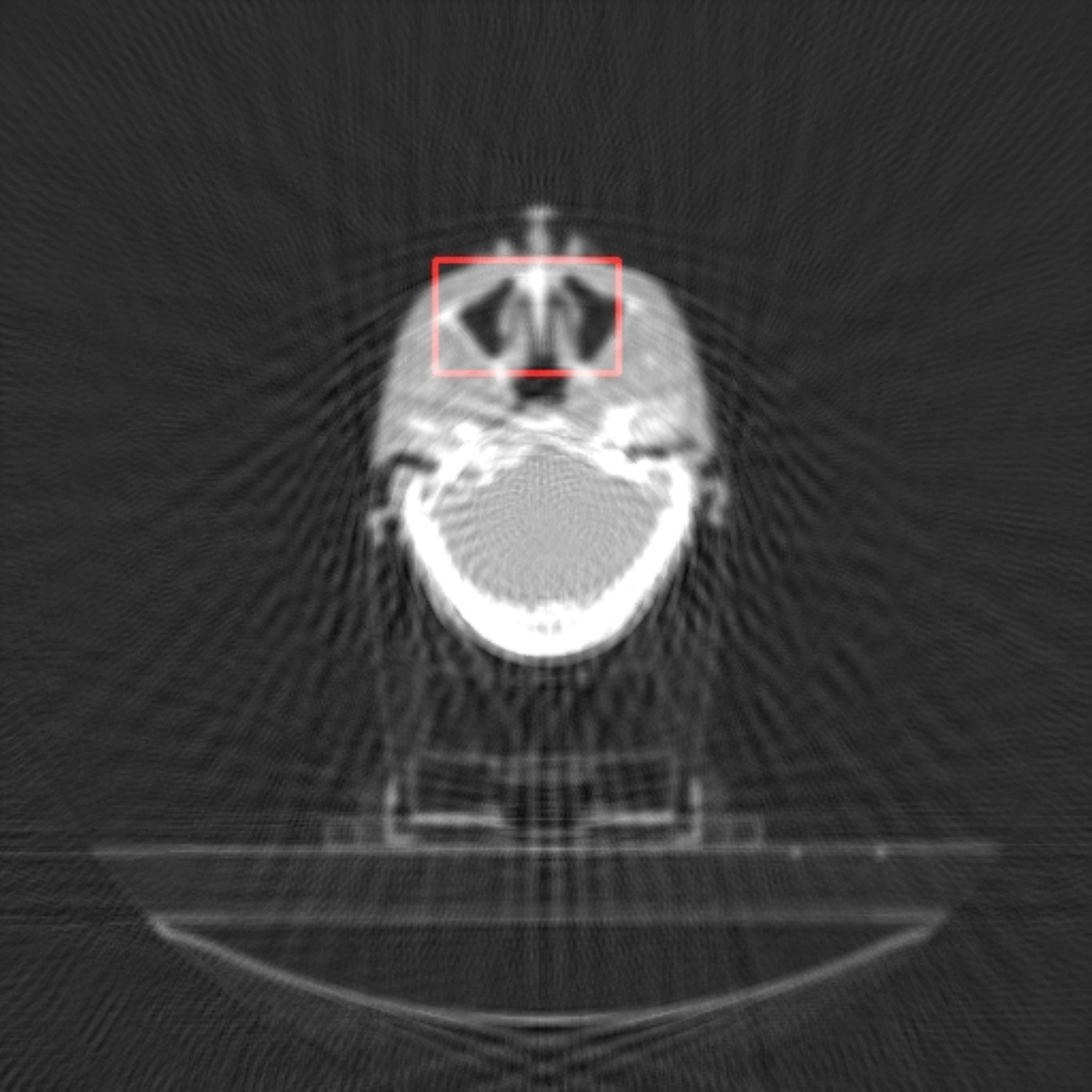}	
				\put(0,0){\color{red}%
					\frame{\includegraphics[scale=0.08]{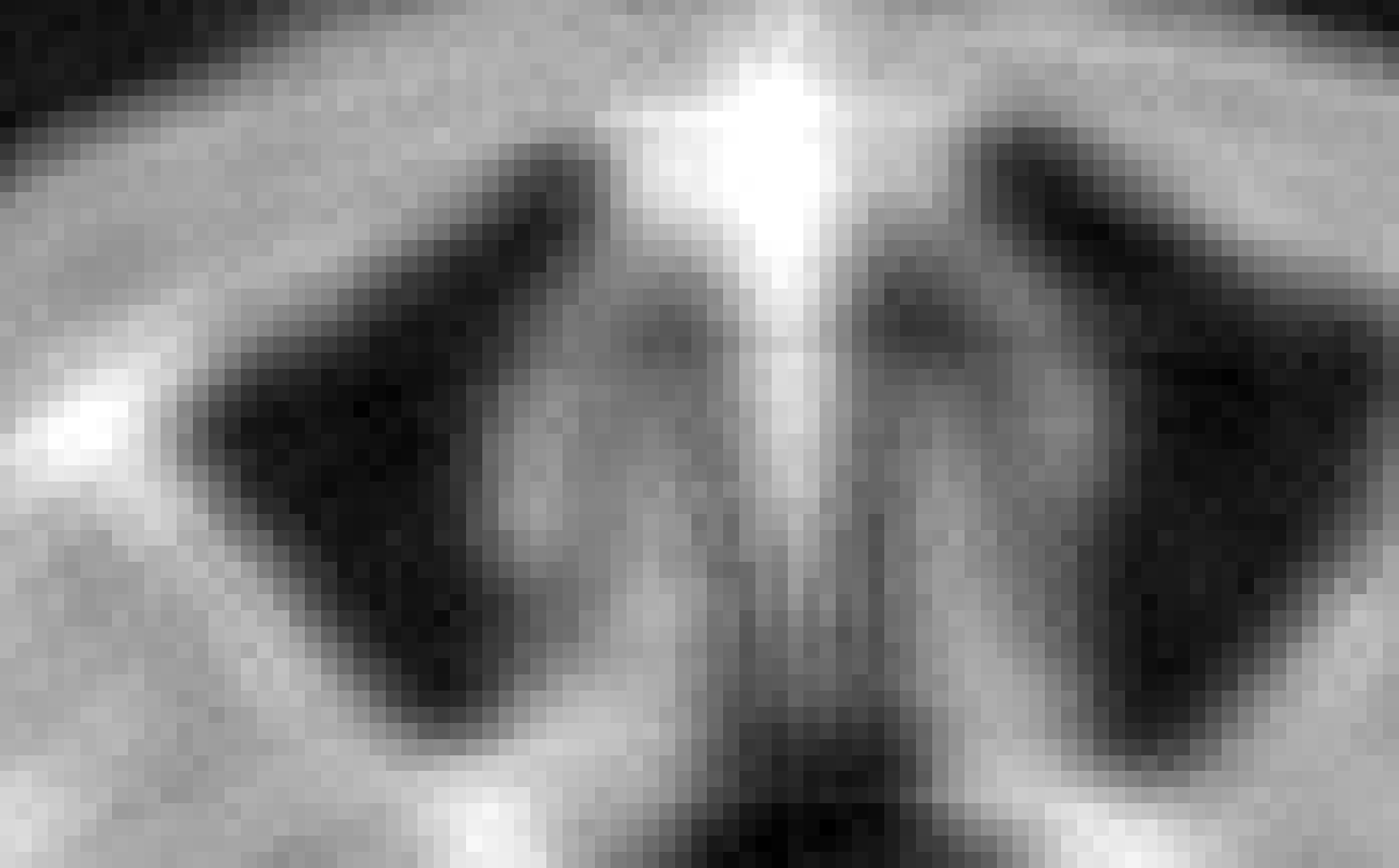}}
				}		
		\end{overpic}}
		\subfloat{
			\begin{overpic}[width=\size\columnwidth]{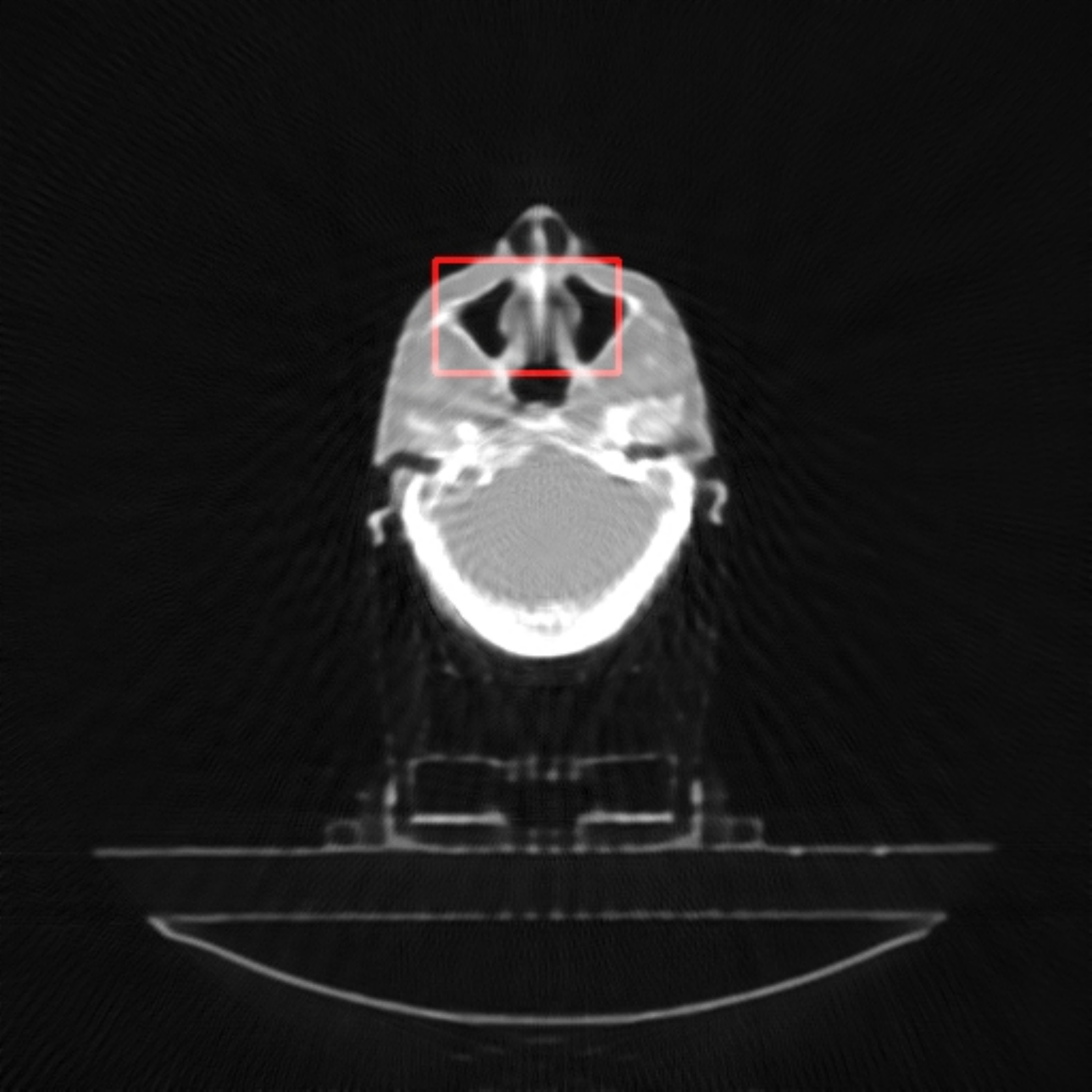}	
				\put(0,0){\color{red}%
					\frame{\includegraphics[scale=0.08]{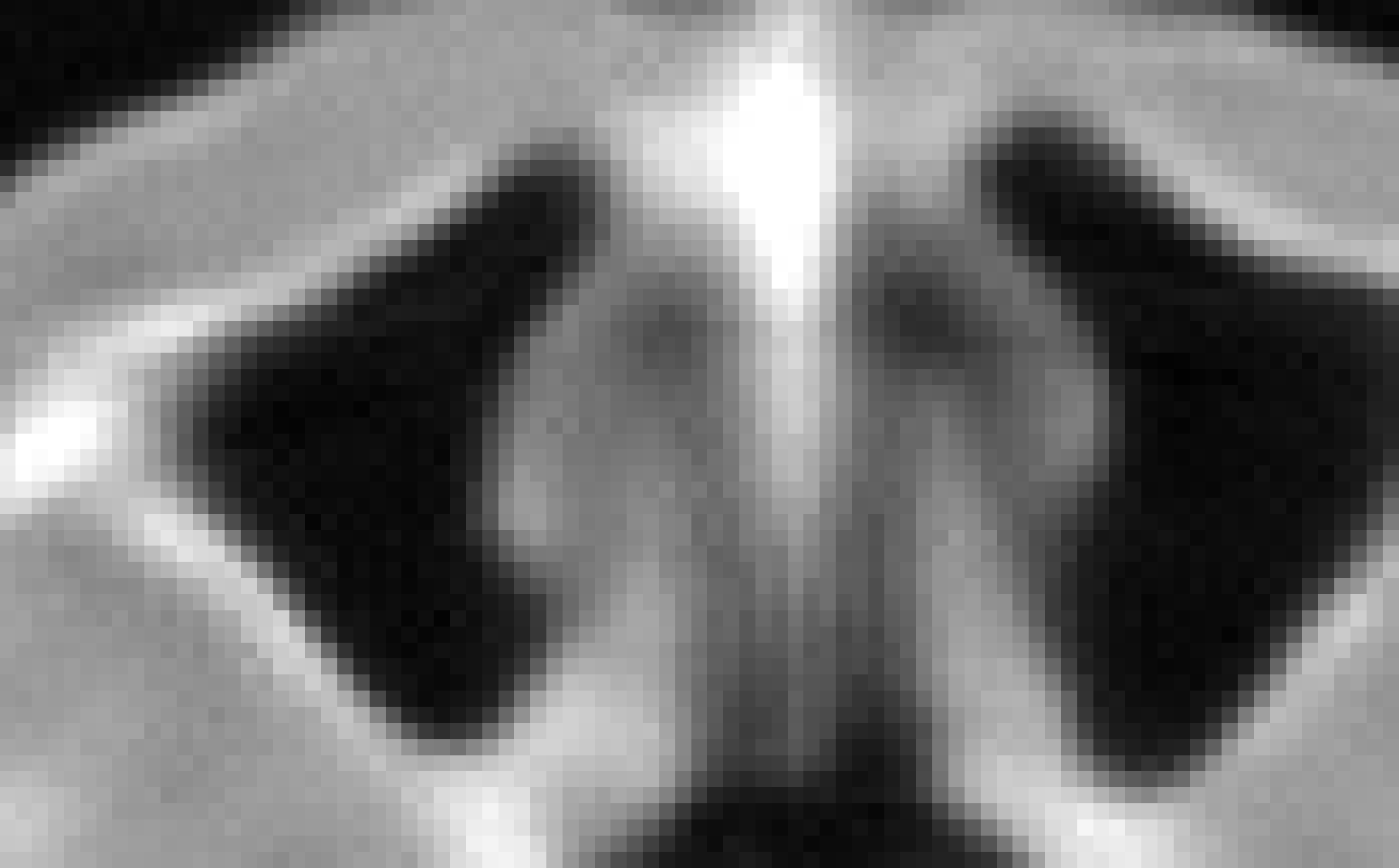}}
				}		
		\end{overpic}}
		\subfloat{
			\begin{overpic}[width=\size\columnwidth]{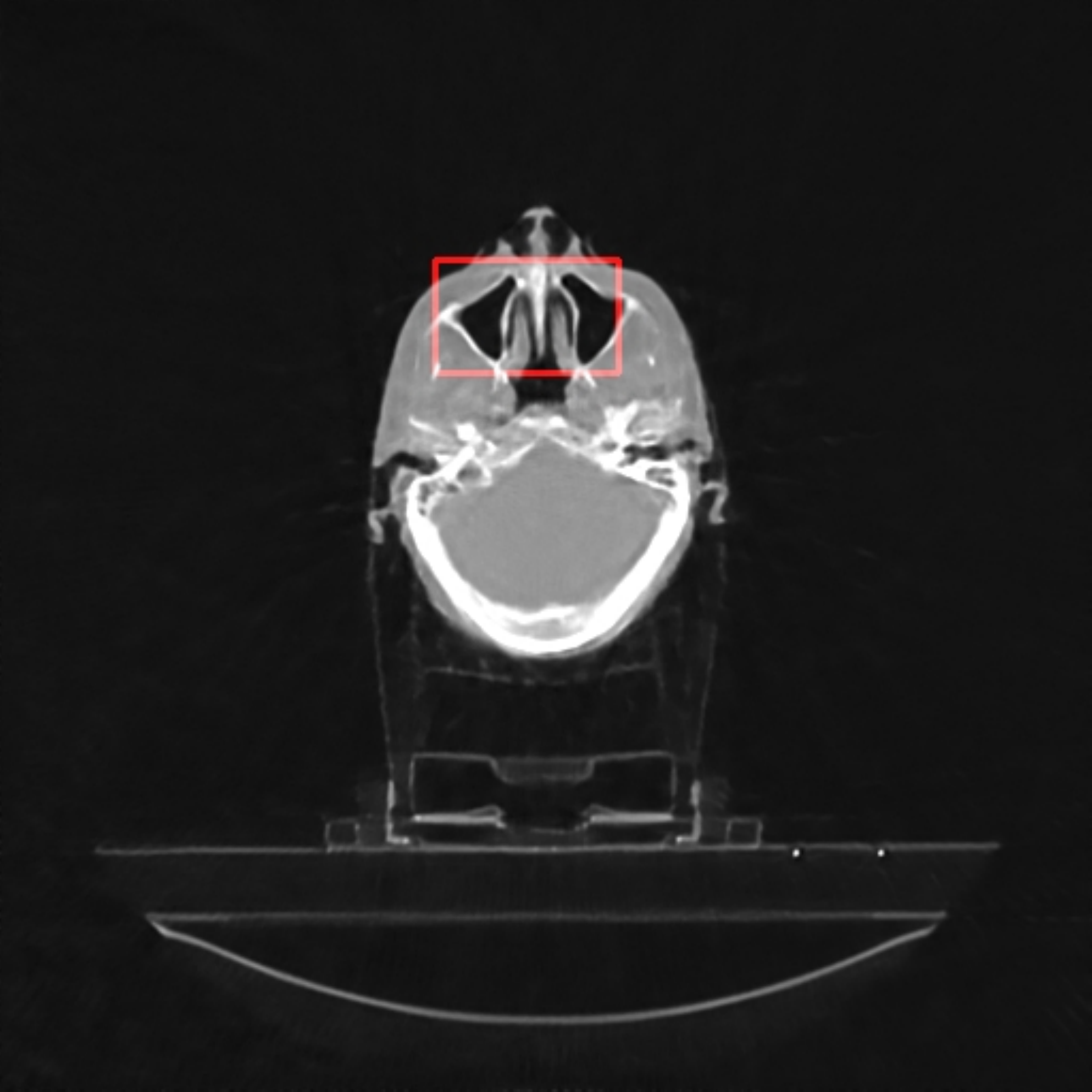}	
				\put(0,0){\color{red}%
					\frame{\includegraphics[scale=0.08]{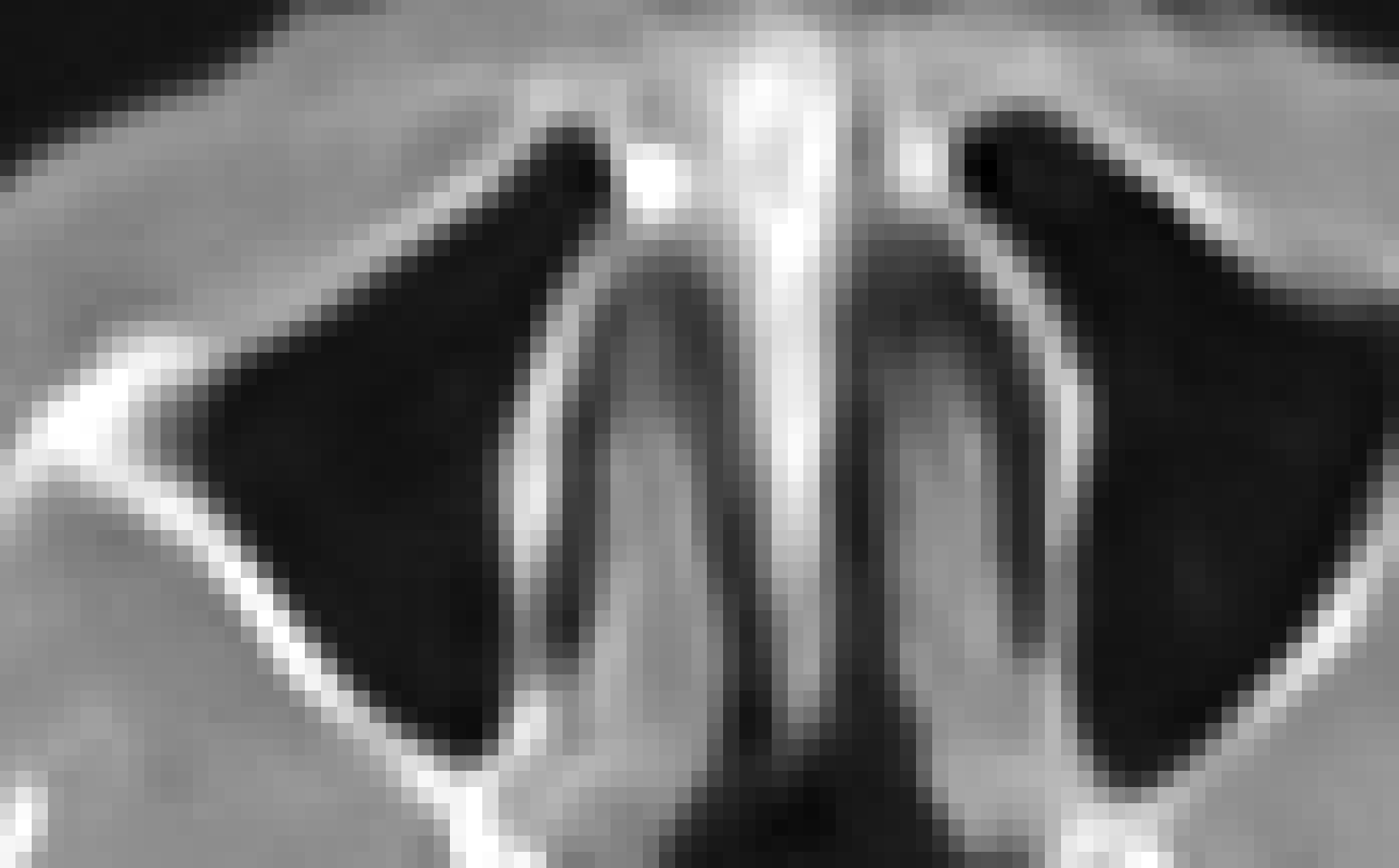}}
				}		
		\end{overpic}}
		
		\vspace{-3mm}
		\renewcommand{\id}{80}
		\subfloat{
			\begin{overpic}[width=\size\columnwidth,percent]{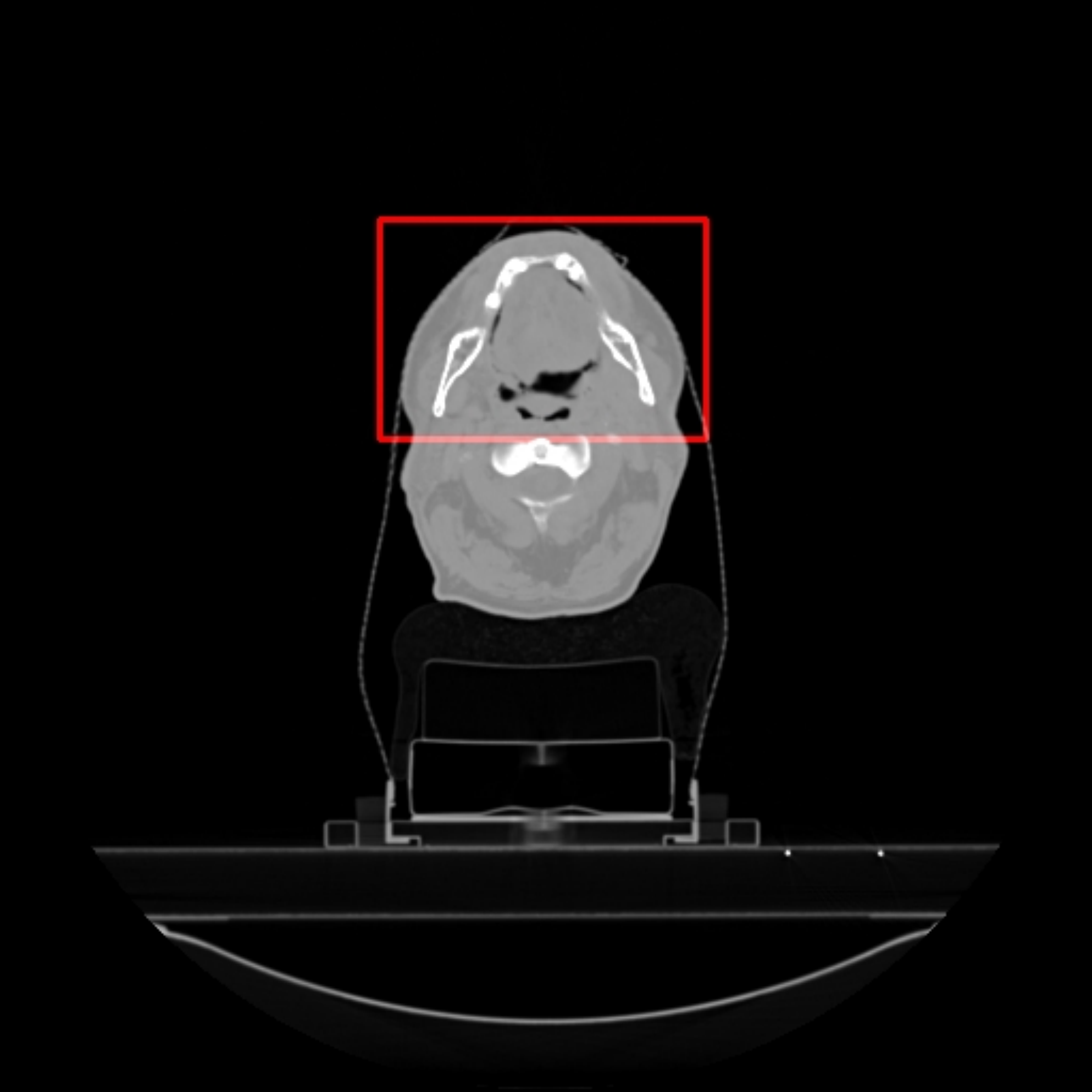}	
				\put(0,0){\color{red}%
					\frame{\includegraphics[scale=0.08]{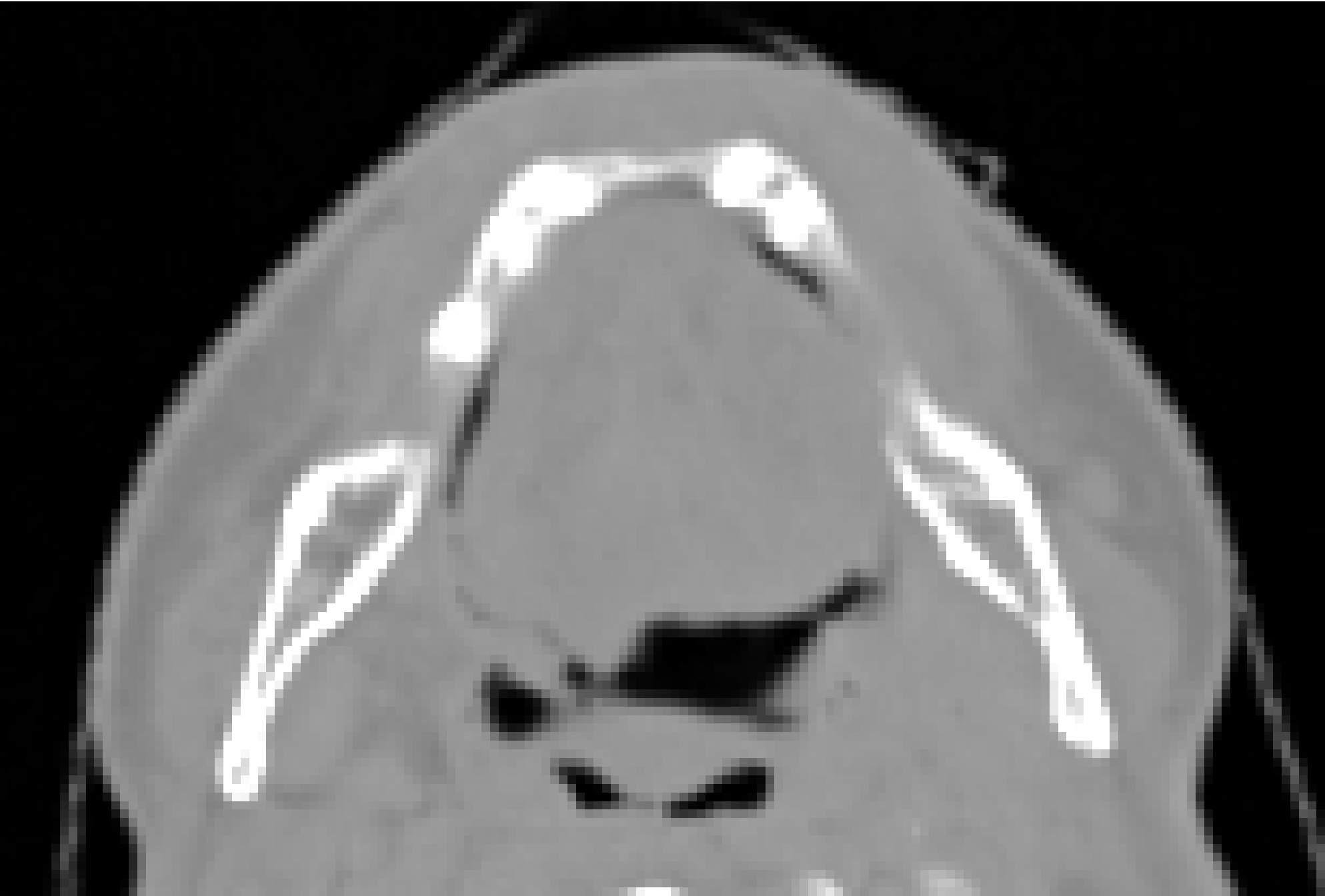}}
				}		
		\end{overpic}}
		\subfloat{
			\begin{overpic}[width=\size\columnwidth]{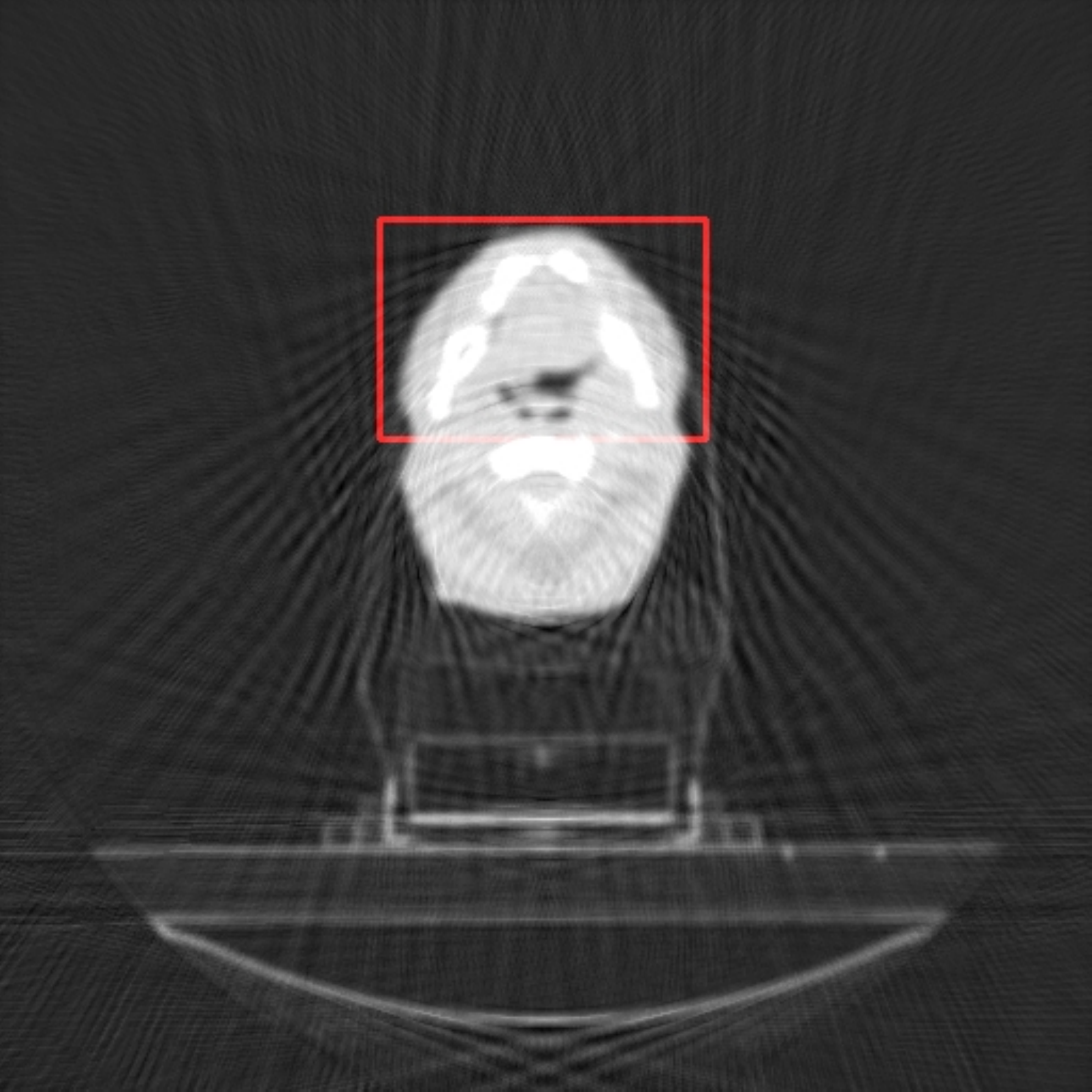}	
				\put(0,0){\color{red}%
					\frame{\includegraphics[scale=0.08]{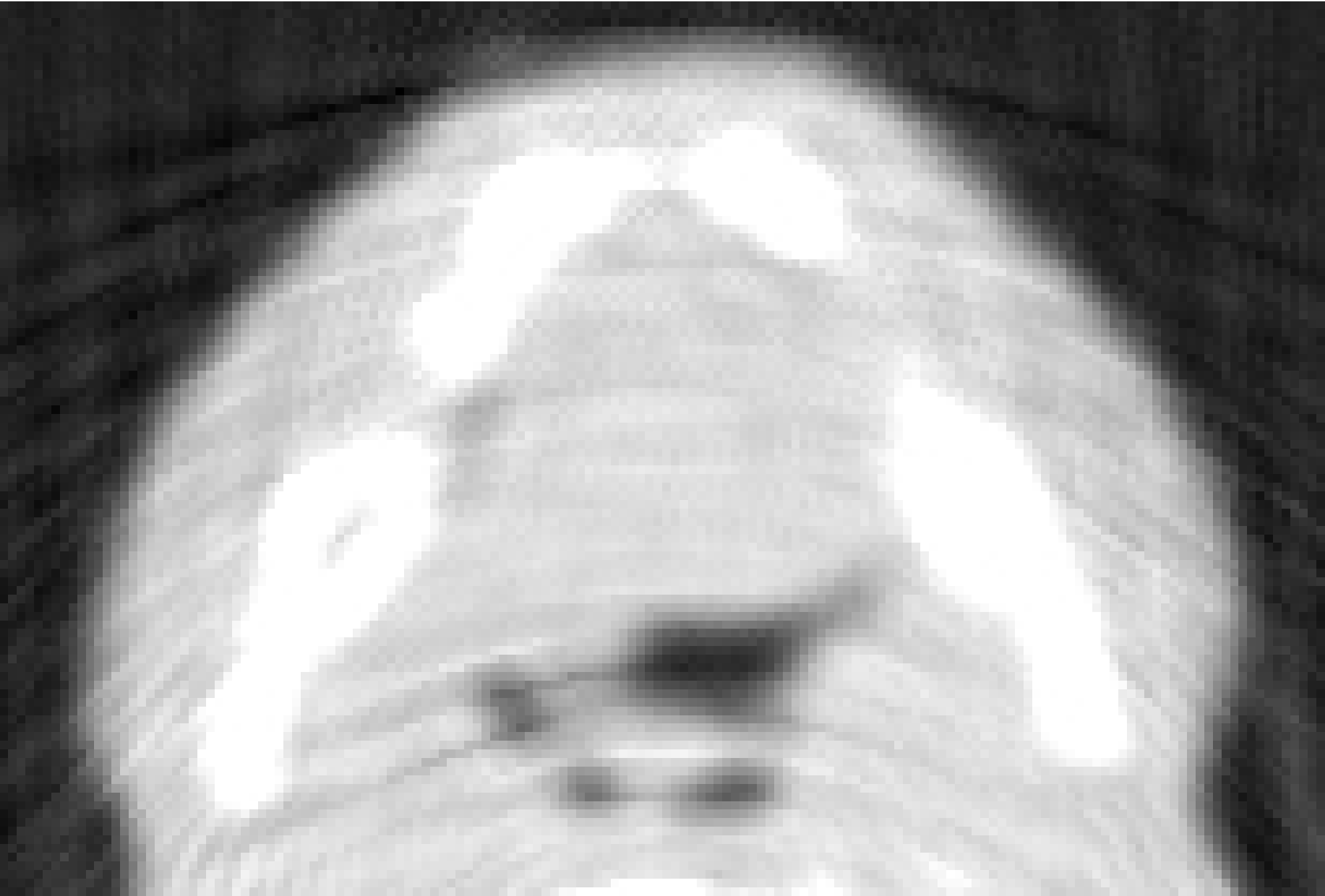}}
				}		
		\end{overpic}}
		\subfloat{
			\begin{overpic}[width=\size\columnwidth]{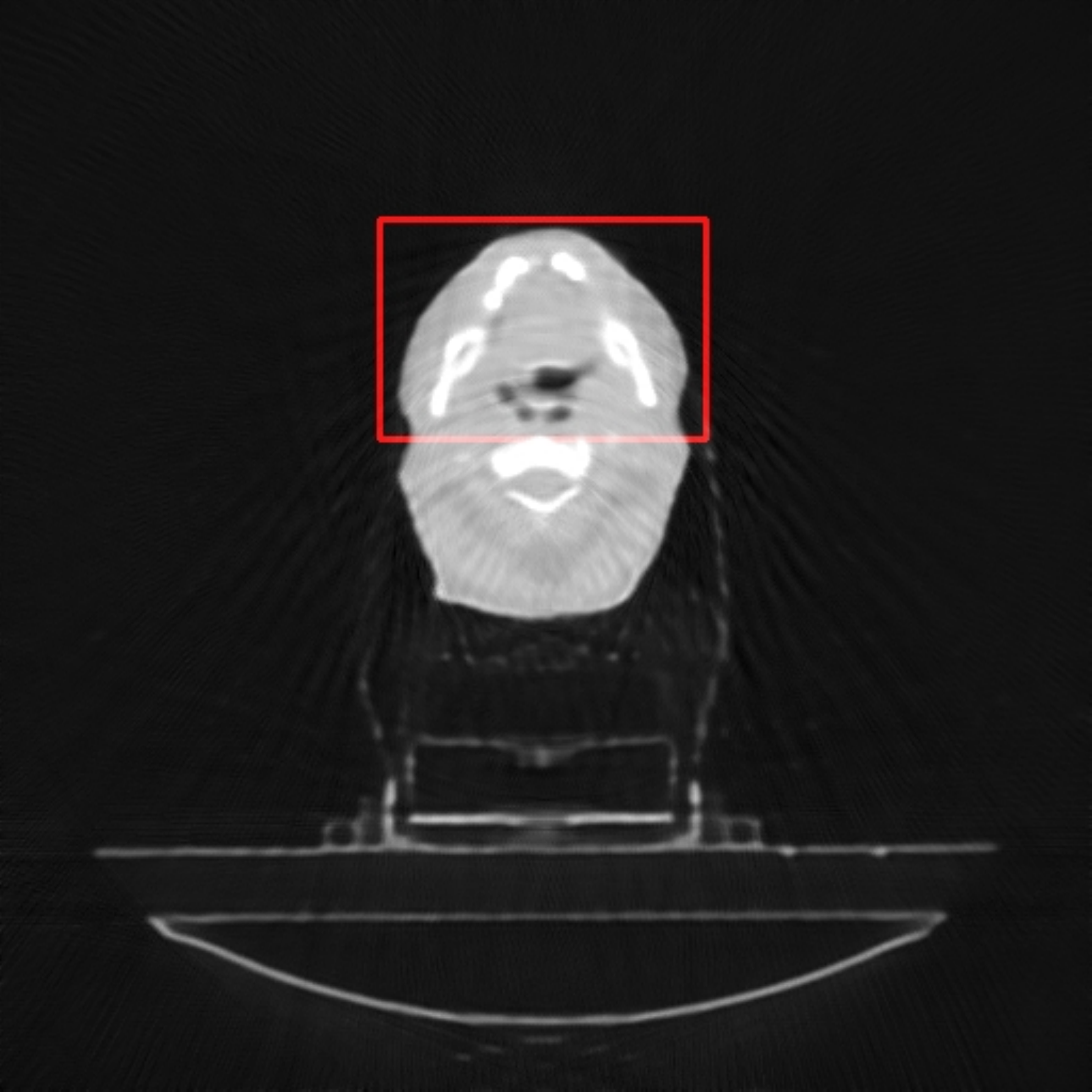}	
				\put(0,0){\color{red}%
					\frame{\includegraphics[scale=0.08]{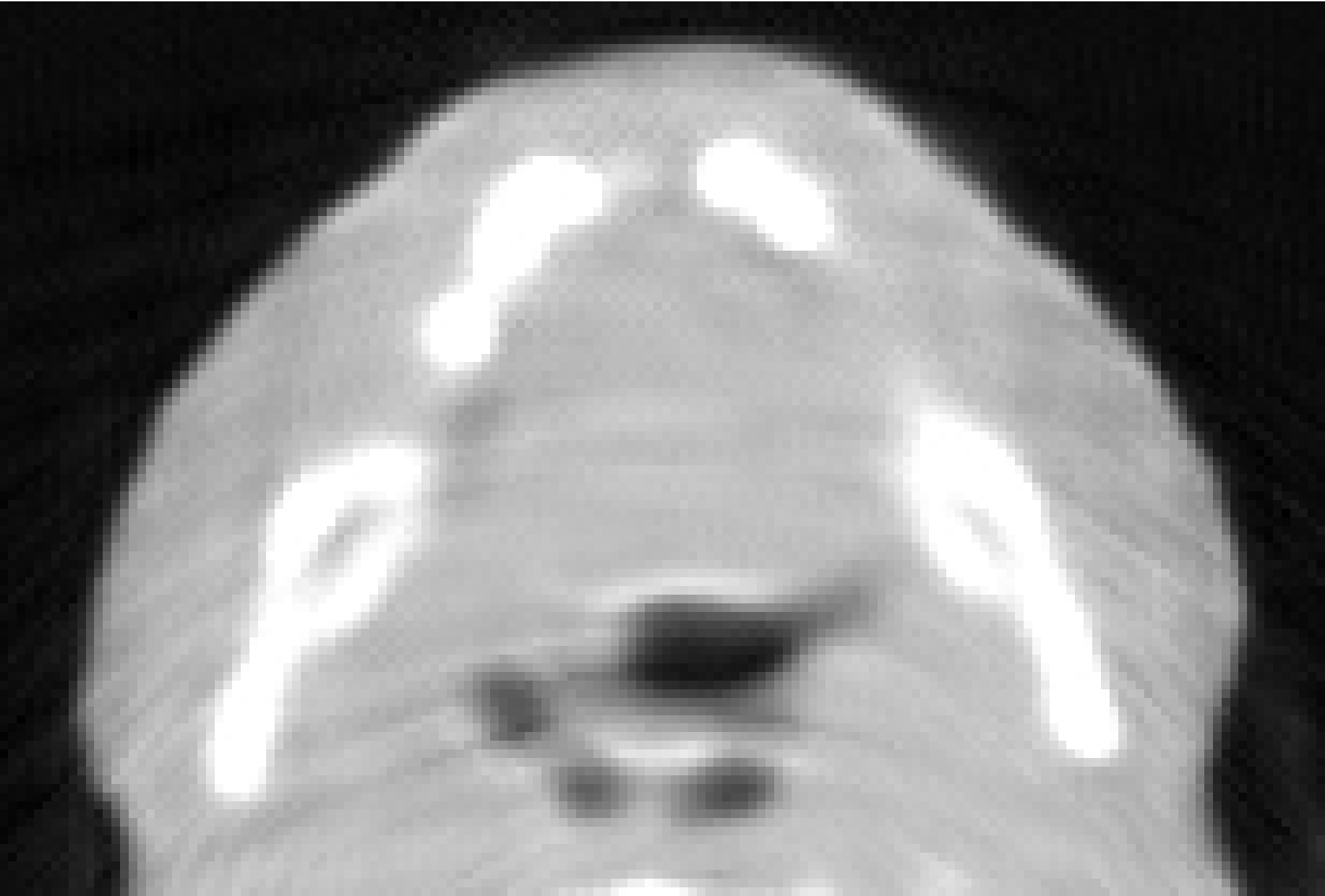}}
				}		
		\end{overpic}}
		\subfloat{
			\begin{overpic}[width=\size\columnwidth]{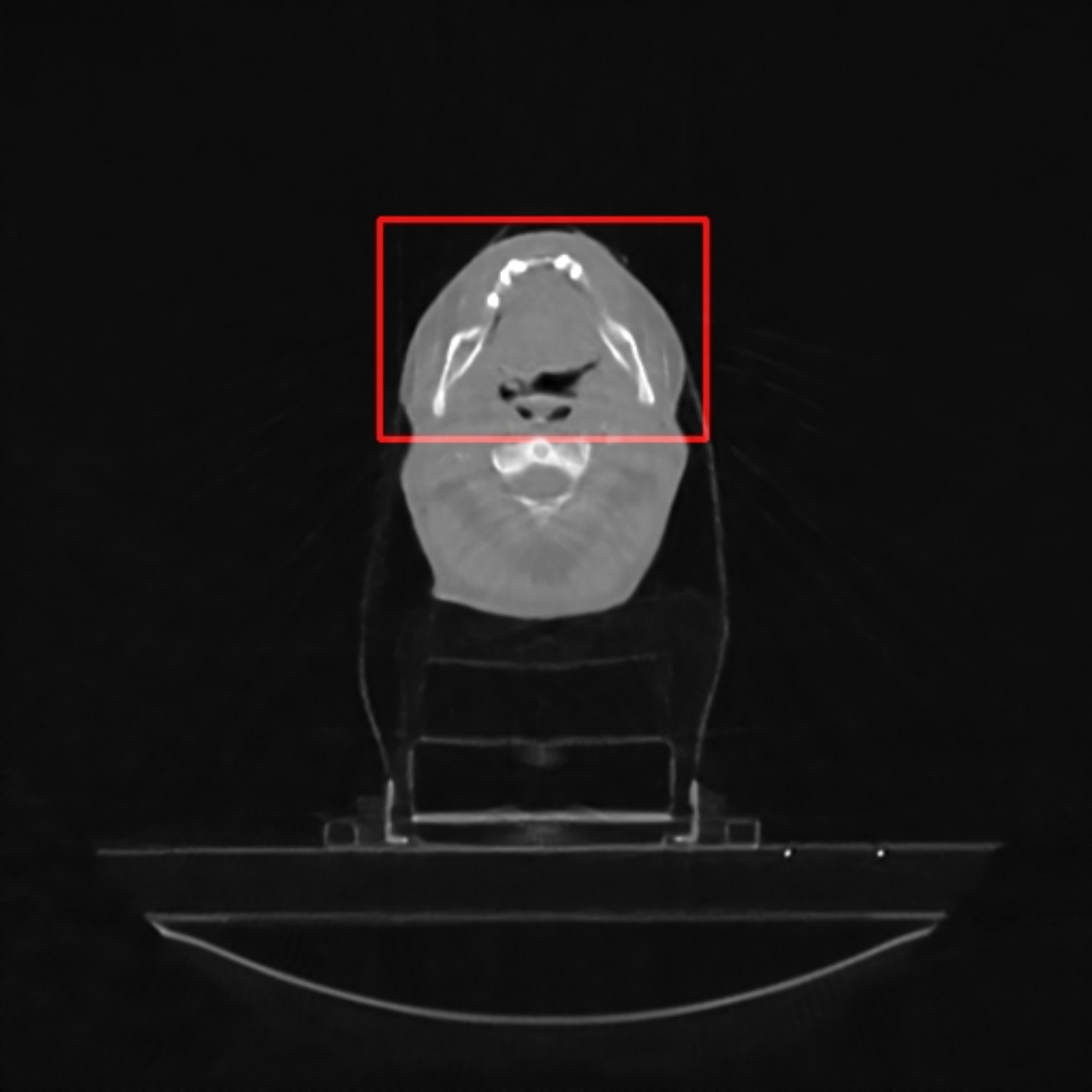}	
				\put(0,0){\color{red}%
					\frame{\includegraphics[scale=0.08]{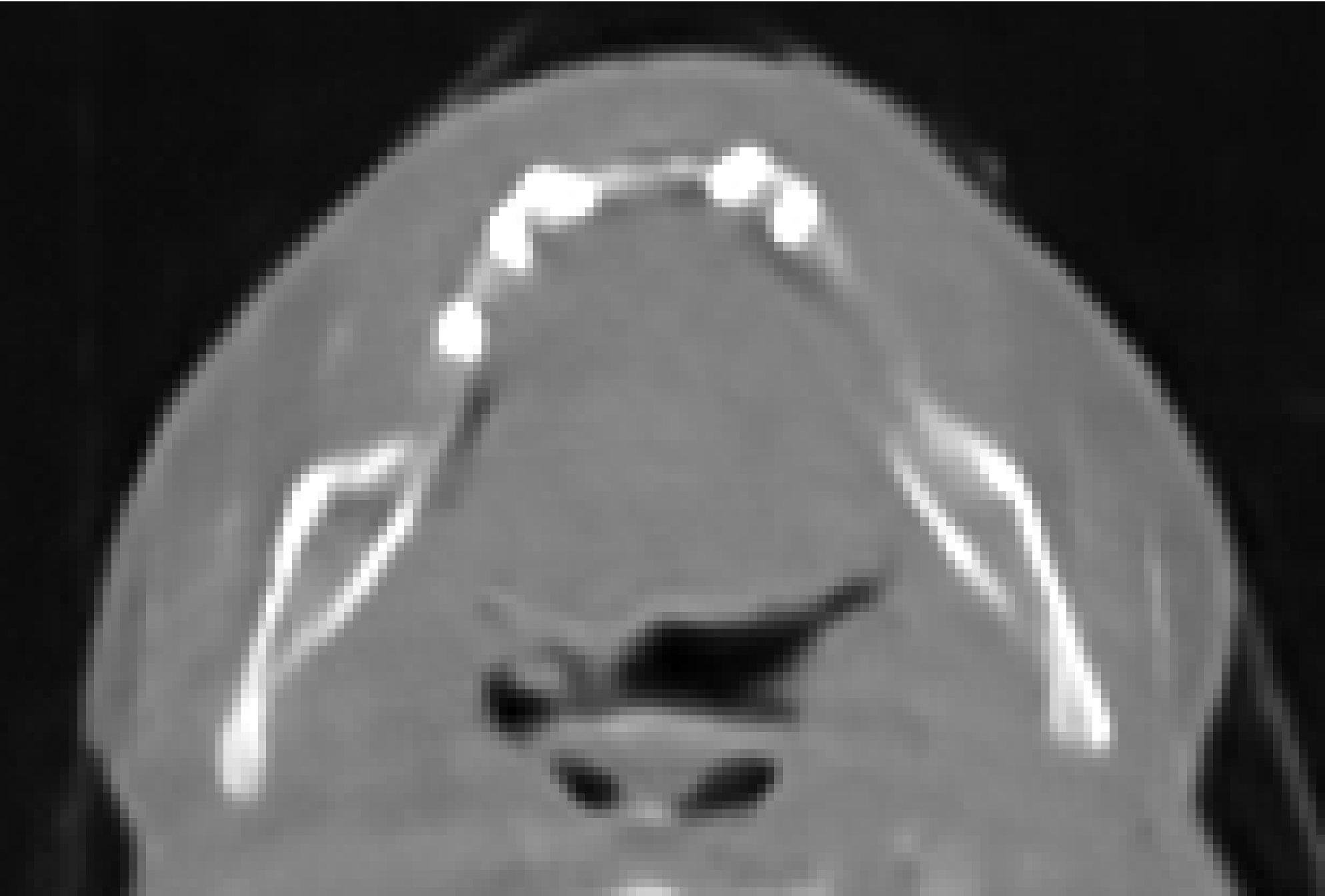}}
				}		
		\end{overpic}}
		\vspace{-3mm}
		
		\renewcommand{\id}{160}
		\subfloat{
			\begin{overpic}[width=\size\columnwidth,percent]{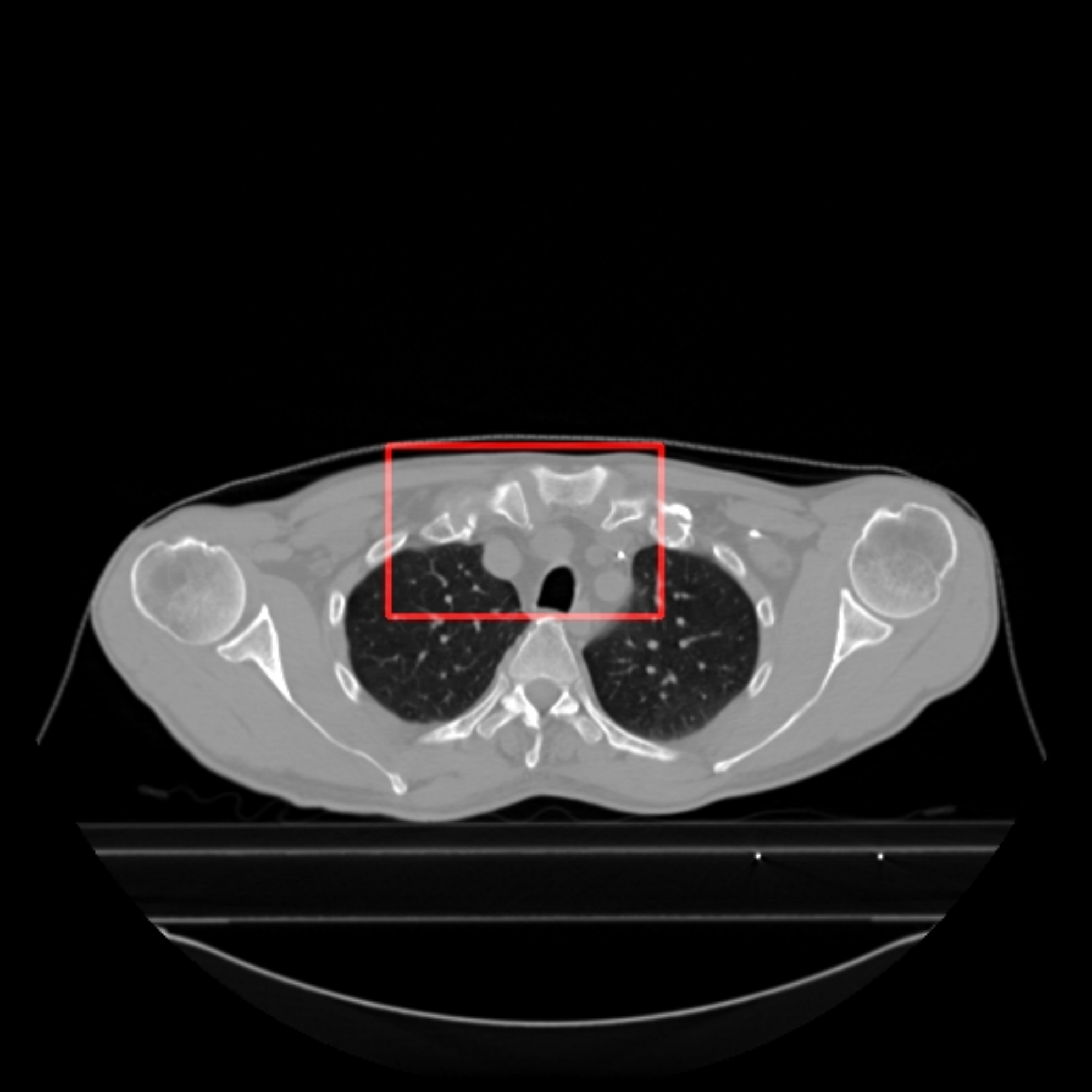}	
				\put(0,0){\color{red}%
					\frame{\includegraphics[scale=0.08]{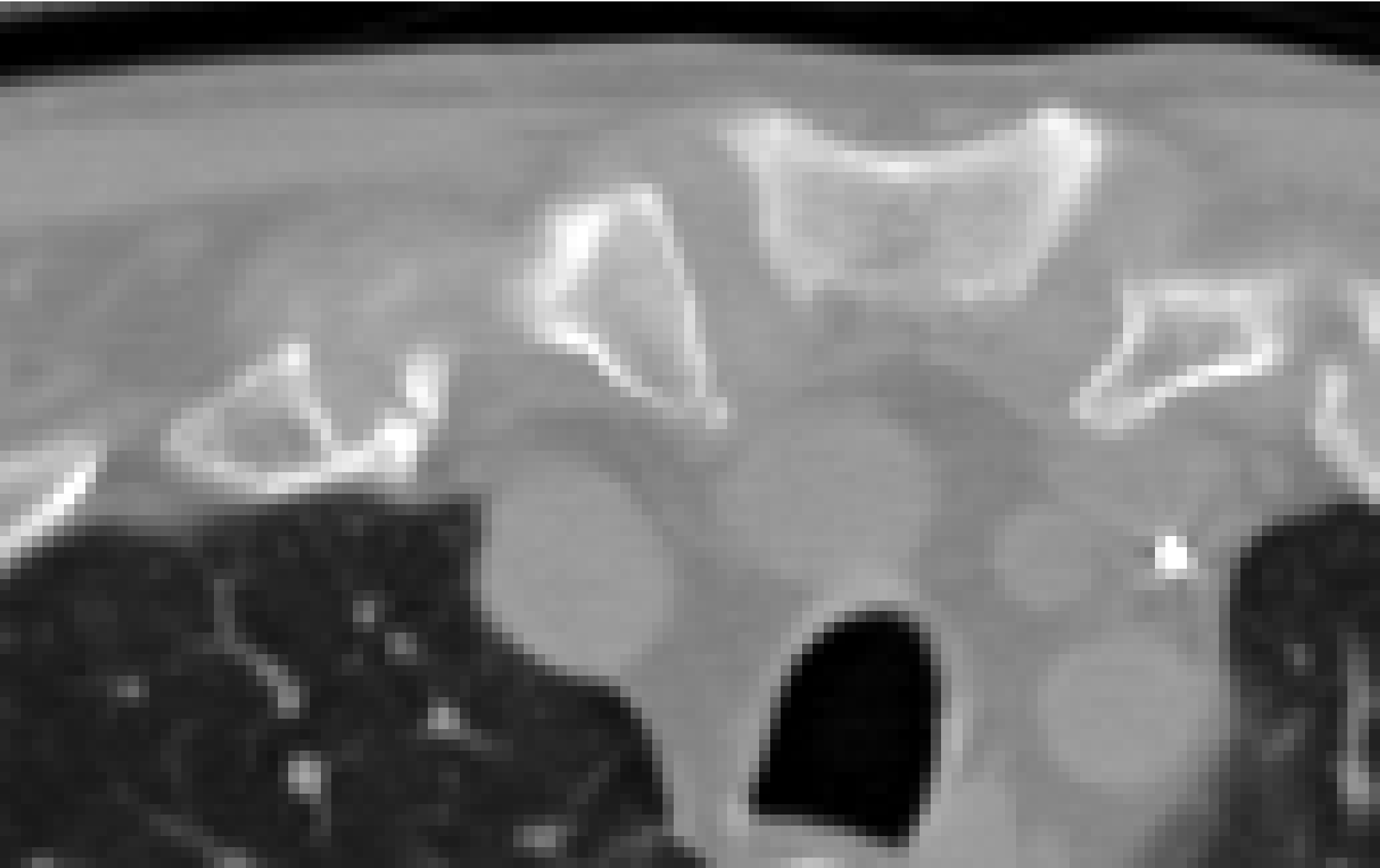}}
				}		
		\end{overpic}}
		\subfloat{
			\begin{overpic}[width=\size\columnwidth]{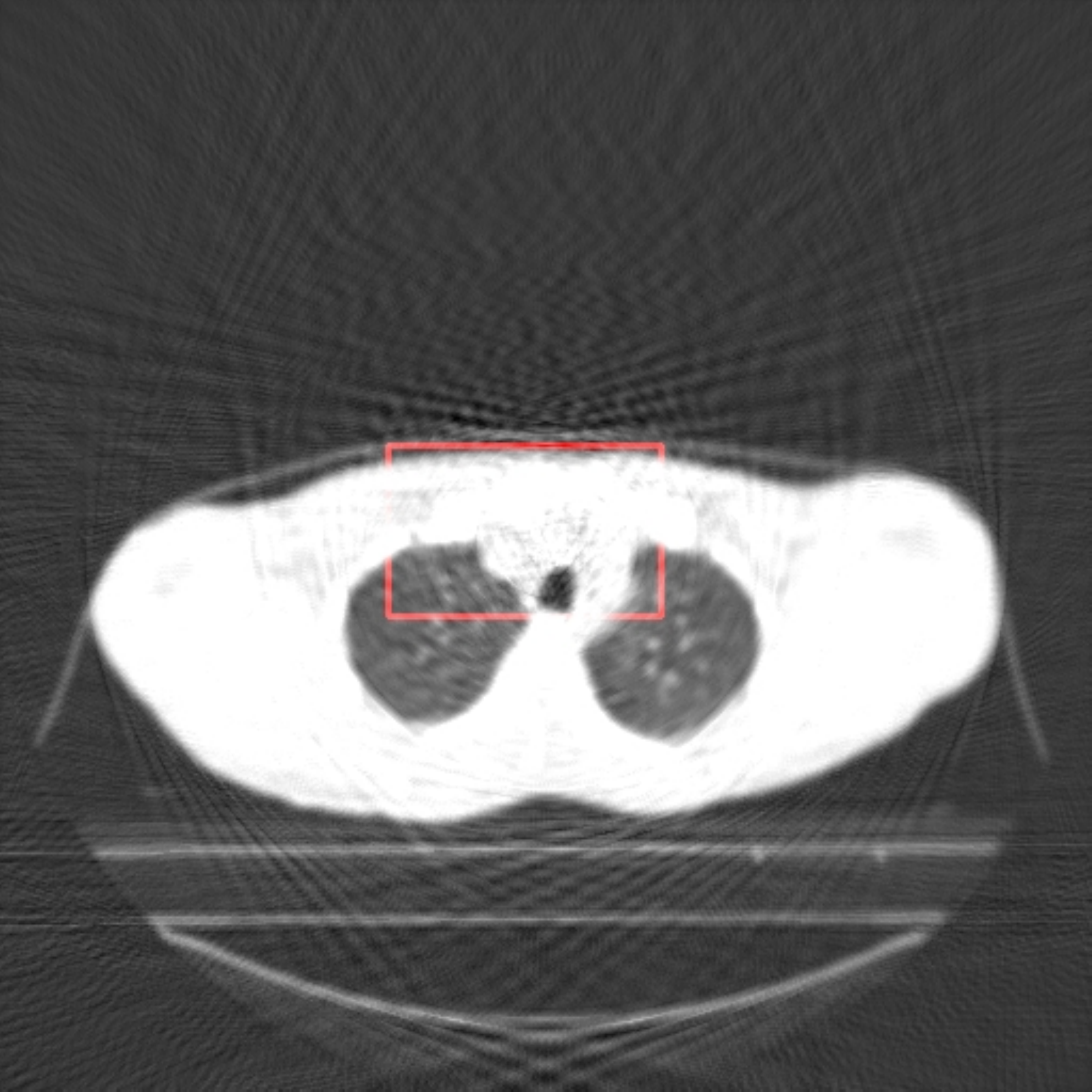}	
				\put(0,0){\color{red}%
					\frame{\includegraphics[scale=0.08]{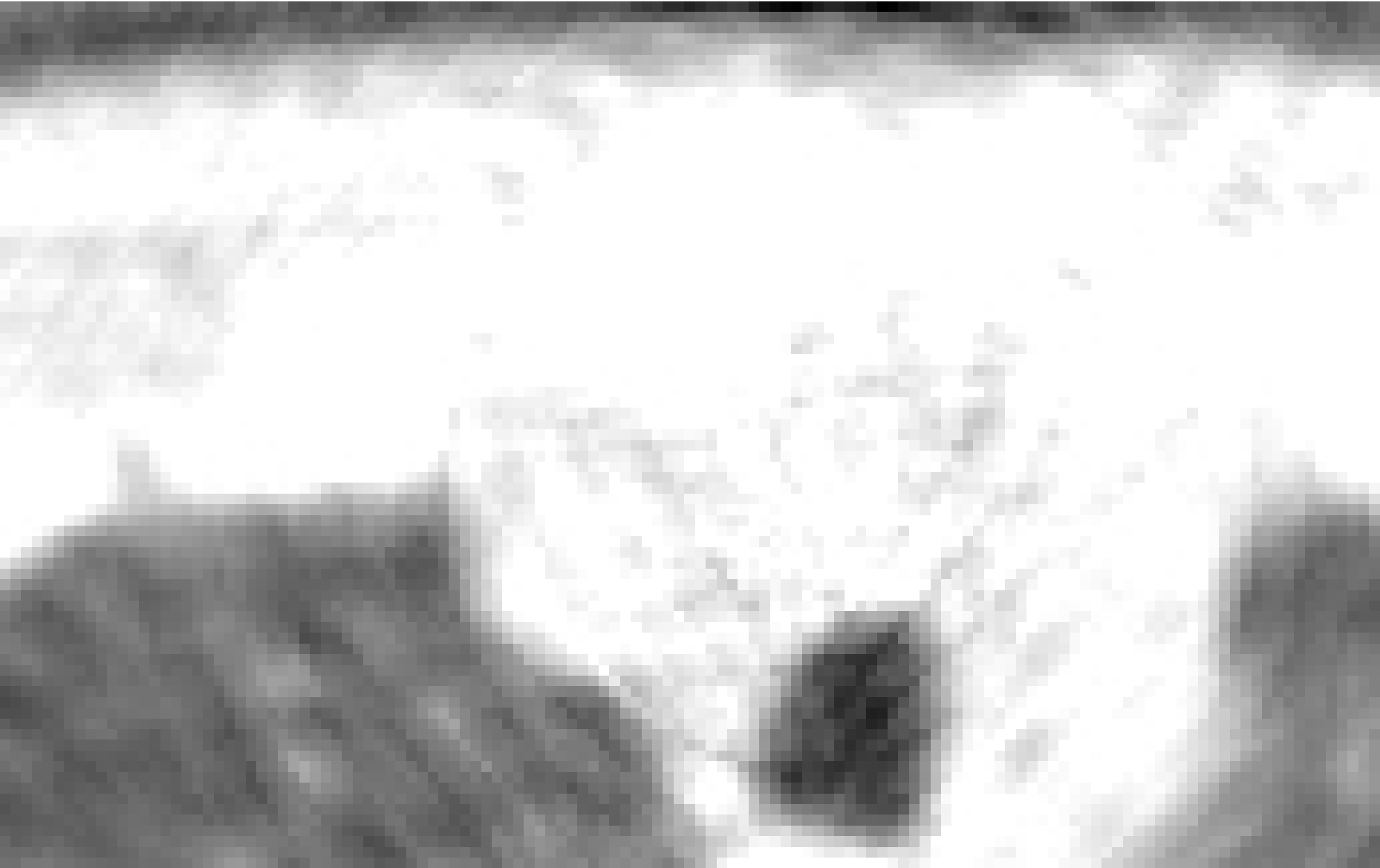}}
				}		
		\end{overpic}}
		\subfloat{
			\begin{overpic}[width=\size\columnwidth]{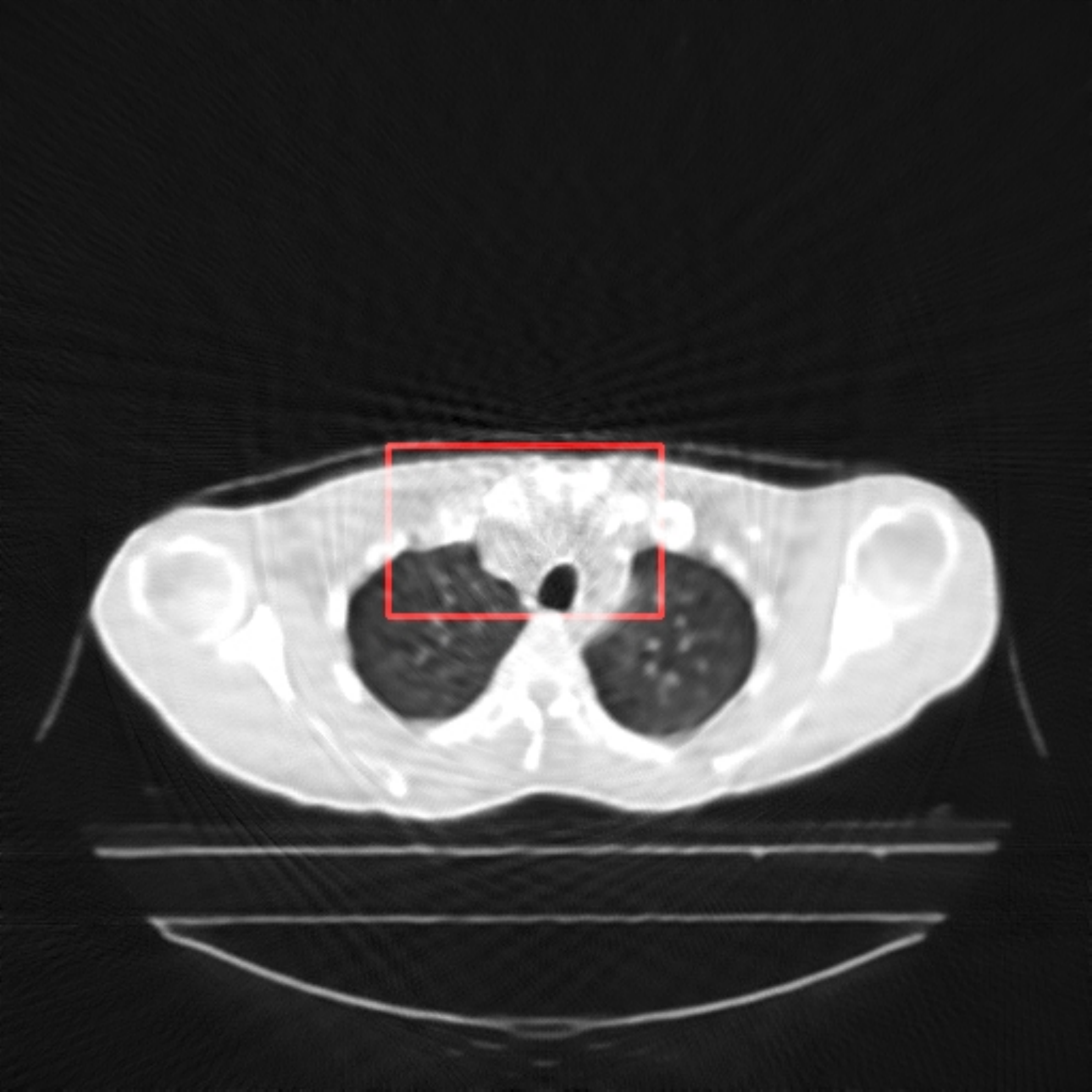}	
				\put(0,0){\color{red}%
					\frame{\includegraphics[scale=0.08]{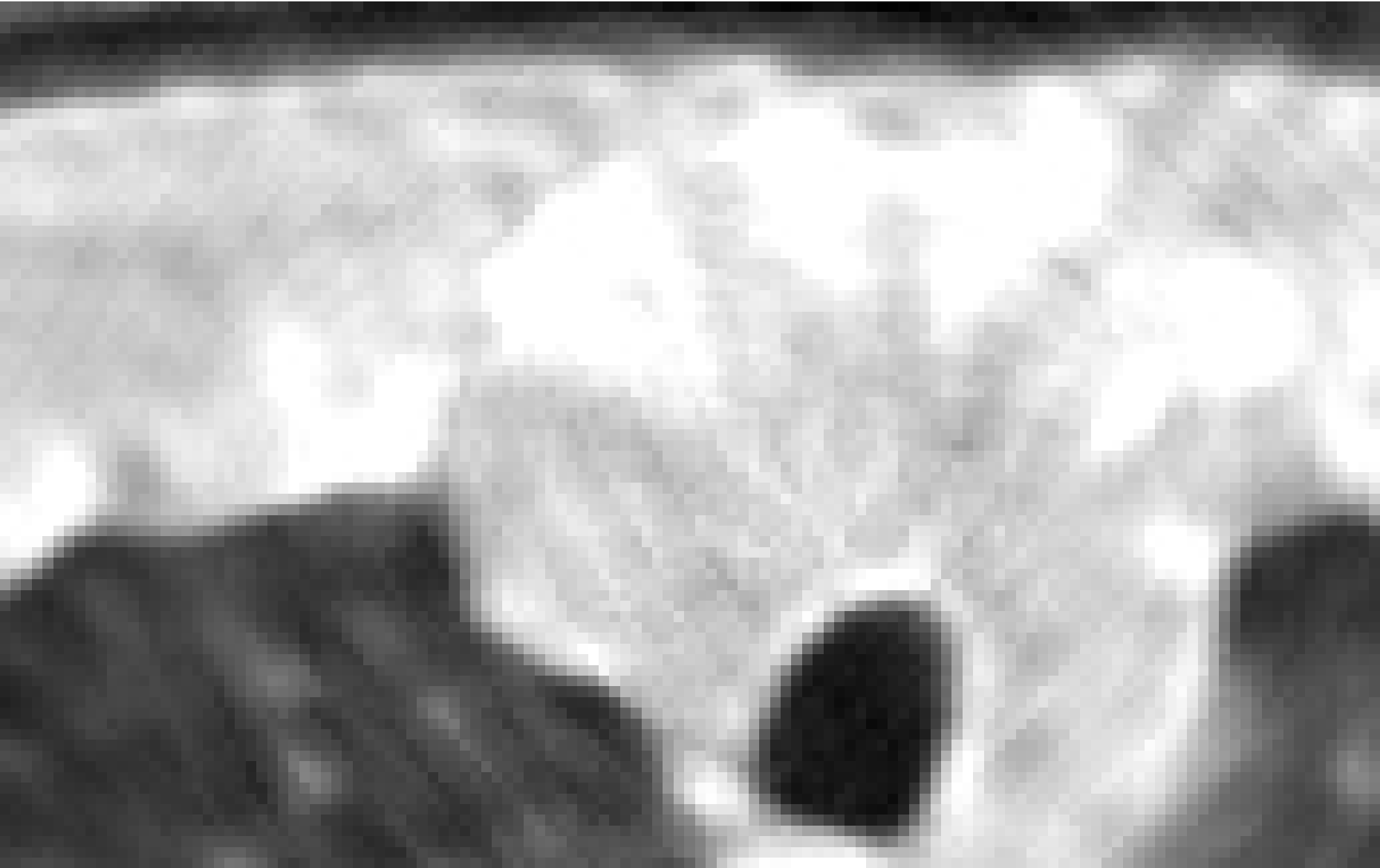}}
				}		
		\end{overpic}}
		\subfloat{
			\begin{overpic}[width=\size\columnwidth]{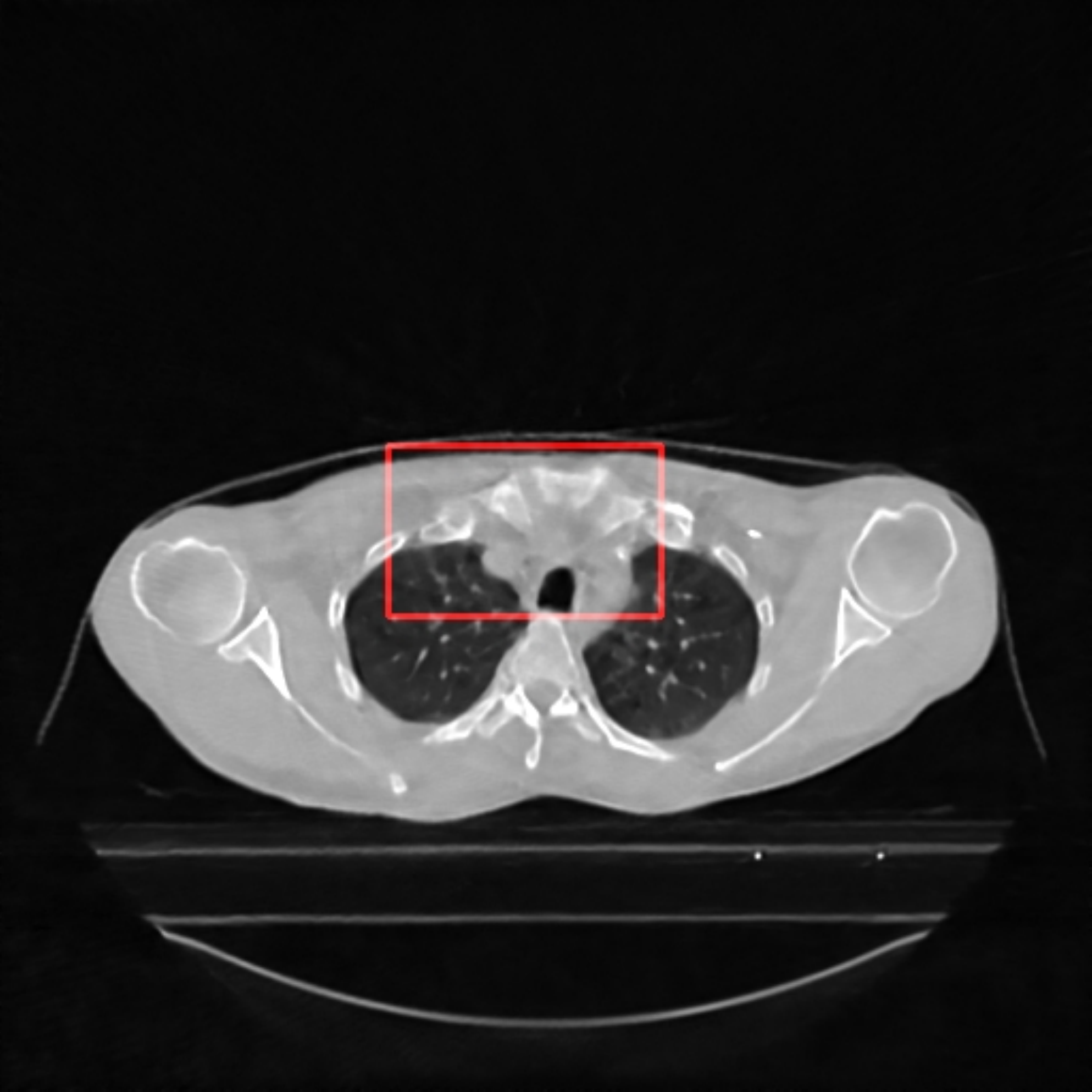}	
				\put(0,0){\color{red}%
					\frame{\includegraphics[scale=0.08]{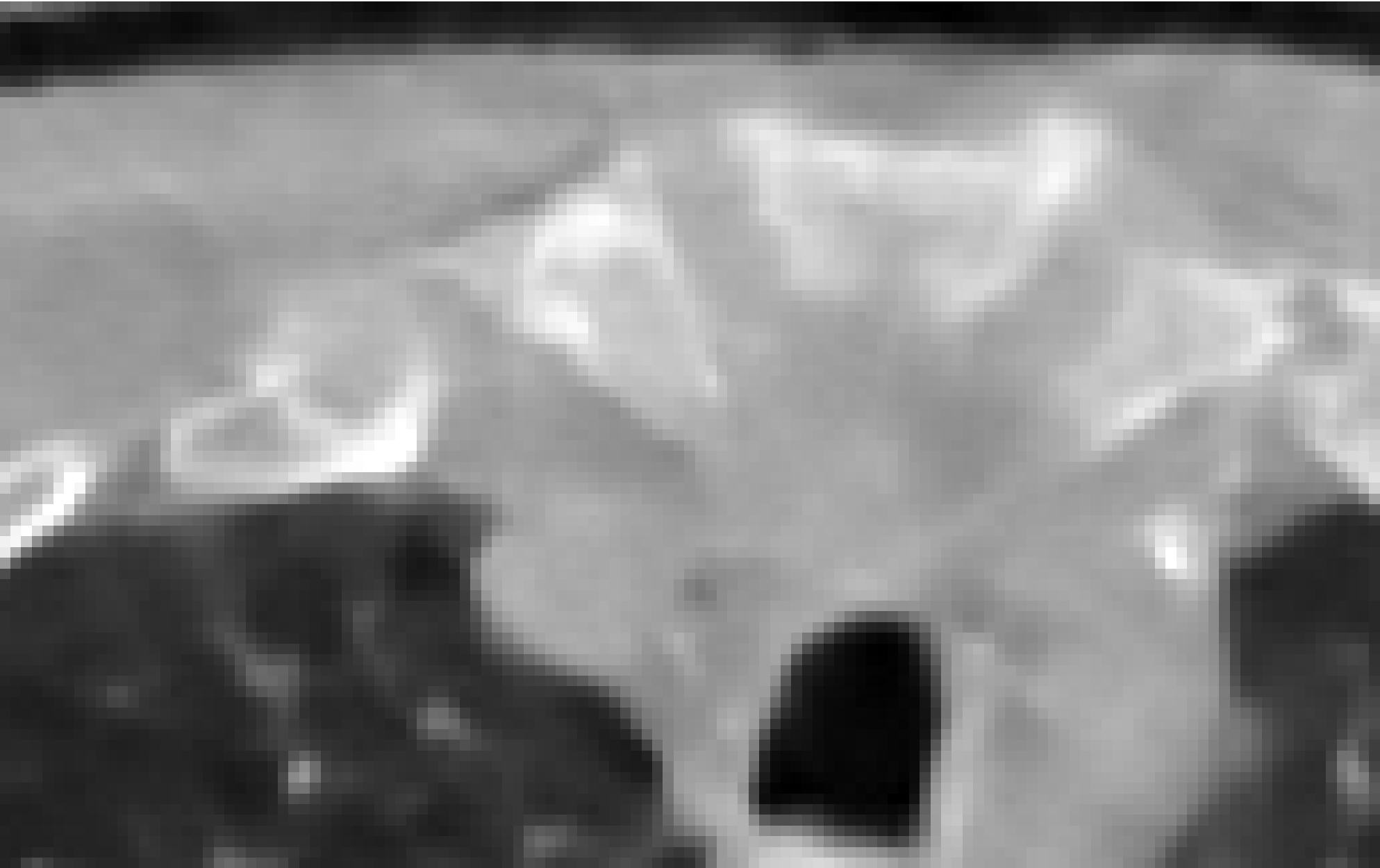}}
				}		
		\end{overpic}}
		\vspace{-3mm}
		\renewcommand{\id}{200}
		\subfloat{
			\begin{overpic}[width=\size\columnwidth,percent]{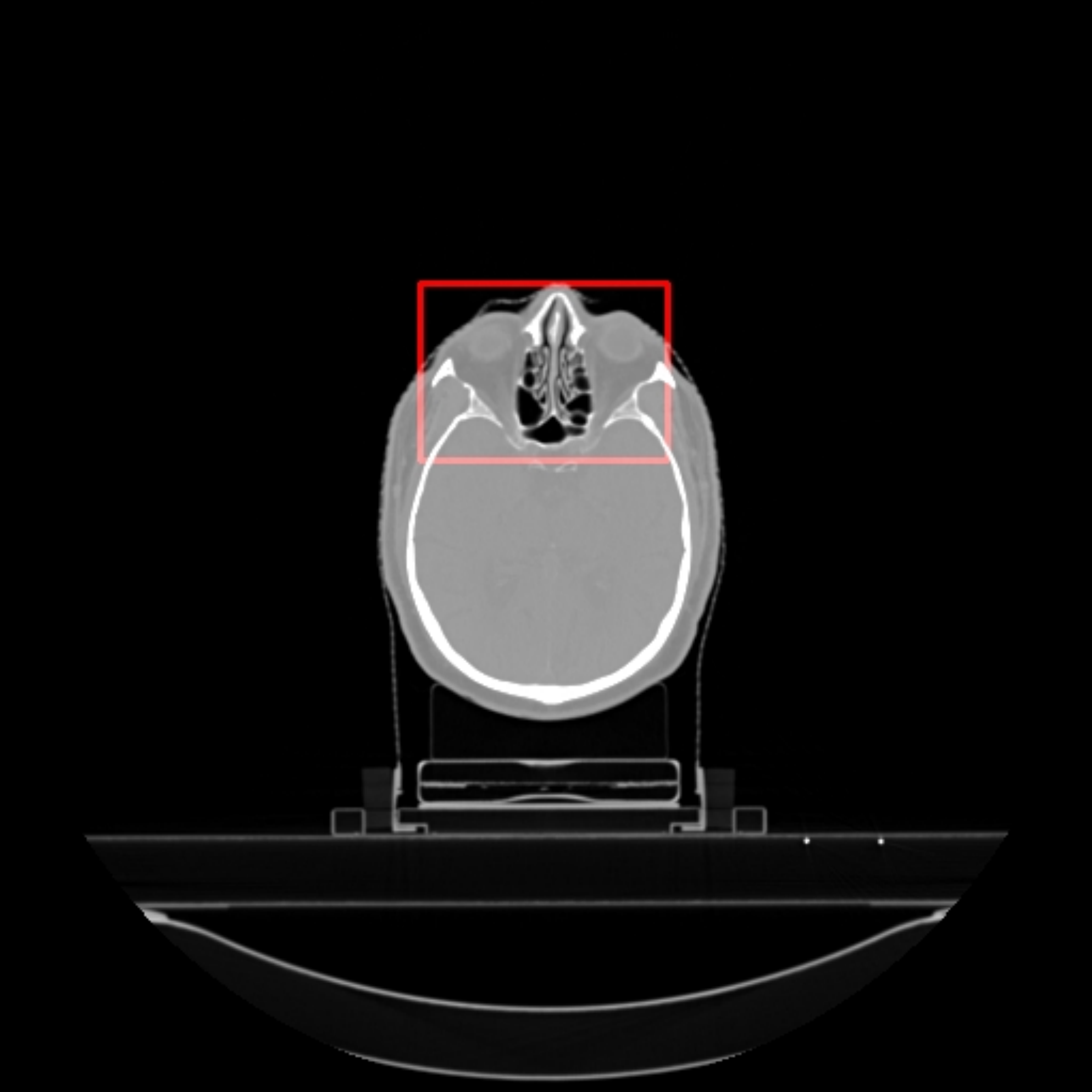}	
				\put(0,0){\color{red}%
					\frame{\includegraphics[scale=0.08]{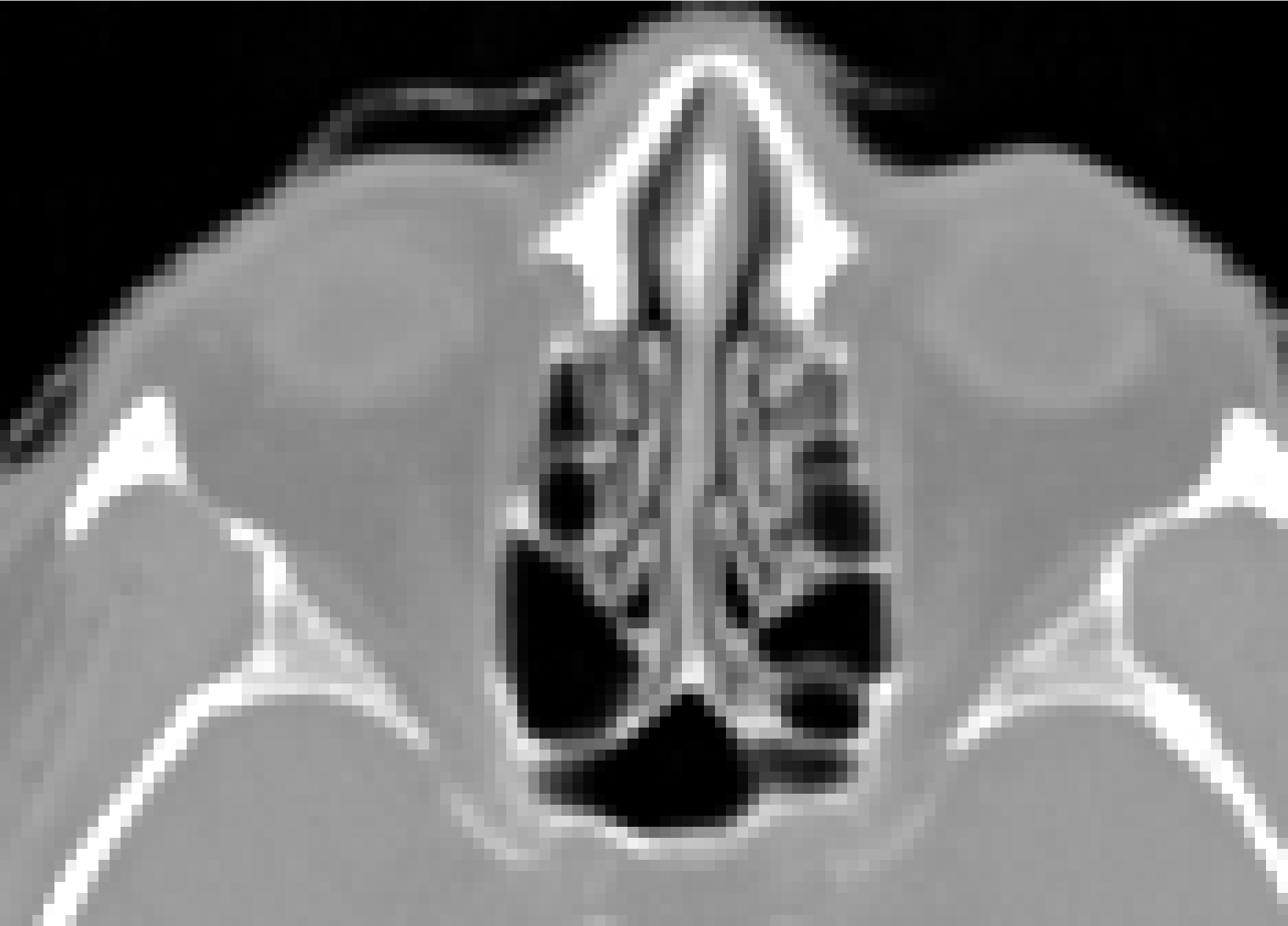}}
				}		
		\end{overpic}}
		\subfloat{
			\begin{overpic}[width=\size\columnwidth]{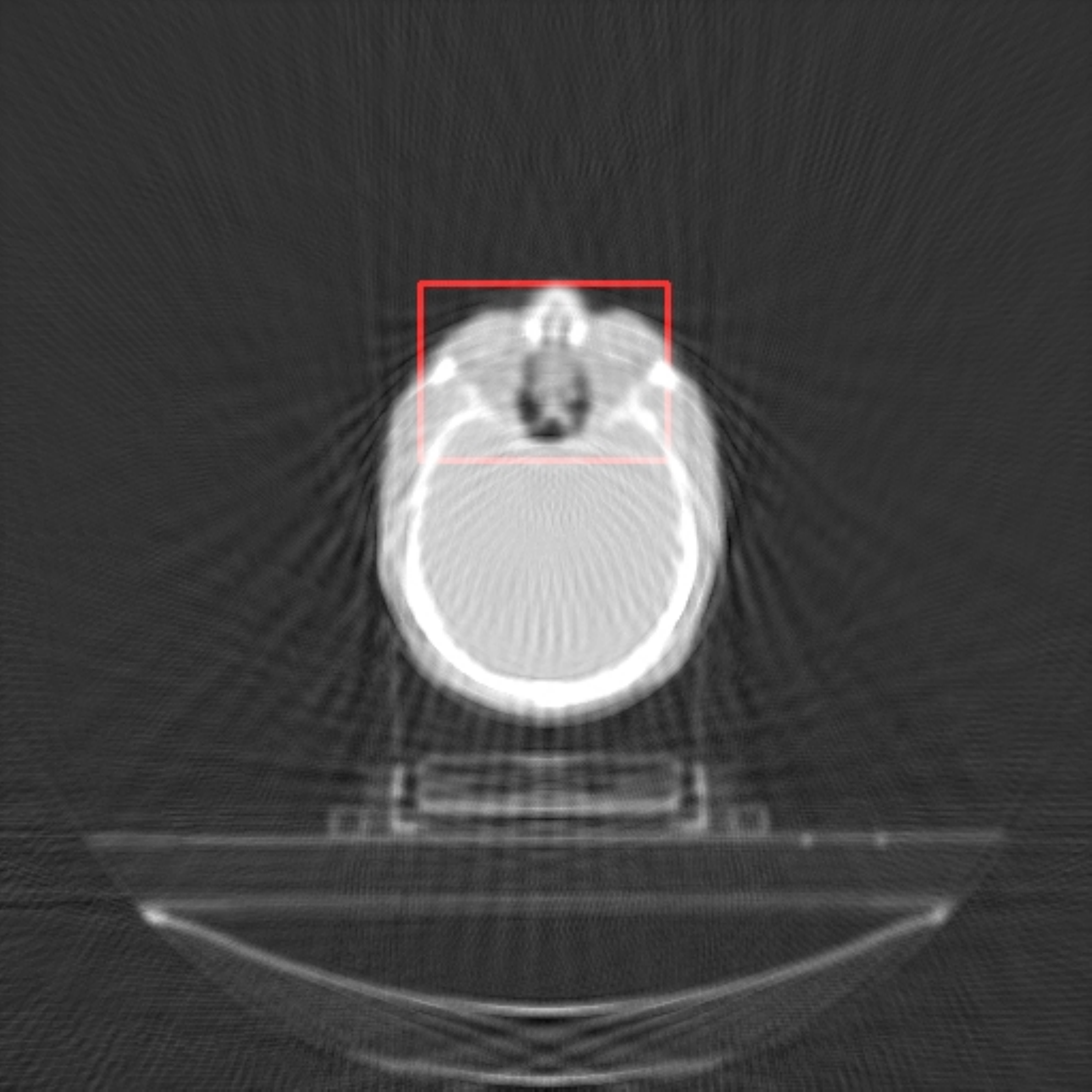}	
				\put(0,0){\color{red}%
					\frame{\includegraphics[scale=0.08]{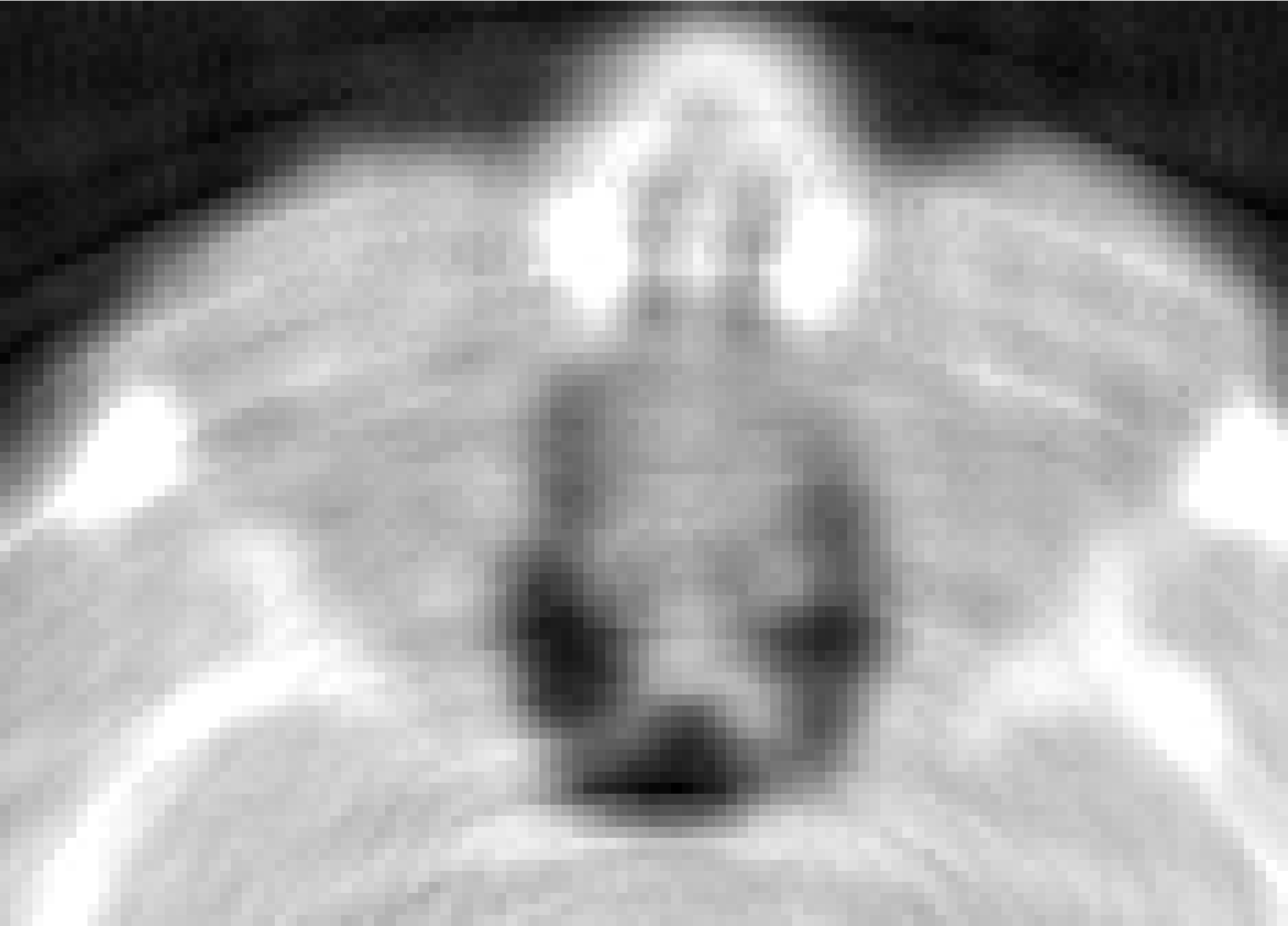}}
				}		
		\end{overpic}}
		\subfloat{
			\begin{overpic}[width=\size\columnwidth]{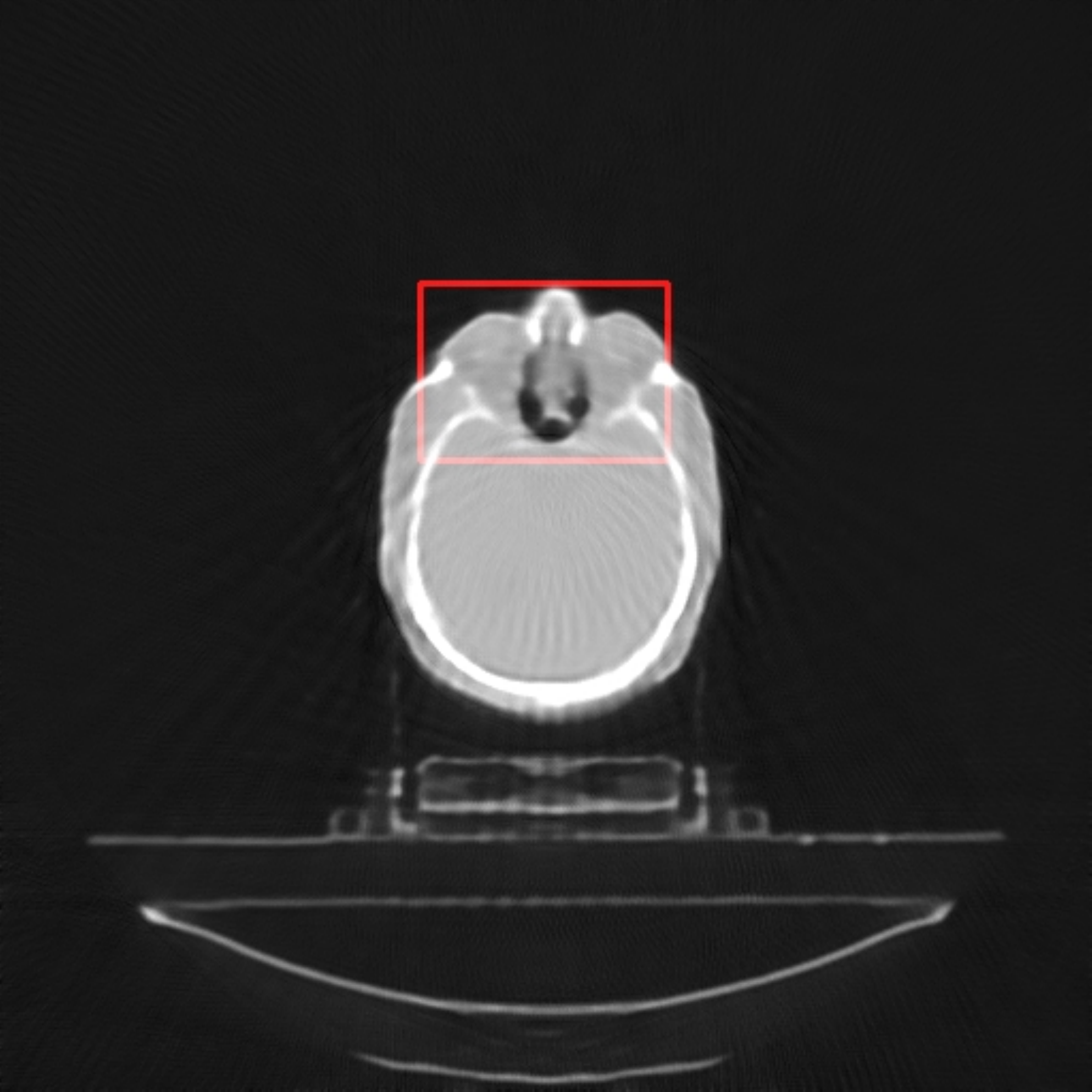}	
				\put(0,0){\color{red}%
					\frame{\includegraphics[scale=0.08]{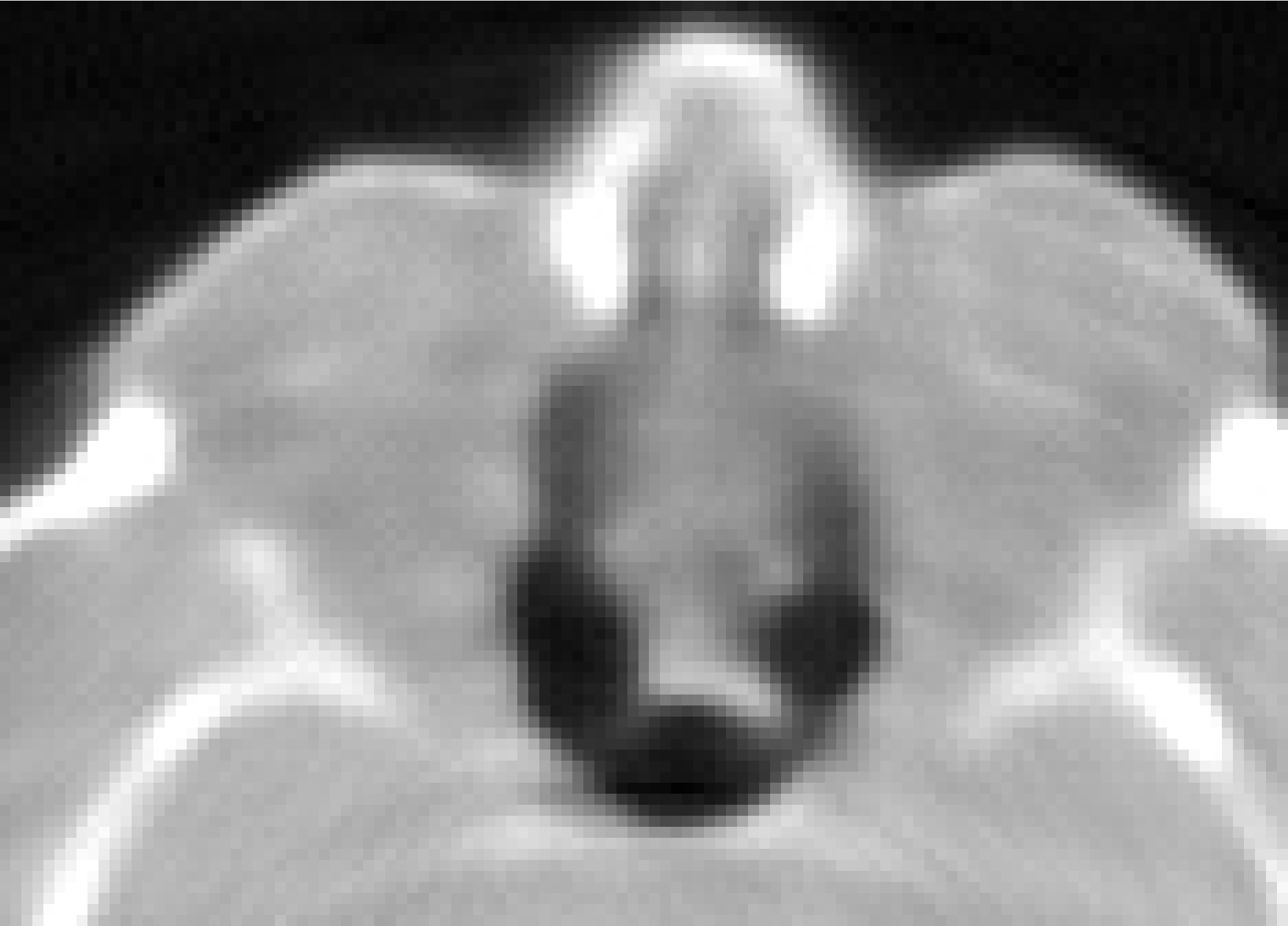}}
				}		
		\end{overpic}}
		\subfloat{
			\begin{overpic}[width=\size\columnwidth]{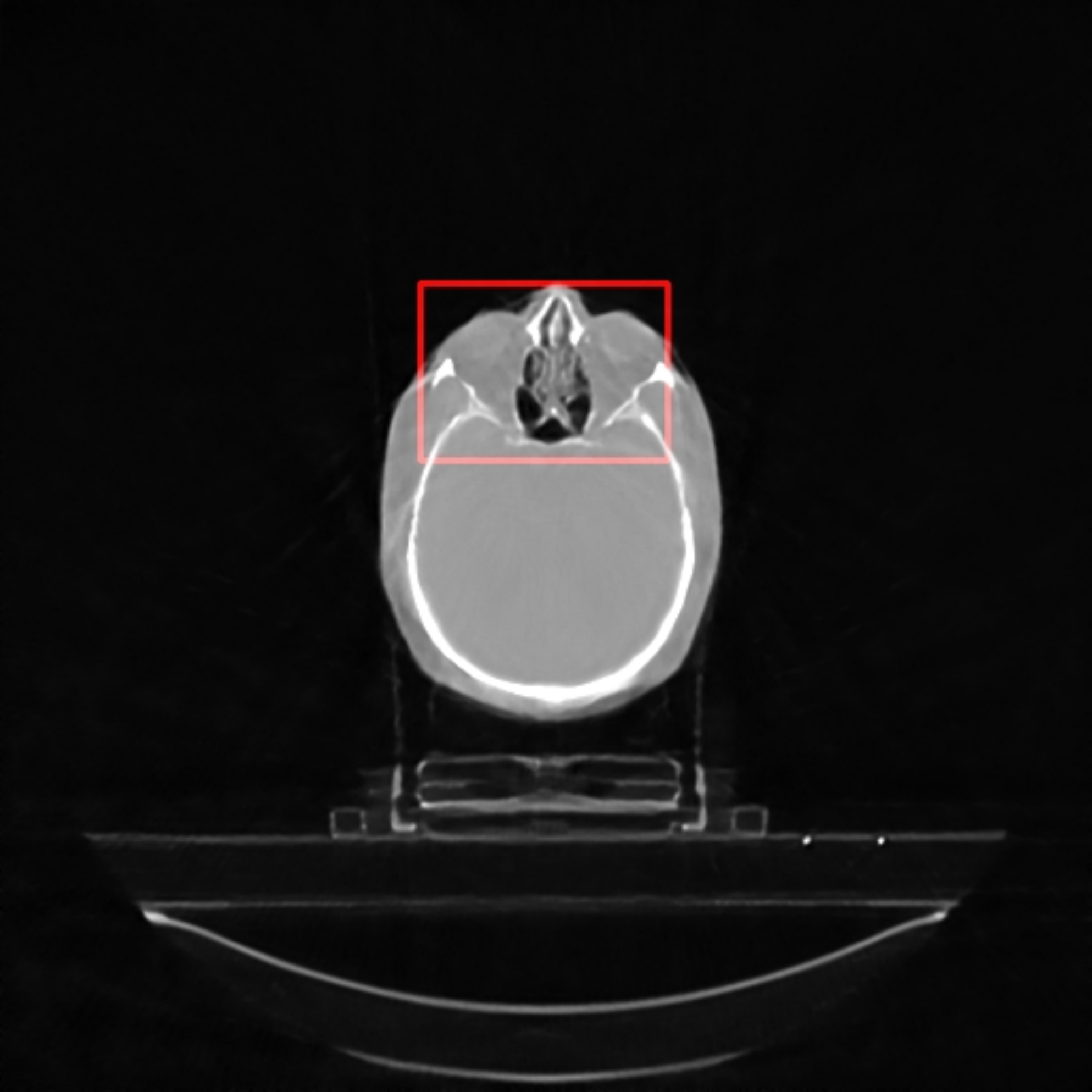}	
				\put(0,0){\color{red}%
					\frame{\includegraphics[scale=0.08]{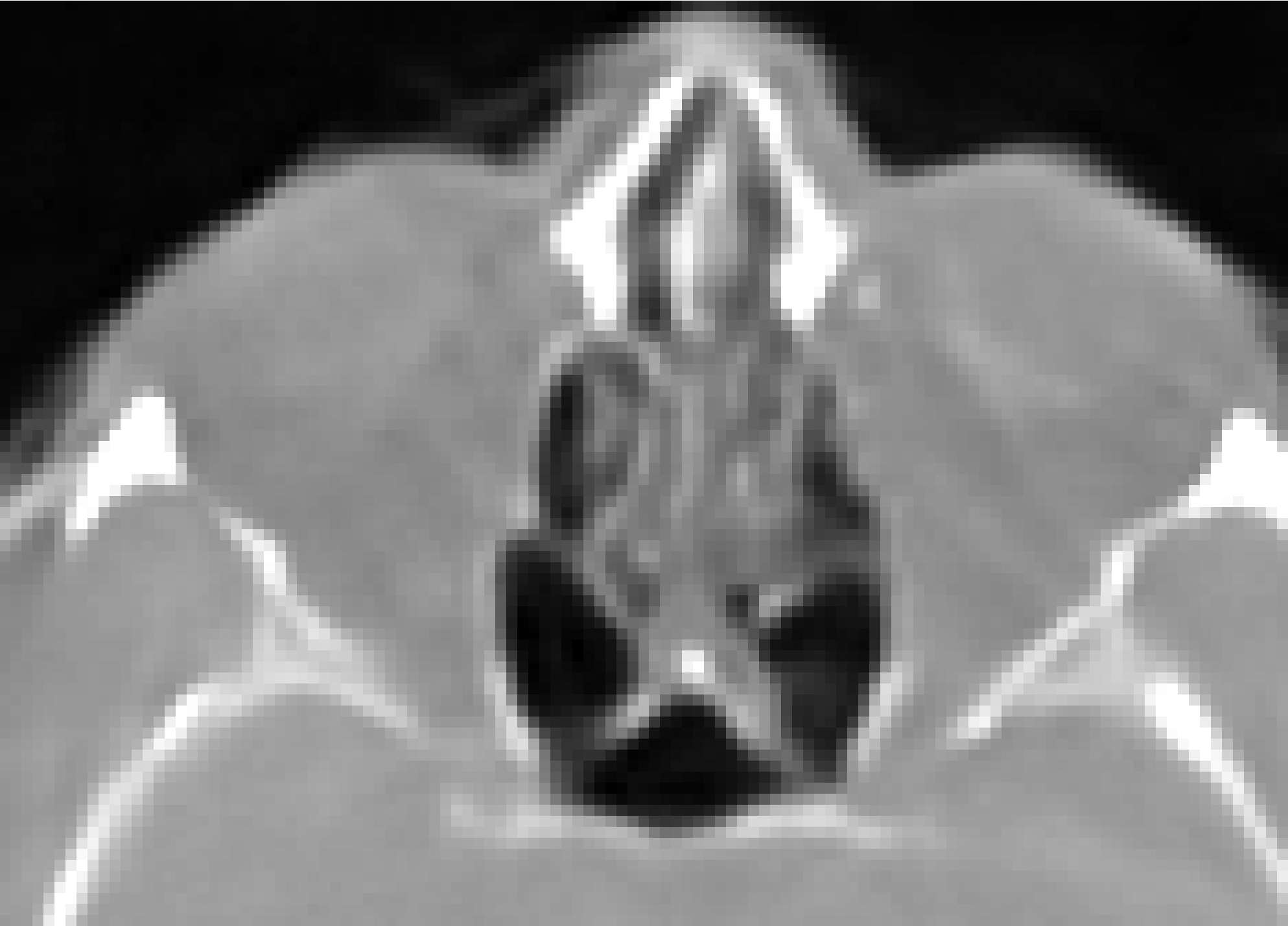}}
				}		
		\end{overpic}}
		\vspace{-3mm}
		\setcounter{subfigure}{0}
		\renewcommand{\id}{240}
		\subfloat[Label]{
			\begin{overpic}[width=\size\columnwidth,percent]{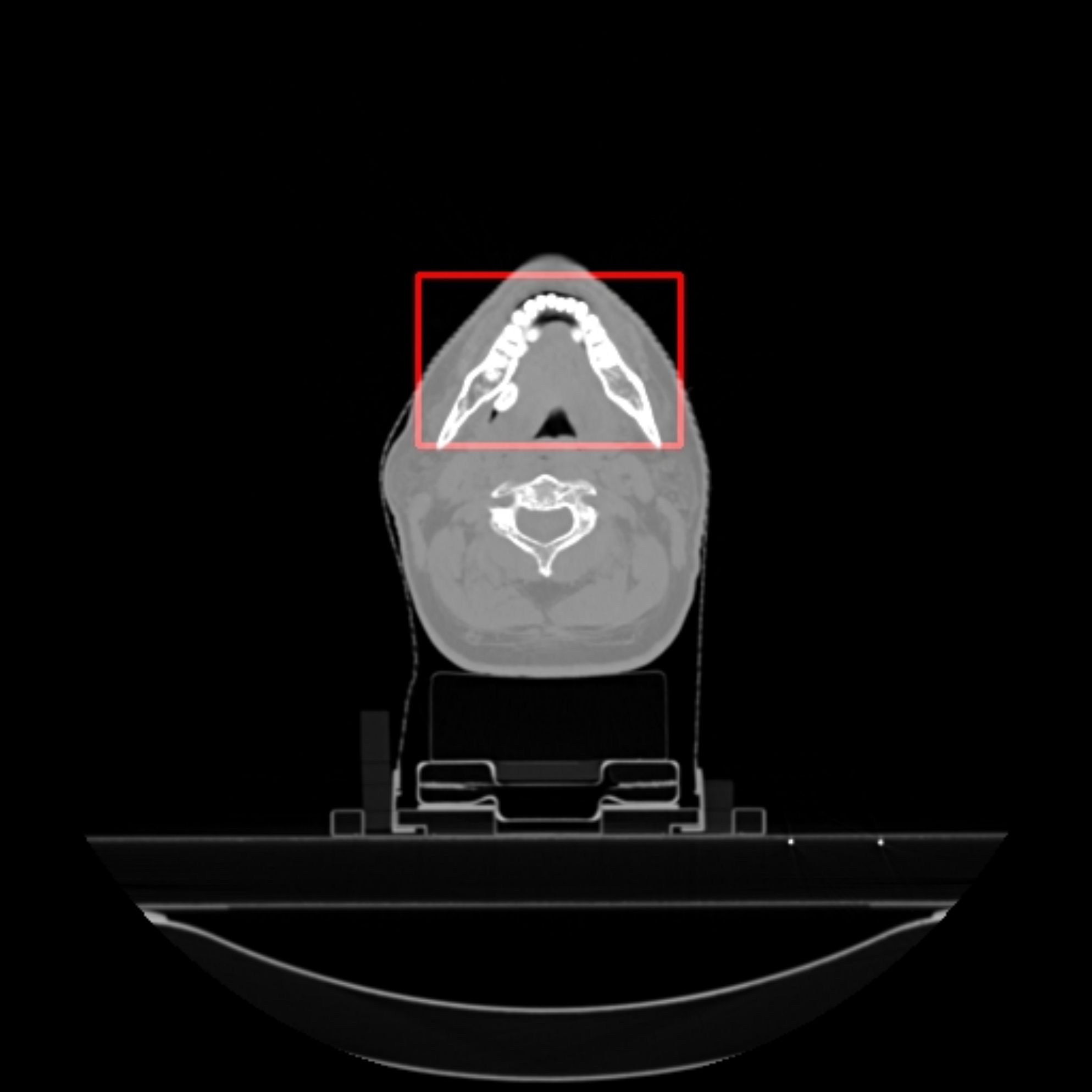}	
				\put(0,0){\color{red}%
					\frame{\includegraphics[scale=0.08]{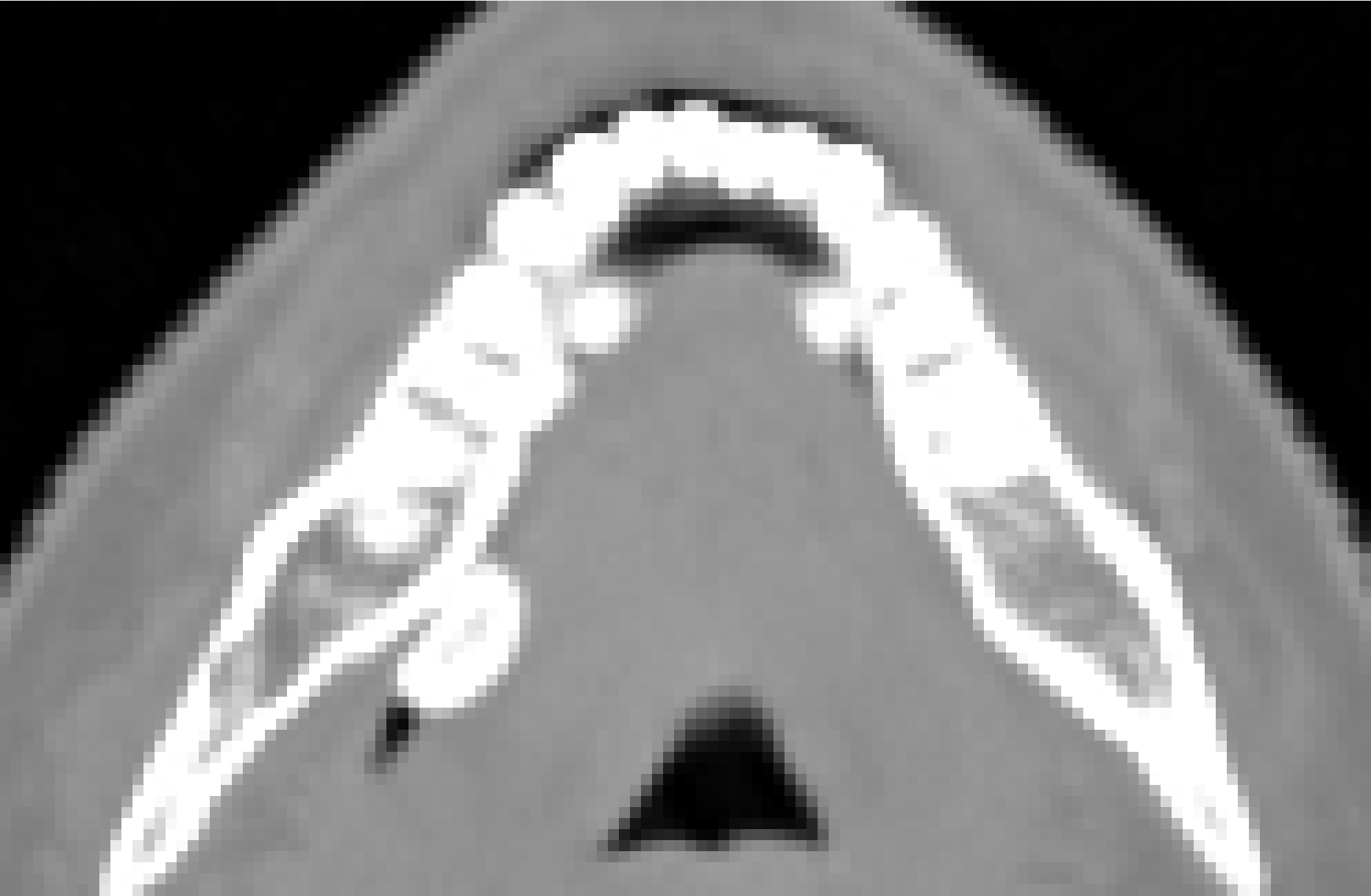}}
				}		
		\end{overpic}}
		\subfloat[Katsevich algorithm]{
			\begin{overpic}[width=\size\columnwidth]{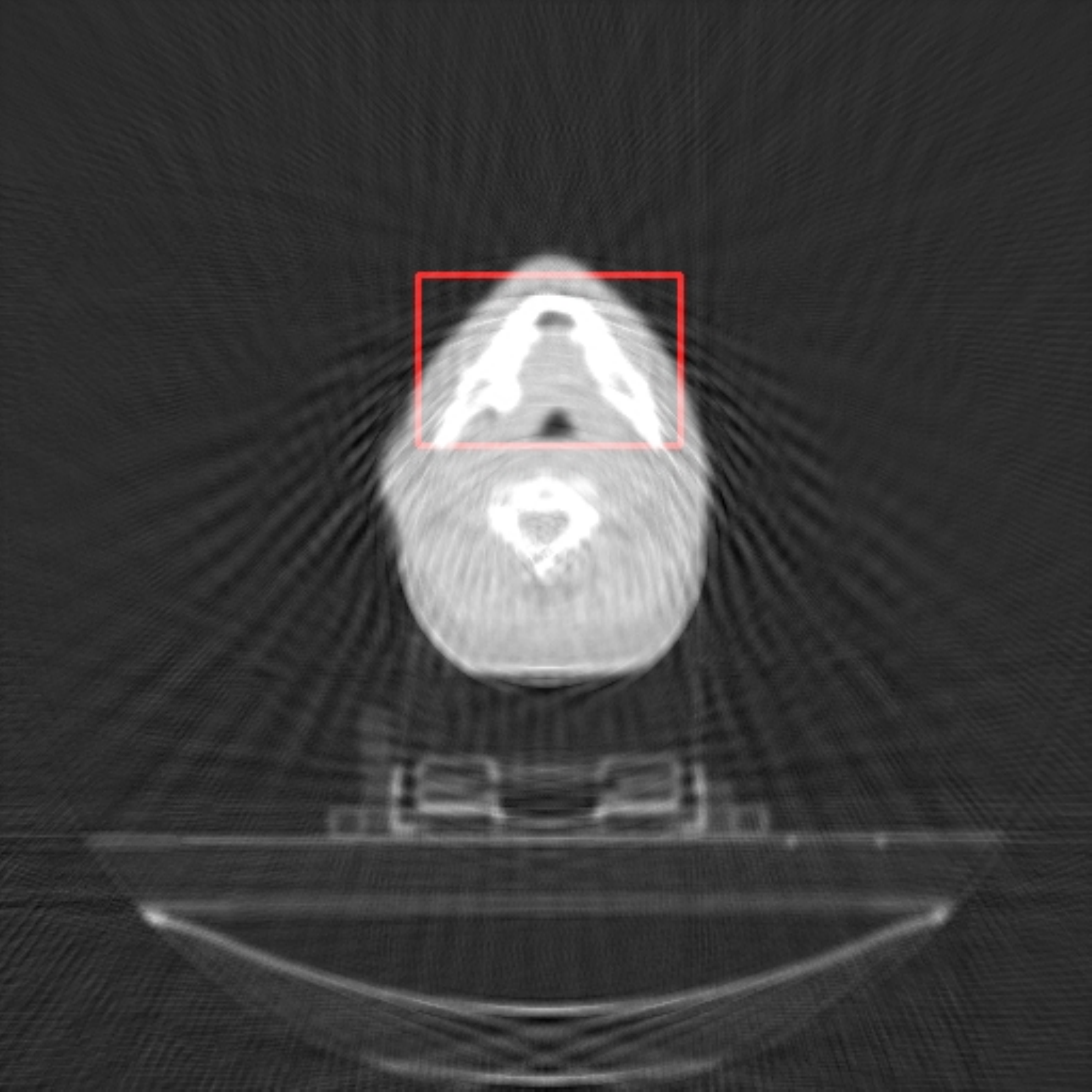}	
				\put(0,0){\color{red}%
					\frame{\includegraphics[scale=0.08]{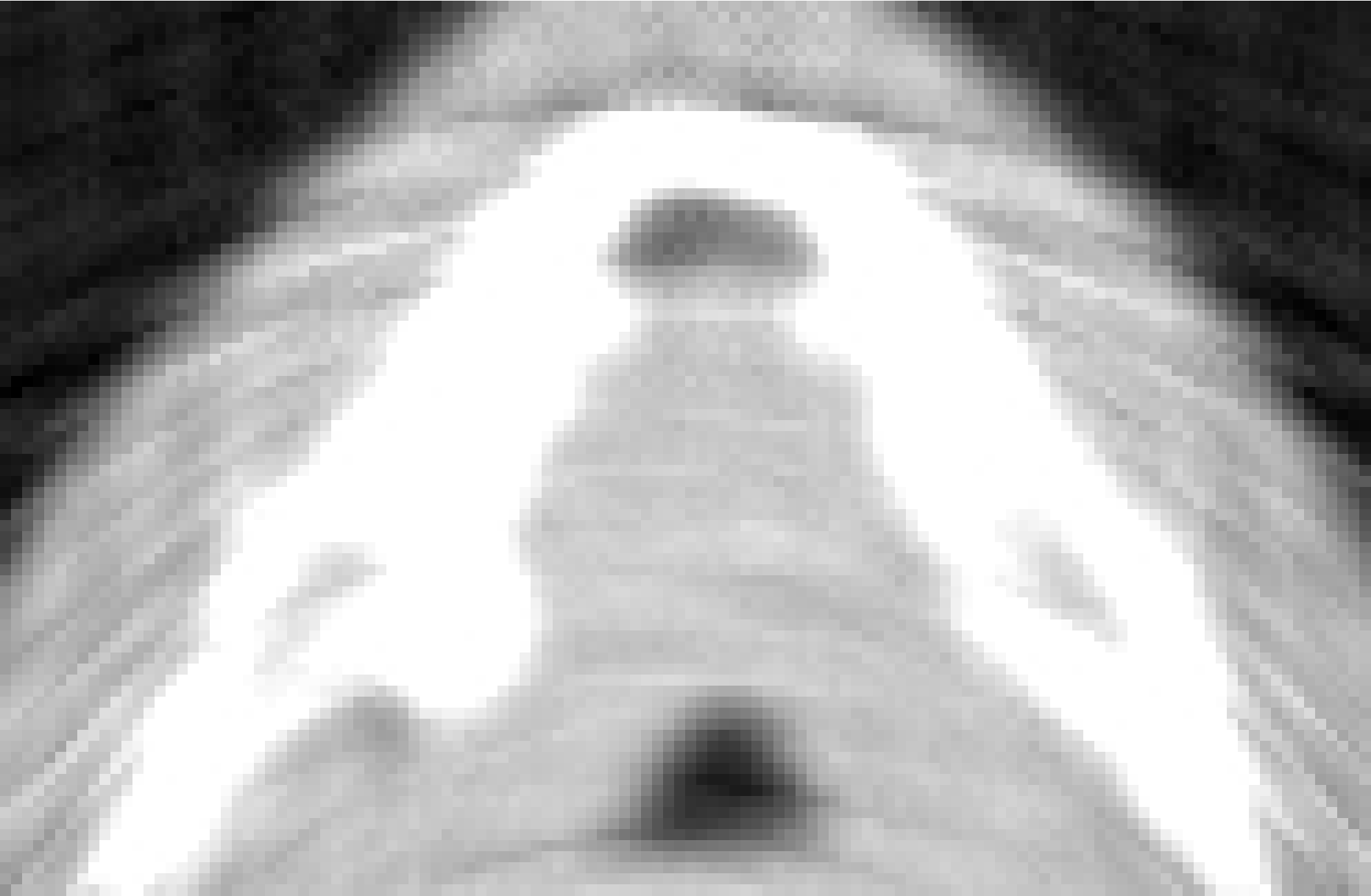}}
				}		
		\end{overpic}}
		\subfloat[FBPConvNet]{
			\begin{overpic}[width=\size\columnwidth]{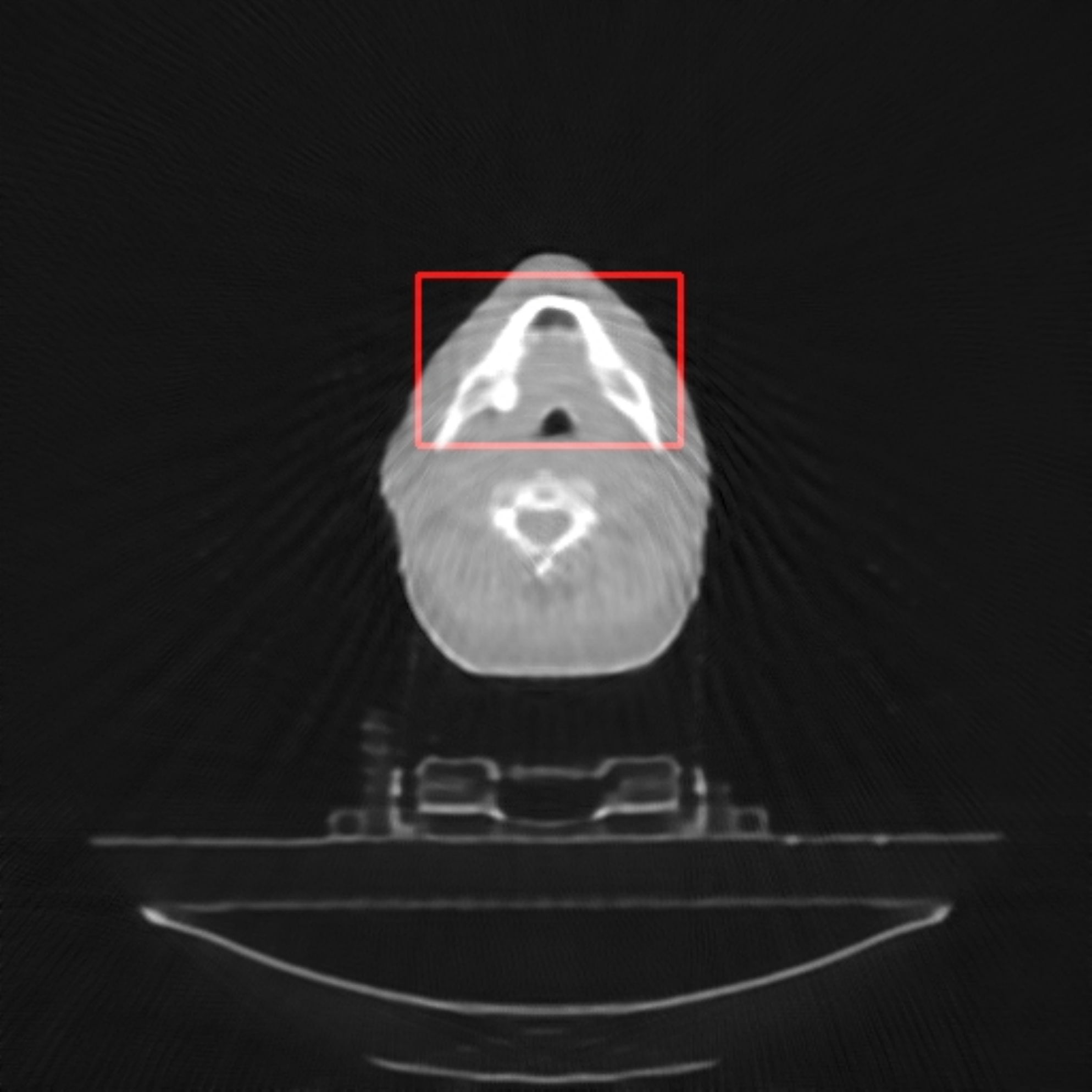}	
				\put(0,0){\color{red}%
					\frame{\includegraphics[scale=0.08]{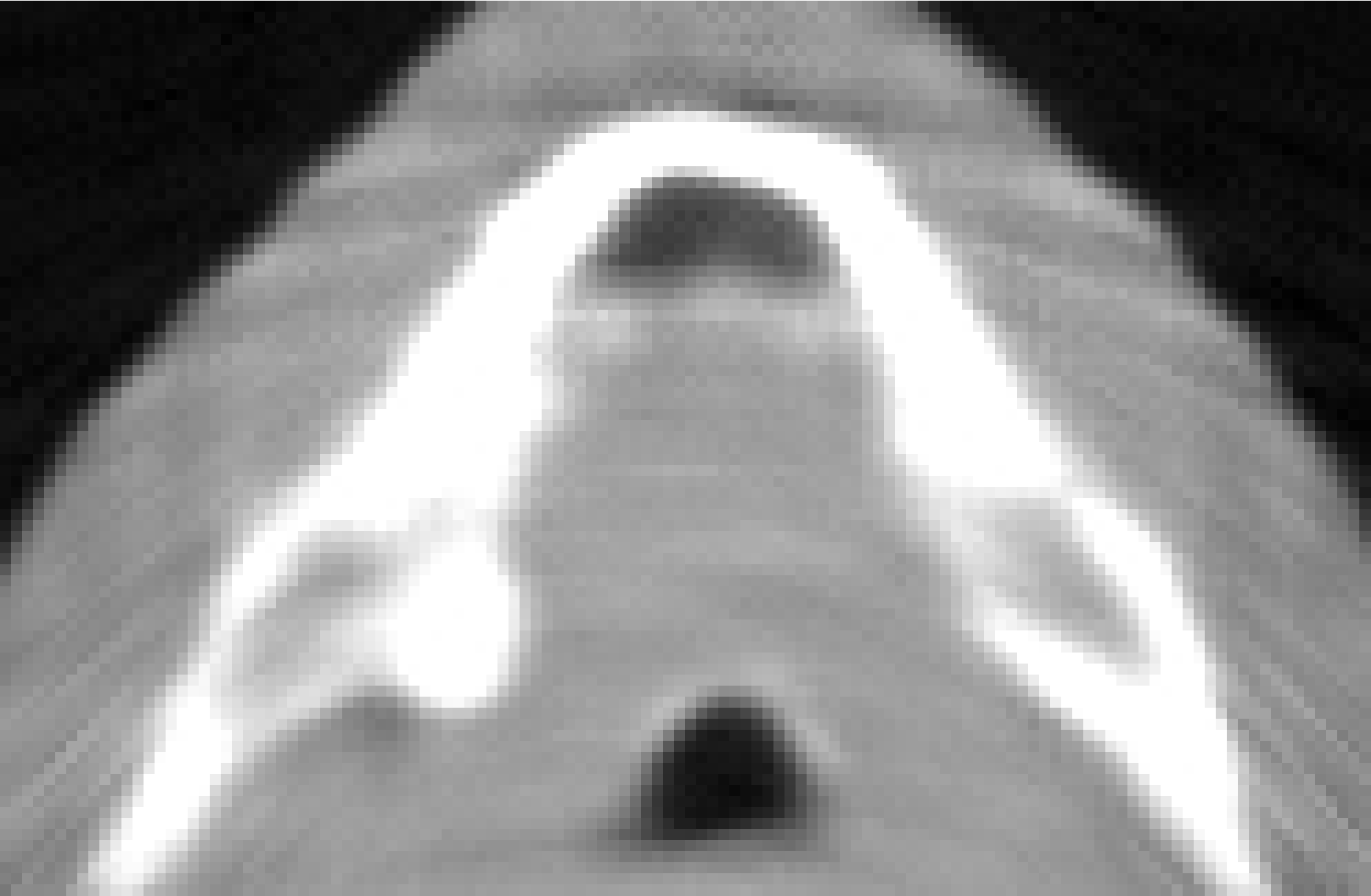}}
				}		
		\end{overpic}}
		\subfloat[Ours]{
			\begin{overpic}[width=\size\columnwidth]{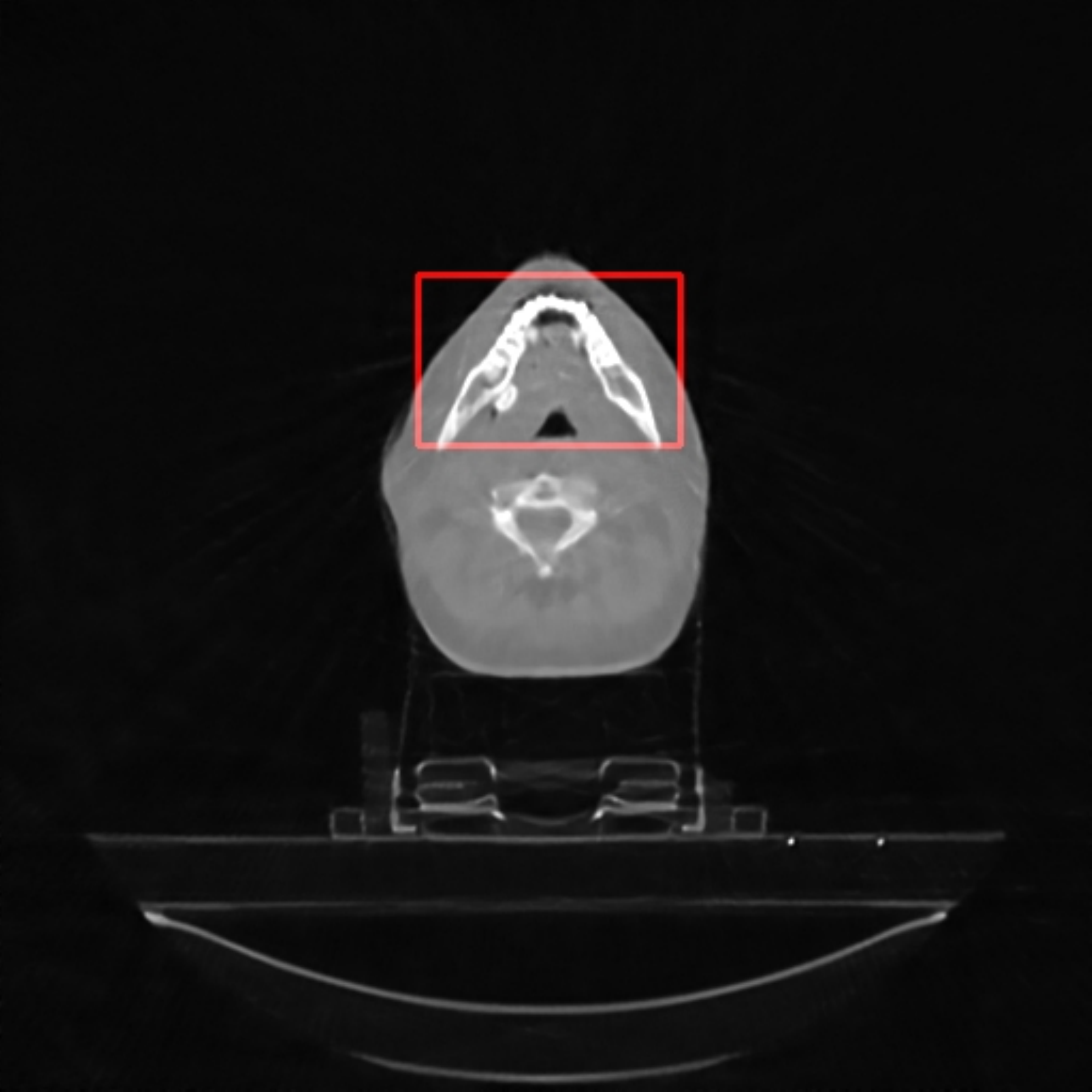}	
				\put(0,0){\color{red}%
					\frame{\includegraphics[scale=0.08]{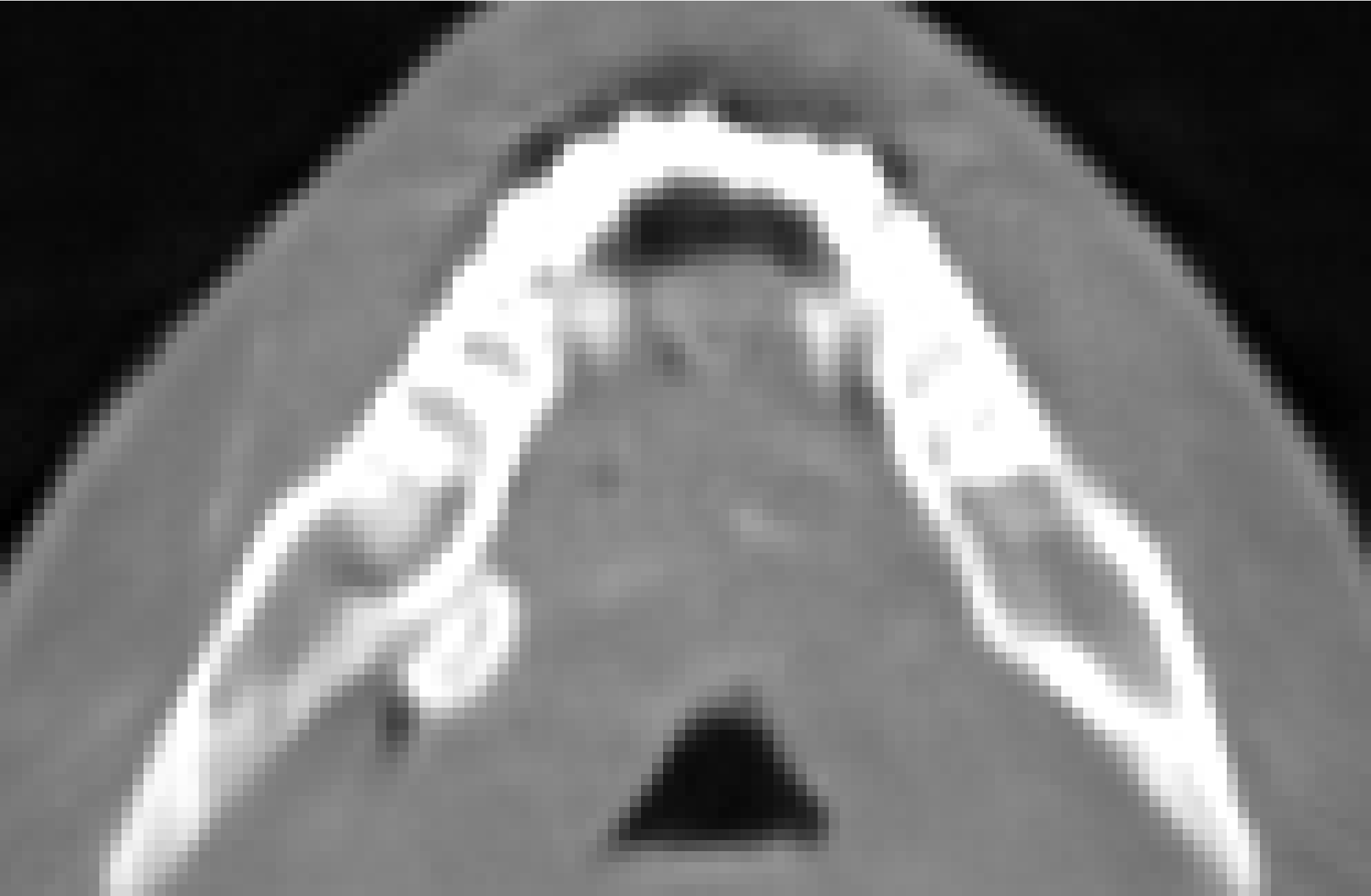}}
				}		
		\end{overpic}}
	}
	\caption{The reconstructed results of head and neck CT data for $p=7\pi$. All images are linearly stretched to [0, 1] and the display window is [0, 0.6].}
	\label{Fig2_neck}
\end{figure*}

\section{Experimental Results}
In this section, we present some simulated experimental results to demonstrate the effectiveness of our method and compare it with the related methods, the nonlocal-TV iterative  algorithm \cite{zhang2013improved} and  the post-processing method FBPConvNet \cite{ISI:000405701500004}, where the convolutional filters in FBPConvNet are extended from 2D to 3D. 

As described in Subsection \ref{Data_preparation}, the CT images in  “the 2016 NIH-AAPM-Mayo Clinic Low Dose CT Grand Challenge” \cite{data} are used to prepare the experimental data, where  the data generated from six patients named 'L192', 'L286', 'L291', 'L310', 'L333' and 'L506' are used as the training data and those from the other four patients named 'L067', 'L096', 'L109' and 'L143' are used as the test data.
The training data involve totally 345 pitches (of size $512\times512\times10$) for $p=7\pi$  and 260 pitches (of size $512\times512\times13$) for $p=14\pi$  and  the test data  involve 215 pitches for $p=7\pi$ and 163 pitches for $p=14\pi$, where $10\%$ of the training data is randomly chosen as the validation.  Note that since the last CT image of one pitch is override with the first CT image of its next pitch, we exclude the last CT image of every pitch in our training and testing data.

For nonlocal-TV iterative  algorithm, we tune its parameters empirically in order to obtain the best results. We use the default parameters of FBPConvNet to train it, where the number of training epoches is 100. 
The CT images reconstructed from the noisy sinograms by the Katsevich algorithm are used as the inputs of 
FBPConvNet, where the batch size is $batch=5$. 
We train our network $60$ epoches with batch size $batch=1$. The size of the learnable parameters of FBPConvNet is about 12.4MB while ours is about 2.3MB. 

\subsection{Experimental Results for $p=7\pi$}
In Fig. \ref{Fig2}, five slices of CT images reconstructed from the test data for $p=7\pi$ are shown. We can observe that the images reconstructed by Katsevich algorithm suffer severe streak artifacts due to the sparsity of the sinogram. The images reconstructed by nonlocal-TV are somewhat over-smoothed and a lot of details are lost. FBPConvNet can reconstruct CT images with some details, but some boundaries in its reconstructed CT images are still blurred. Compared to nonlocal-TV and FBPConvNet, the CT images reconstructed by our method can reconstruct more details and boundaries as shown in Fig. \ref{Fig2_our}. To better observe the differences, the areas marked by the red rectangles in the reconstructed images are enlarged and presented on the left-down positions.

\begin{table}[!t]
	\renewcommand{\arraystretch}{1.3}
	\caption{The RMSE and SSIM of the four methods for NIH-AAPM-Mayo test data.}
	\label{T2}
	\centering
	\begin{tabular}{cccc}
		\hline
		Pitch & Method & RMSE & SSIM \\
		& Katsevich & 97.616$\pm$7.549 & 0.777$\pm$0.028 \\
		$p=7\pi$ & Nonlocal-TV &93.174$\pm$7.271  & 0.815$\pm$0.028 \\
		& FBPConvNet &  70.329$\pm$8.169& 0.845$\pm$0.028 \\
		& Ours & \textbf{59.068}$\pm$7.028 &\textbf{0.867}$\pm$0.027  \\
		& Katsevich & 97.783$\pm$7.493 & 0.776$\pm$0.028 \\
		$p=14\pi$ & Nonlocal-TV &93.362$\pm$7.217  &0.814$\pm$0.028  \\
		& FBPConvNet &71.609$\pm$8.140  &0.842$\pm$0.028  \\
		& Ours & \textbf{60.028}$\pm$6.988 & \textbf{0.865}$\pm$0.027 \\
		\hline
	\end{tabular}
\end{table}%

To quantitatively evaluate the performances of the compared methods, two metrics, the root-mean-square-error (RMSE) and the structural-similarity (SSIM) are  used to measure the similarity of the reconstructed images and the labels. The definition of RMSE is
\begin{equation}
	\operatorname{RMSE}=\sqrt{\frac{\|f-f_{GT}\|_{2}^{2}}{n}},
\end{equation}
where $f$ and $f_{GT}$ are, respectively, the reconstructed CT image and label image, $n$ is the number of pixels.
The value of SSIM is defined as
\begin{equation}
	\operatorname{SSIM}=\frac{\left(2 \mu_{1} \mu_{2}+c_{1}\right)\left(2 \sigma_{1,2}+c_{2}\right)}{\left(\mu_{1}^{2}+\mu_{2}^{2}+c_{1}\right)\left(\sigma_{1}^{2}+\sigma_{2}^{2}+c_{2}\right)},
\end{equation}
where 
\begin{equation}
	\begin{aligned}
		C_{1}&=(0.01 \times(\max (f_{GT})-\min (f_{GT})))^{2}\\
		C_{2}&=(0.03 \times(\max (f_{GT})-\min (f_{GT})))^{2},
	\end{aligned}
\end{equation}
$\mu_{1}$, $\sigma_1$ are the mean and variance of the reconstructed CT image $f$,  $\mu_{2}$,  $\sigma_2$ are the mean and variance of the  referenced image $f_{GT}$, respectively, $\sigma_{1,2}$ is the covariance between $f$ and $f_{GT}$,

In Table \ref{T2}, the average  RMSE and SSIM of the CT images reconstructed by the four methods  are listed. It can be observed that our network  gains the lowest RMSE and highest SSIM in average, which coincides with our subjective evaluation.

\begin{figure*}[htbp]	
	\centering{	
		\newcommand{\id}{10}
		\subfloat{
			\begin{overpic}[width=\size\columnwidth,percent]{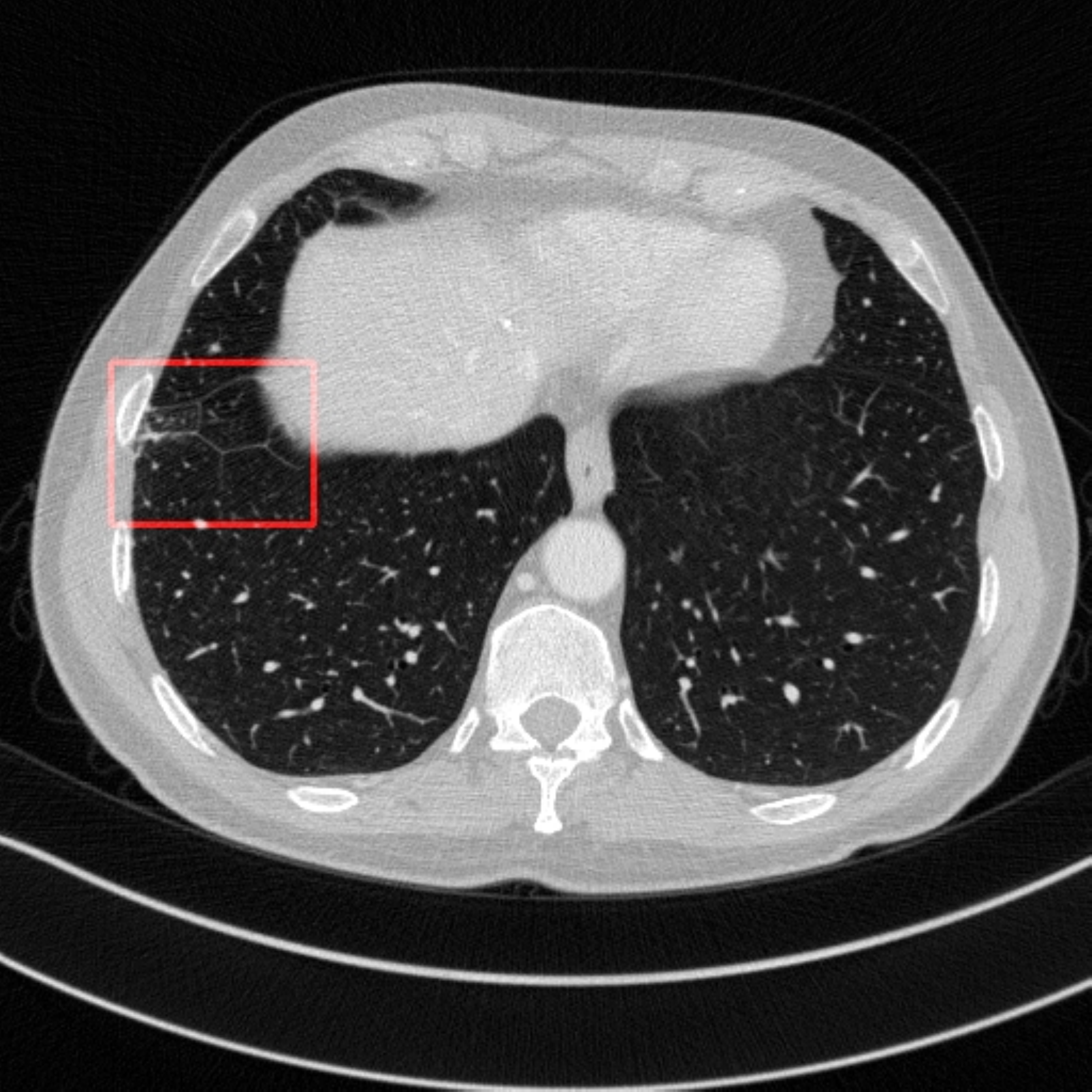}	
				\put(0,0){\color{red}%
					\frame{\includegraphics[scale=0.08]{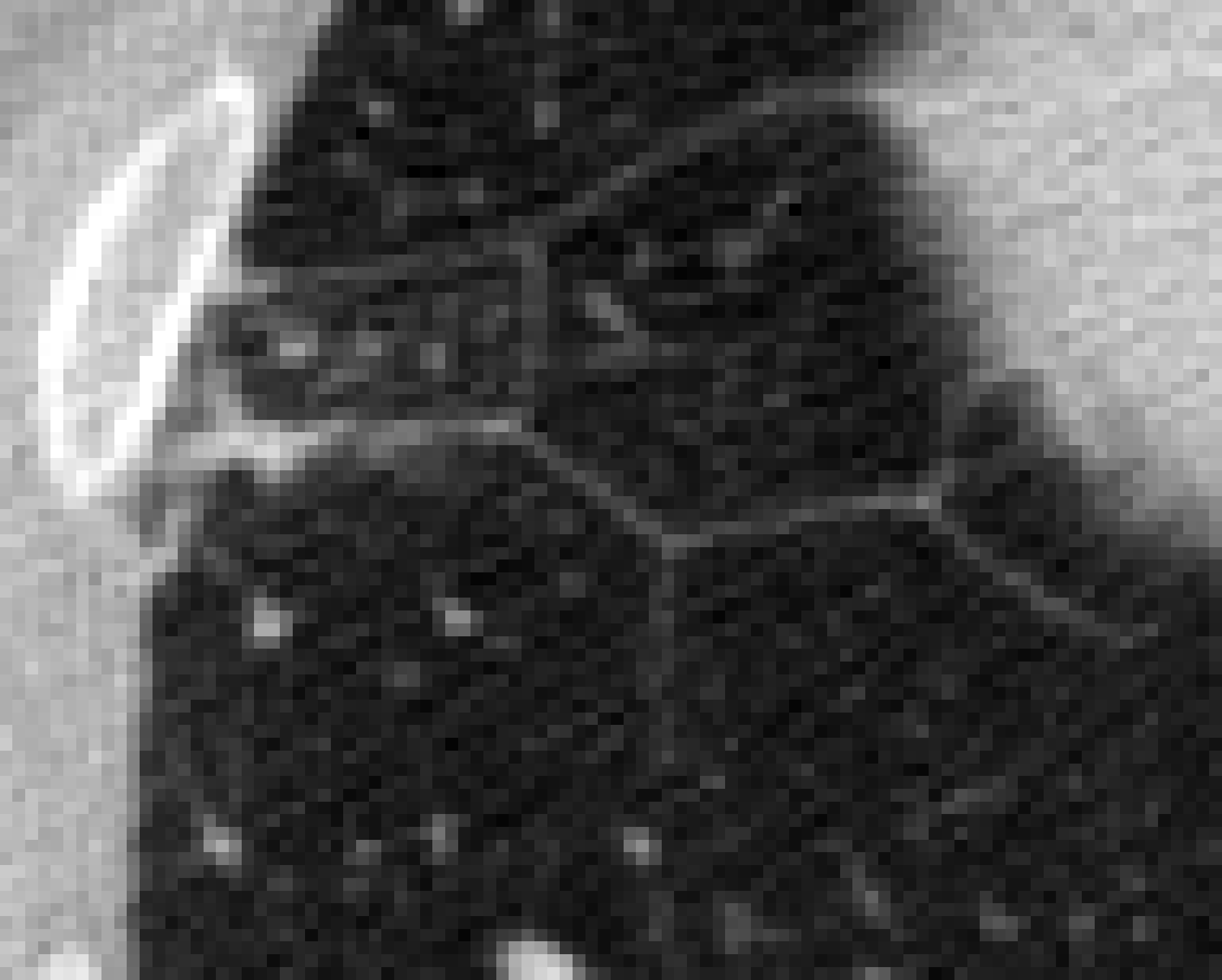}}
				}		
		\end{overpic}}
		\subfloat{
			\begin{overpic}[width=\size\columnwidth]{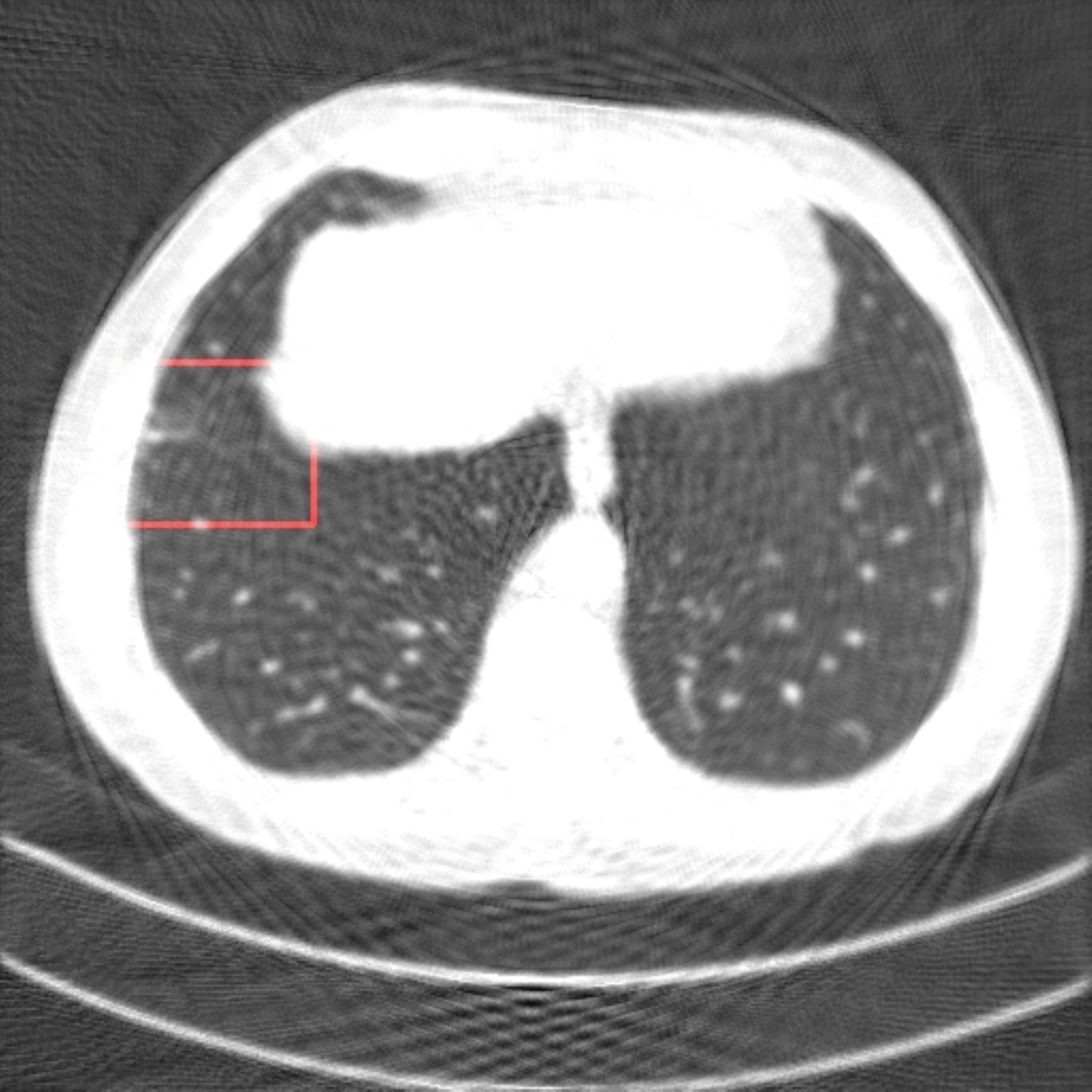}	
				\put(0,0){\color{red}%
					\frame{\includegraphics[scale=0.08]{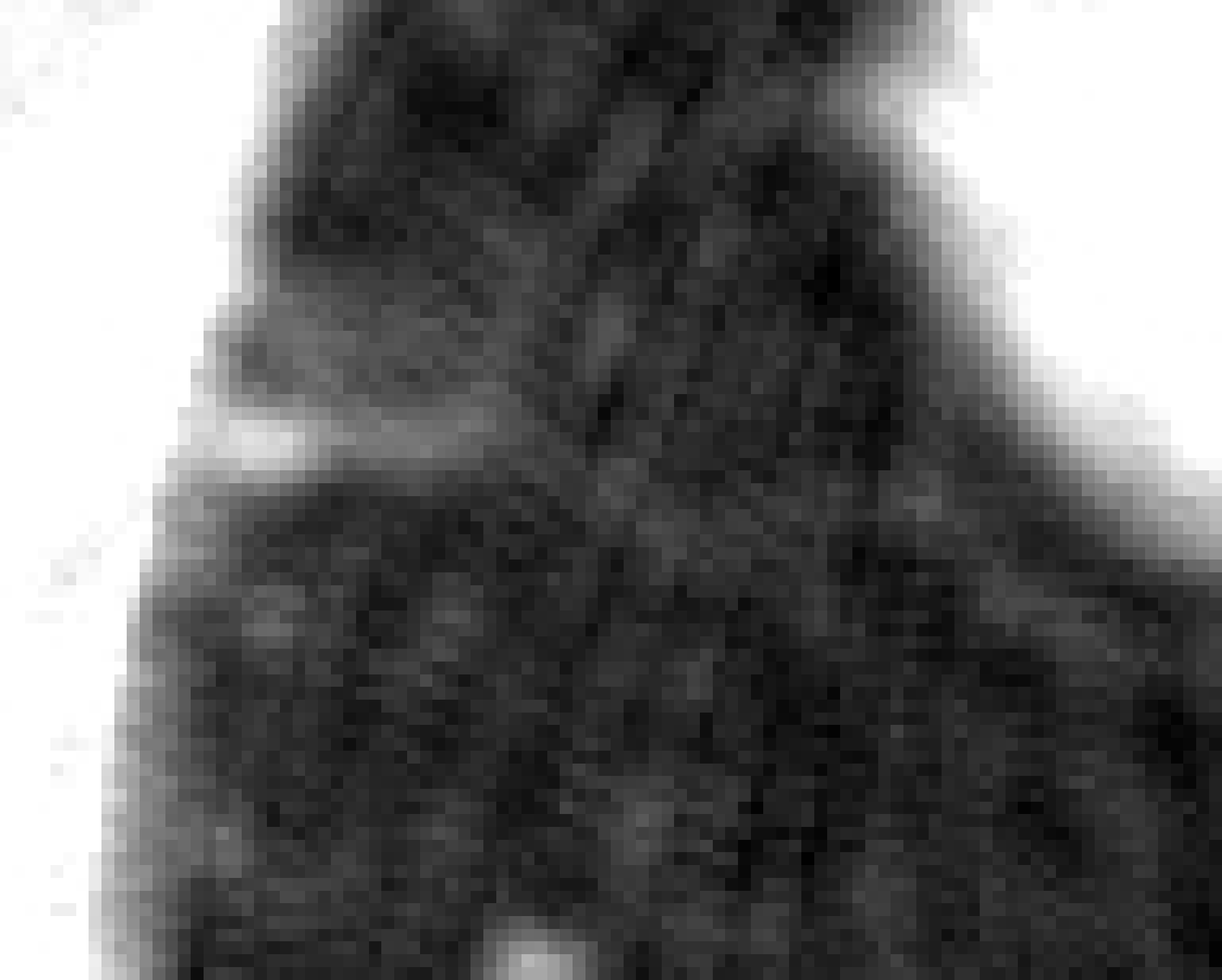}}
				}		
		\end{overpic}}
		\subfloat{
			\begin{overpic}[width=\size\columnwidth]{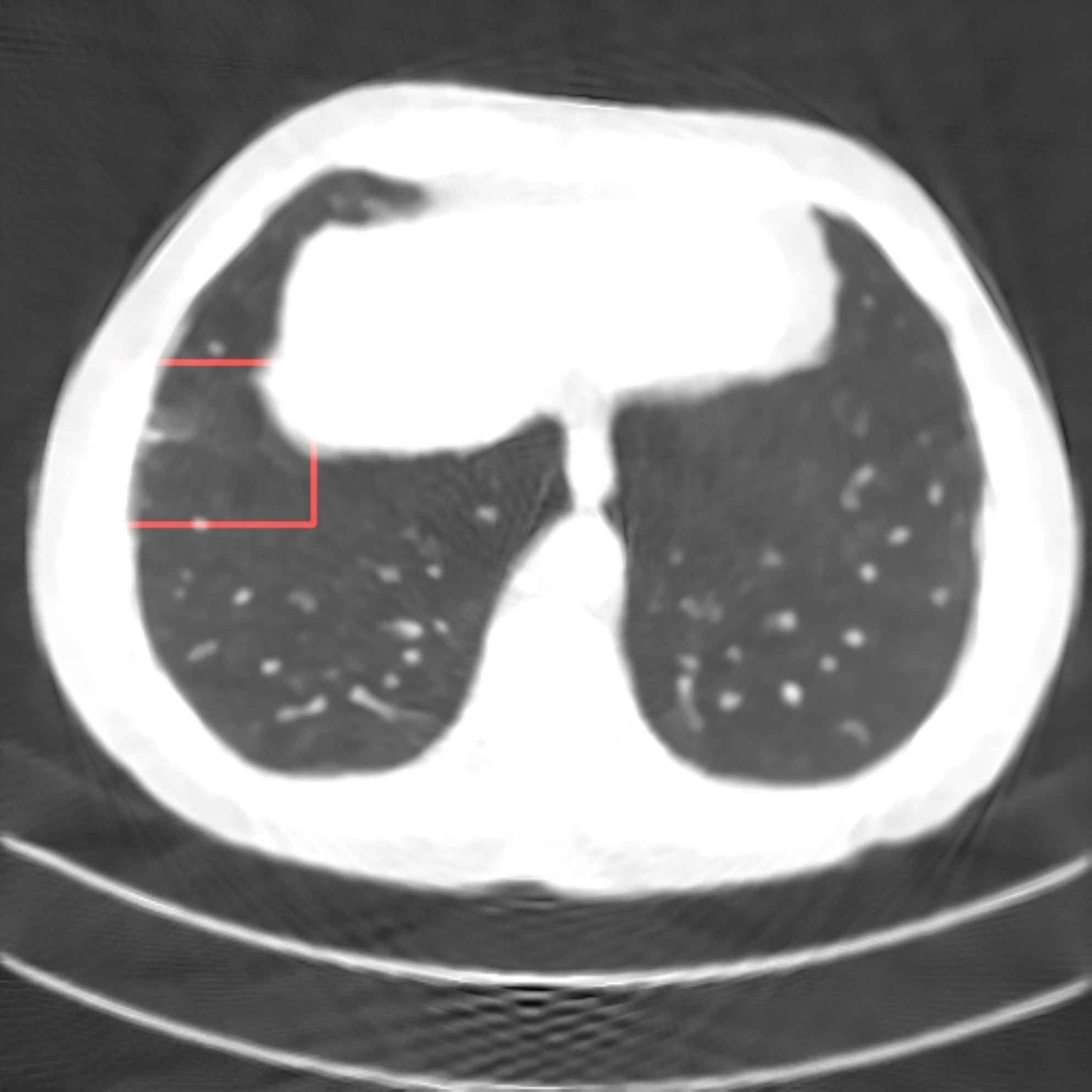}	
				\put(0,0){\color{red}%
					\frame{\includegraphics[scale=0.08]{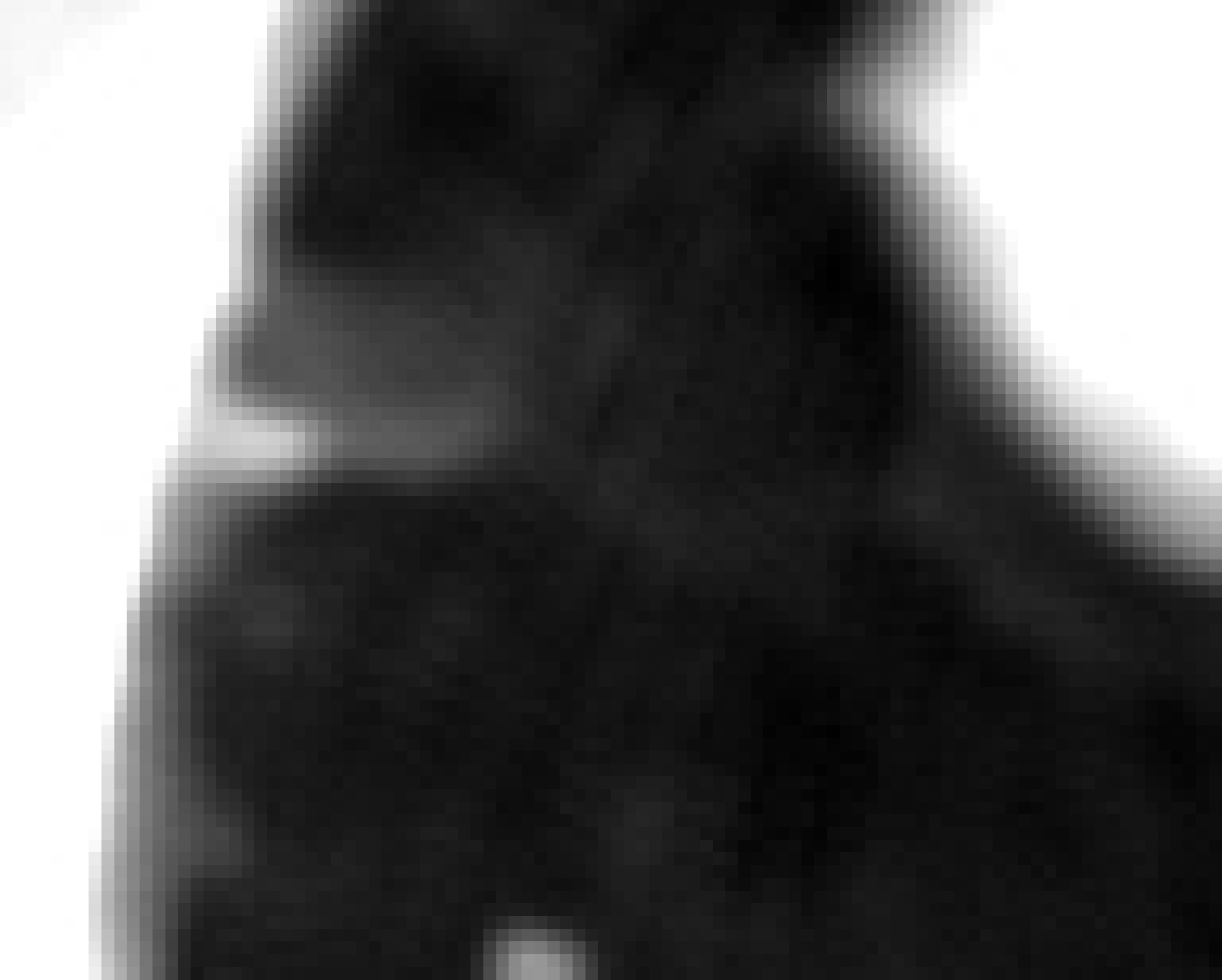}}
				}		
		\end{overpic}}
		\subfloat{
			\begin{overpic}[width=\size\columnwidth]{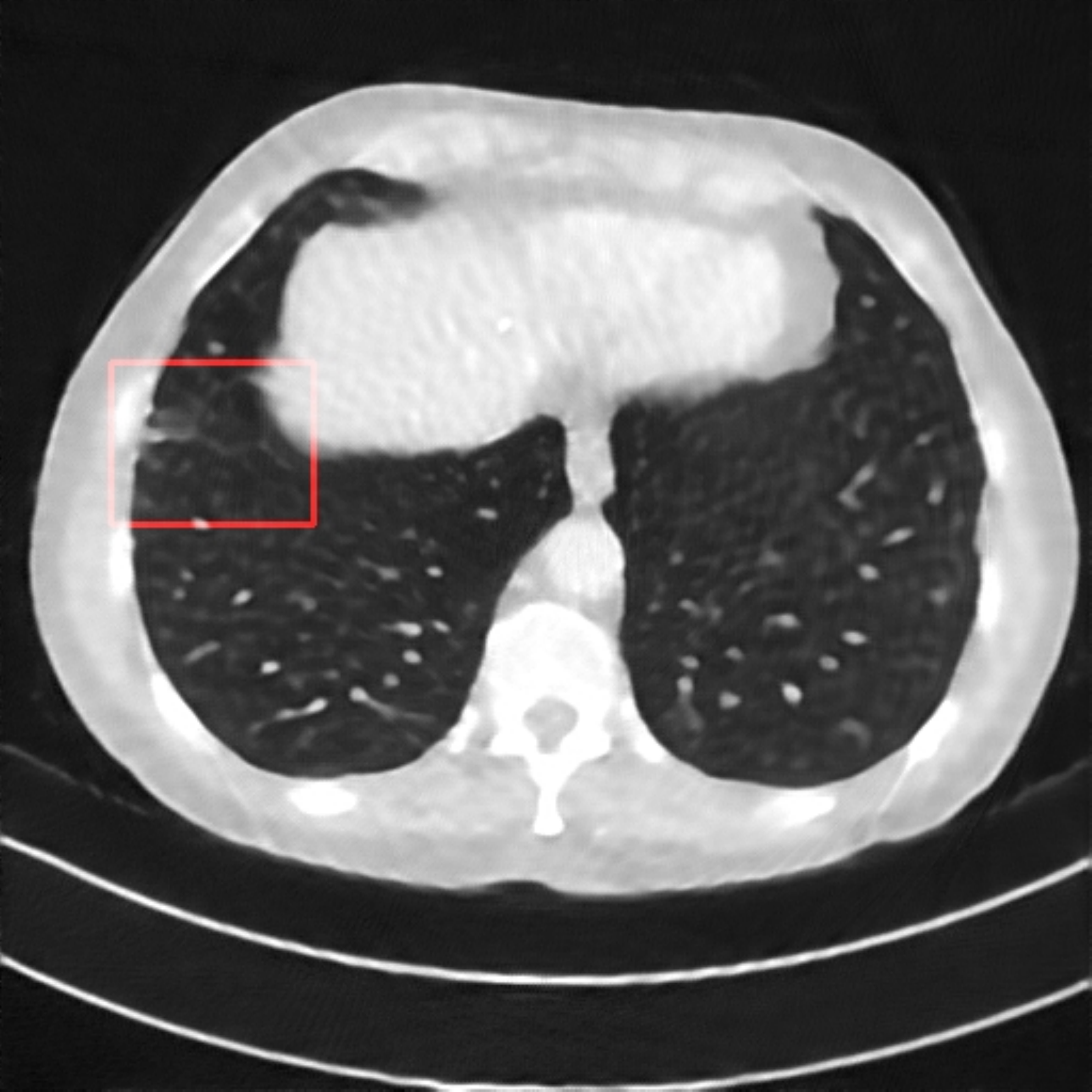}	
				\put(0,0){\color{red}%
					\frame{\includegraphics[scale=0.08]{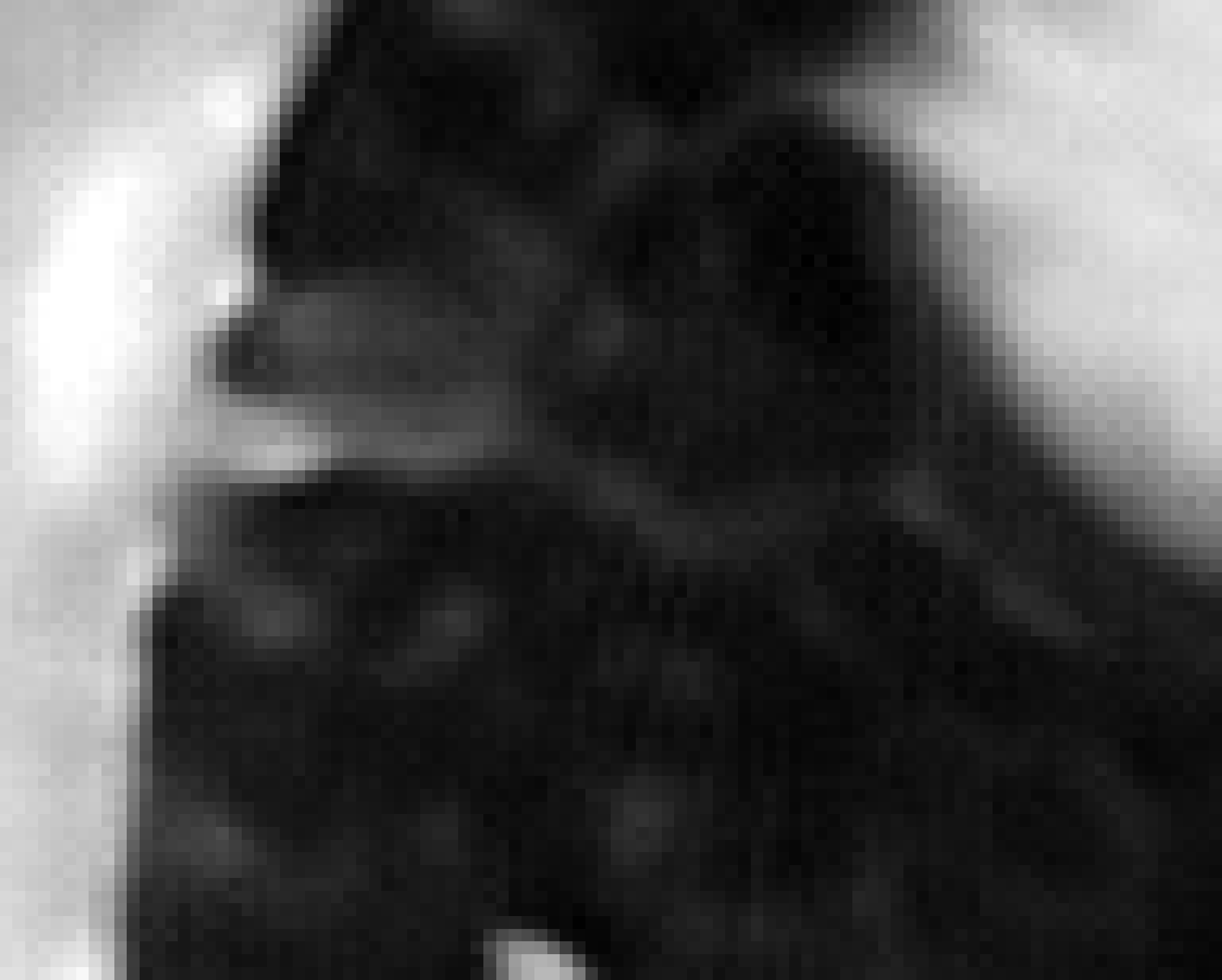}}
				}		
		\end{overpic}}
		\subfloat{
			\begin{overpic}[width=\size\columnwidth]{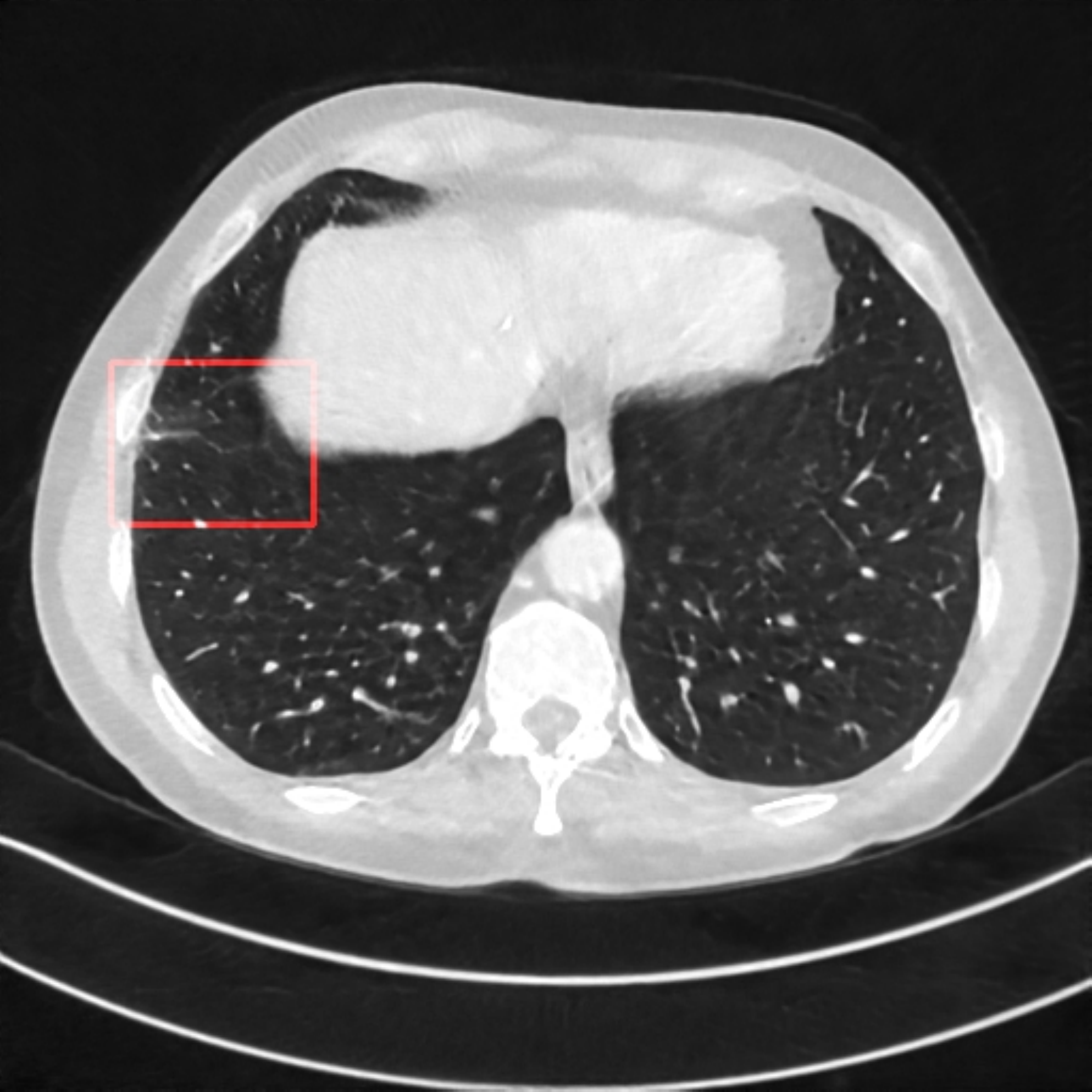}	
				\put(0,0){\color{red}%
					\frame{\includegraphics[scale=0.08]{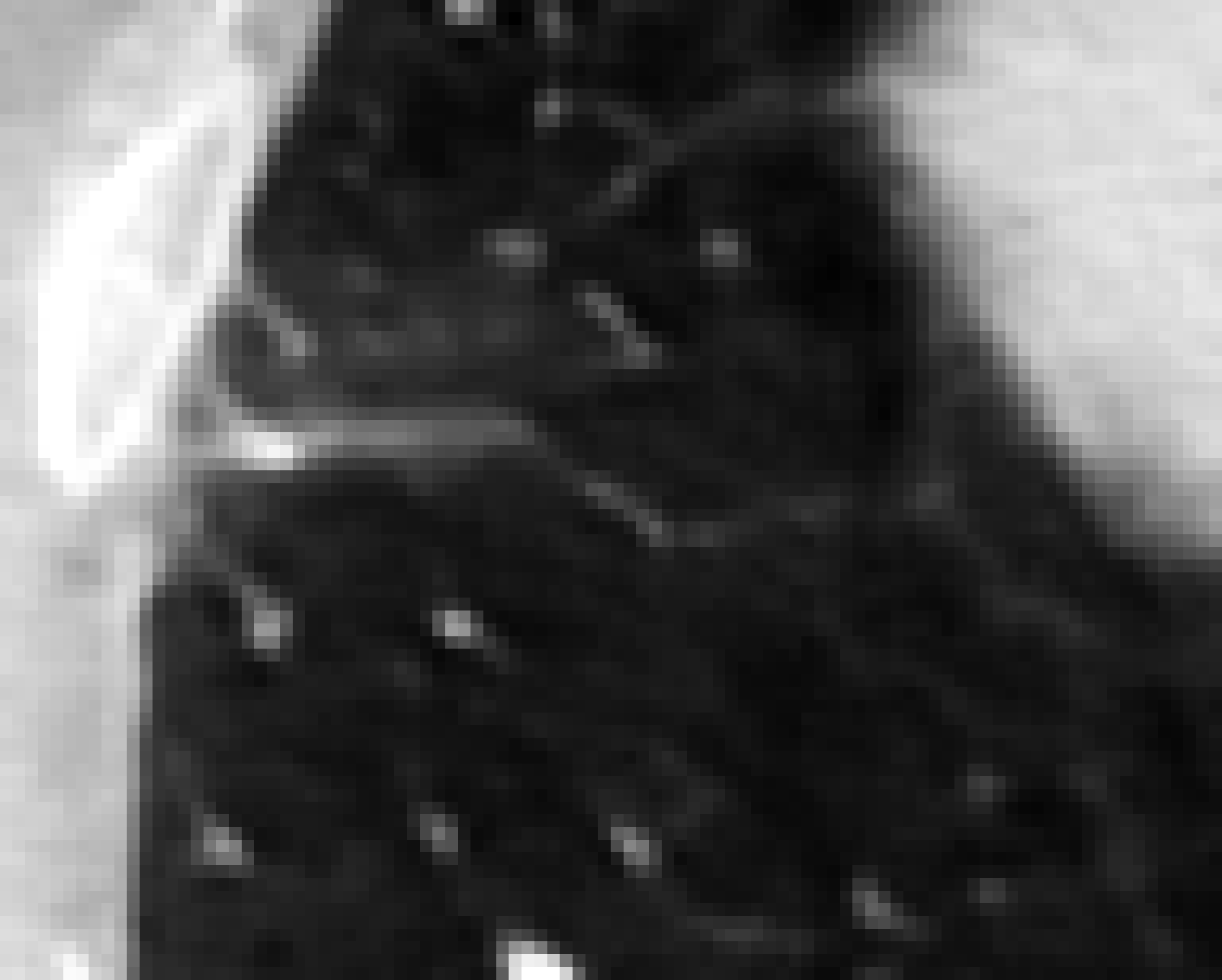}}
				}		
		\end{overpic}}
		\vspace{-3mm}
		\renewcommand{\id}{200}
		\subfloat{
			\begin{overpic}[width=\size\columnwidth,percent]{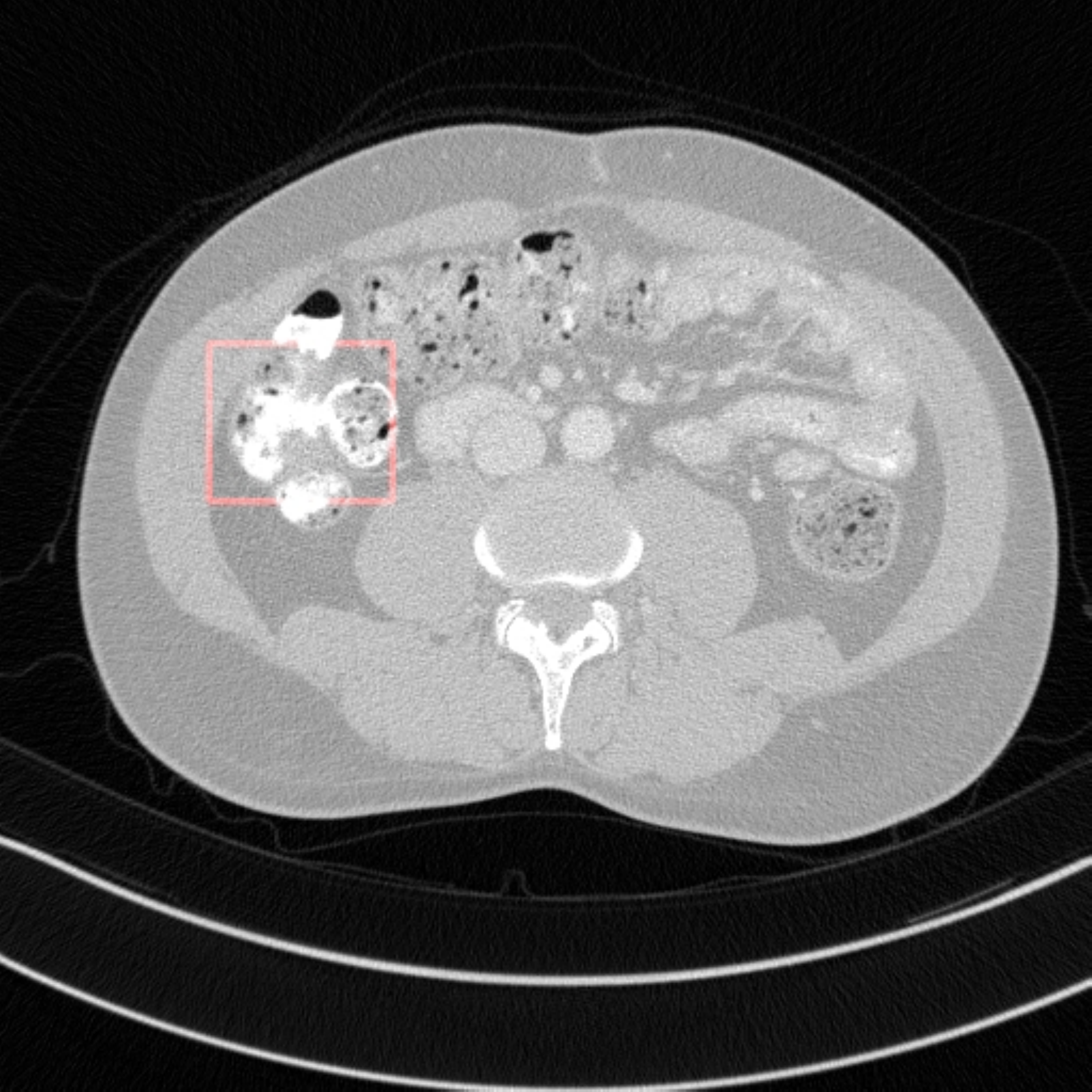}	
				\put(0,0){\color{red}%
					\frame{\includegraphics[scale=0.08]{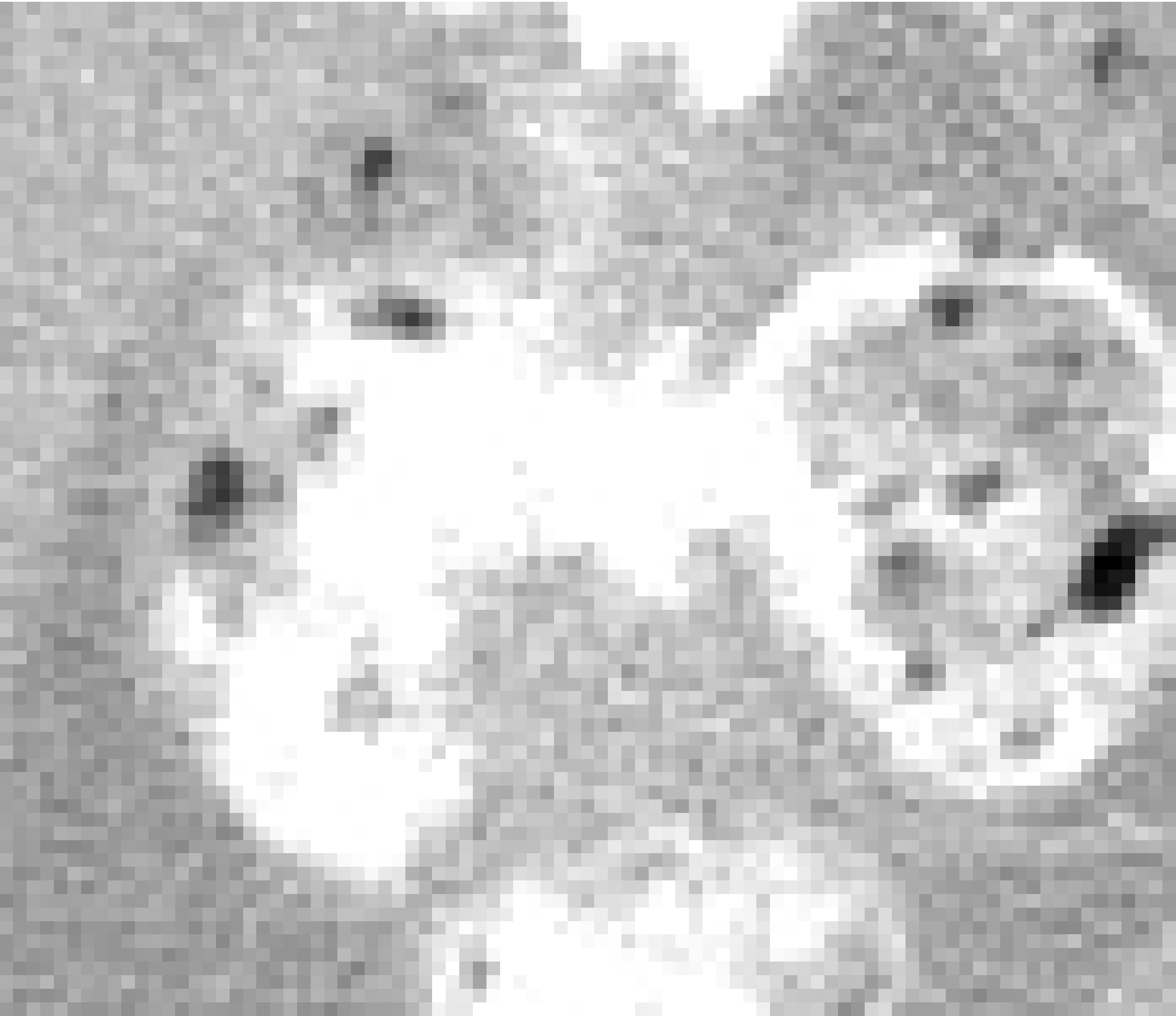}}
				}		
		\end{overpic}}
		\subfloat{
			\begin{overpic}[width=\size\columnwidth]{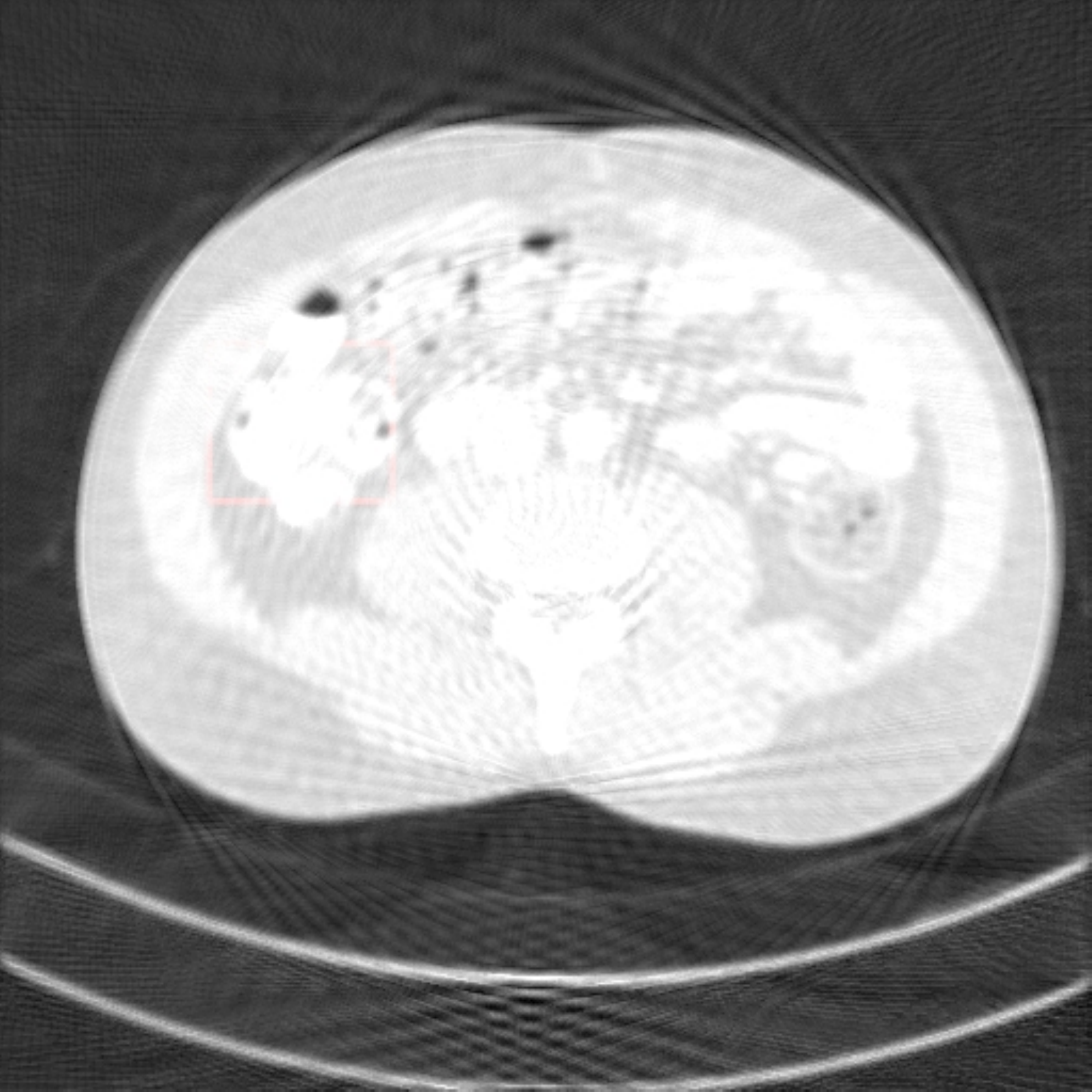}	
				\put(0,0){\color{red}%
					\frame{\includegraphics[scale=0.08]{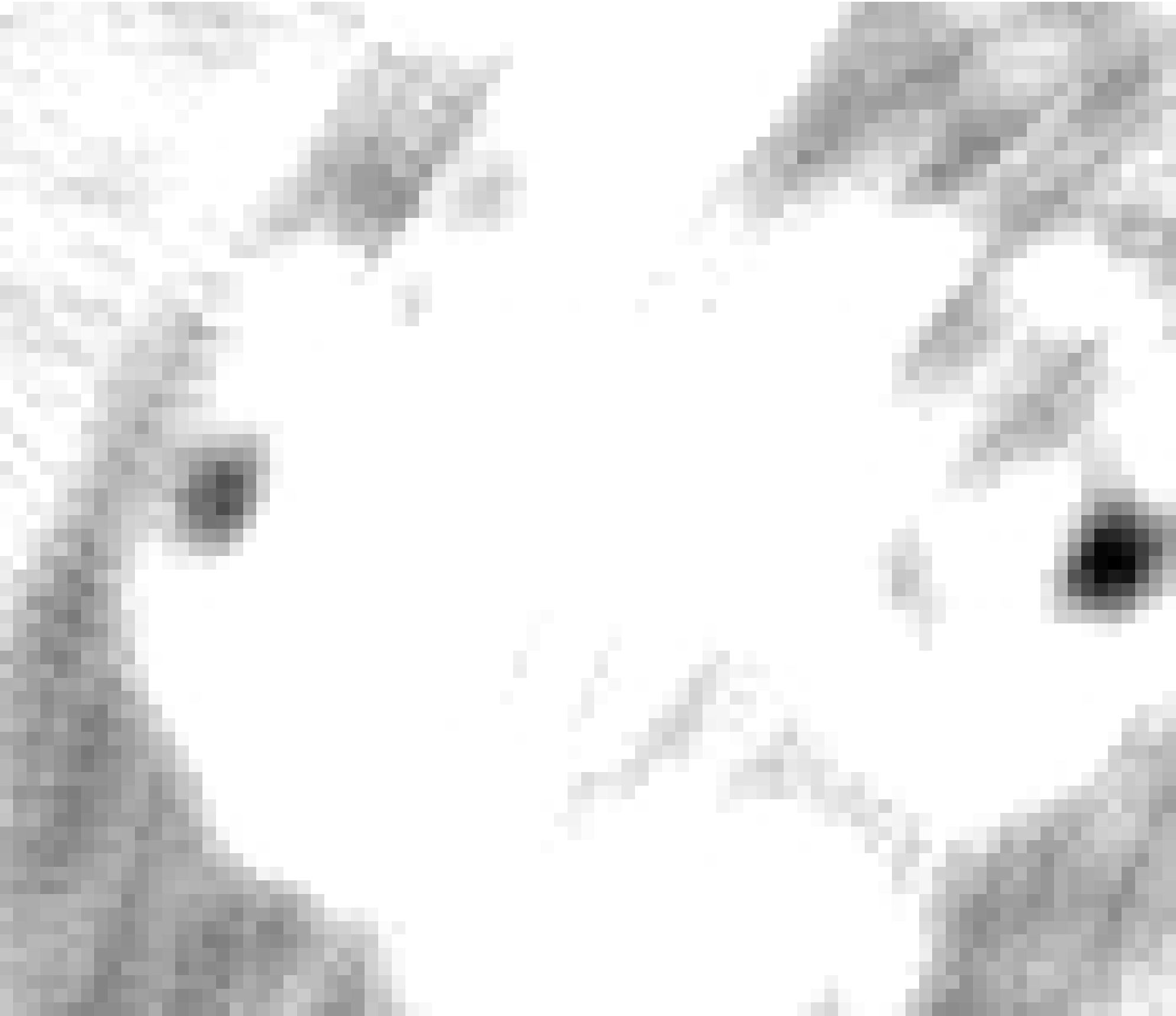}}
				}		
		\end{overpic}}
		\subfloat{
			\begin{overpic}[width=\size\columnwidth]{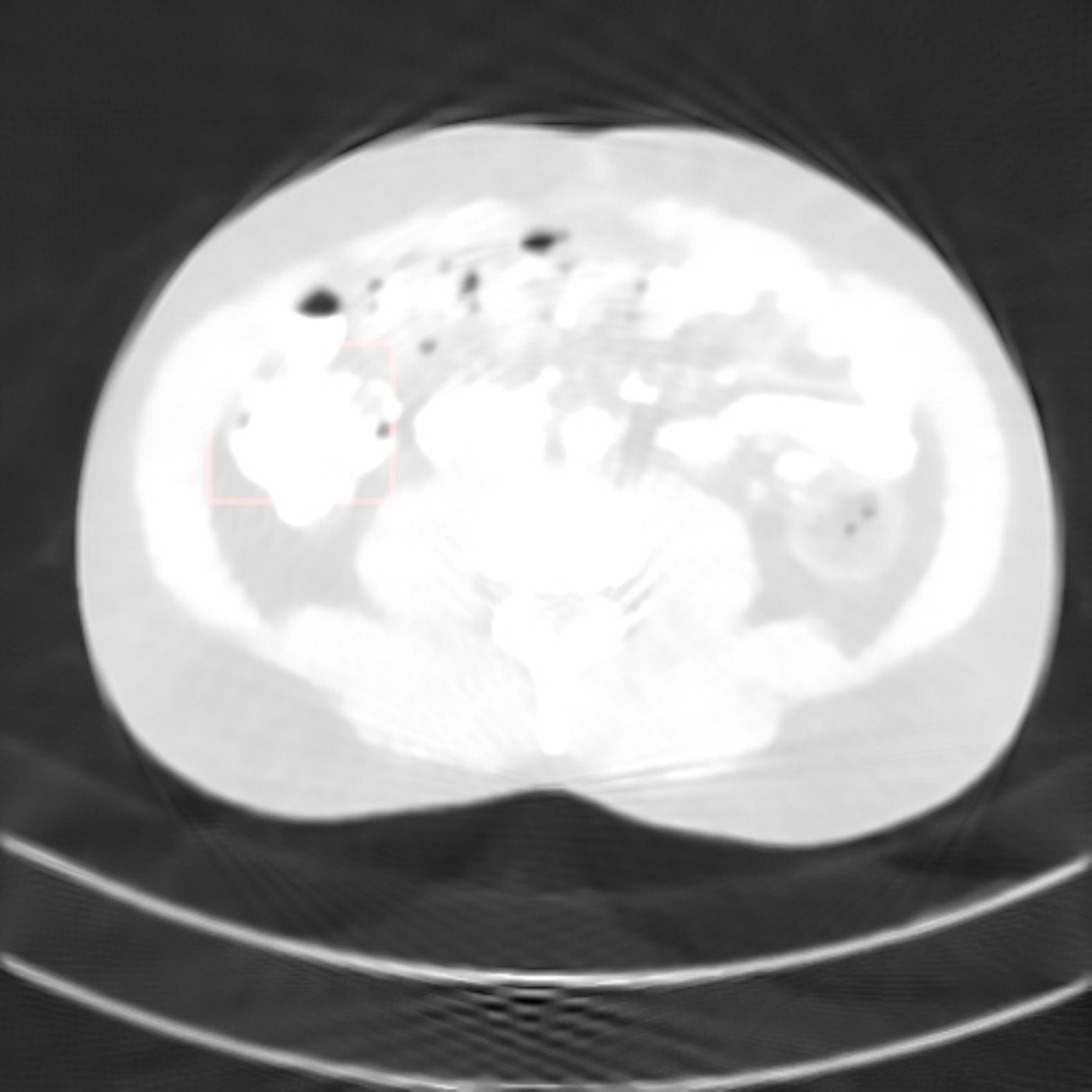}	
				\put(0,0){\color{red}%
					\frame{\includegraphics[scale=0.08]{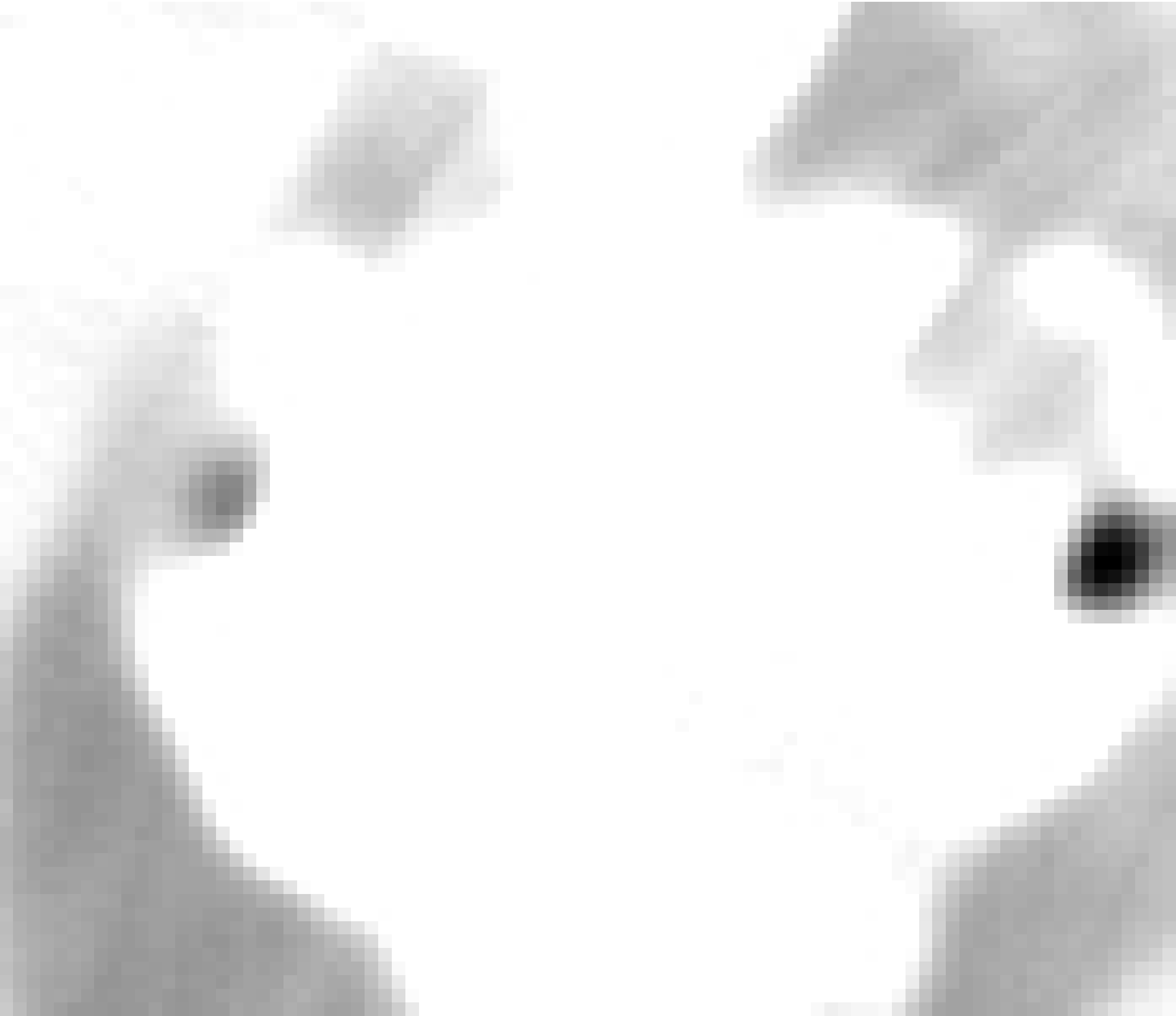}}
				}		
		\end{overpic}}
		\subfloat{
			\begin{overpic}[width=\size\columnwidth]{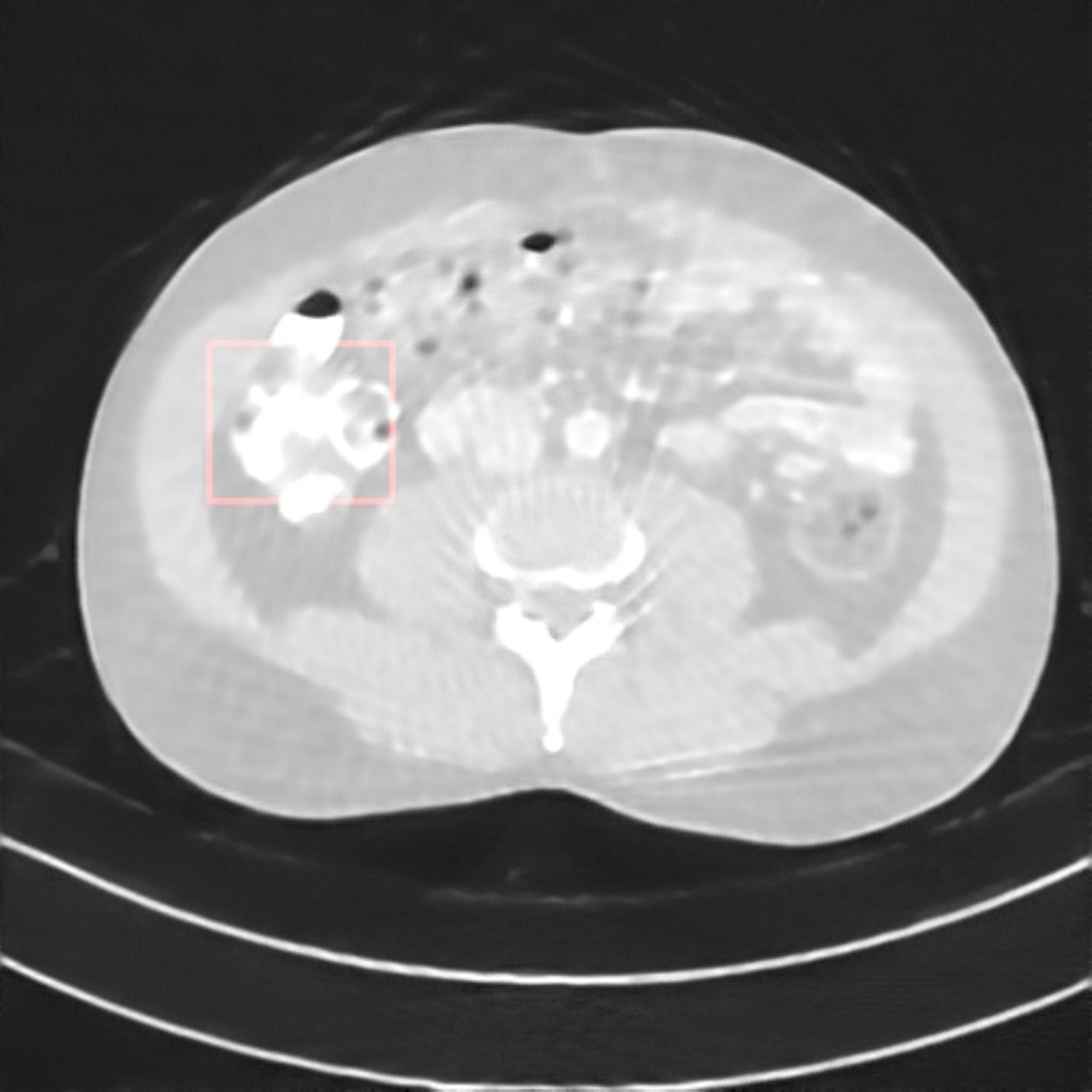}	
				\put(0,0){\color{red}%
					\frame{\includegraphics[scale=0.08]{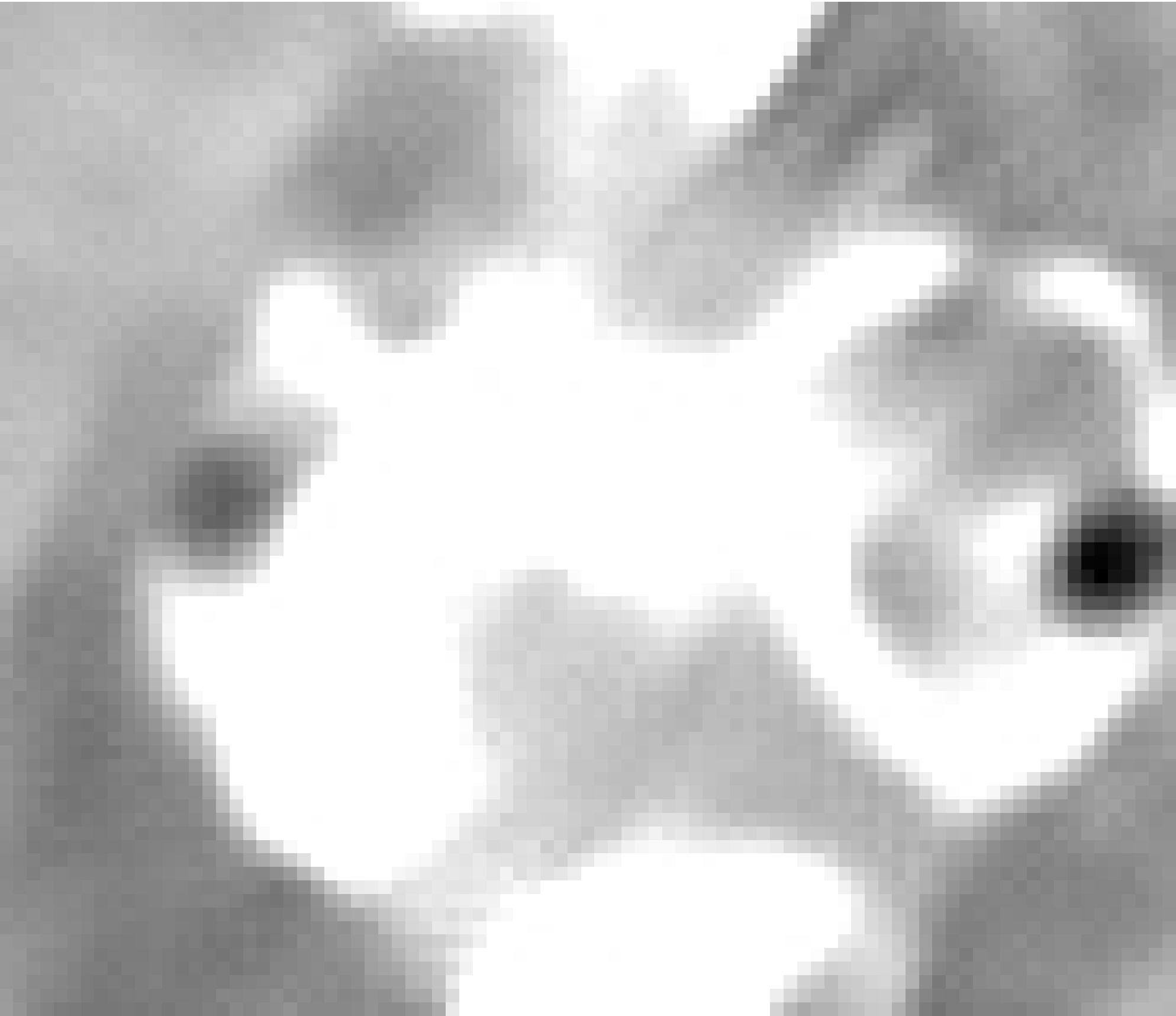}}
				}		
		\end{overpic}}
		\subfloat{
			\begin{overpic}[width=\size\columnwidth]{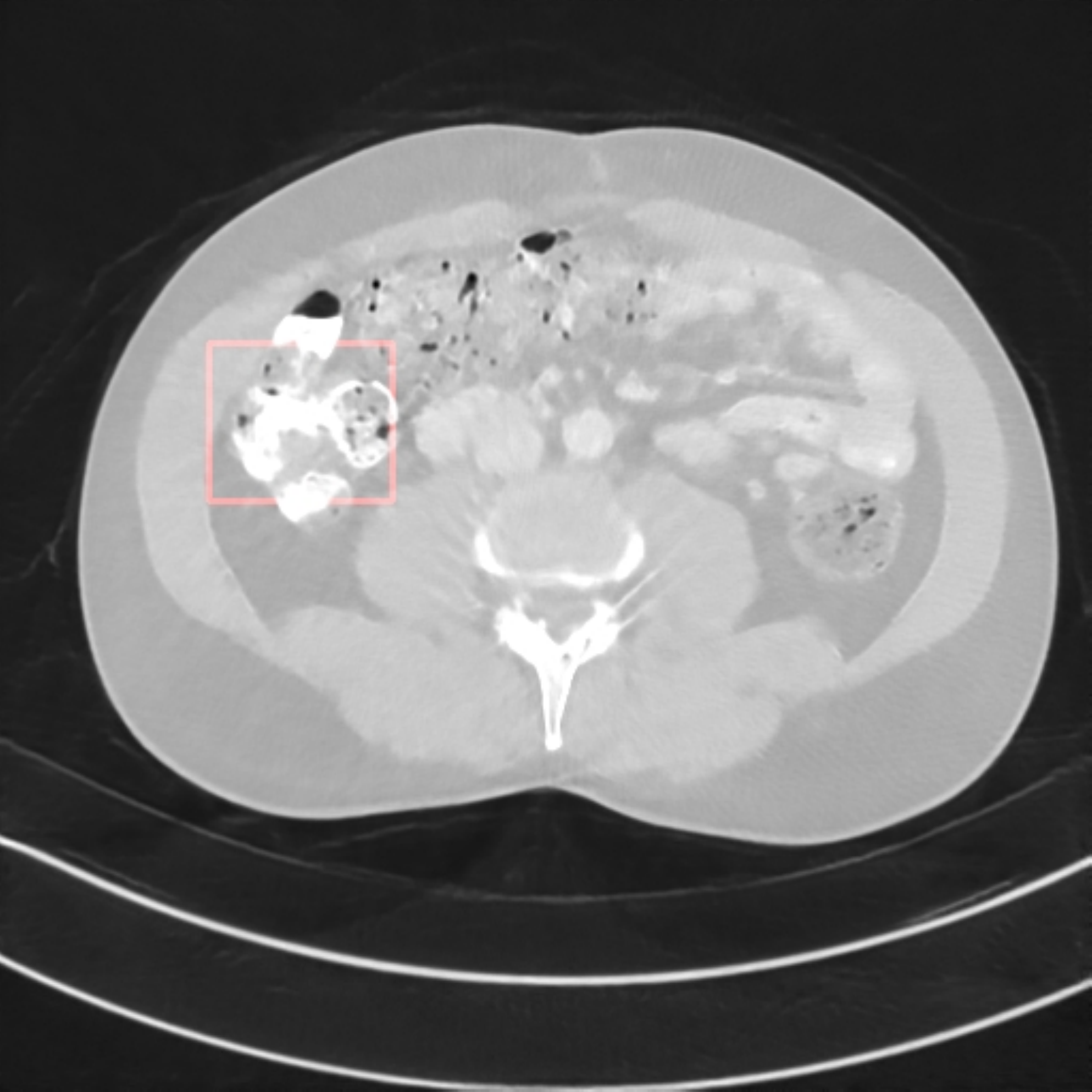}	
				\put(0,0){\color{red}%
					\frame{\includegraphics[scale=0.08]{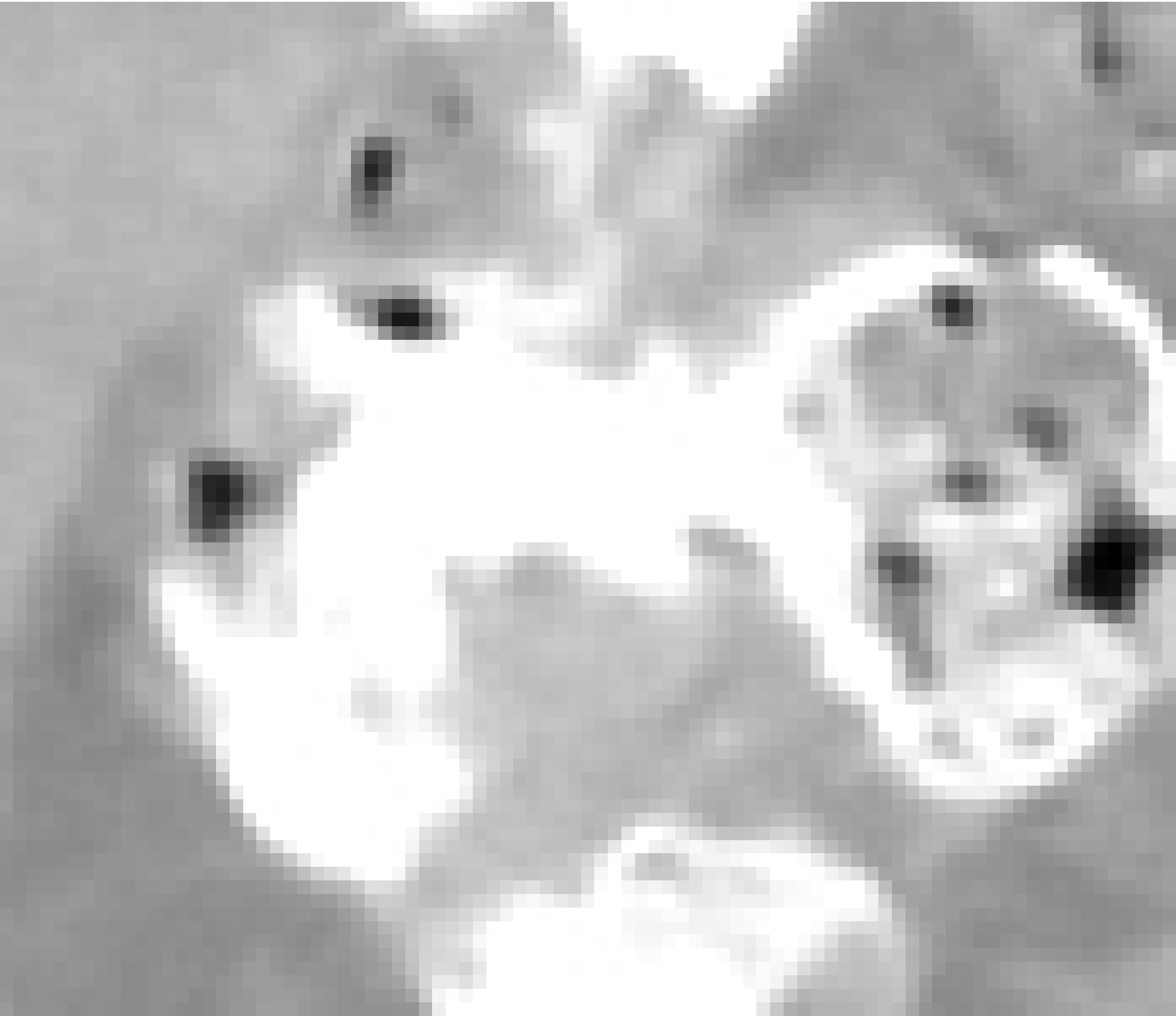}}
				}		
		\end{overpic}}
		\vspace{-3mm}
		\renewcommand{\id}{400}
		\subfloat{
			\begin{overpic}[width=\size\columnwidth,percent]{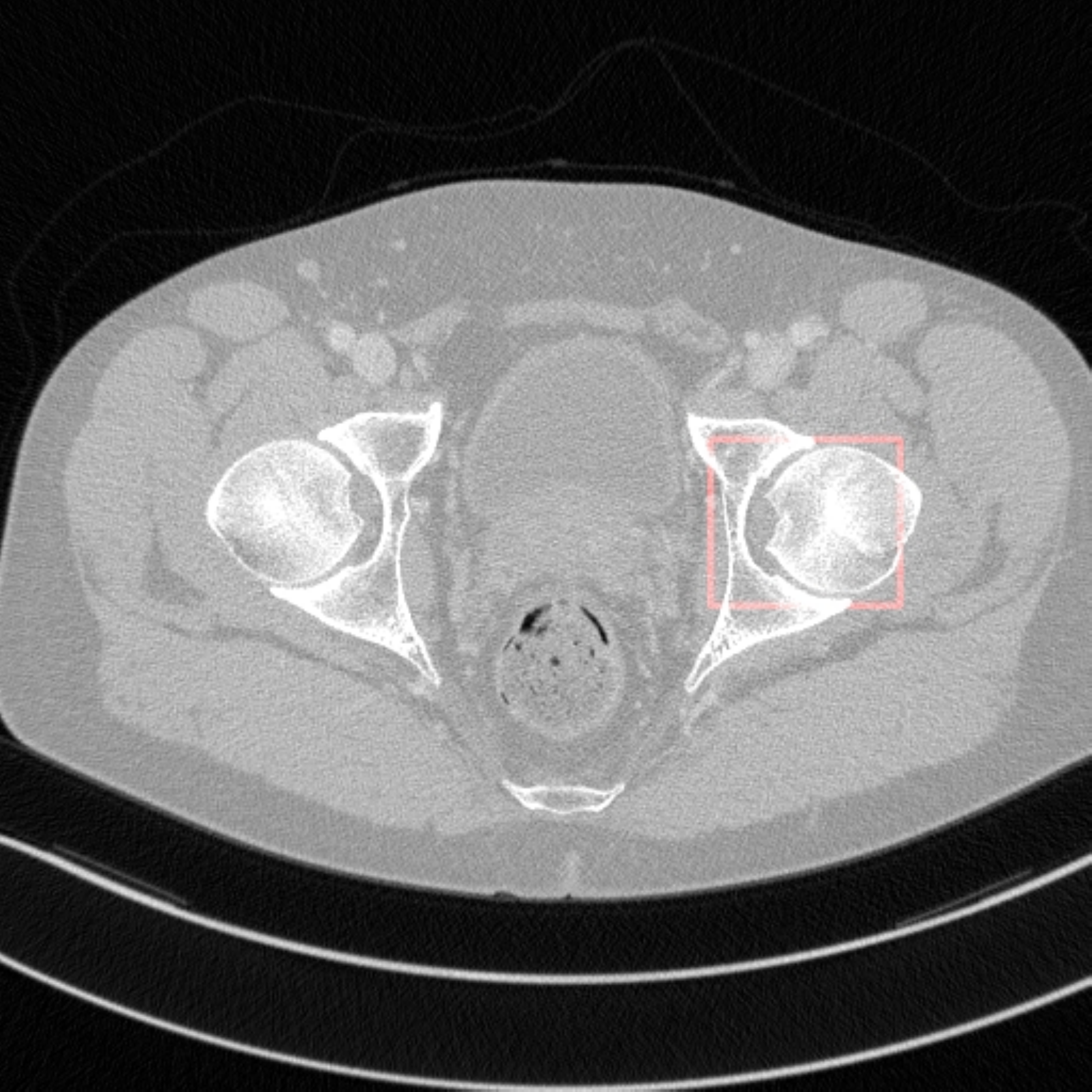}	
				\put(0,0){\color{red}%
					\frame{\includegraphics[scale=0.08]{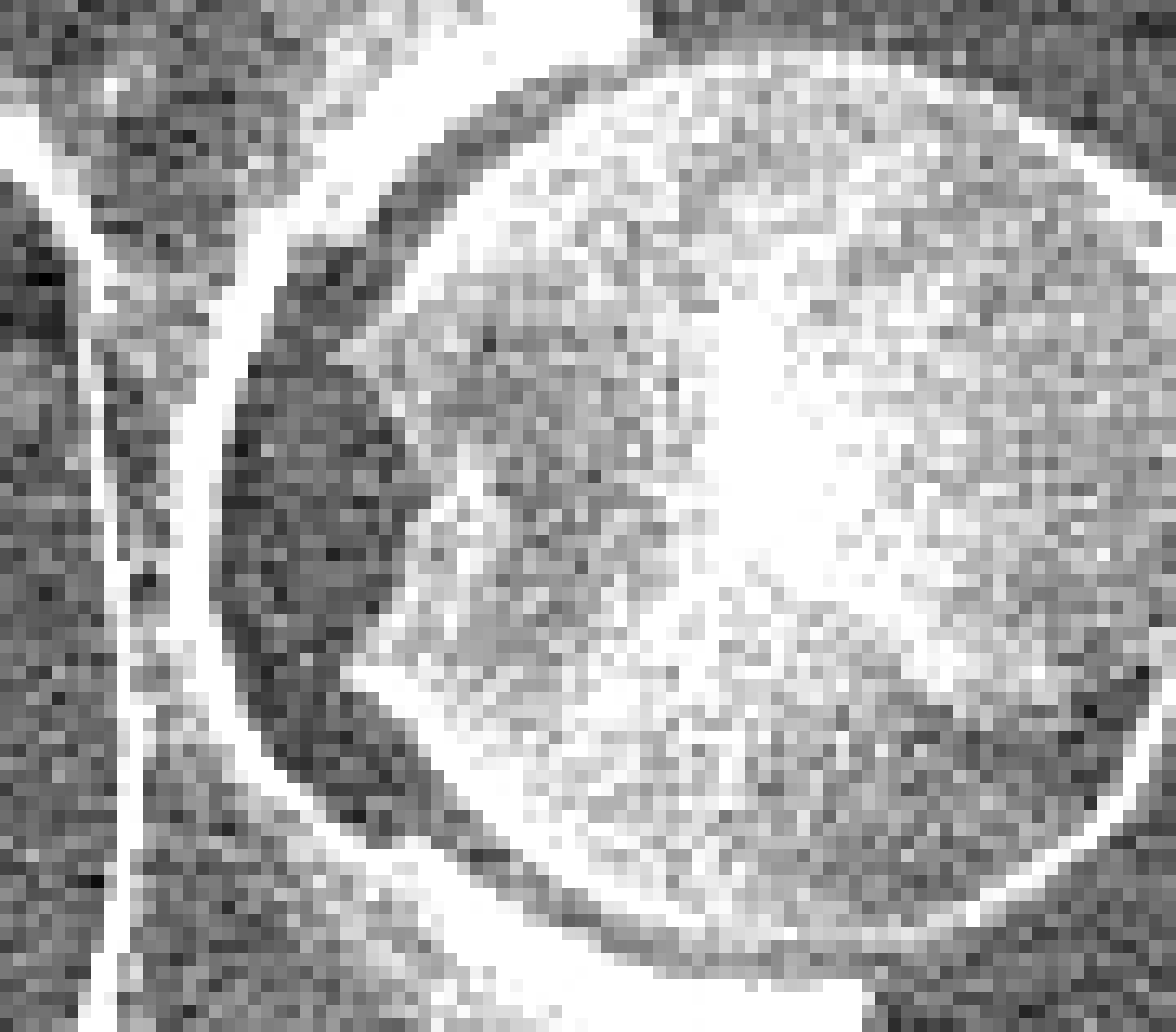}}
				}		
		\end{overpic}}
		\subfloat{
			\begin{overpic}[width=\size\columnwidth]{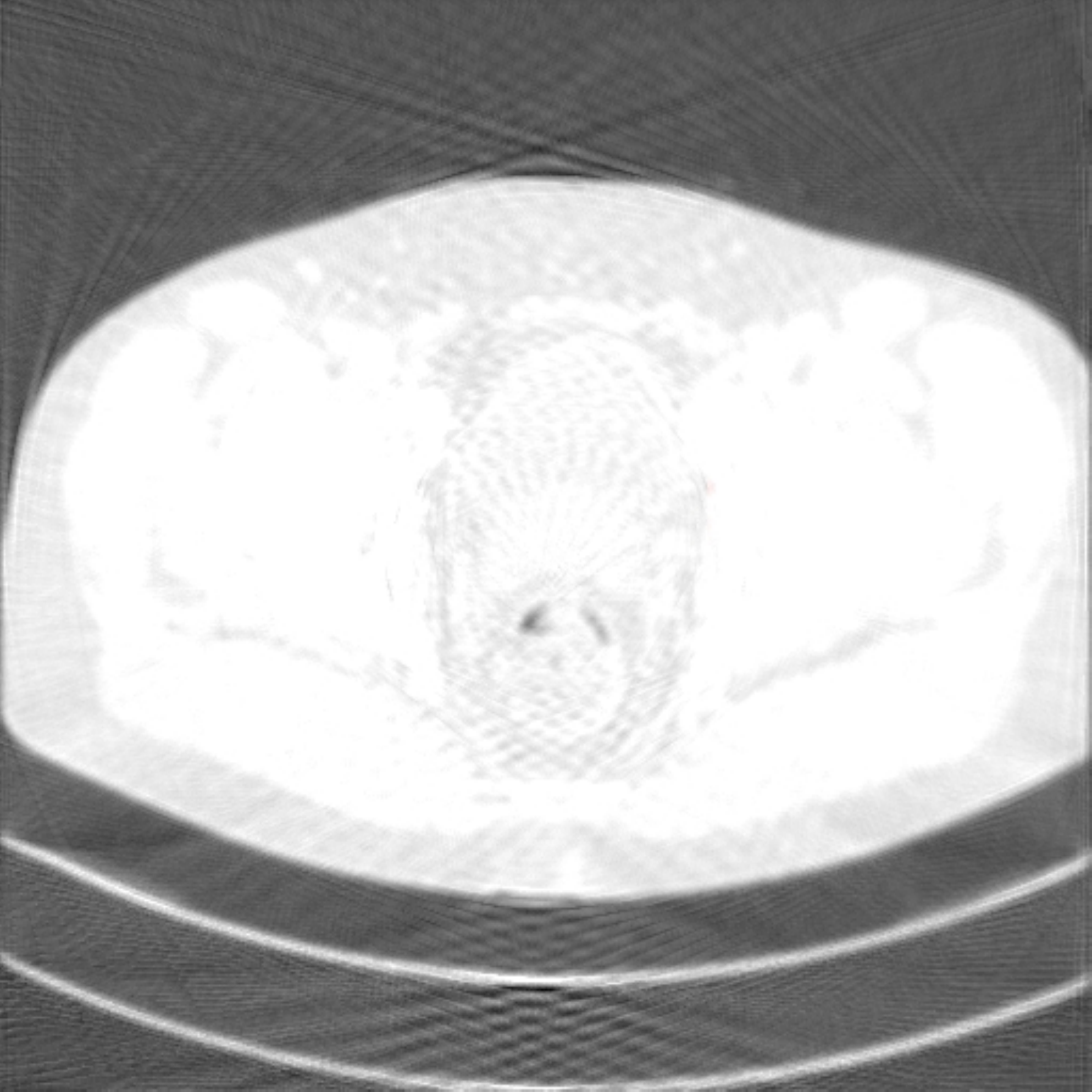}	
				\put(0,0){\color{red}%
					\frame{\includegraphics[scale=0.08]{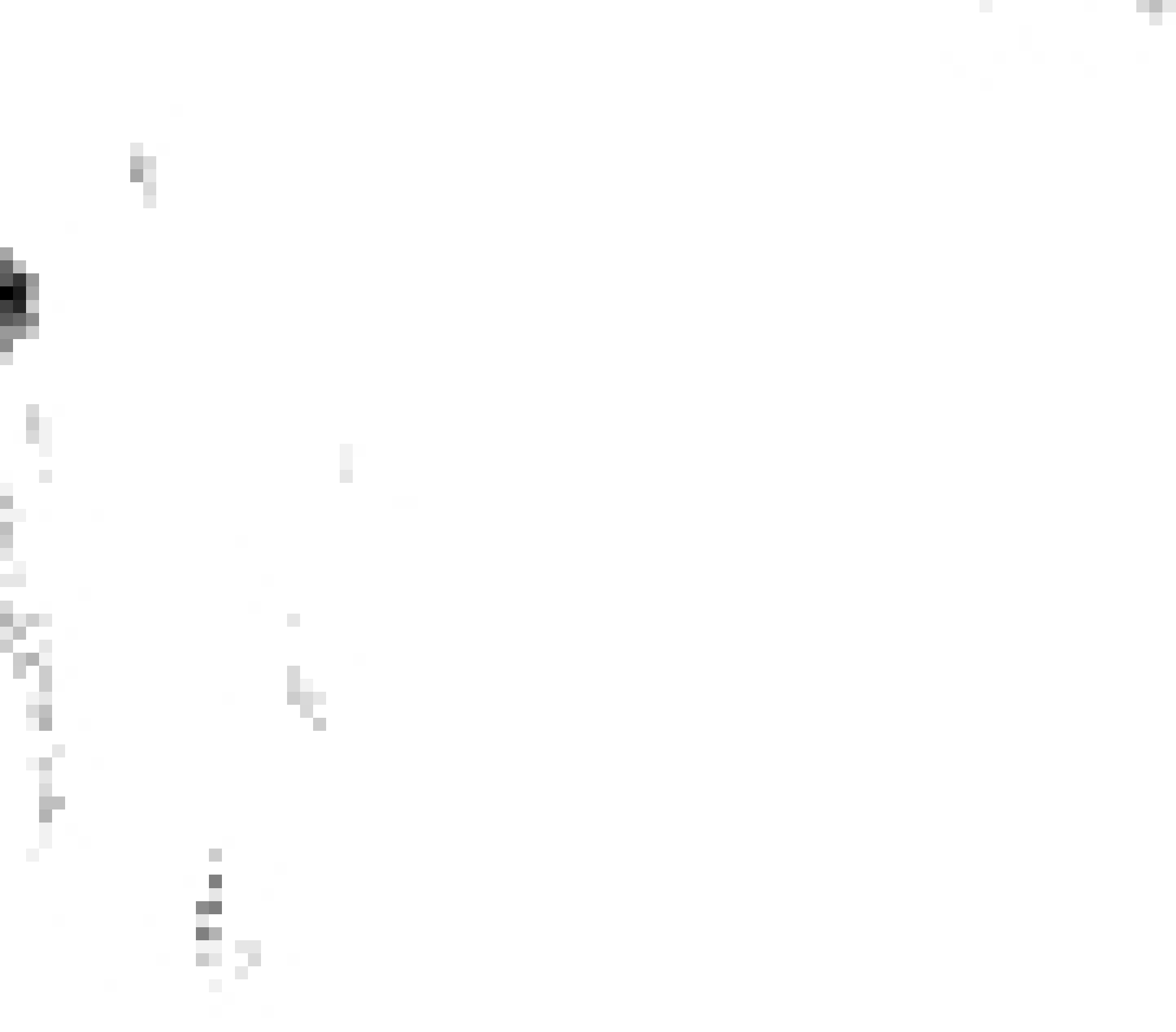}}
				}		
		\end{overpic}}
		\subfloat{
			\begin{overpic}[width=\size\columnwidth]{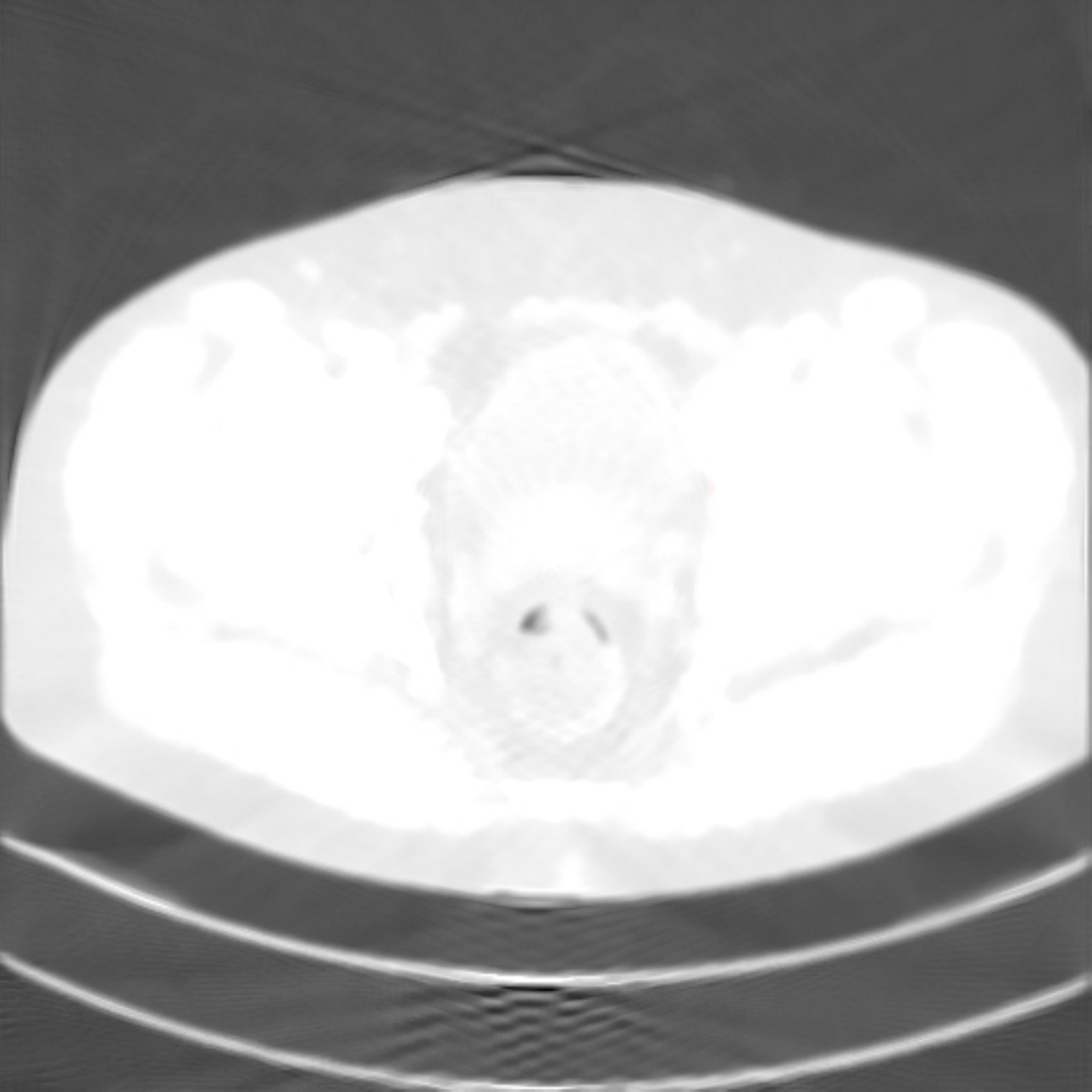}	
				\put(0,0){\color{red}%
					\frame{\includegraphics[scale=0.08]{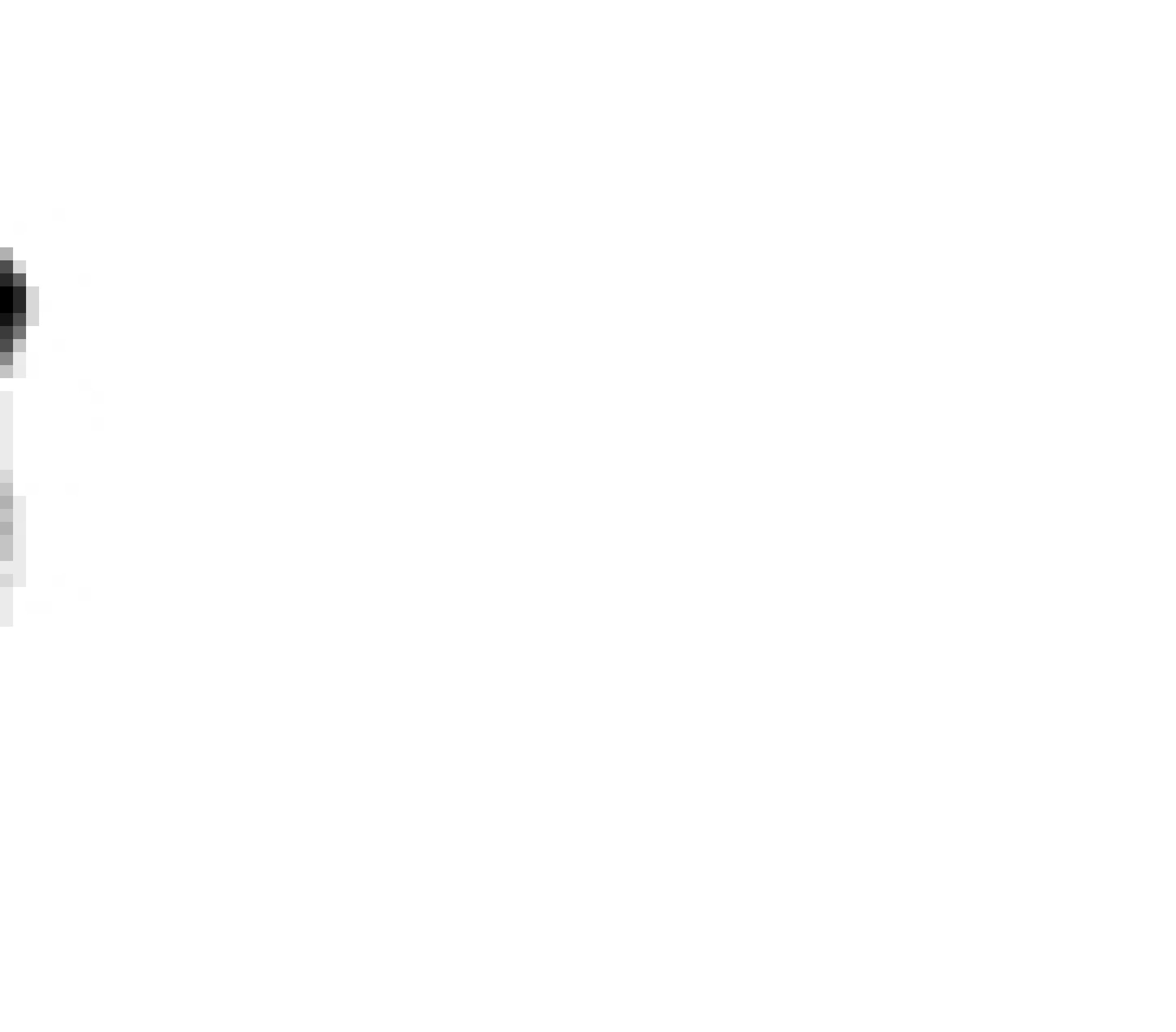}}
				}		
		\end{overpic}}
		\subfloat{
			\begin{overpic}[width=\size\columnwidth]{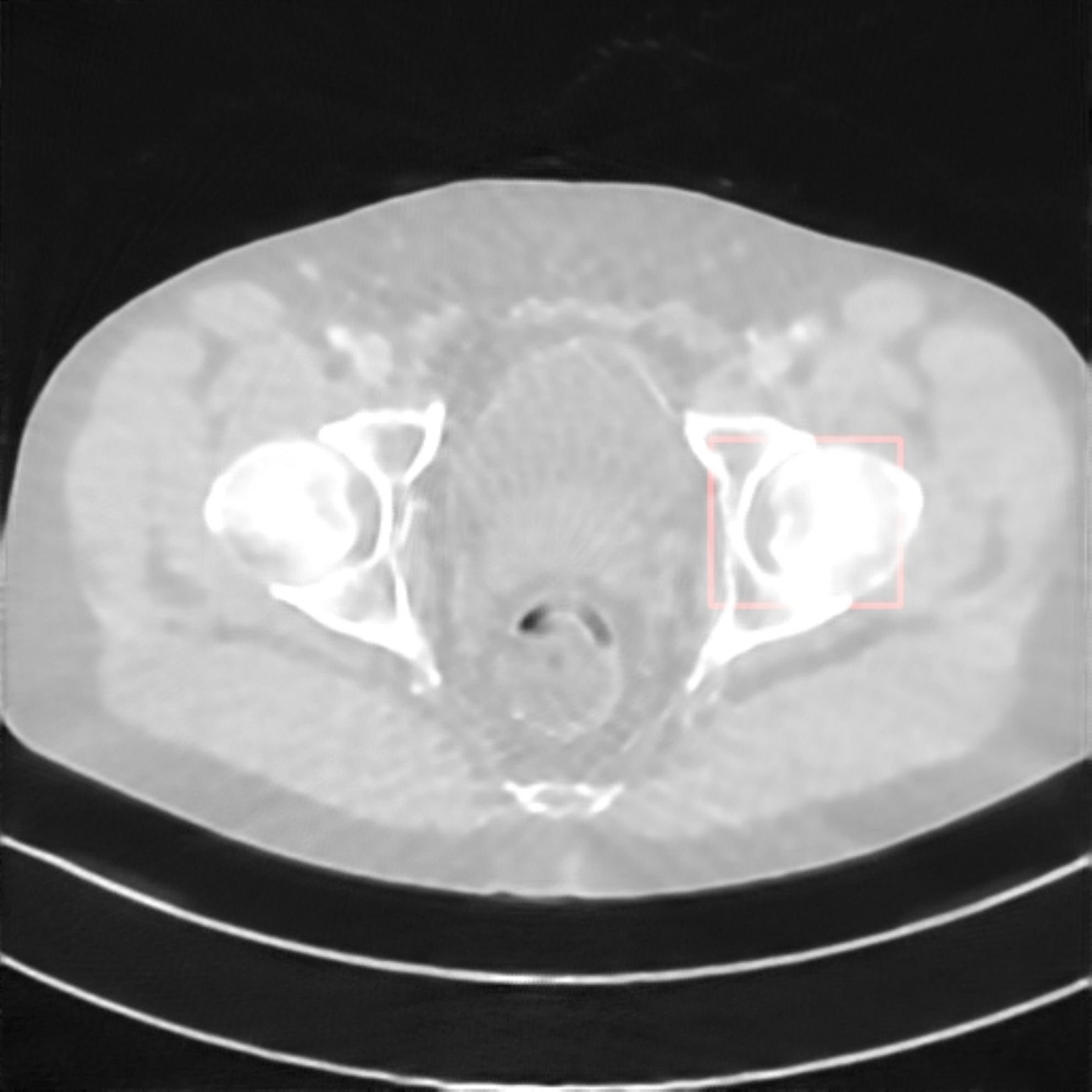}	
				\put(0,0){\color{red}%
					\frame{\includegraphics[scale=0.08]{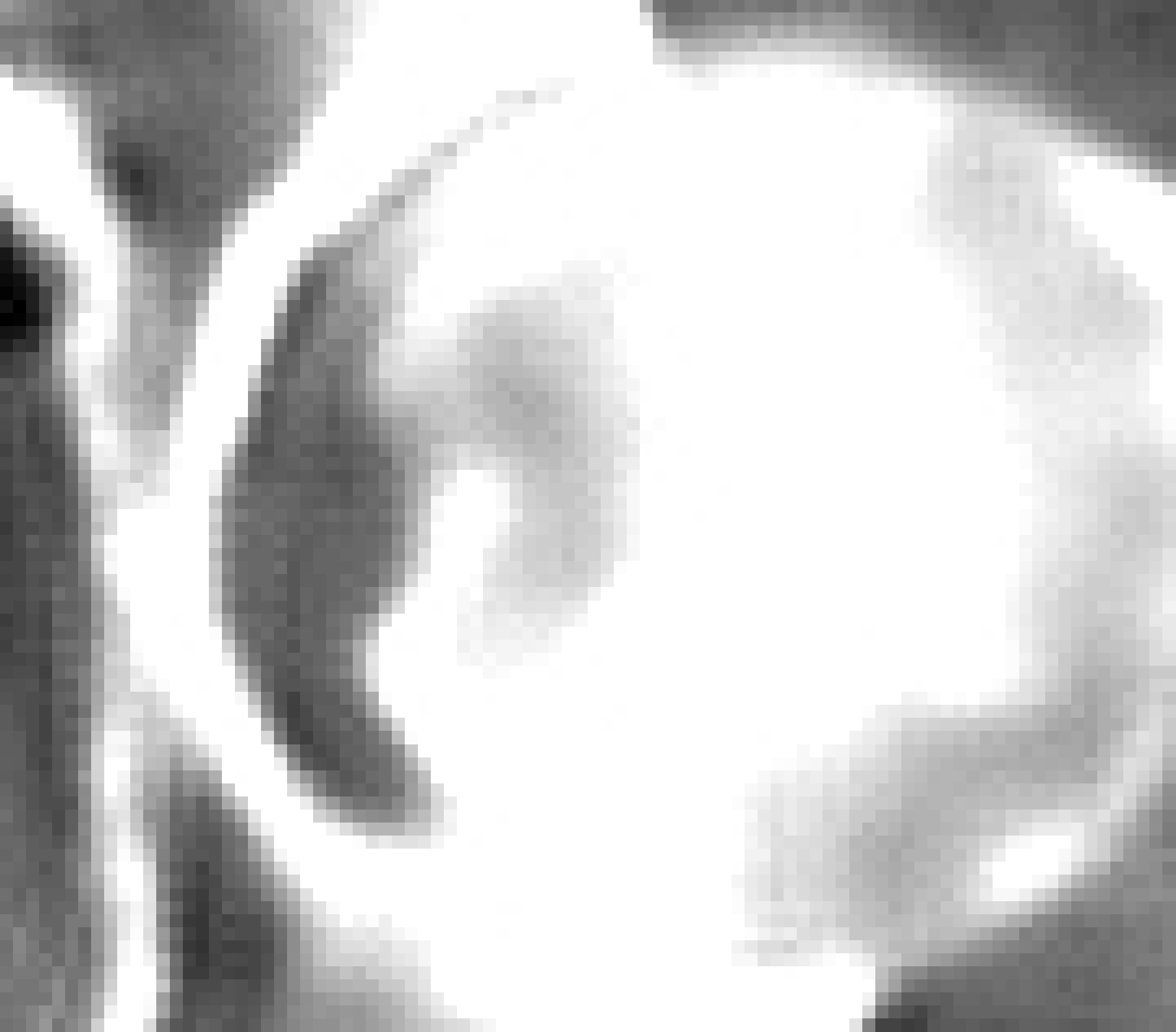}}
				}		
		\end{overpic}}
		\subfloat{
			\begin{overpic}[width=\size\columnwidth]{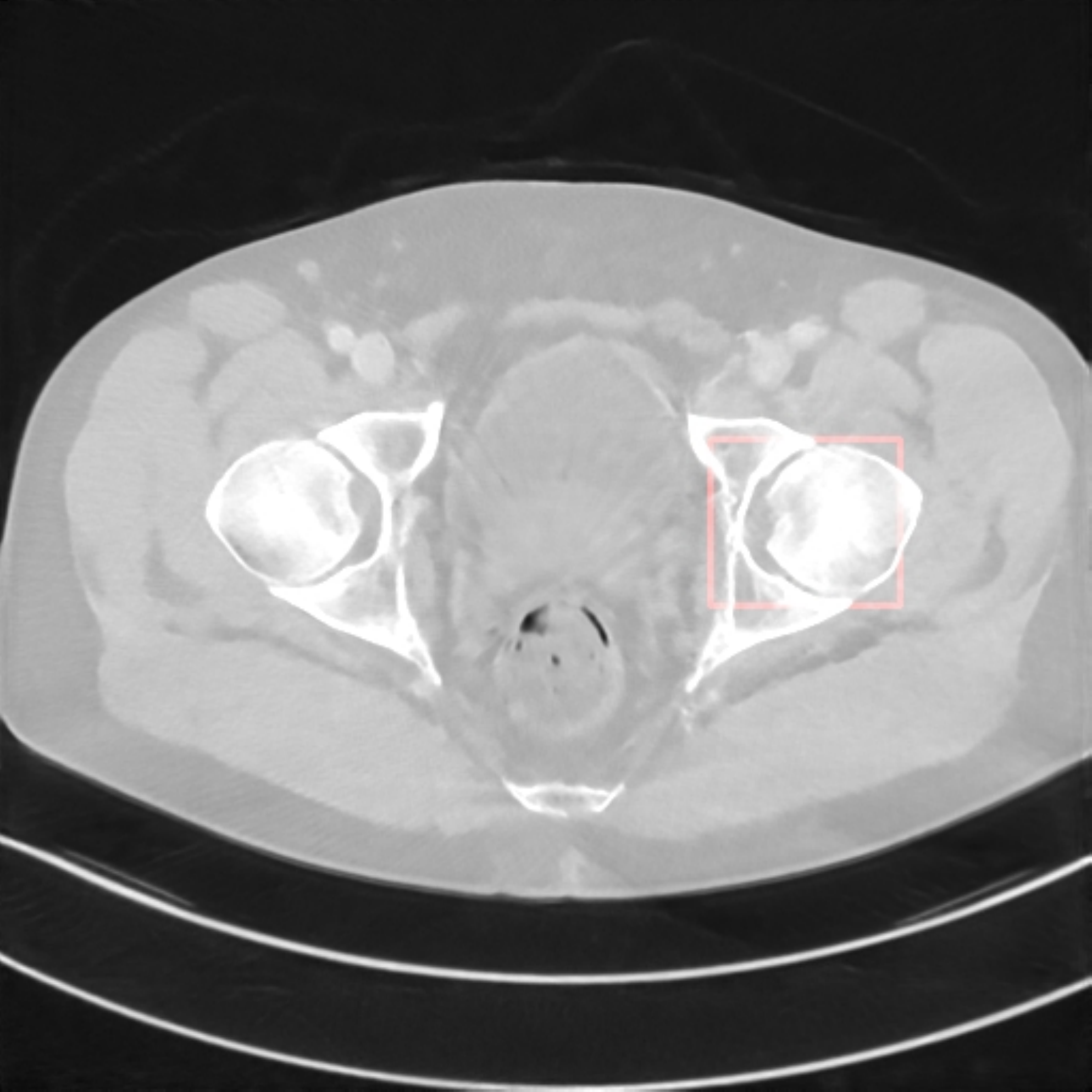}	
				\put(0,0){\color{red}%
					\frame{\includegraphics[scale=0.08]{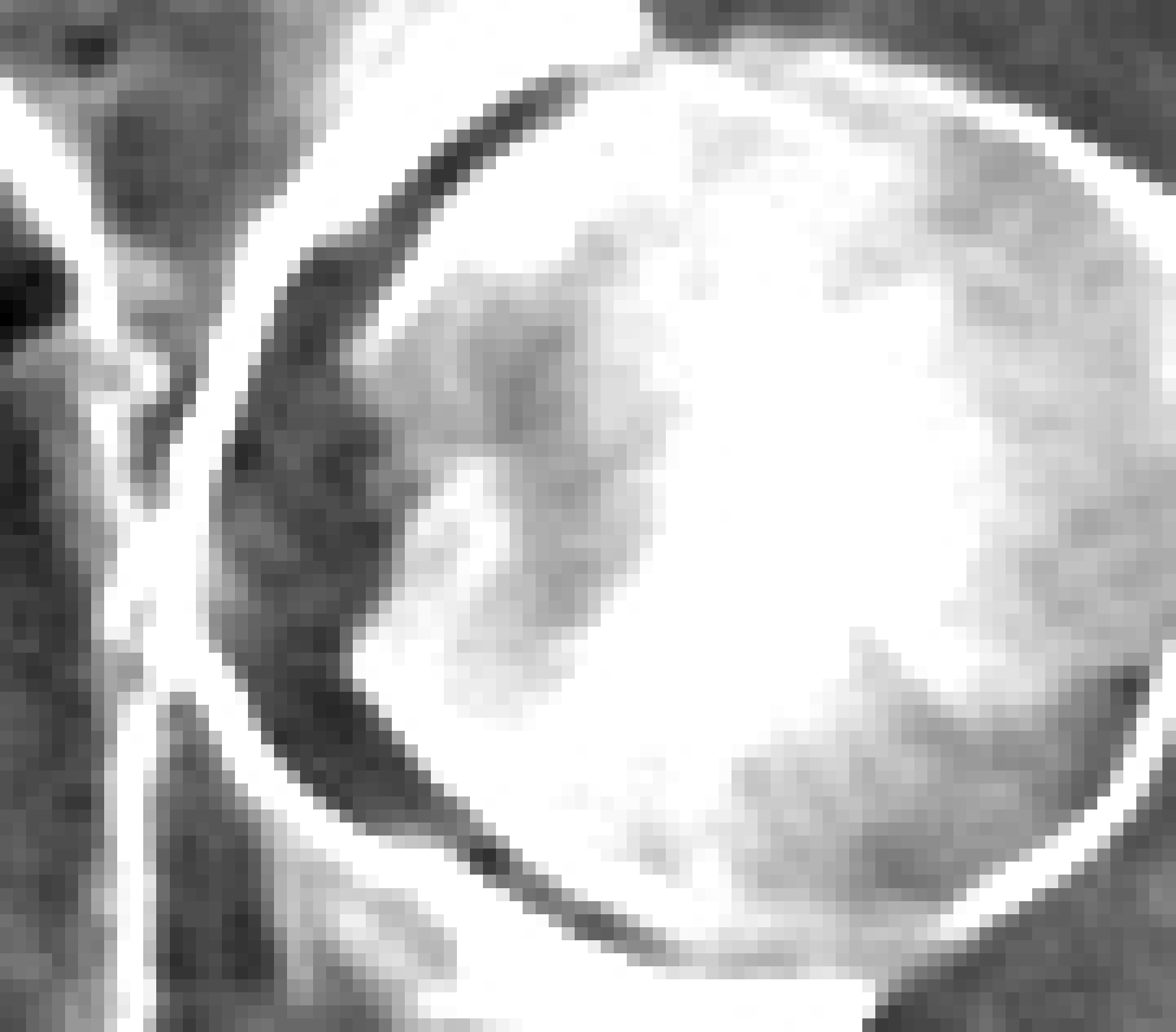}}
				}		
		\end{overpic}}
		\vspace{-3mm}
		\renewcommand{\id}{800}
		\subfloat{
			\begin{overpic}[width=\size\columnwidth,percent]{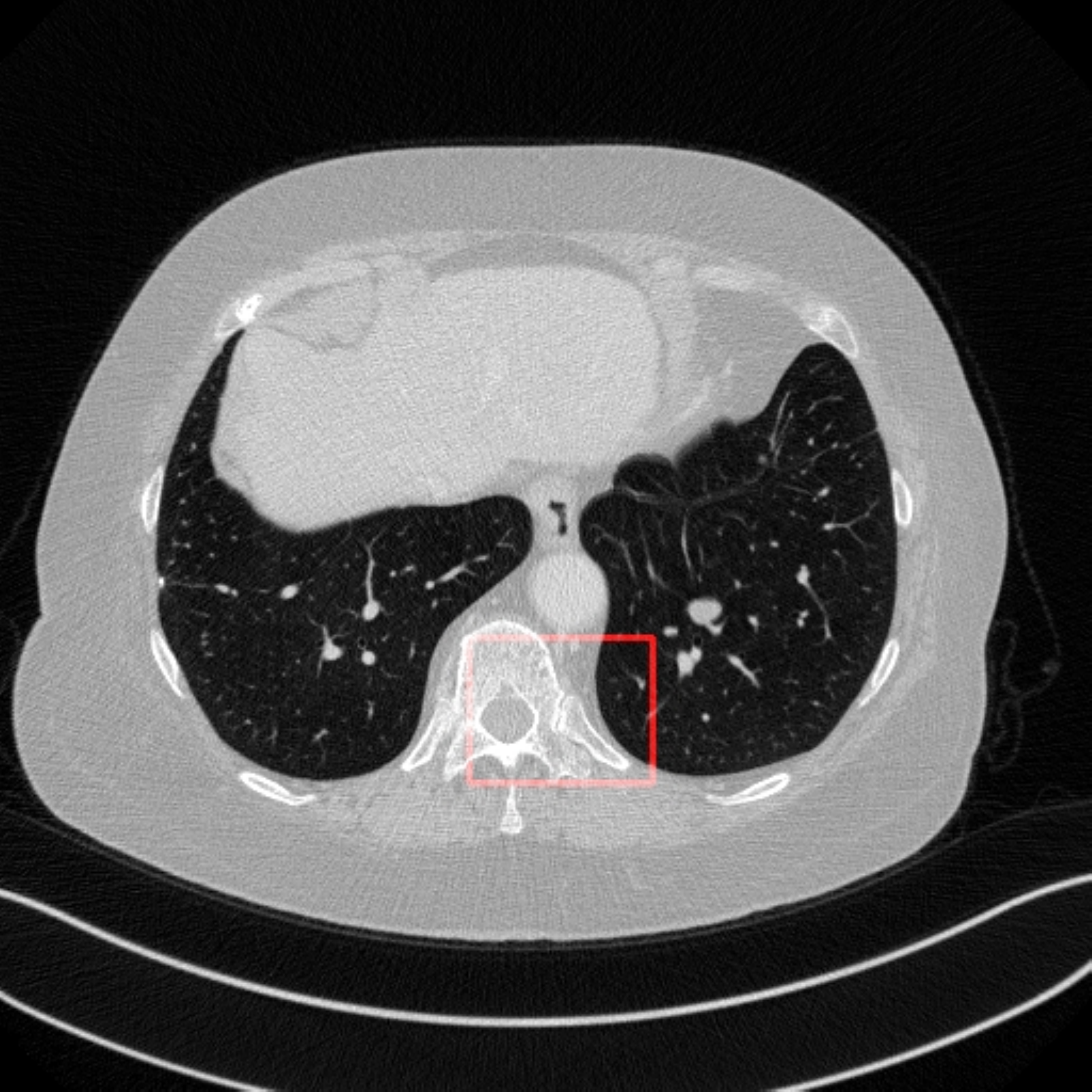}	
				\put(0,0){\color{red}%
					\frame{\includegraphics[scale=0.08]{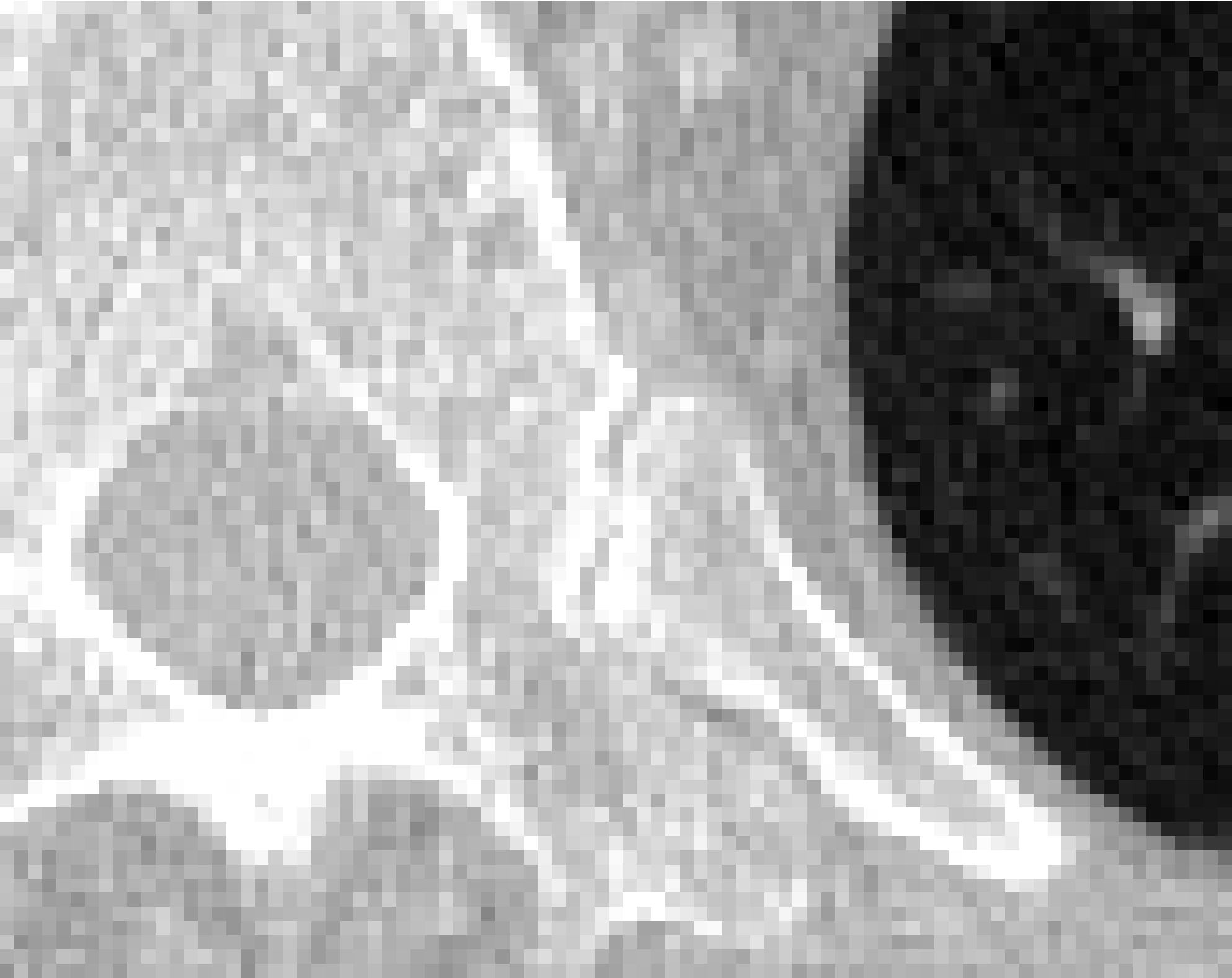}}
				}		
		\end{overpic}}
		\subfloat{
			\begin{overpic}[width=\size\columnwidth]{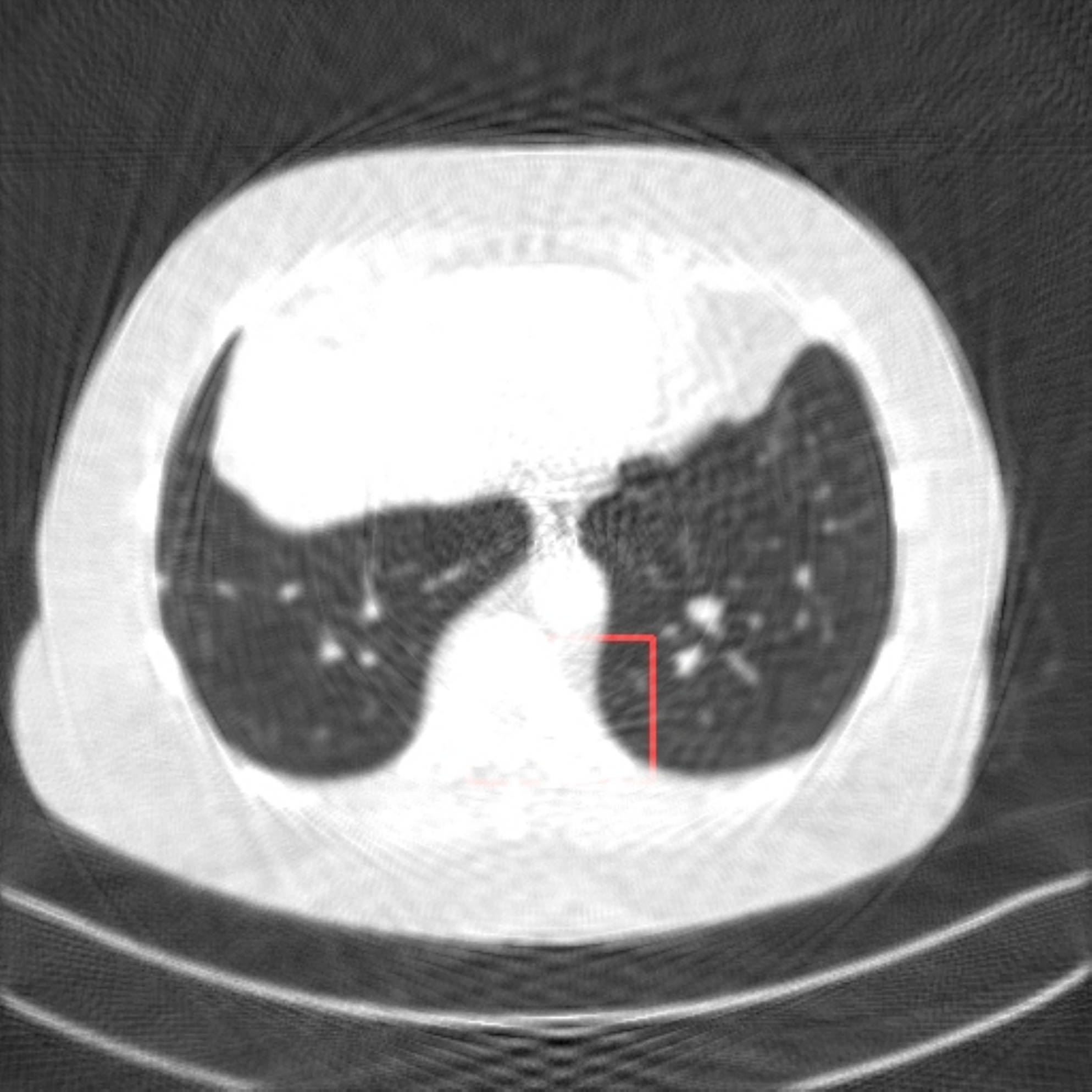}	
				\put(0,0){\color{red}%
					\frame{\includegraphics[scale=0.08]{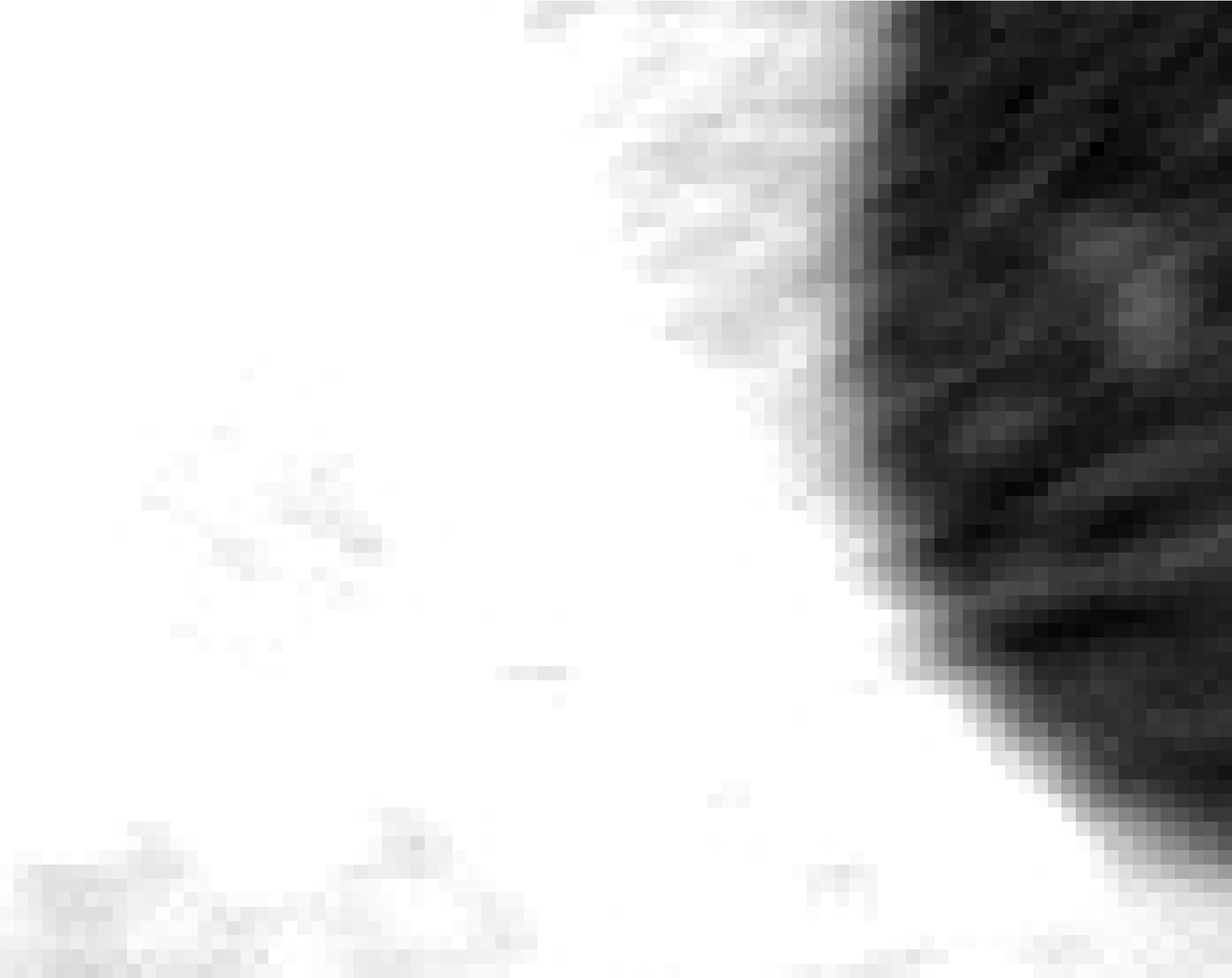}}
				}		
		\end{overpic}}
		\subfloat{
			\begin{overpic}[width=\size\columnwidth]{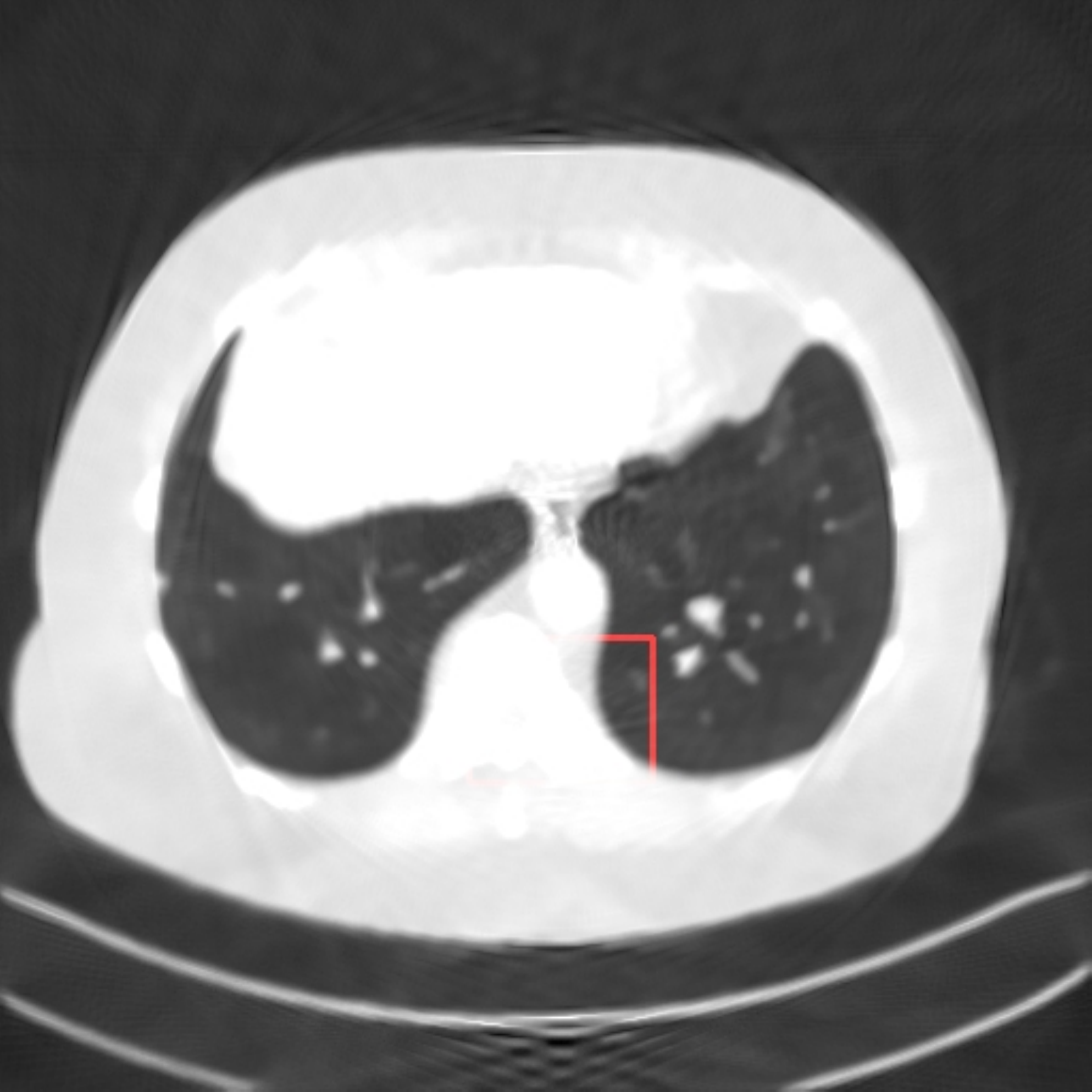}	
				\put(0,0){\color{red}%
					\frame{\includegraphics[scale=0.08]{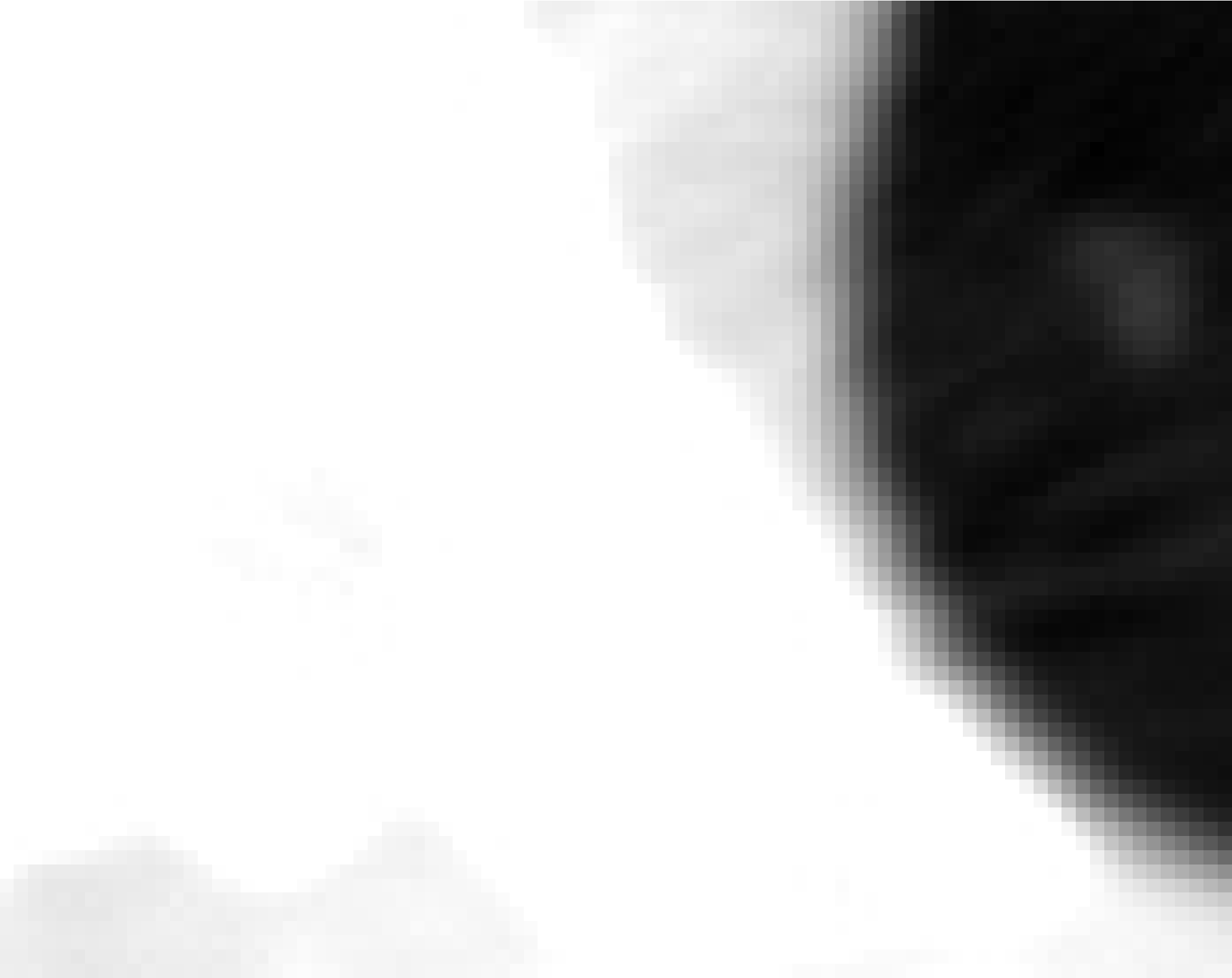}}
				}		
		\end{overpic}}
		\subfloat{
			\begin{overpic}[width=\size\columnwidth]{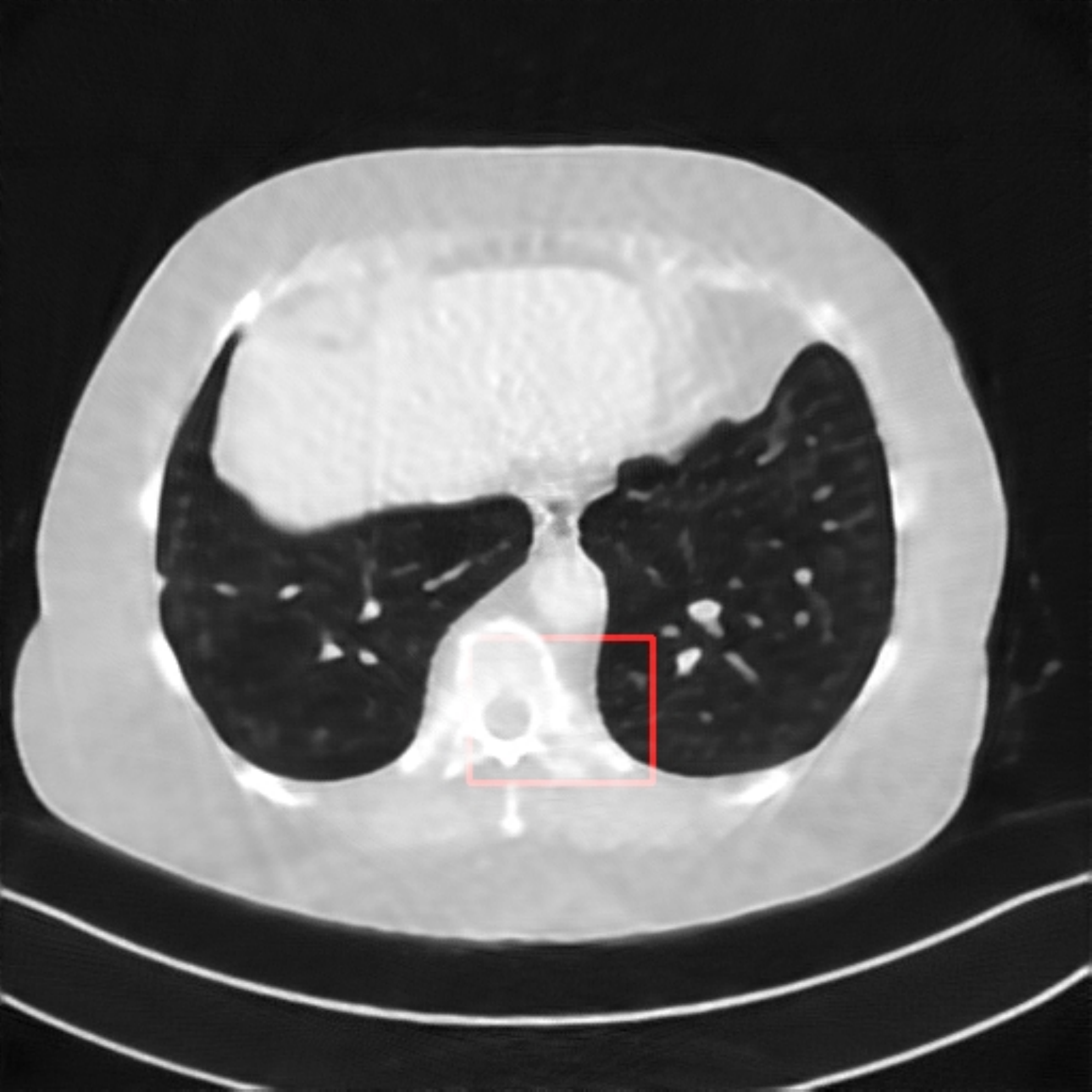}	
				\put(0,0){\color{red}%
					\frame{\includegraphics[scale=0.08]{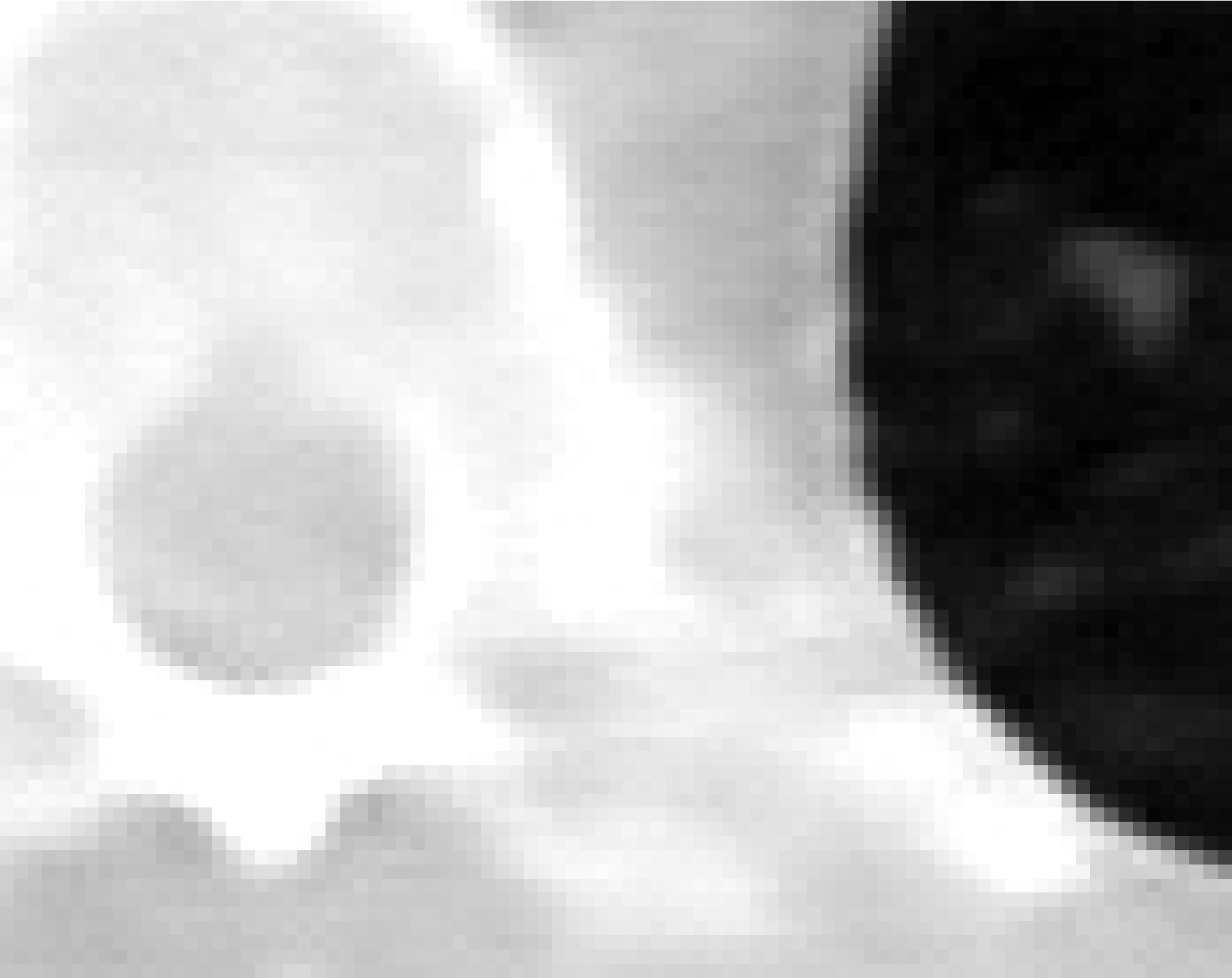}}
				}		
		\end{overpic}}
		\subfloat{
			\begin{overpic}[width=\size\columnwidth]{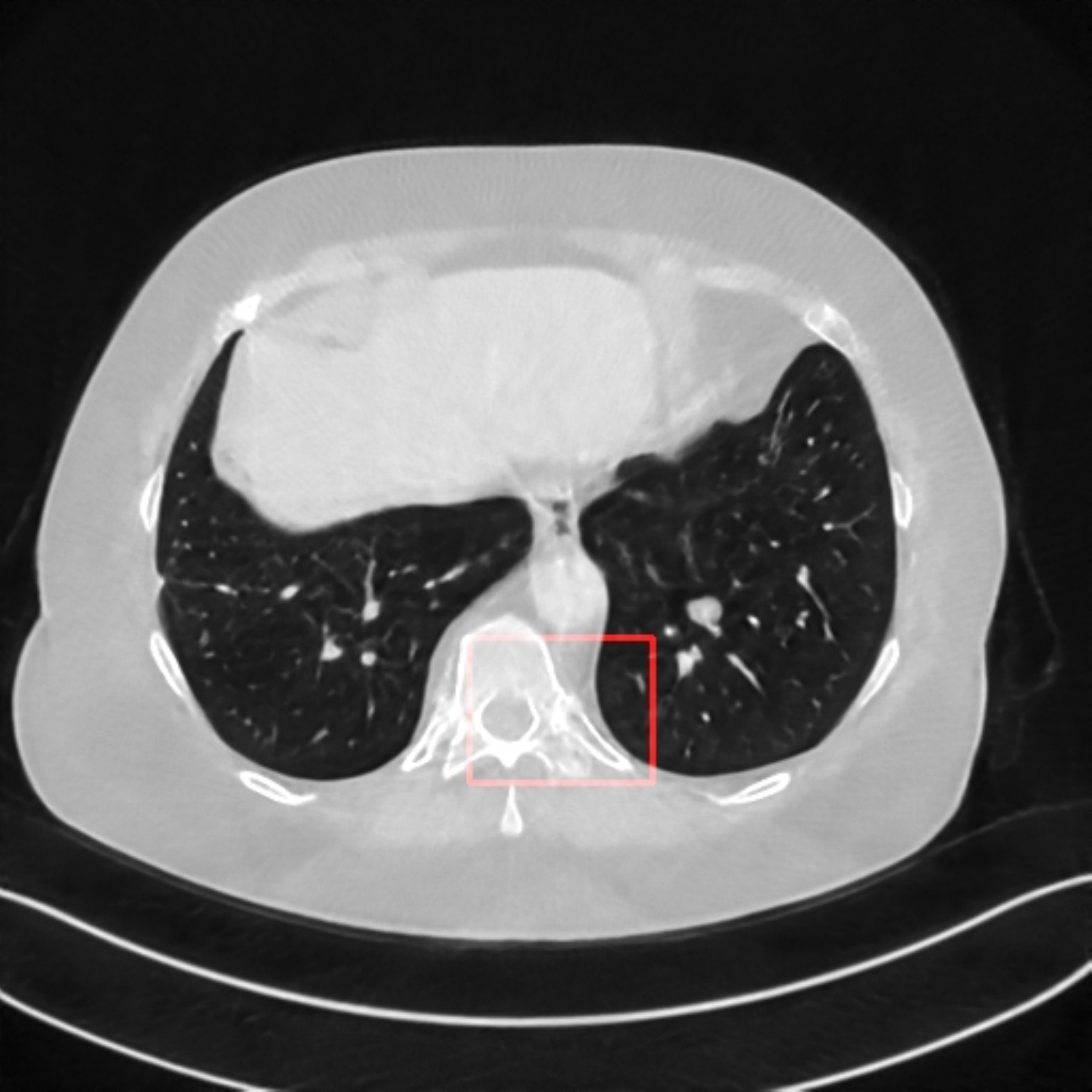}	
				\put(0,0){\color{red}%
					\frame{\includegraphics[scale=0.08]{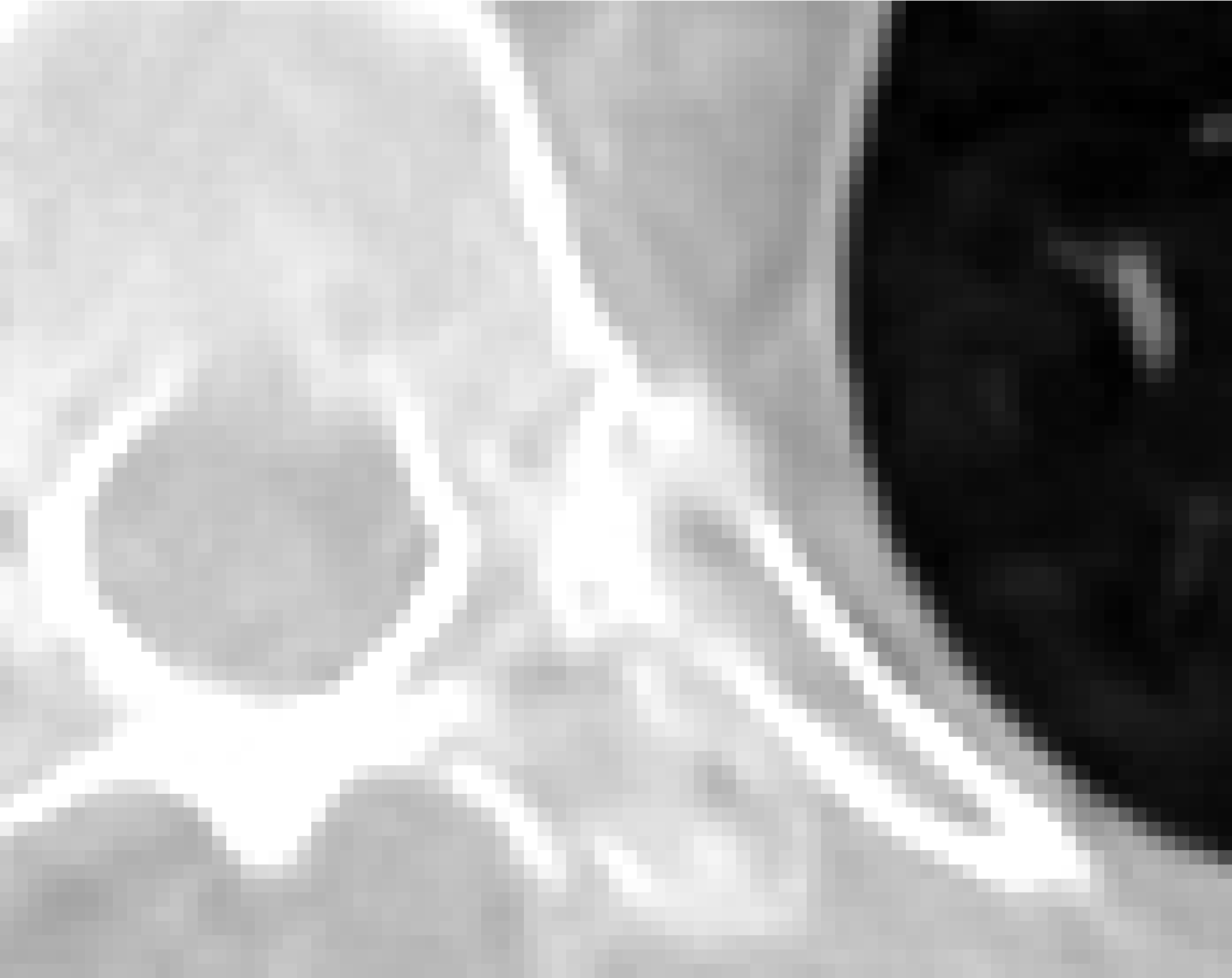}}
				}		
		\end{overpic}}
		\vspace{-3mm}
		\setcounter{subfigure}{0}
		\renewcommand{\id}{2000}
		\subfloat[Label]{
			\begin{overpic}[width=\size\columnwidth,percent]{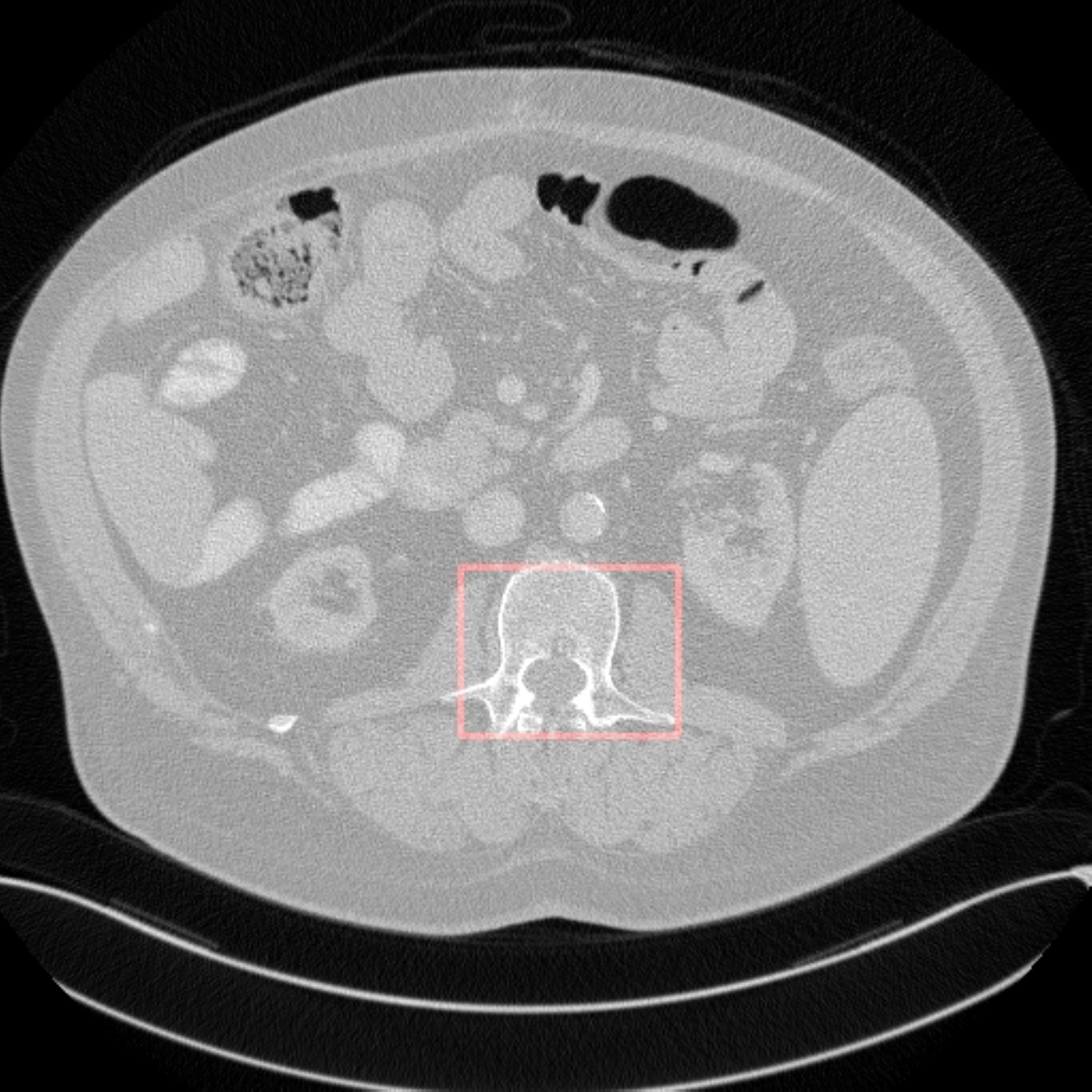}	
				\put(0,0){\color{red}%
					\frame{\includegraphics[scale=0.08]{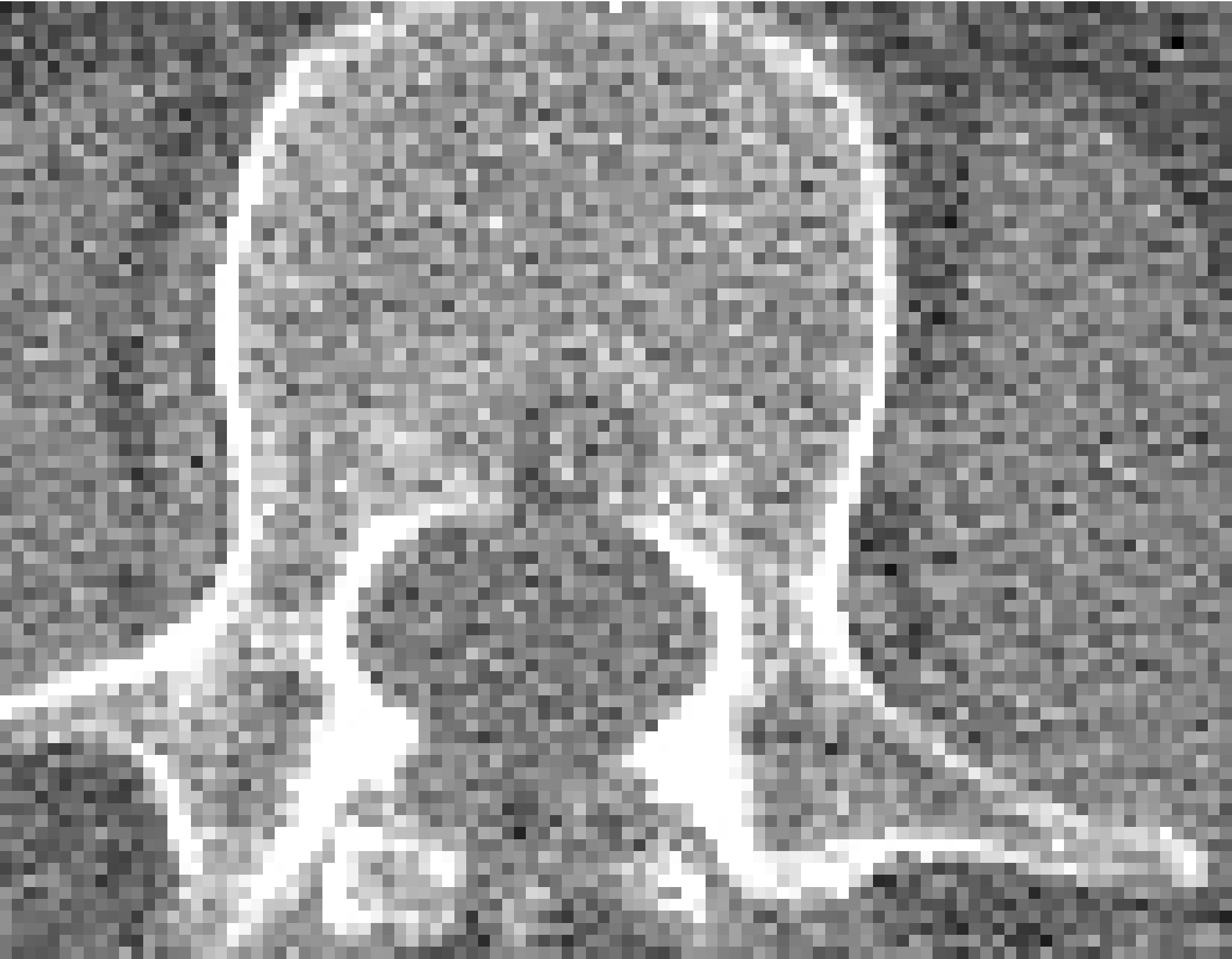}}
				}		
		\end{overpic}}
		\subfloat[Katsevich algorithm]{
			\begin{overpic}[width=\size\columnwidth]{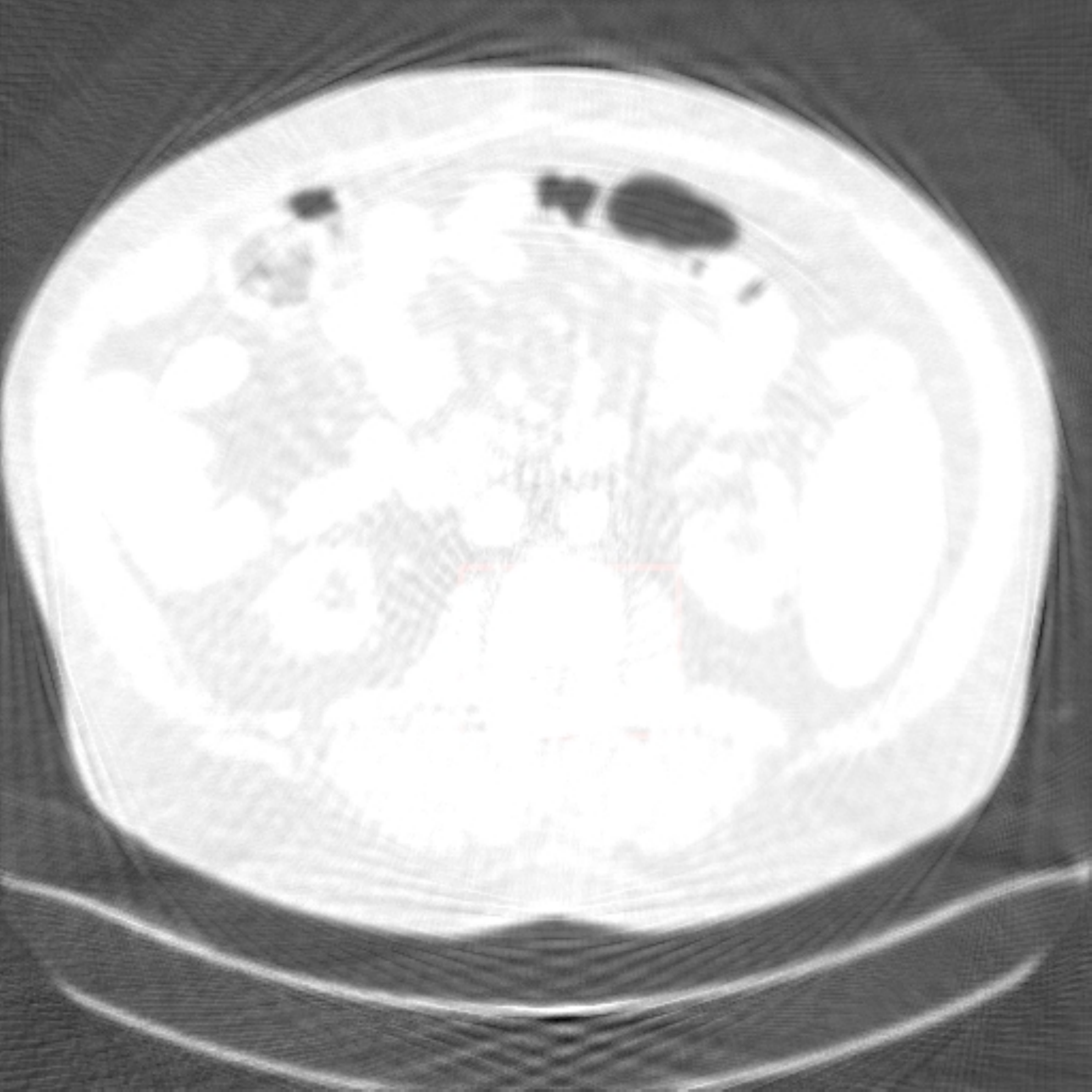}	
				\put(0,0){\color{red}%
					\frame{\includegraphics[scale=0.08]{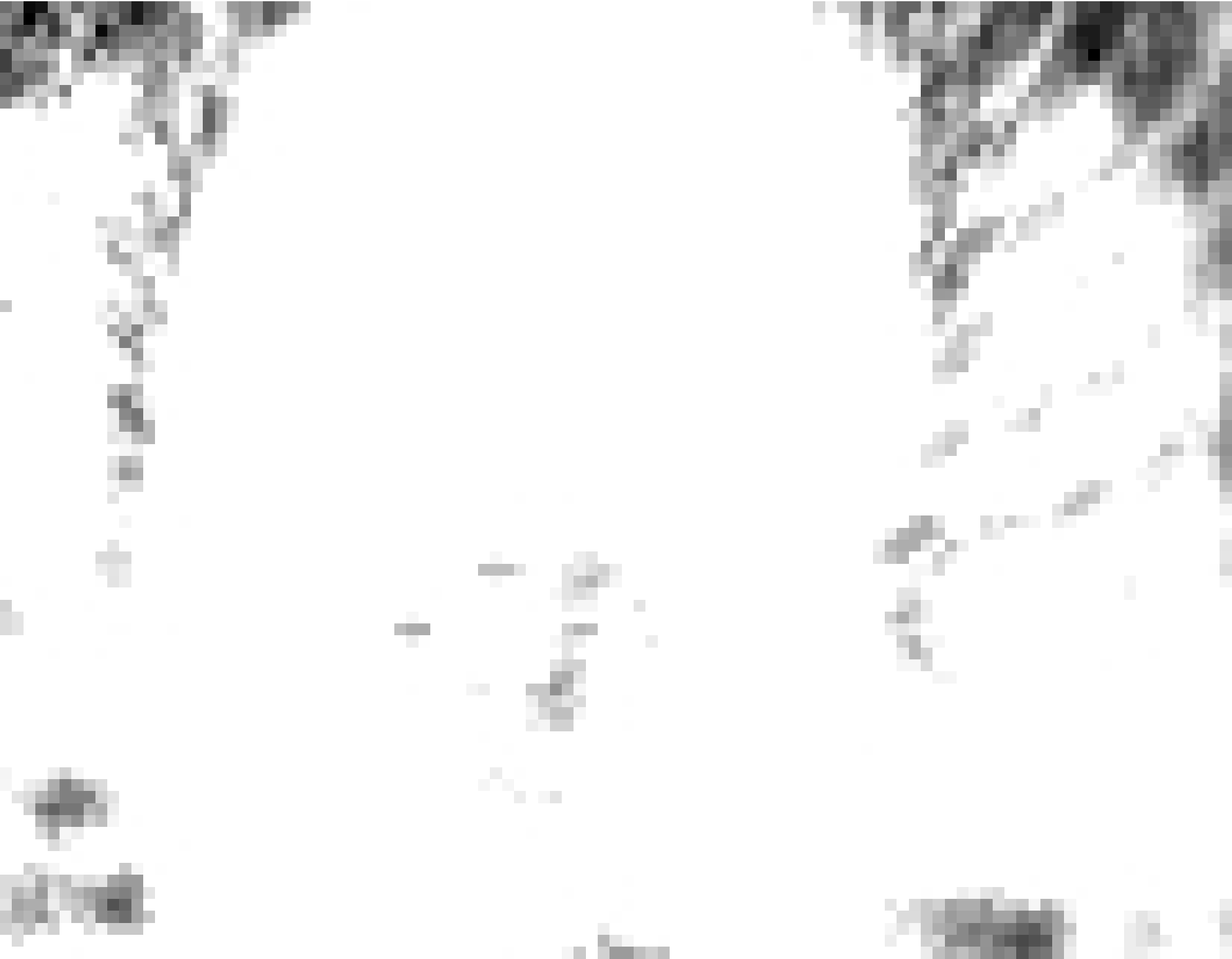}}
				}		
		\end{overpic}}
		\subfloat[Nonlocal-TV]{
			\begin{overpic}[width=\size\columnwidth]{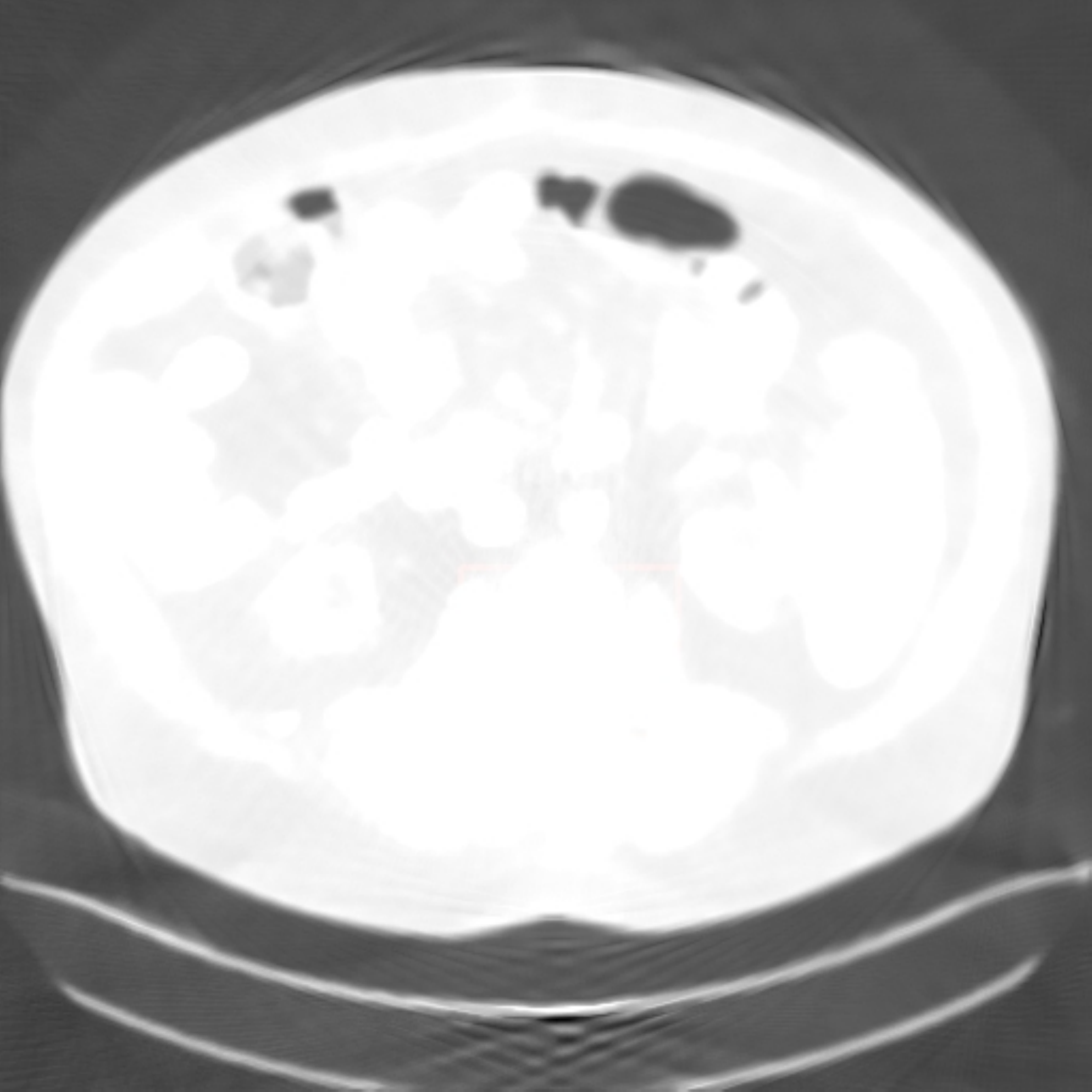}	
				\put(0,0){\color{red}%
					\frame{\includegraphics[scale=0.08]{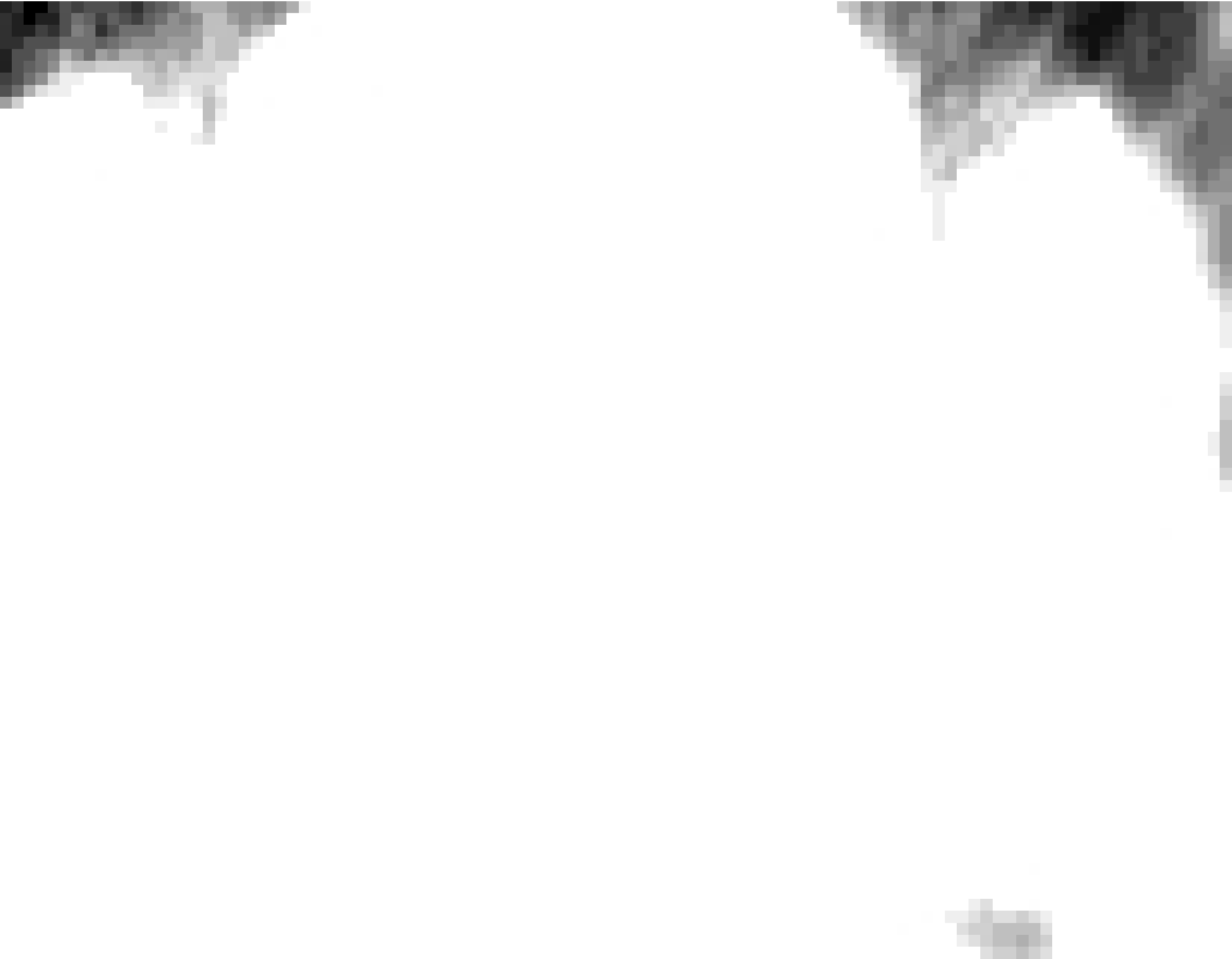}}
				}		
		\end{overpic}}
		\subfloat[FBPConvNet]{
			\begin{overpic}[width=\size\columnwidth]{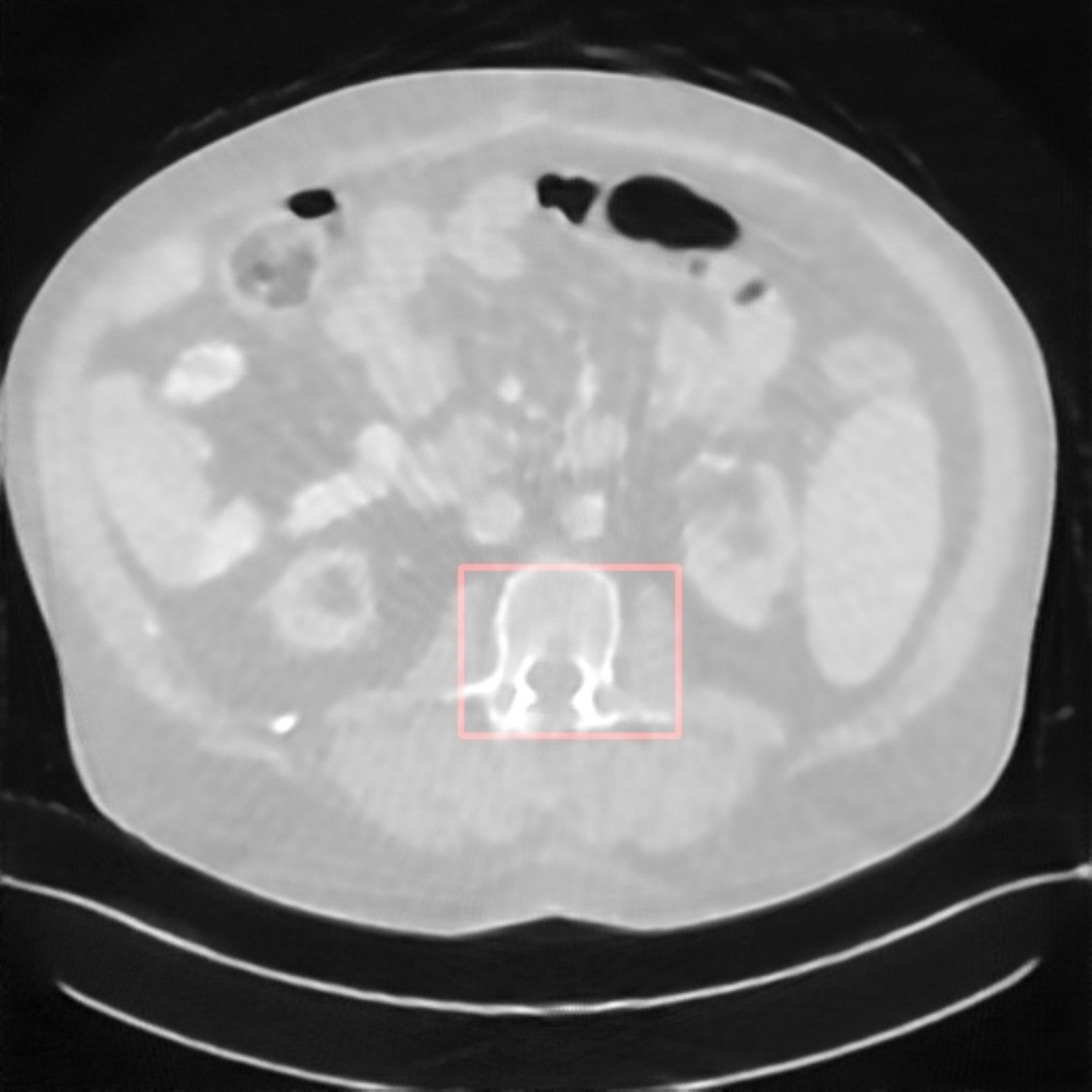}	
				\put(0,0){\color{red}%
					\frame{\includegraphics[scale=0.08]{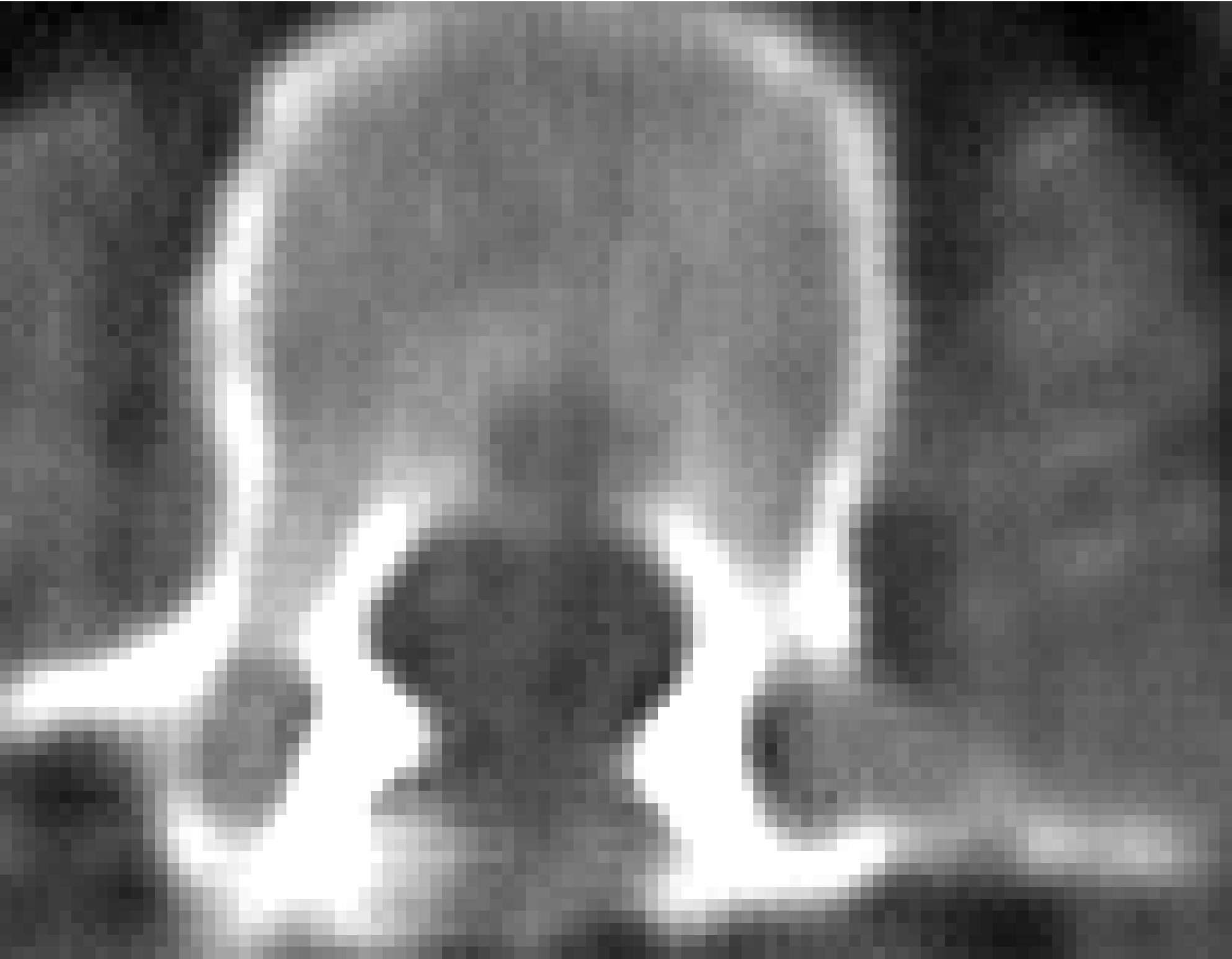}}
				}		
		\end{overpic}}
		\subfloat[Ours]{
			\begin{overpic}[width=\size\columnwidth]{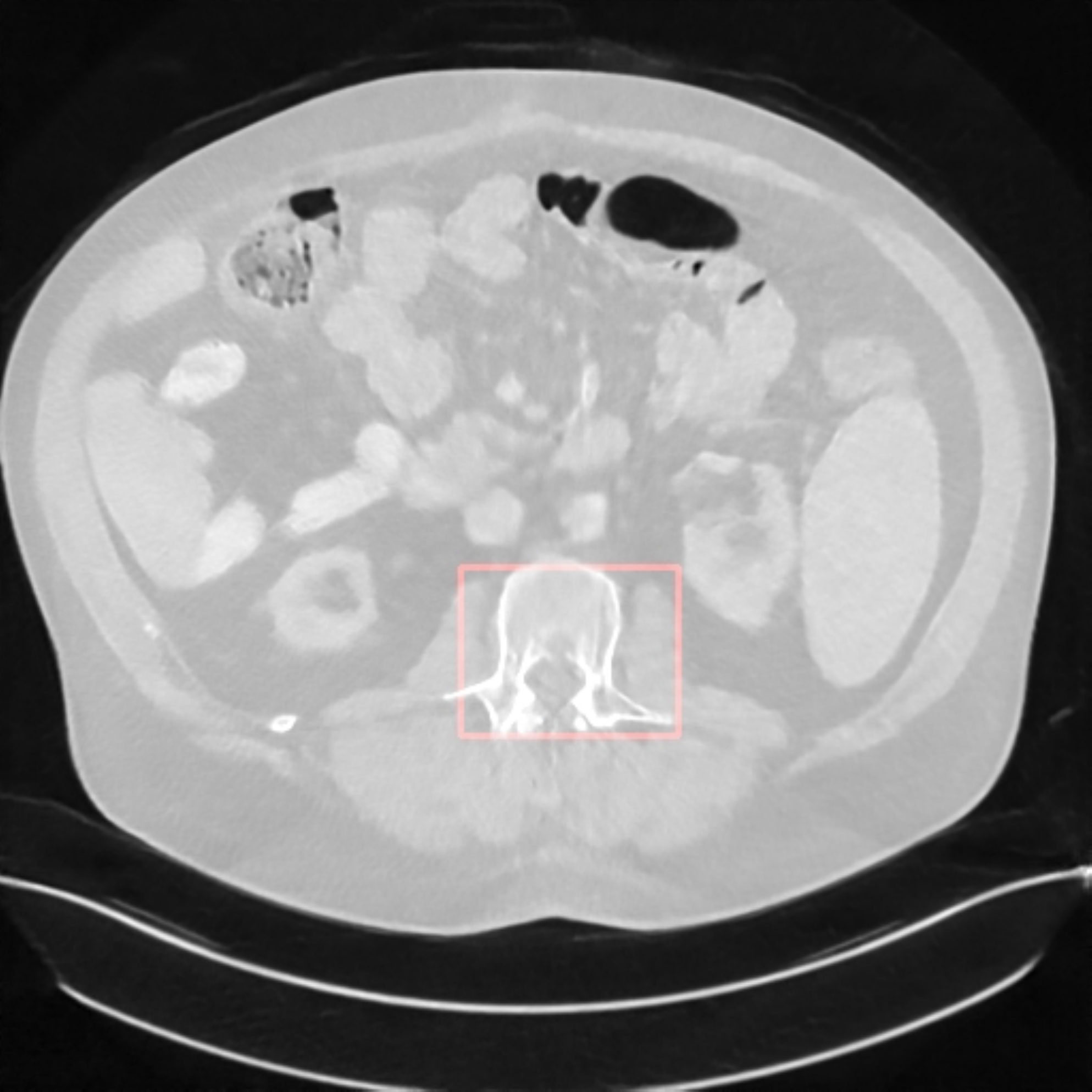}	
				\put(0,0){\color{red}%
					\frame{\includegraphics[scale=0.08]{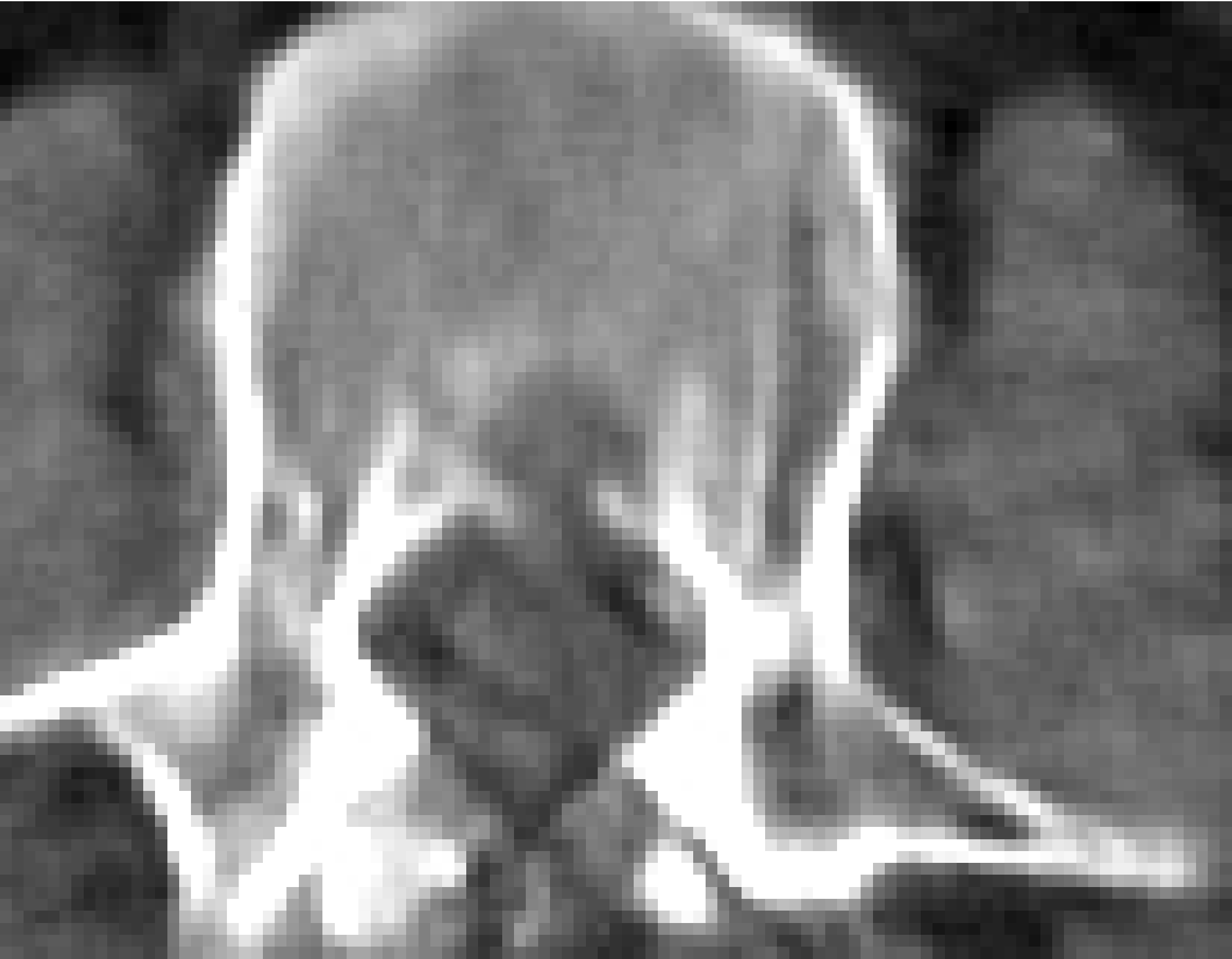}}
				}		
		\end{overpic}}
	}
	\caption{The reconstructed CT NIH-AAPM-Mayo data for $p=14\pi$. All images are linearly stretched to [0, 1] and the display window is [0, 0.6].}
	\label{Fig4}
\end{figure*}

\begin{figure*}[htbp]	
	\centering{	
		\newcommand{\id}{40}
		\subfloat{
			\begin{overpic}[width=\size\columnwidth,percent]{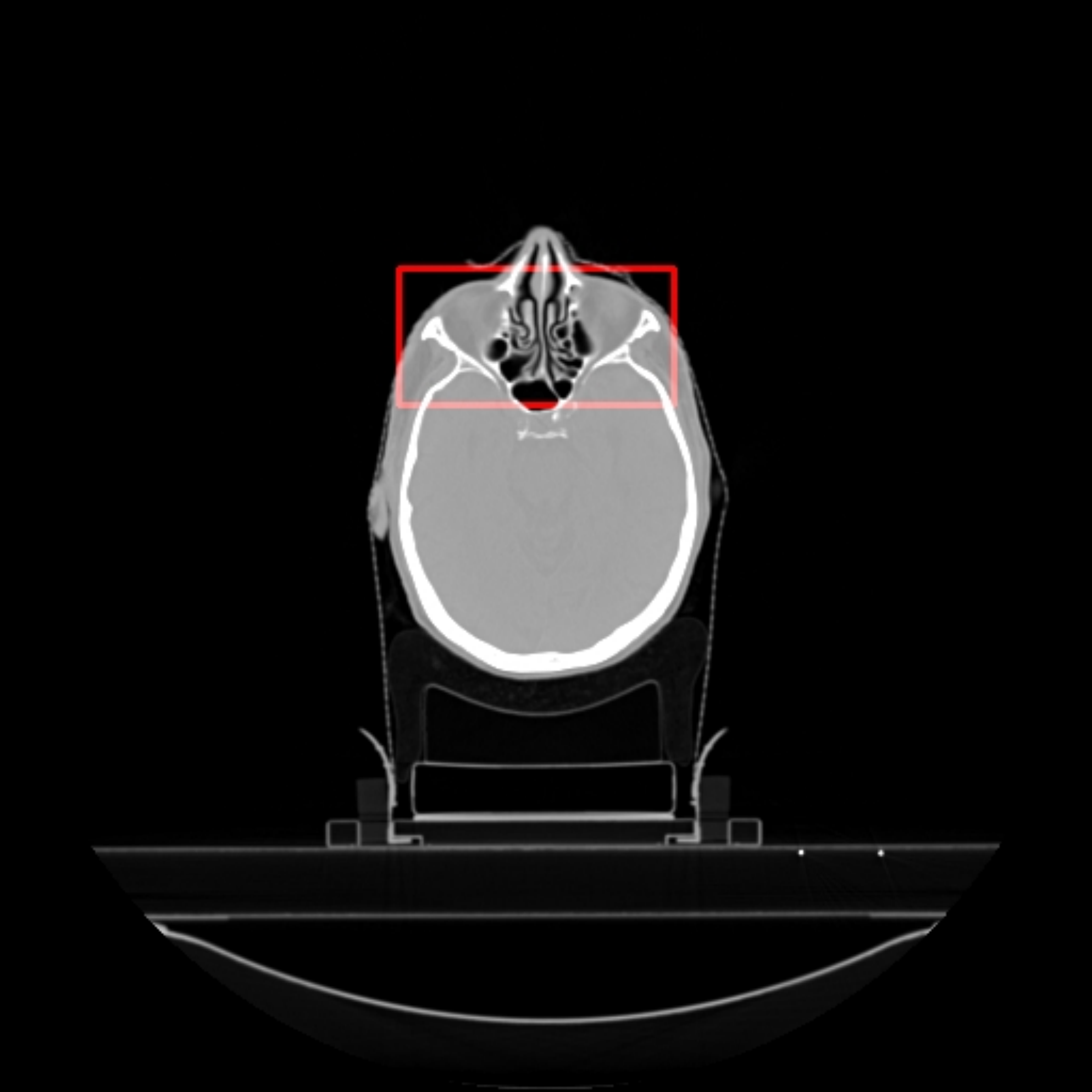}	
				\put(0,0){\color{red}%
					\frame{\includegraphics[scale=0.08]{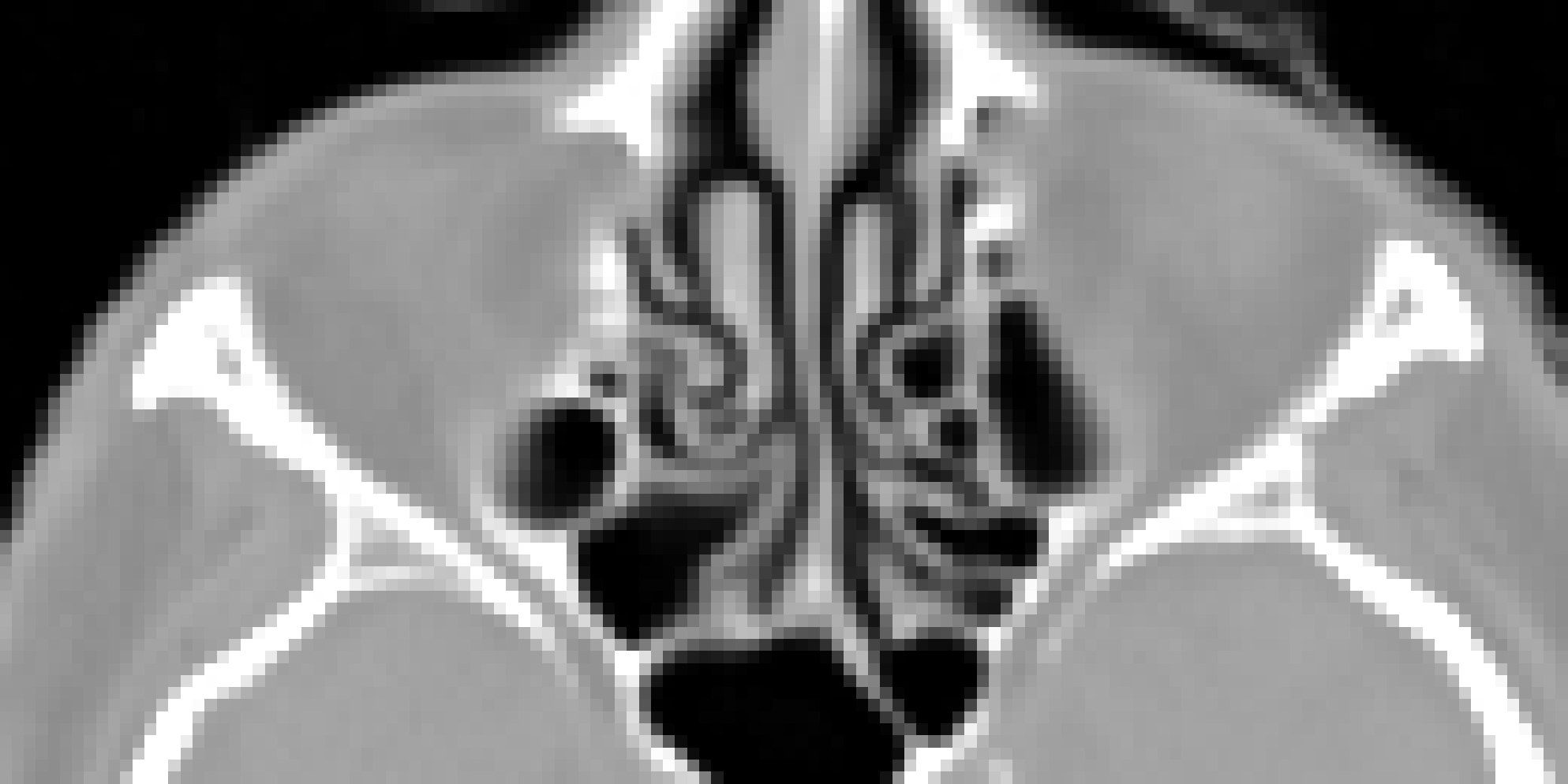}}
				}		
		\end{overpic}}
		\subfloat{
			\begin{overpic}[width=\size\columnwidth]{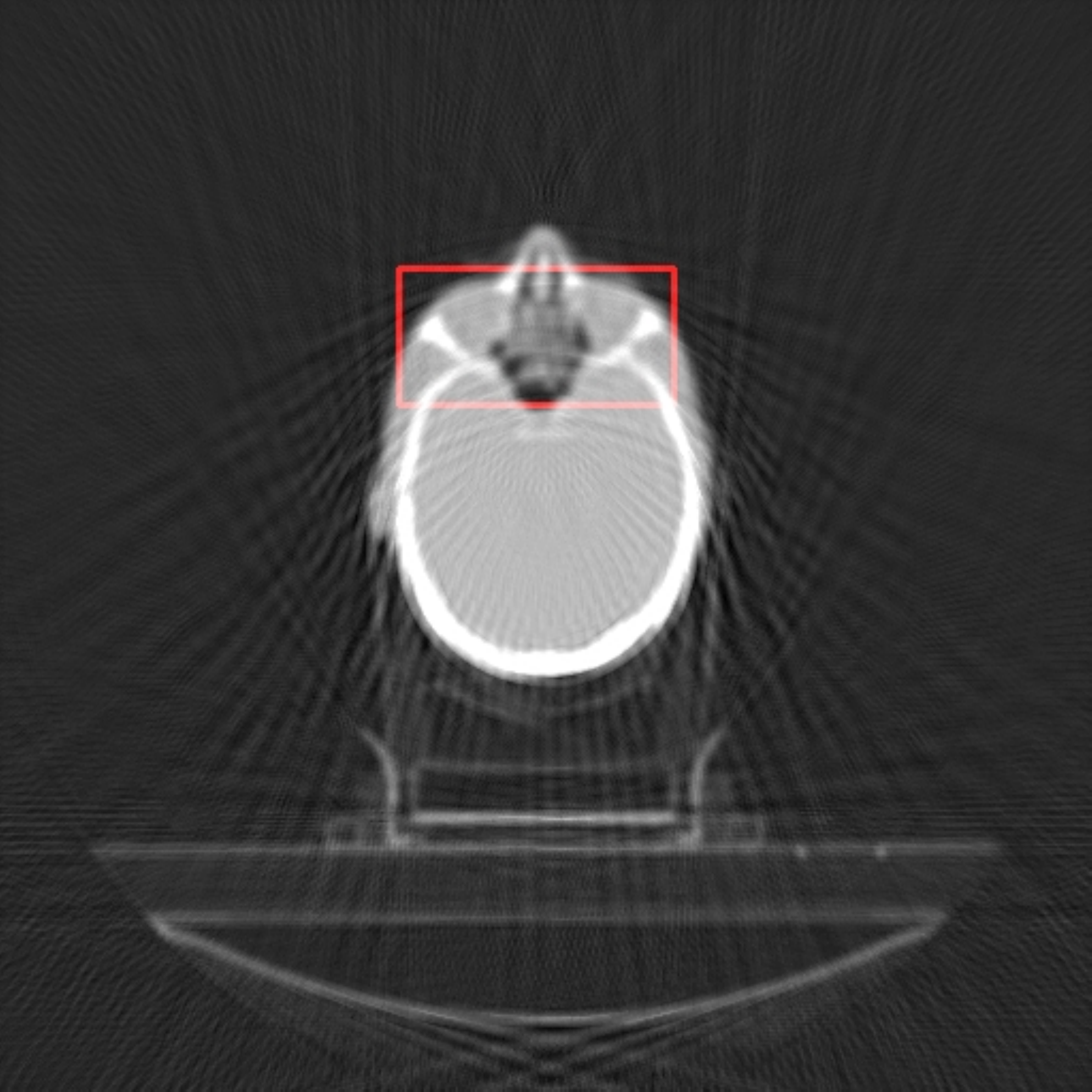}	
				\put(0,0){\color{red}%
					\frame{\includegraphics[scale=0.08]{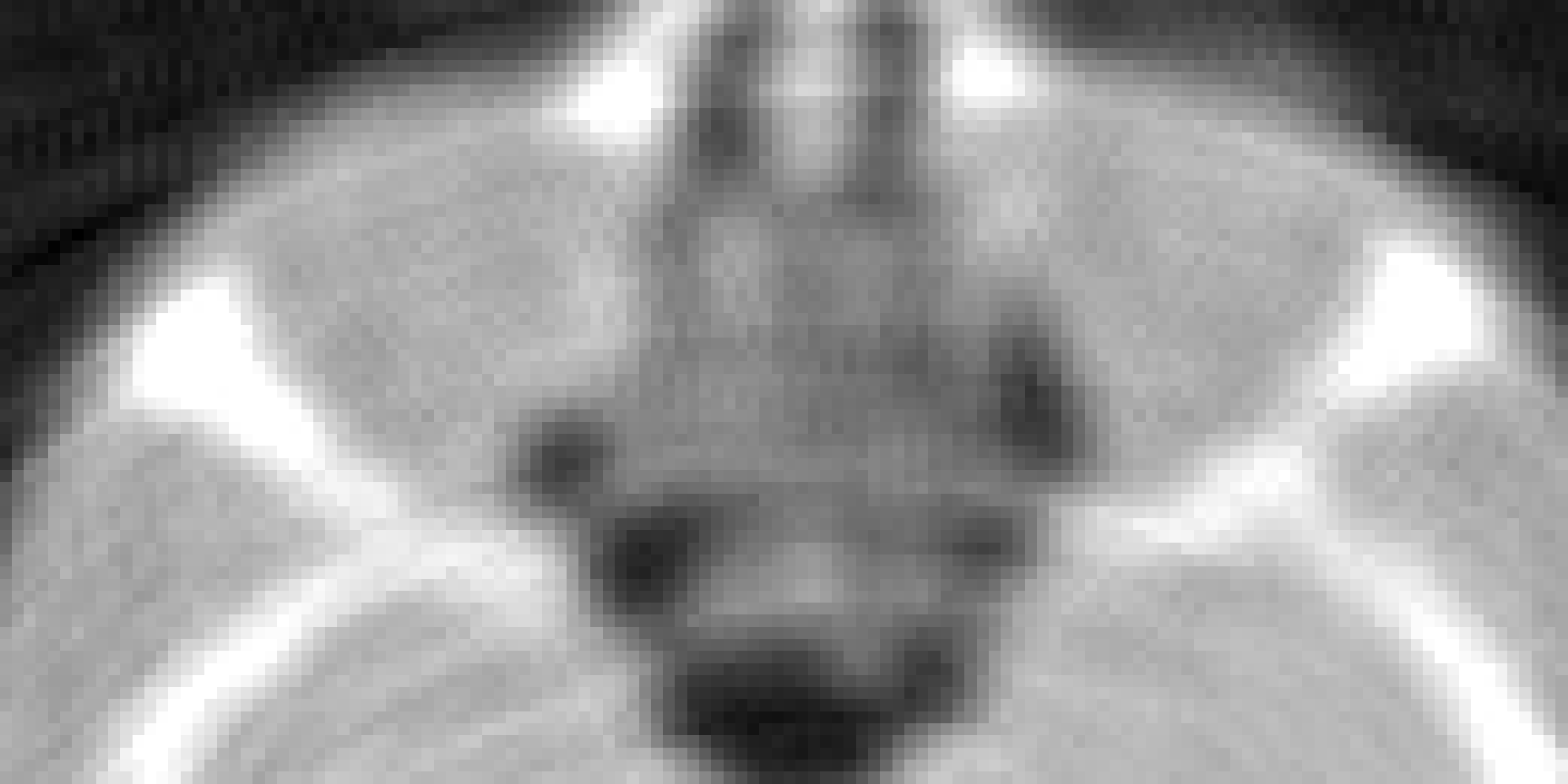}}
				}		
		\end{overpic}}
		\subfloat{
			\begin{overpic}[width=\size\columnwidth]{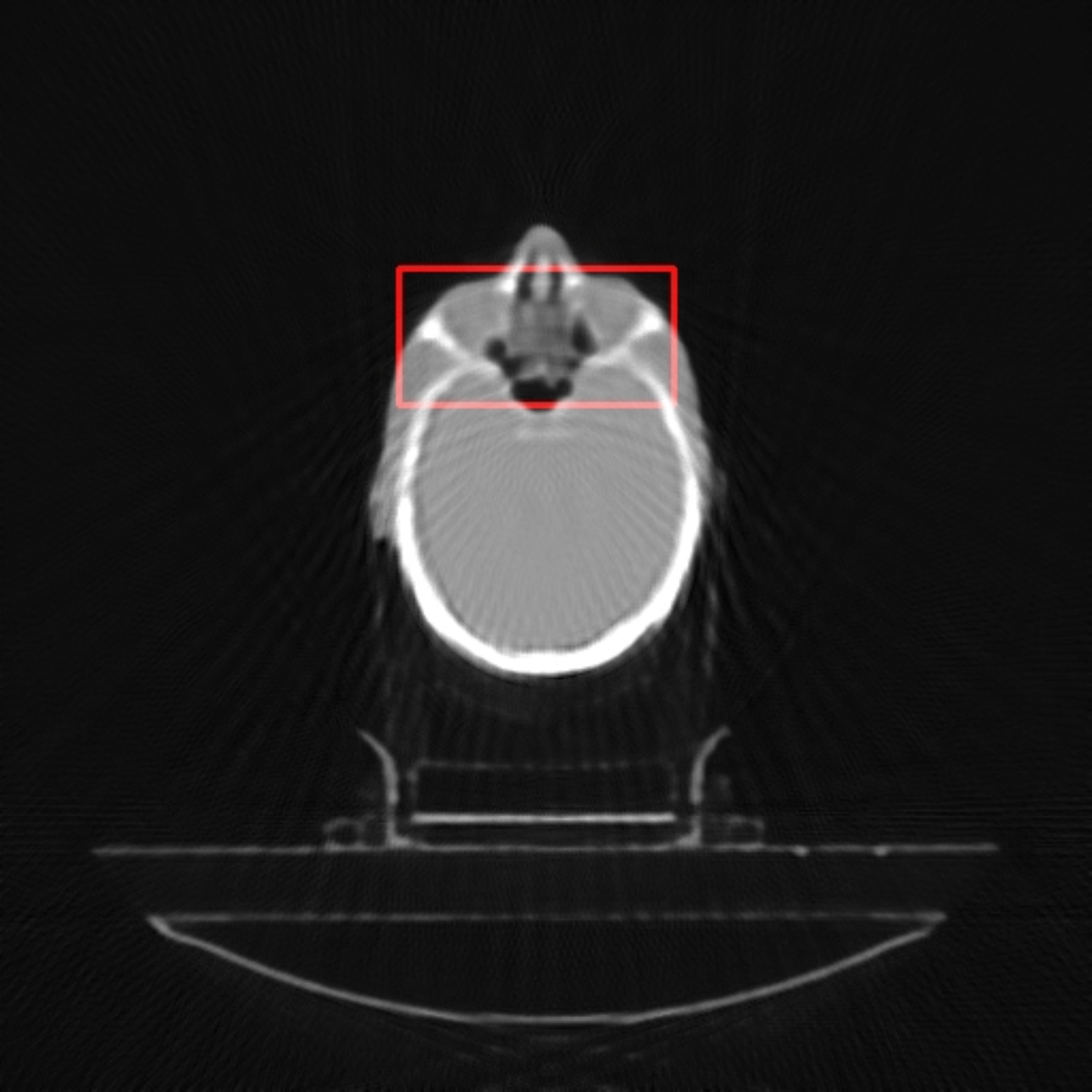}	
				\put(0,0){\color{red}%
					\frame{\includegraphics[scale=0.08]{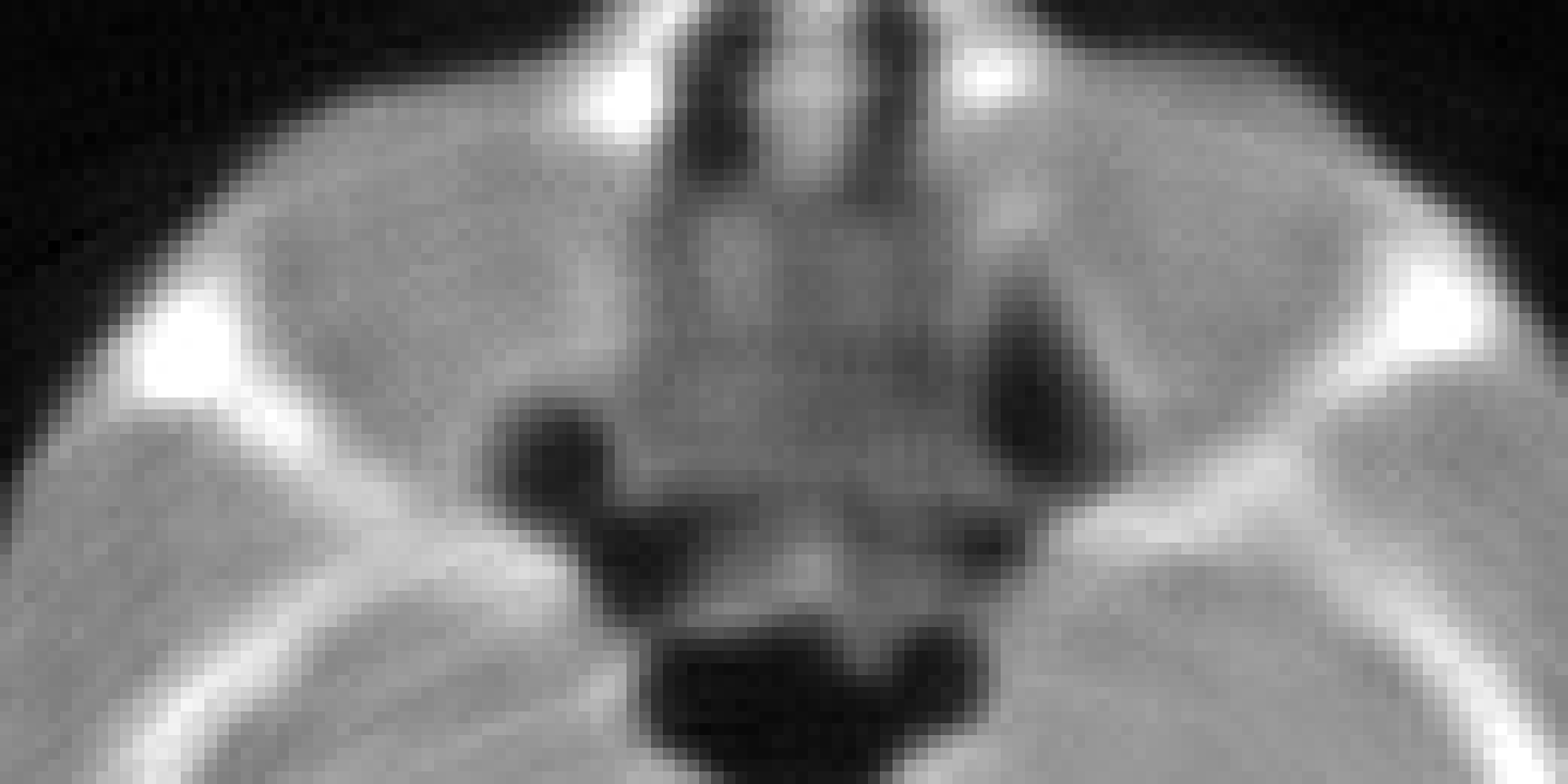}}
				}		
		\end{overpic}}
		\subfloat{
			\begin{overpic}[width=\size\columnwidth]{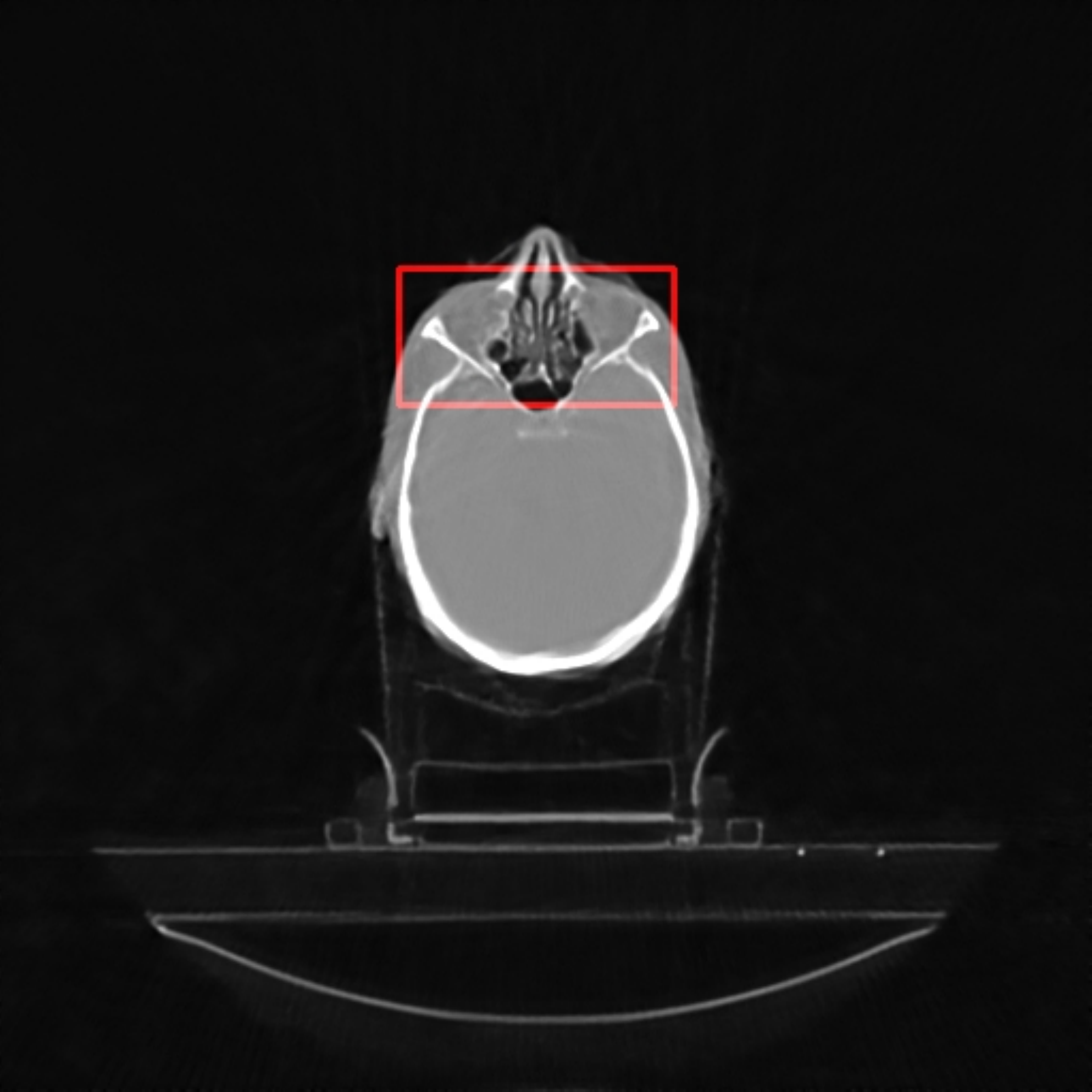}	
				\put(0,0){\color{red}%
					\frame{\includegraphics[scale=0.08]{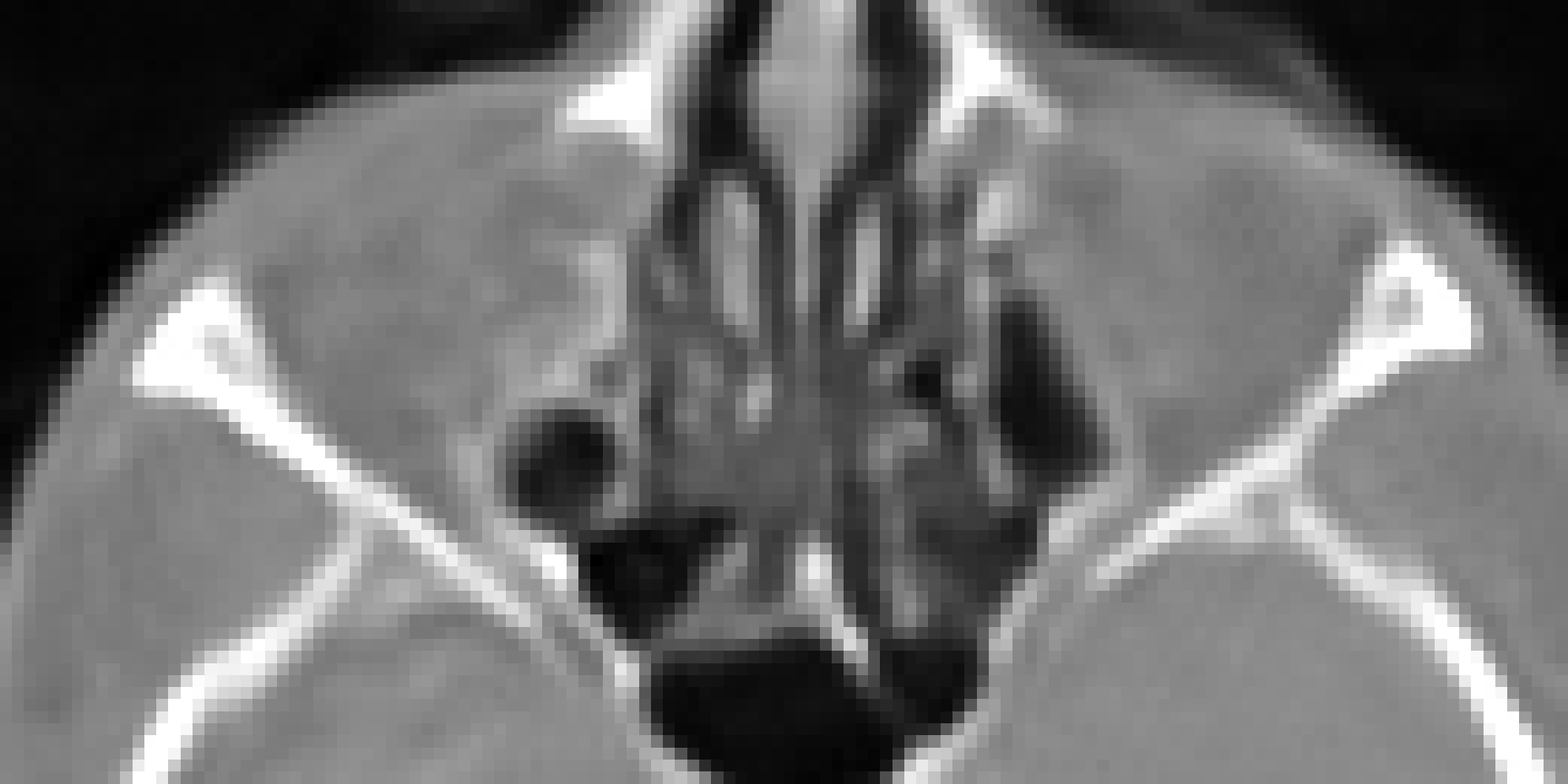}}
				}		
		\end{overpic}}
		
		\vspace{-3mm}
		\renewcommand{\id}{60}
		\subfloat{
			\begin{overpic}[width=\size\columnwidth,percent]{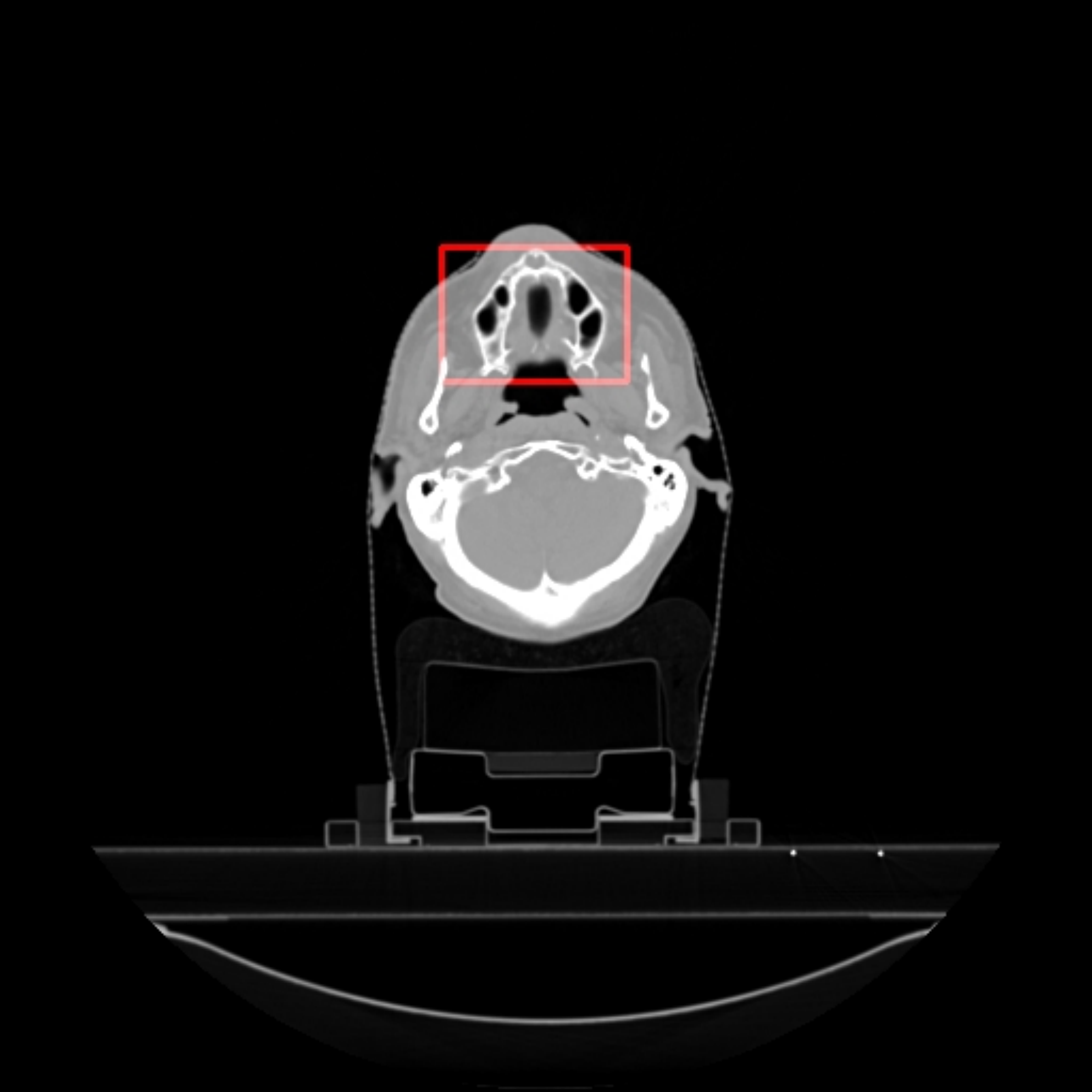}	
				\put(0,0){\color{red}%
					\frame{\includegraphics[scale=0.08]{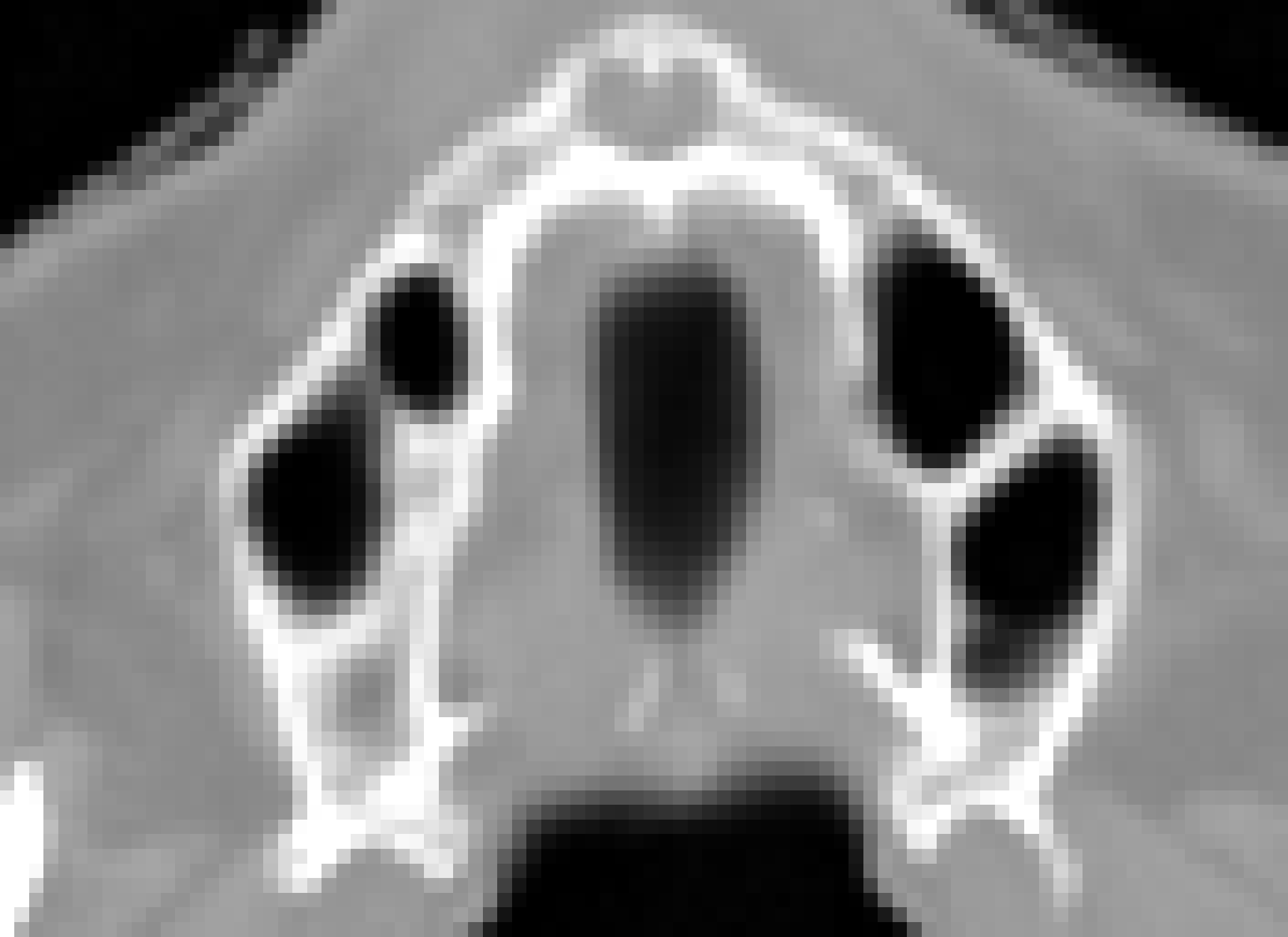}}
				}		
		\end{overpic}}
		\subfloat{
			\begin{overpic}[width=\size\columnwidth]{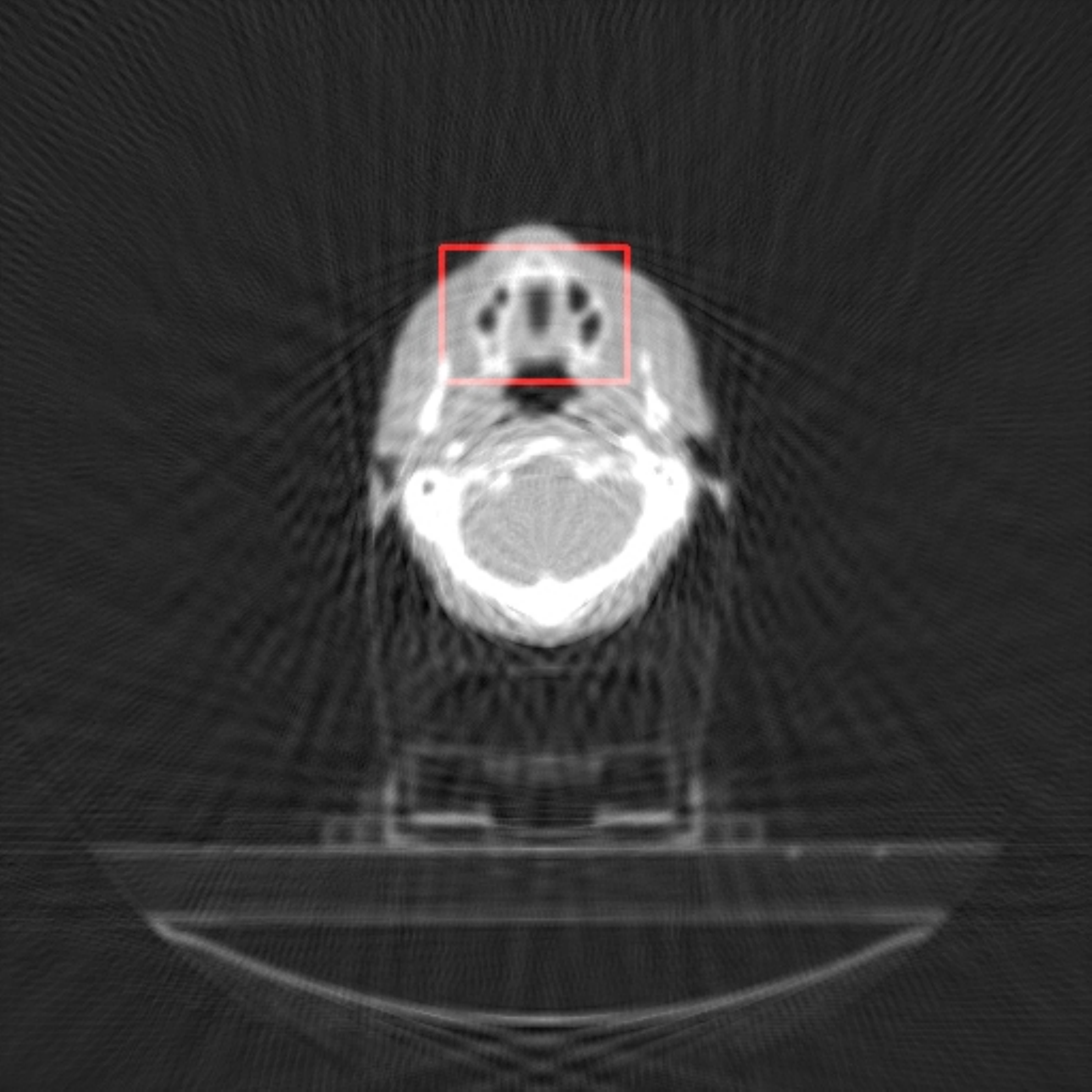}	
				\put(0,0){\color{red}%
					\frame{\includegraphics[scale=0.08]{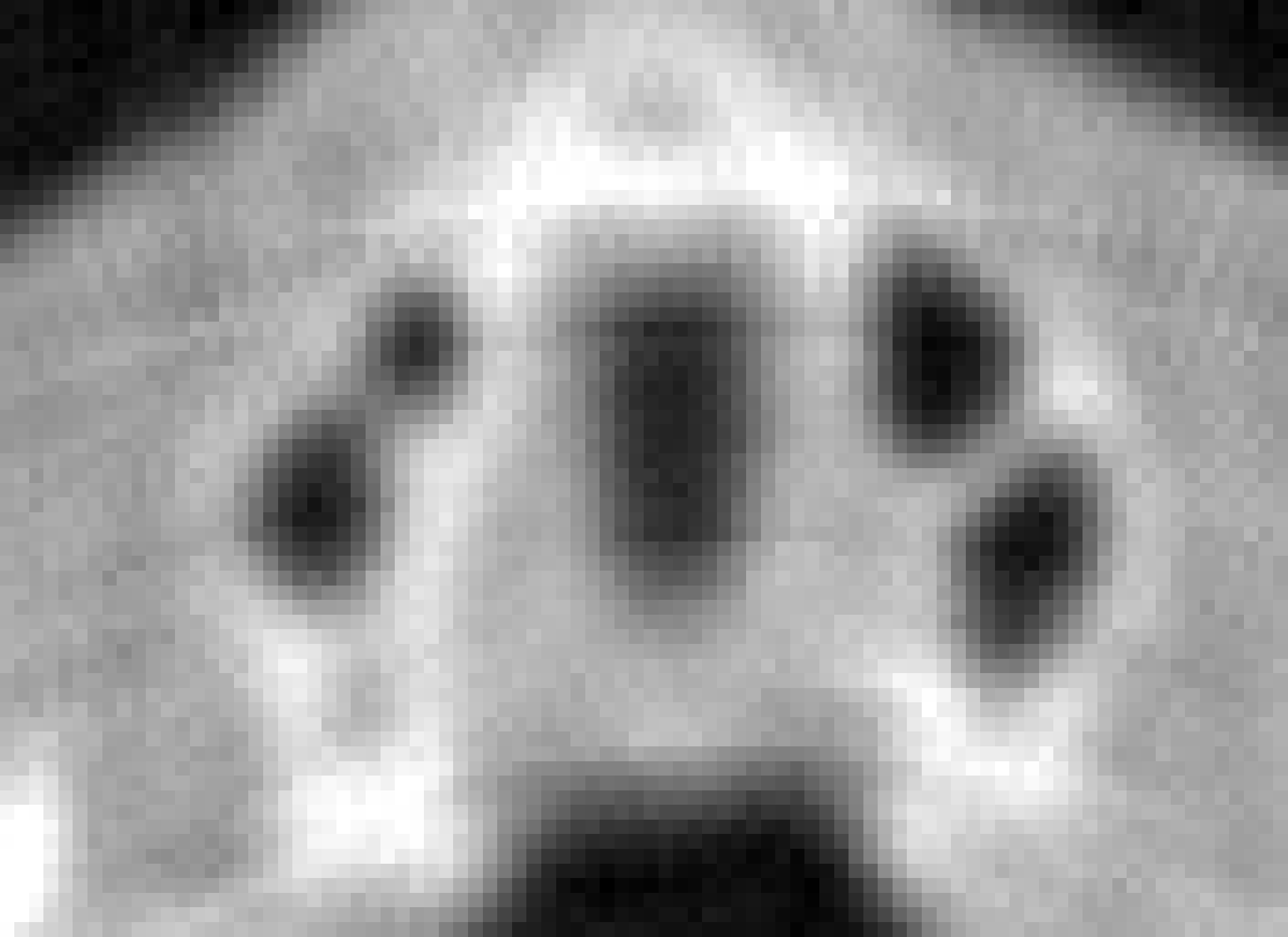}}
				}		
		\end{overpic}}
		\subfloat{
			\begin{overpic}[width=\size\columnwidth]{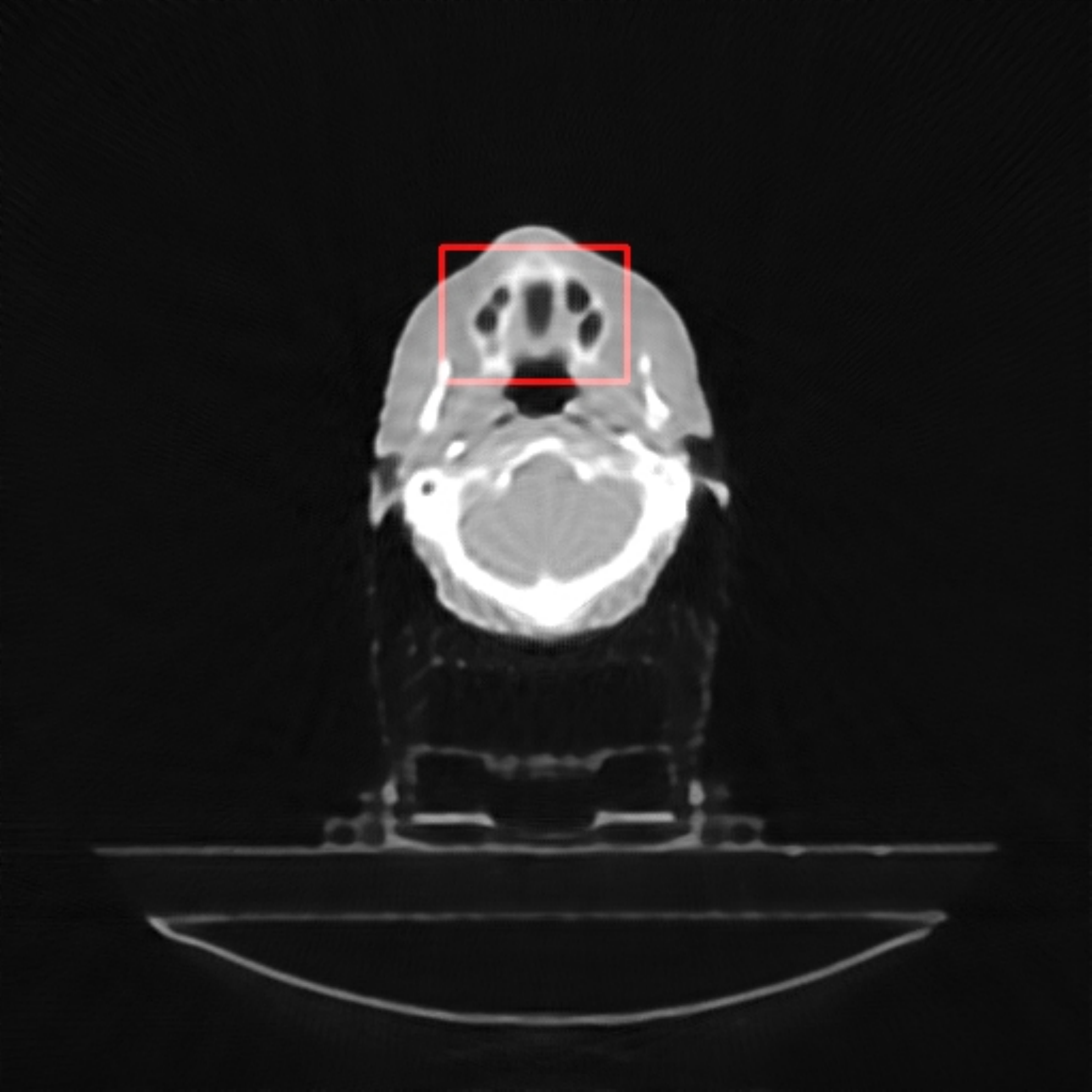}	
				\put(0,0){\color{red}%
					\frame{\includegraphics[scale=0.08]{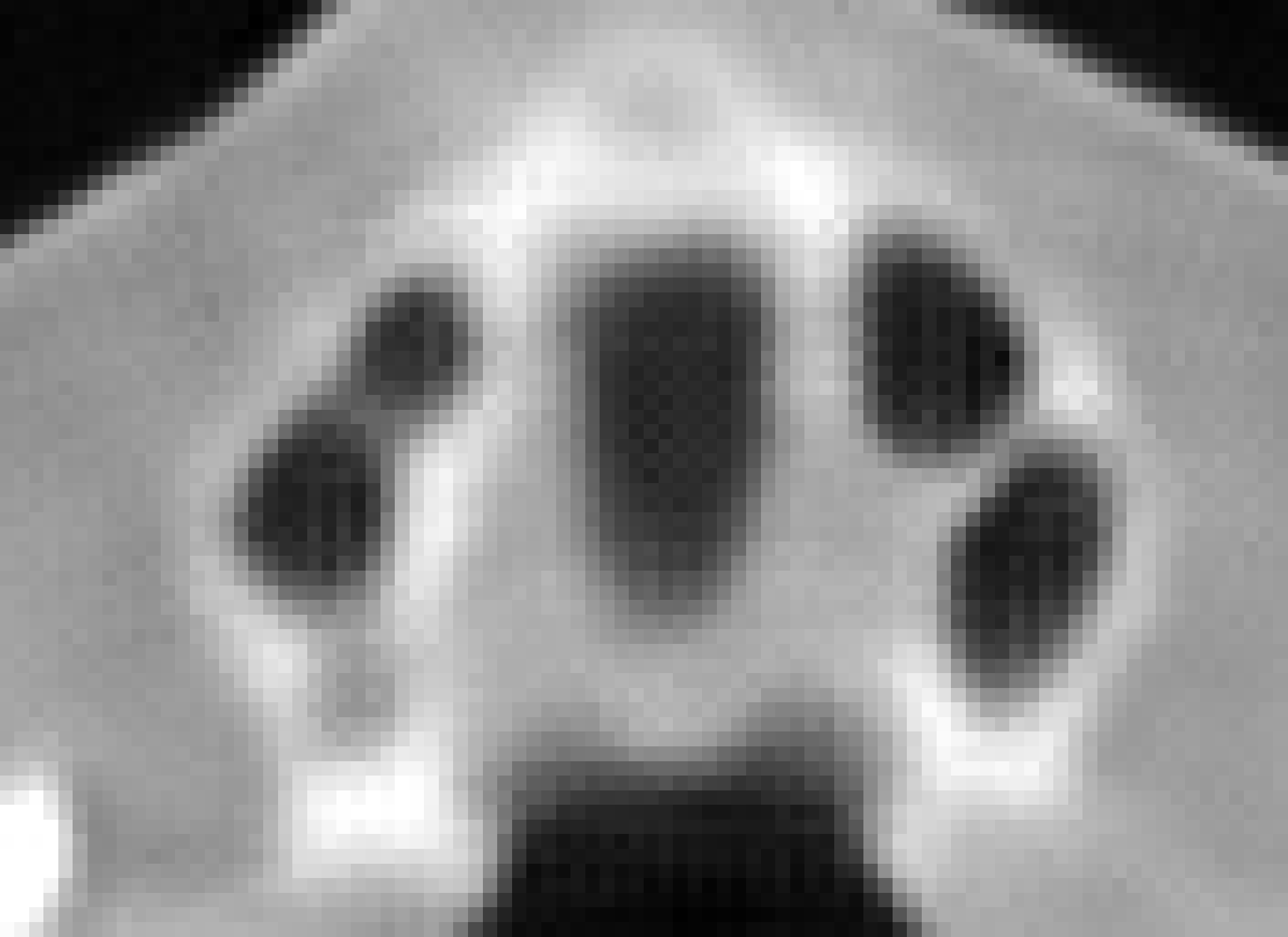}}
				}		
		\end{overpic}}
		\subfloat{
			\begin{overpic}[width=\size\columnwidth]{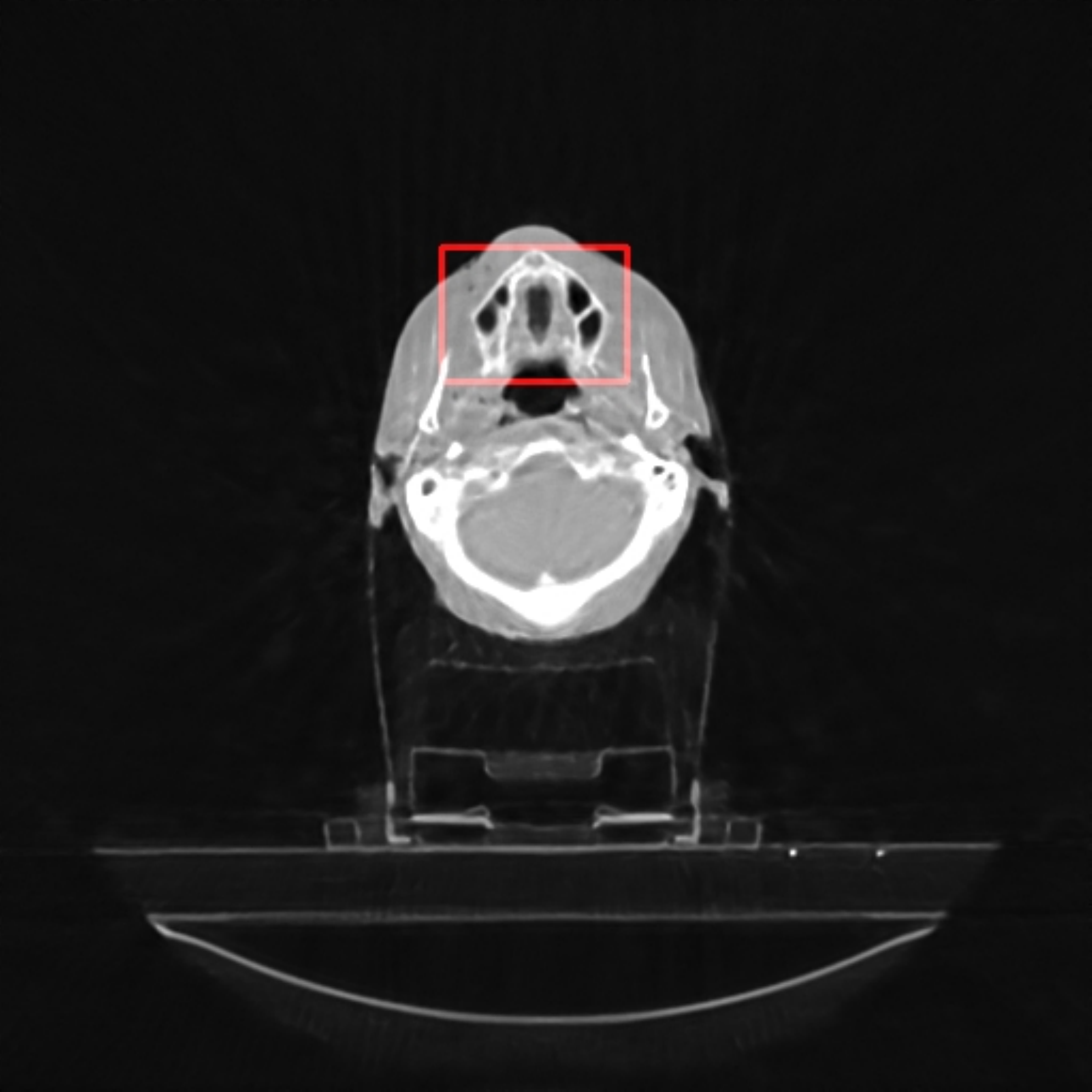}	
				\put(0,0){\color{red}%
					\frame{\includegraphics[scale=0.08]{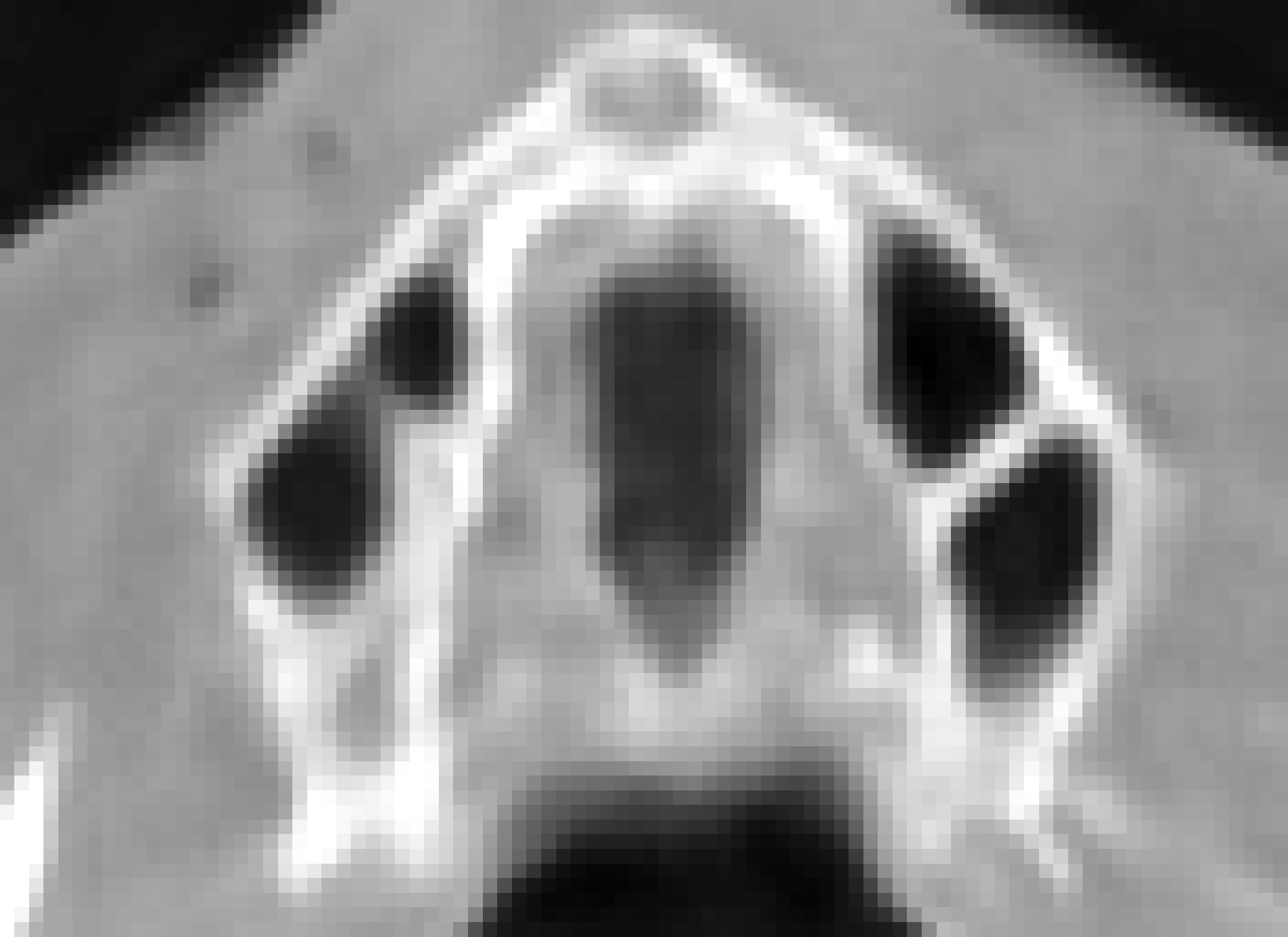}}
				}		
		\end{overpic}}
		\vspace{-3mm}
		
		\renewcommand{\id}{80}
		\subfloat{
			\begin{overpic}[width=\size\columnwidth,percent]{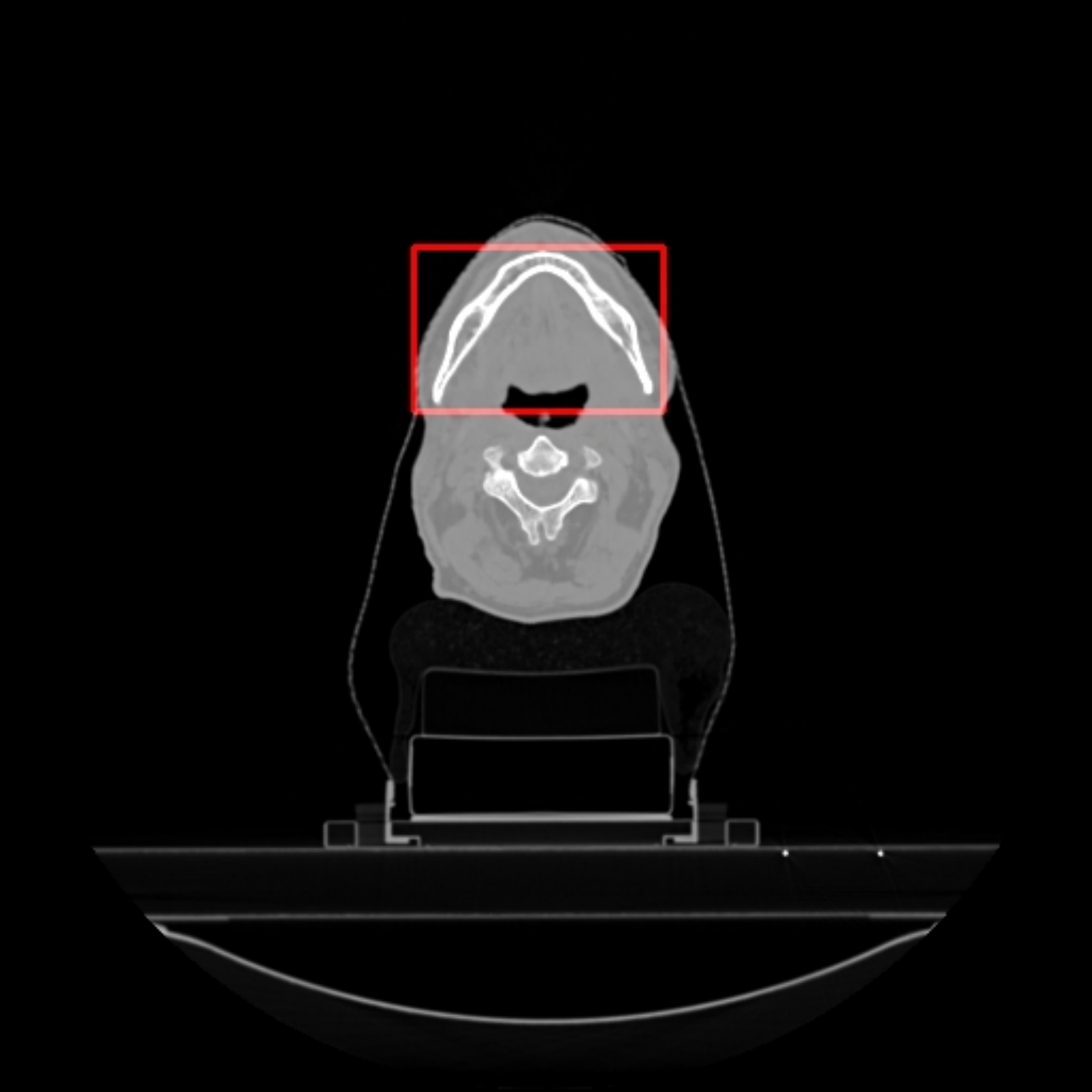}	
				\put(0,0){\color{red}%
					\frame{\includegraphics[scale=0.08]{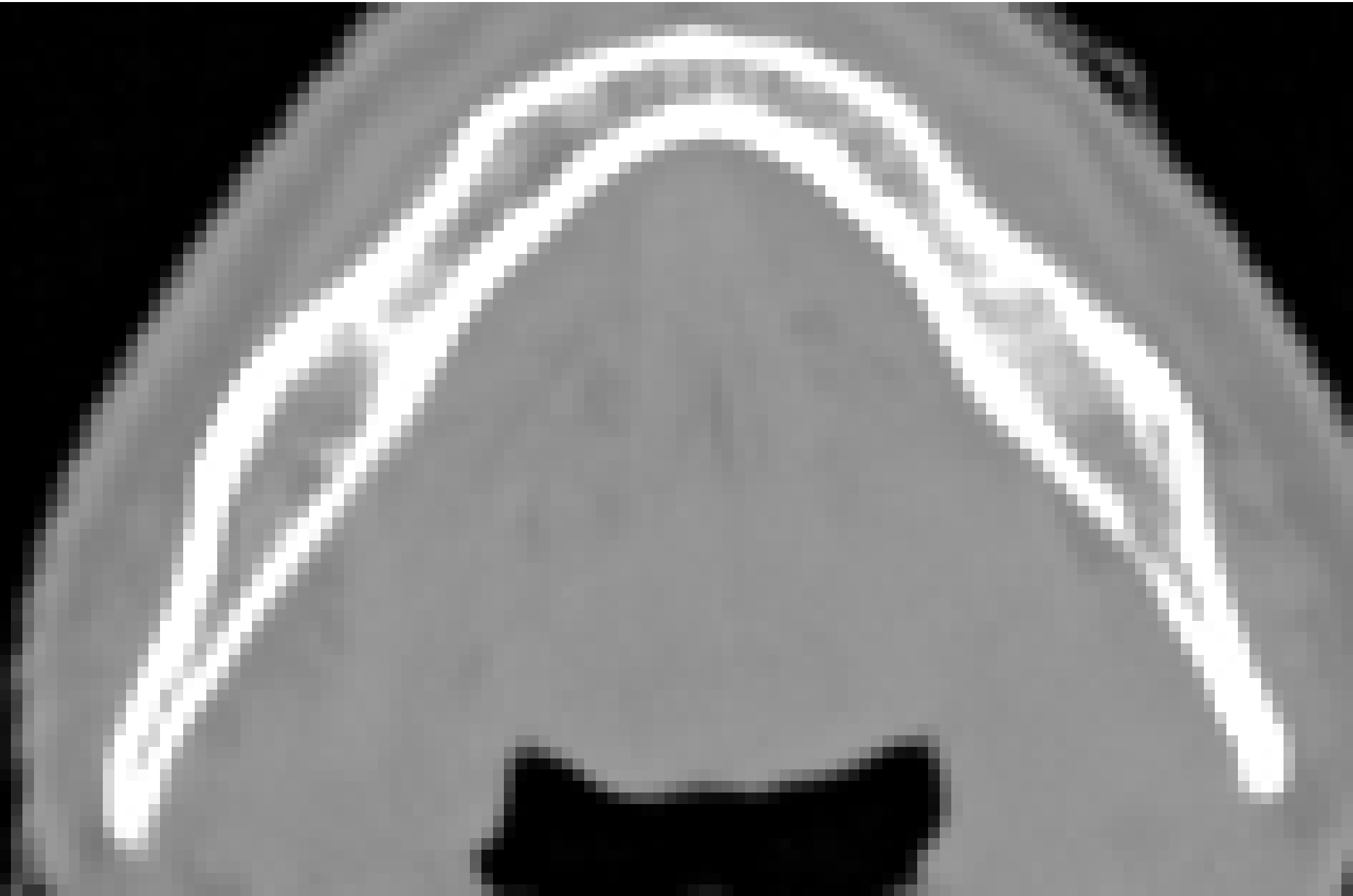}}
				}		
		\end{overpic}}
		\subfloat{
			\begin{overpic}[width=\size\columnwidth]{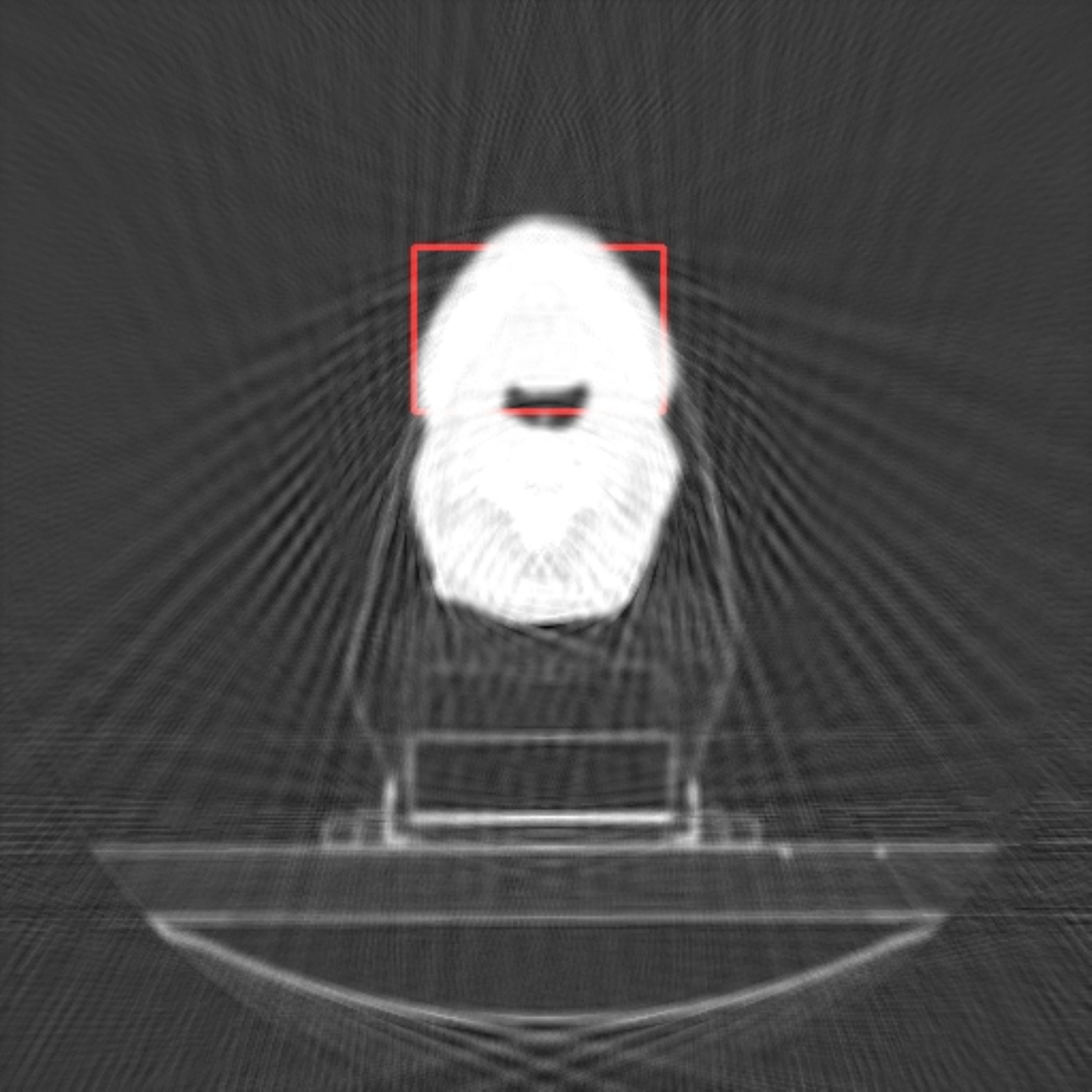}	
				\put(0,0){\color{red}%
					\frame{\includegraphics[scale=0.08]{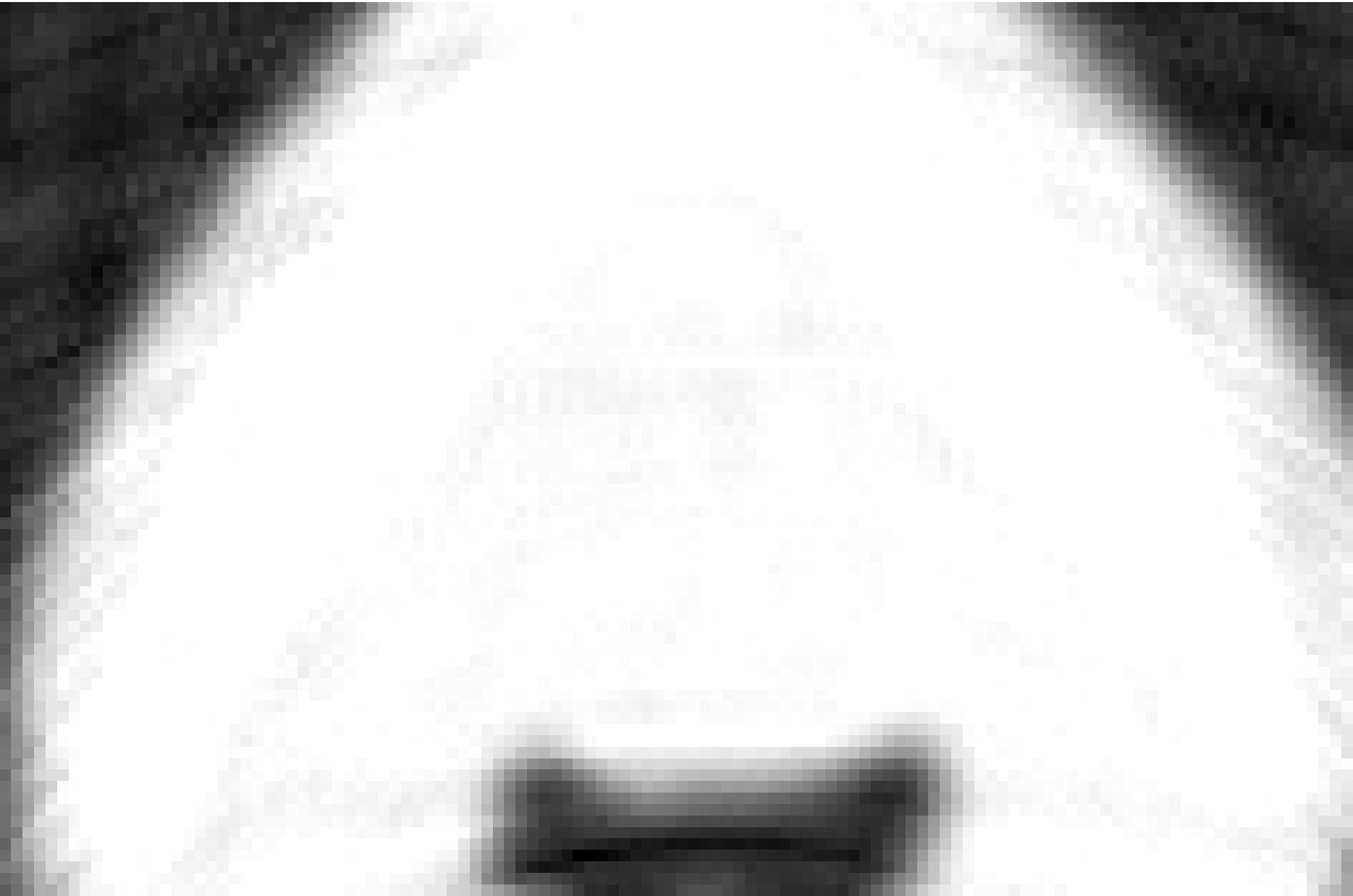}}
				}		
		\end{overpic}}
		\subfloat{
			\begin{overpic}[width=\size\columnwidth]{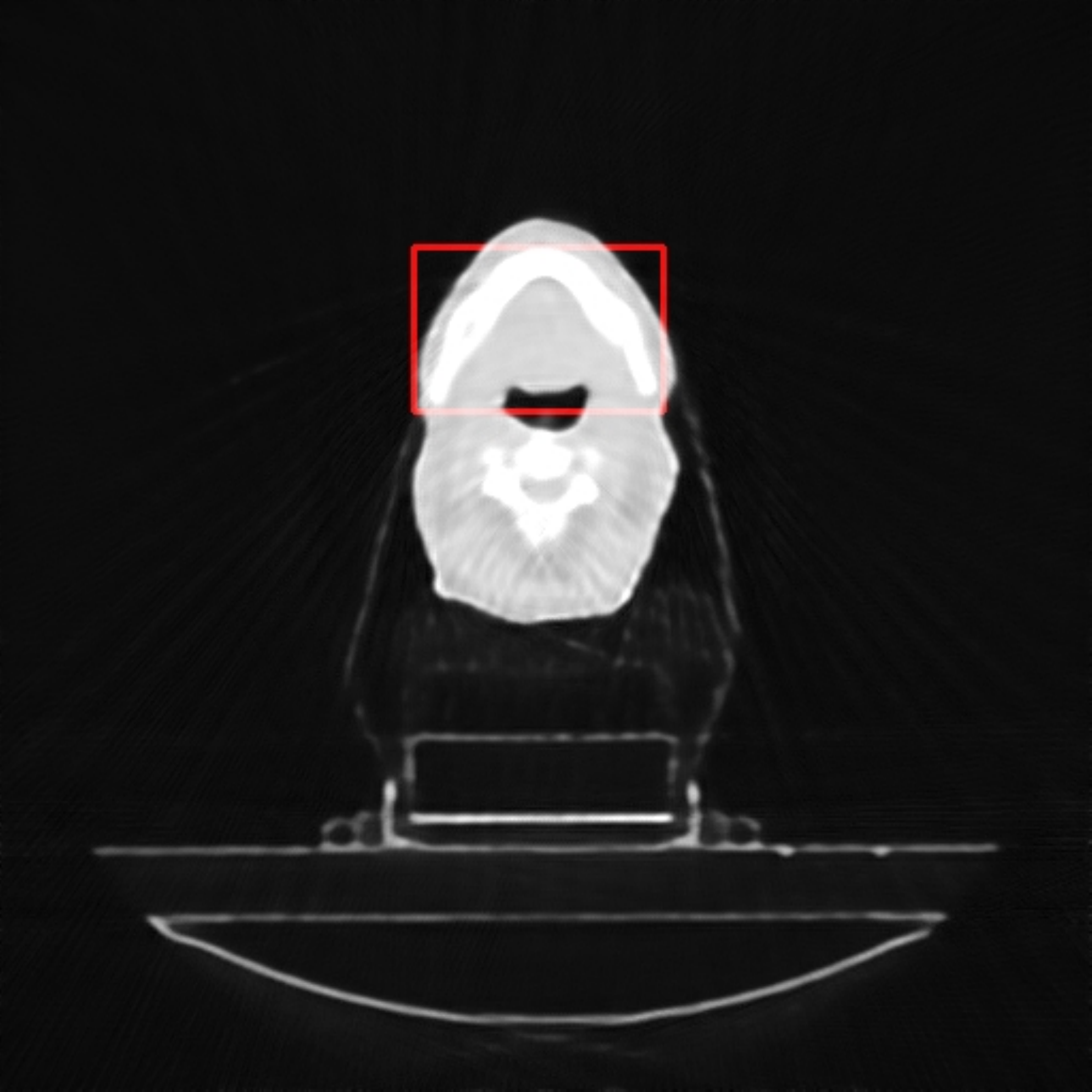}	
				\put(0,0){\color{red}%
					\frame{\includegraphics[scale=0.08]{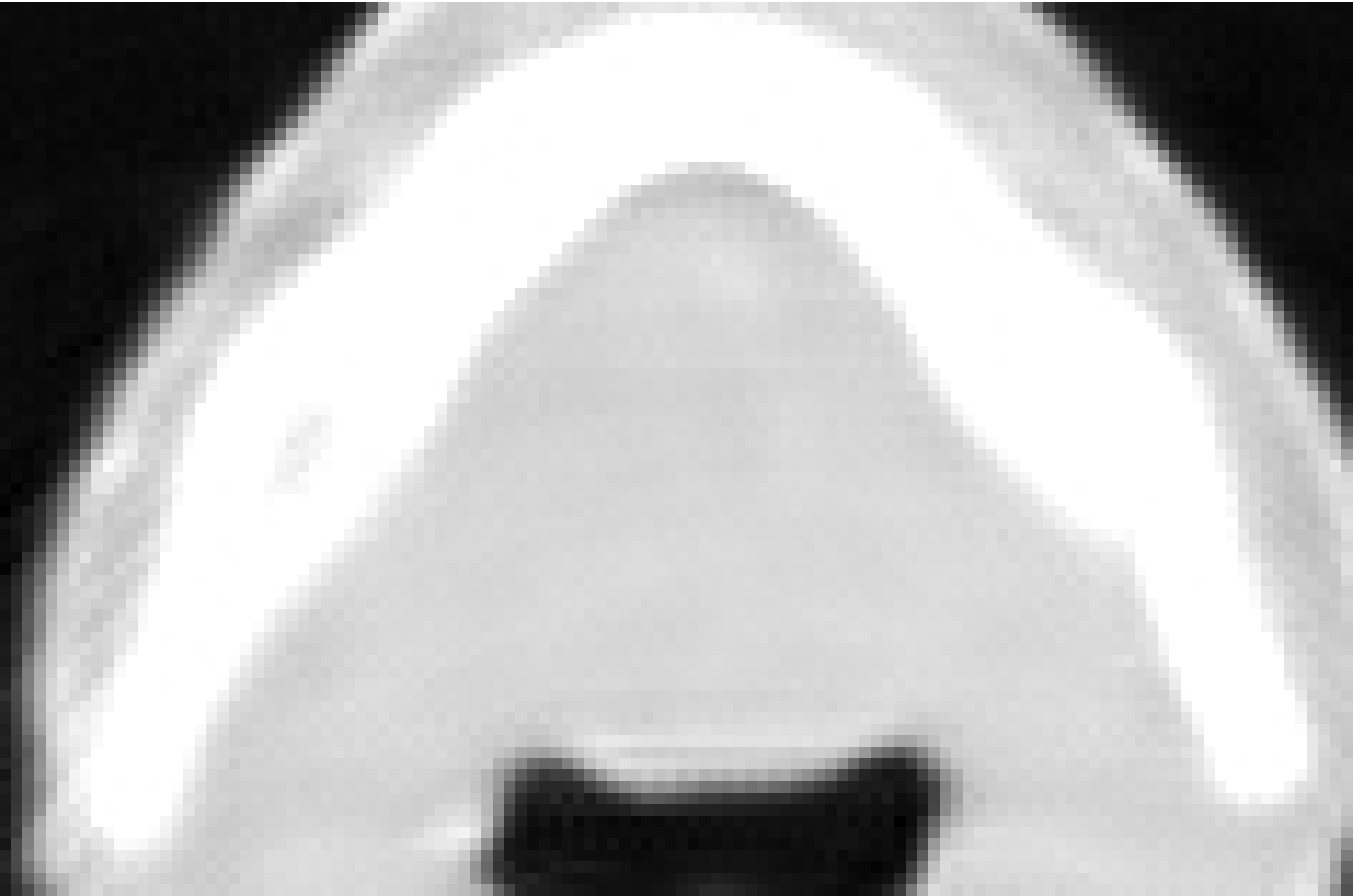}}
				}		
		\end{overpic}}
		\subfloat{
			\begin{overpic}[width=\size\columnwidth]{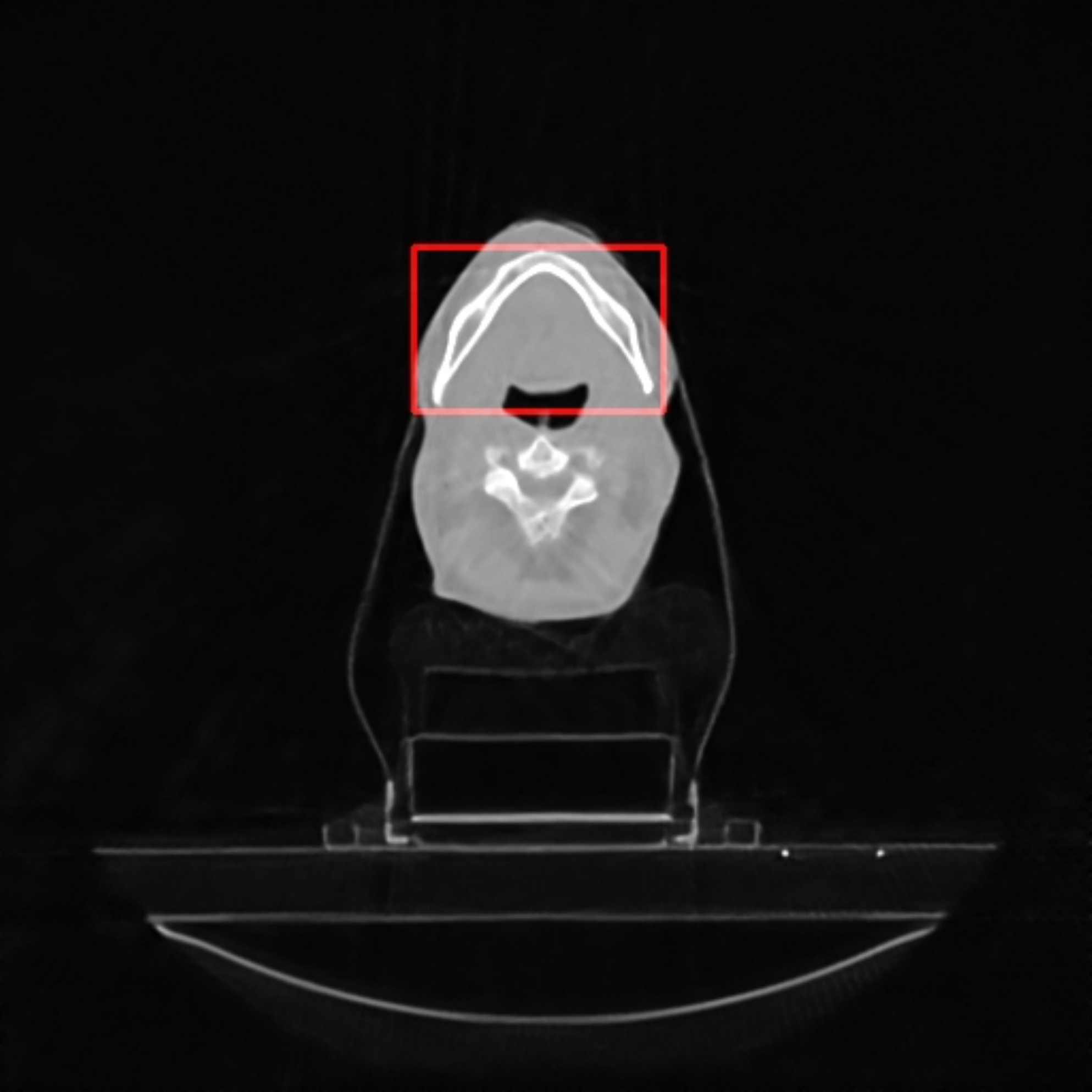}	
				\put(0,0){\color{red}%
					\frame{\includegraphics[scale=0.08]{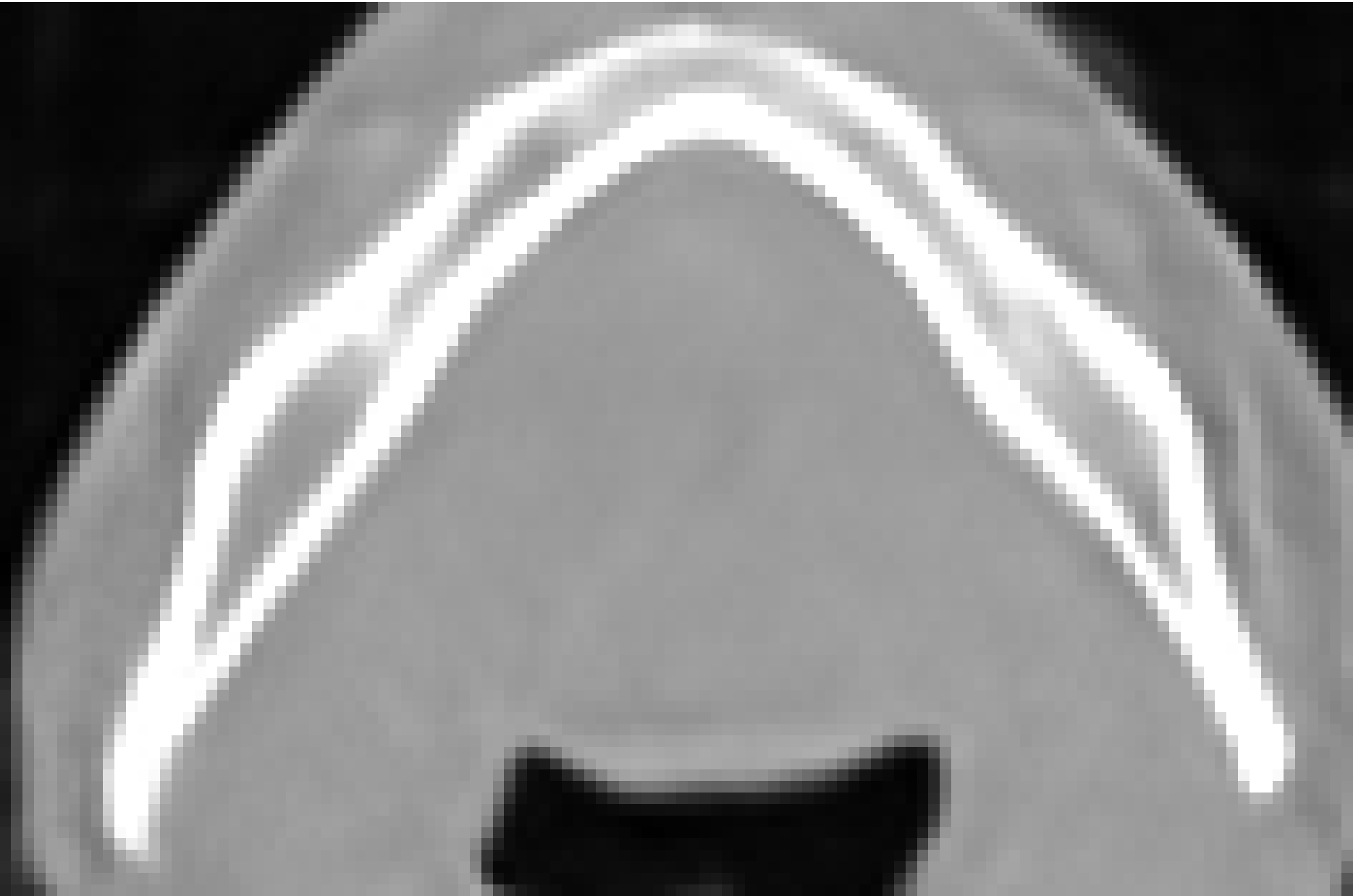}}
				}		
		\end{overpic}}
		\vspace{-3mm}
		\renewcommand{\id}{140}
		\subfloat{
			\begin{overpic}[width=\size\columnwidth,percent]{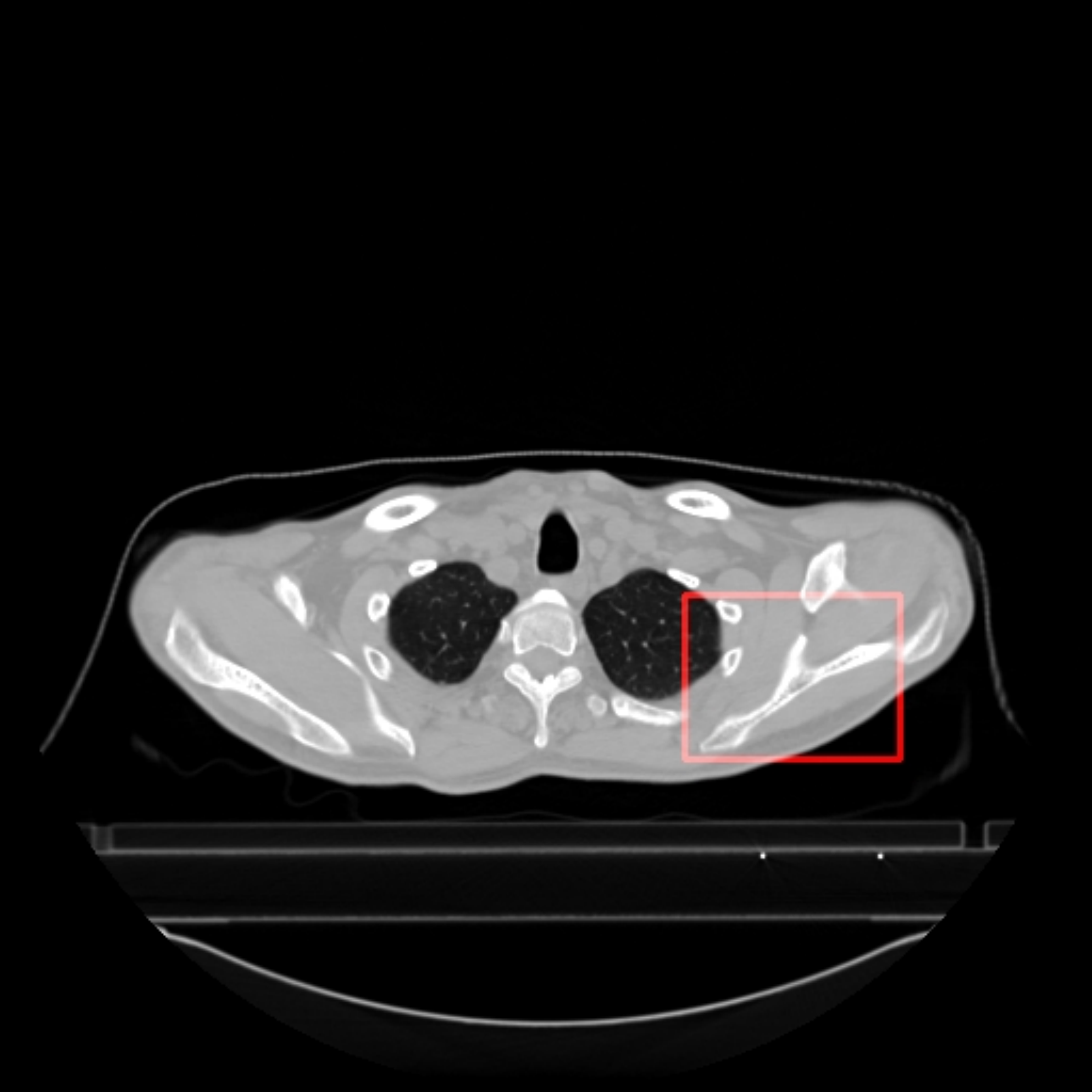}	
				\put(0,0){\color{red}%
					\frame{\includegraphics[scale=0.08]{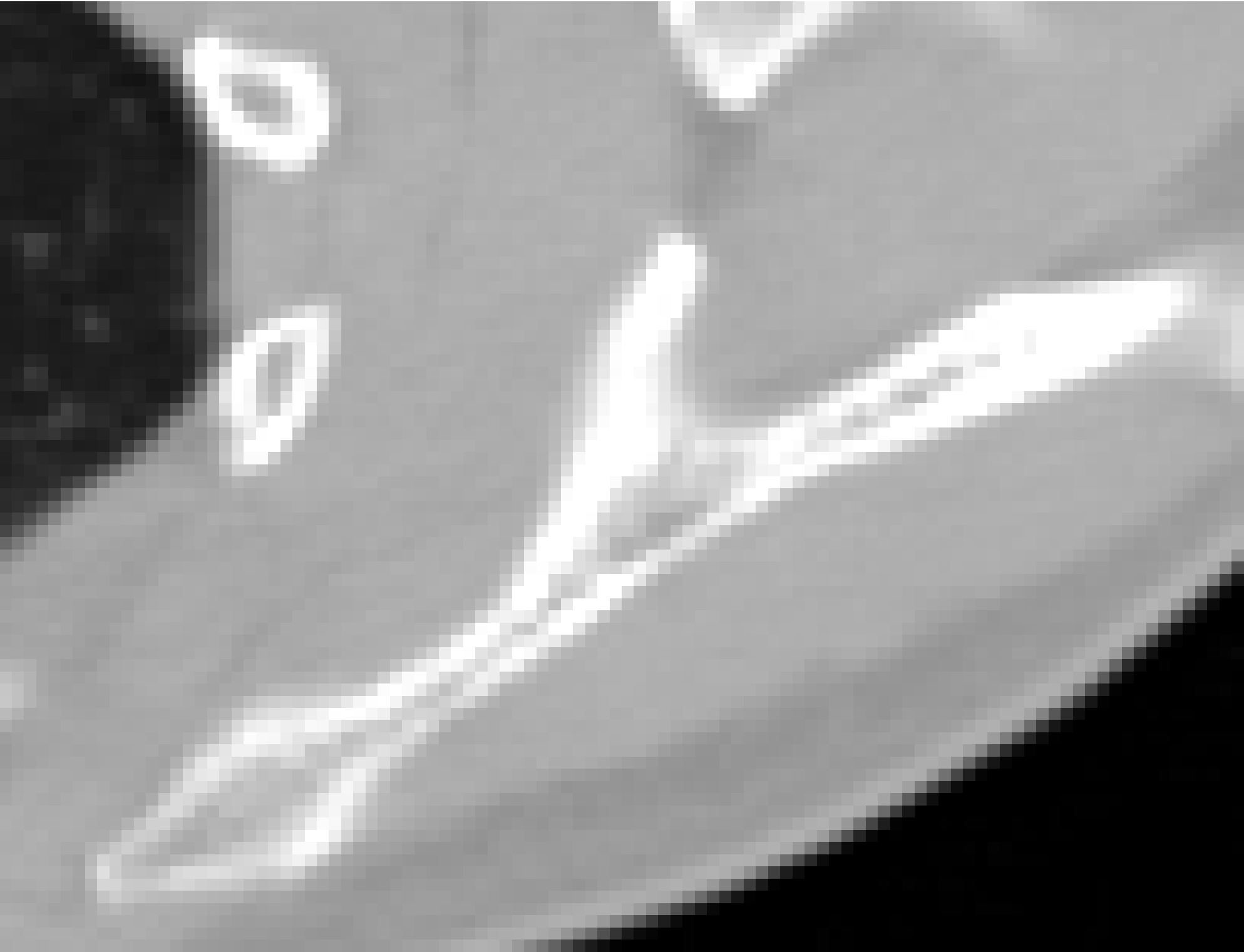}}
				}		
		\end{overpic}}
		\subfloat{
			\begin{overpic}[width=\size\columnwidth]{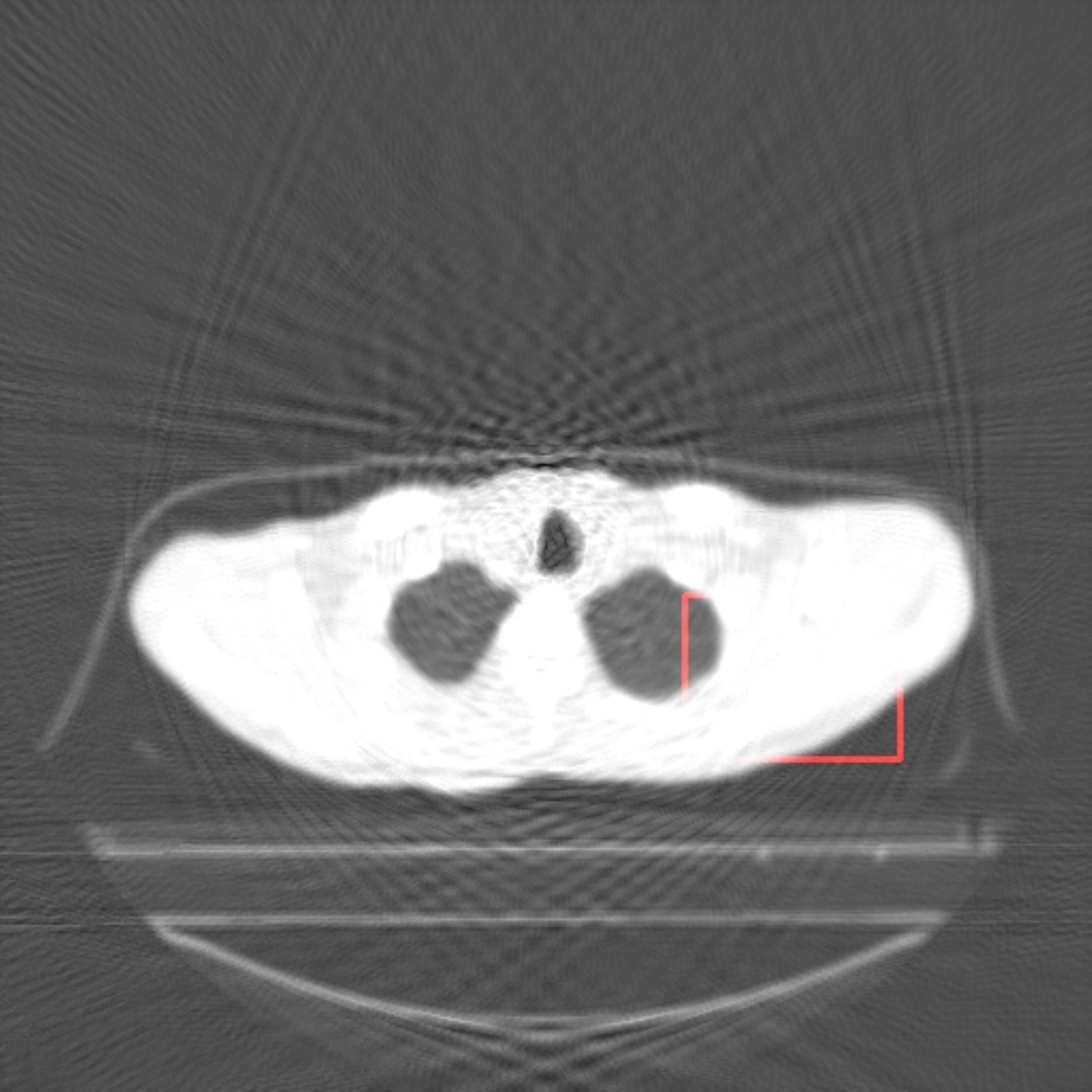}	
				\put(0,0){\color{red}%
					\frame{\includegraphics[scale=0.08]{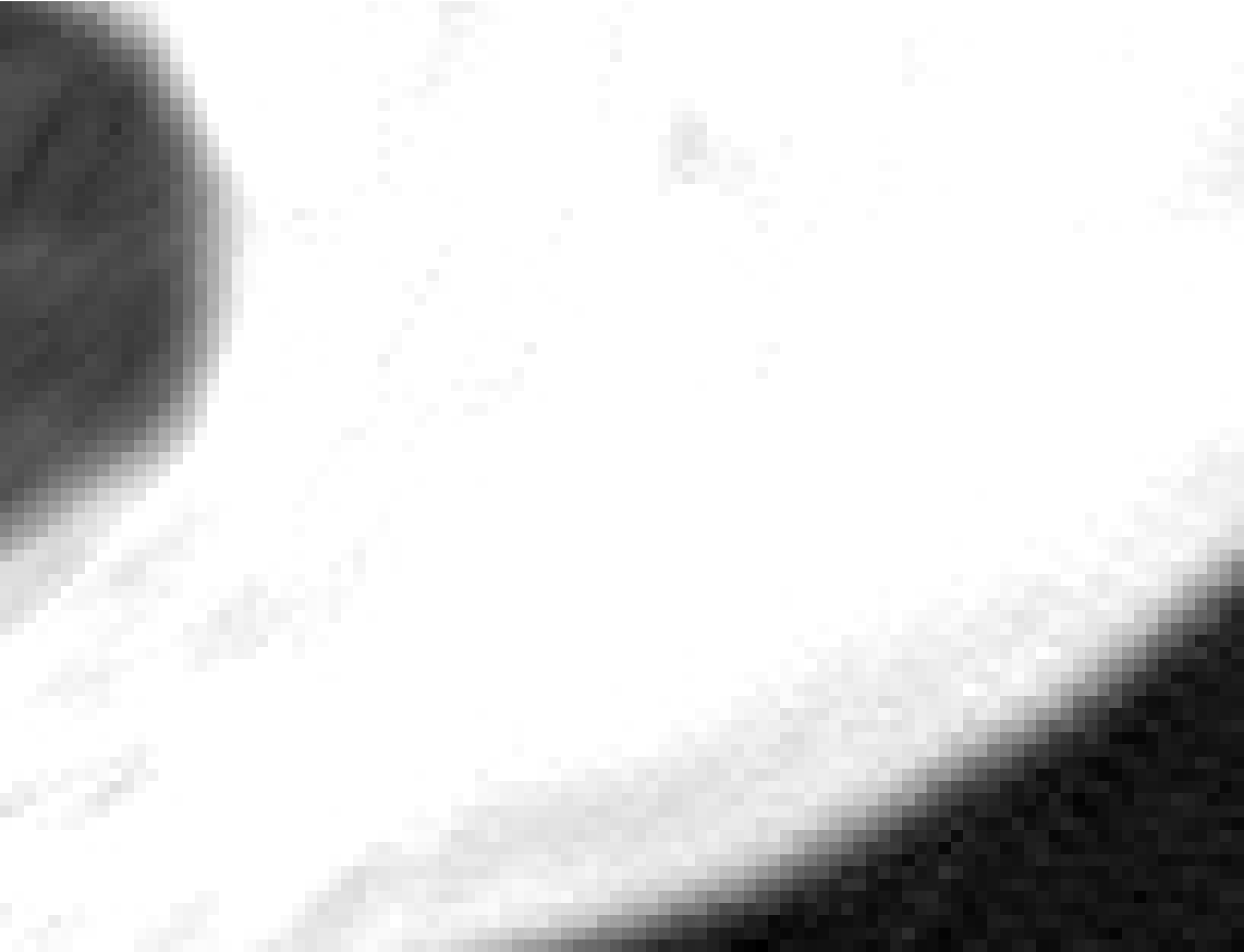}}
				}		
		\end{overpic}}
		\subfloat{
			\begin{overpic}[width=\size\columnwidth]{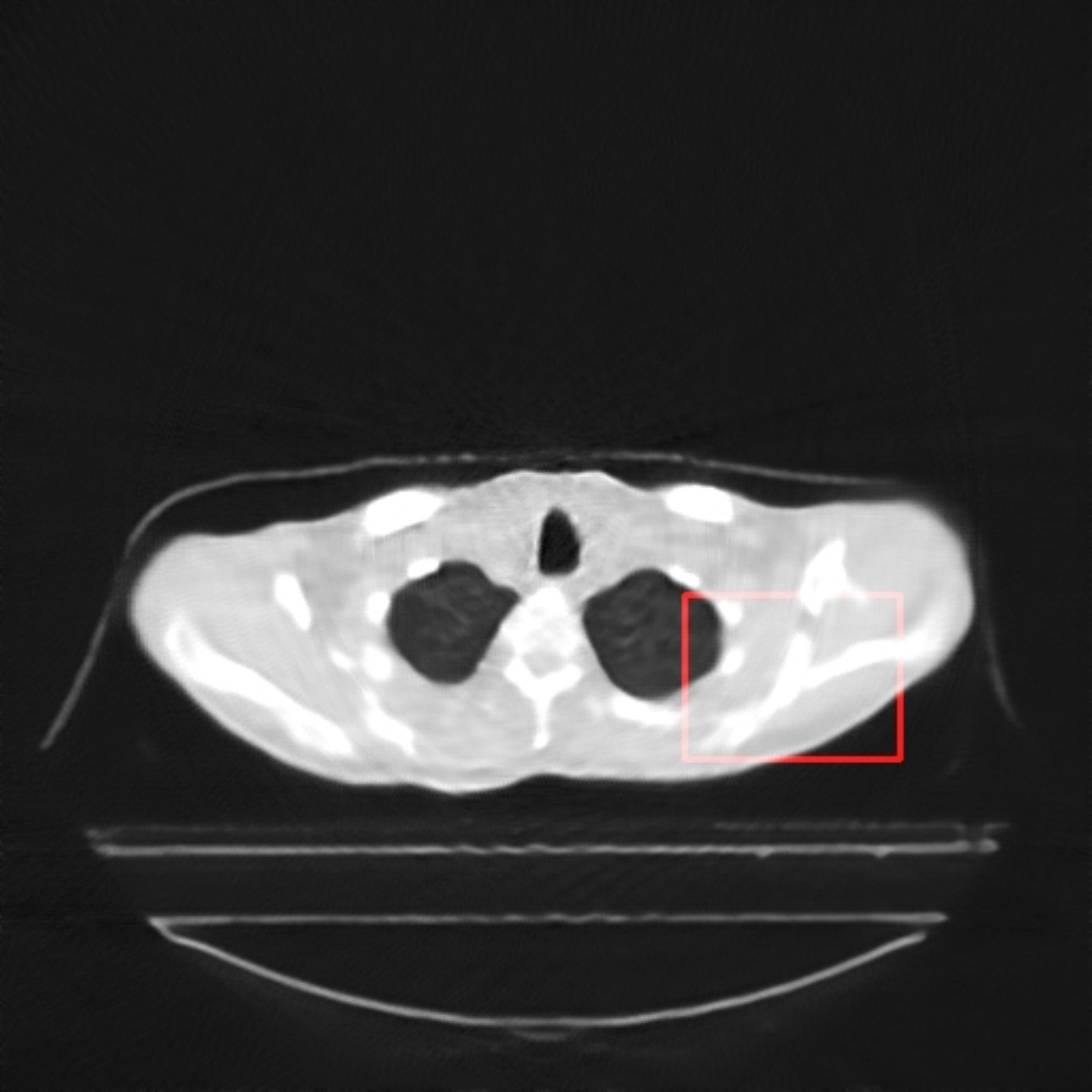}	
				\put(0,0){\color{red}%
					\frame{\includegraphics[scale=0.08]{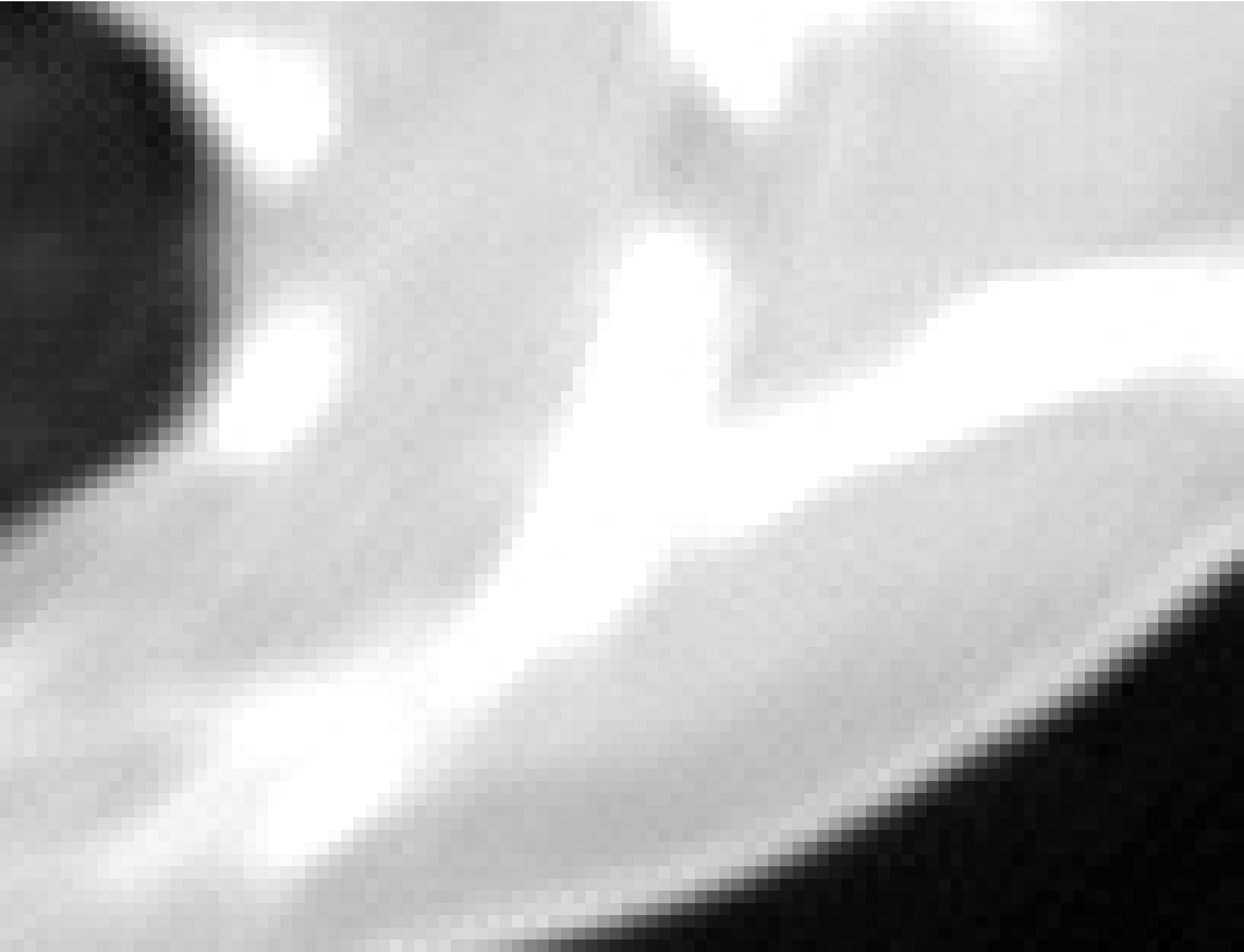}}
				}		
		\end{overpic}}
		\subfloat{
			\begin{overpic}[width=\size\columnwidth]{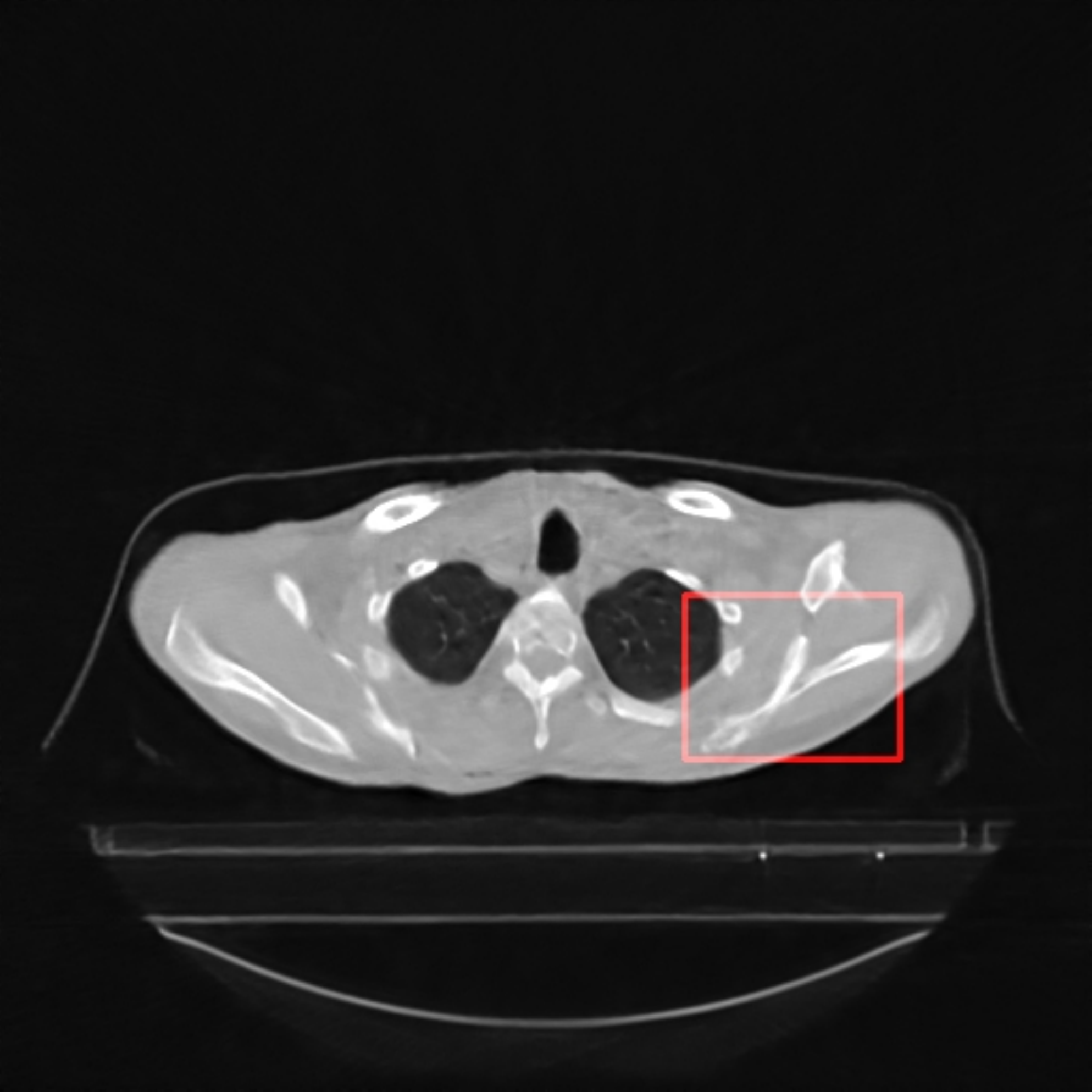}	
				\put(0,0){\color{red}%
					\frame{\includegraphics[scale=0.08]{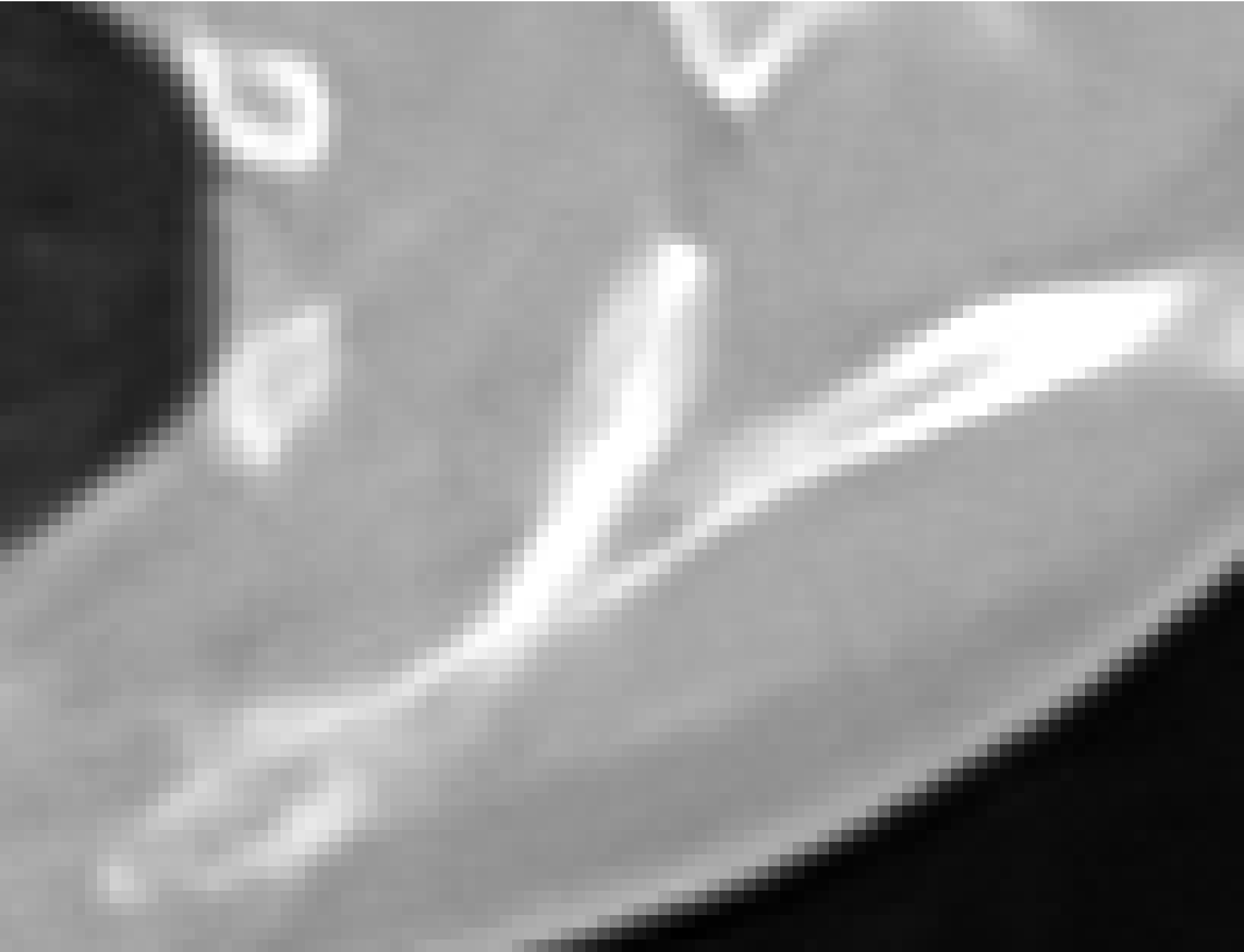}}
				}		
		\end{overpic}}
		\vspace{-3mm}
		\setcounter{subfigure}{0}
		\renewcommand{\id}{200}
		\subfloat[Label]{
			\begin{overpic}[width=\size\columnwidth,percent]{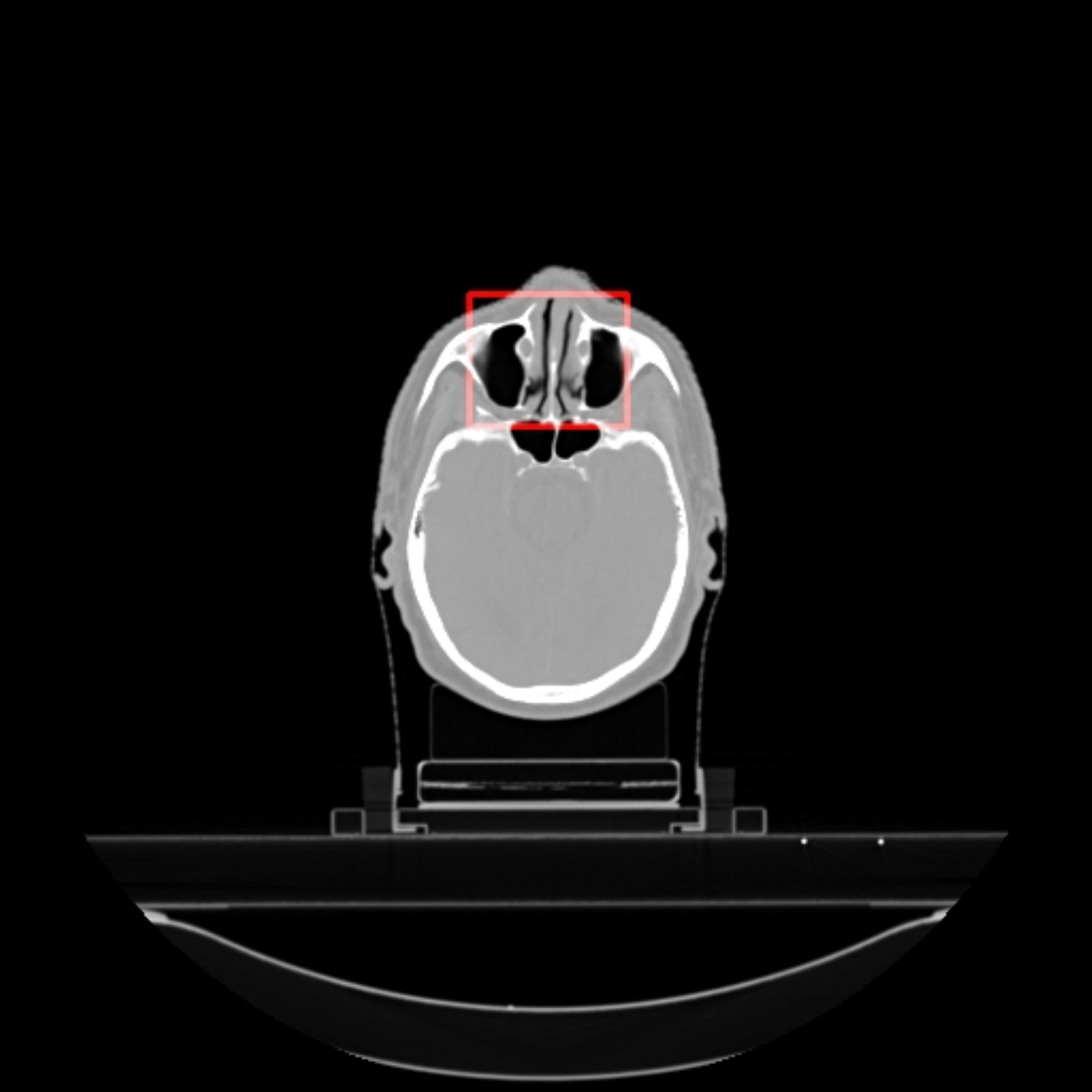}	
				\put(0,0){\color{red}%
					\frame{\includegraphics[scale=0.08]{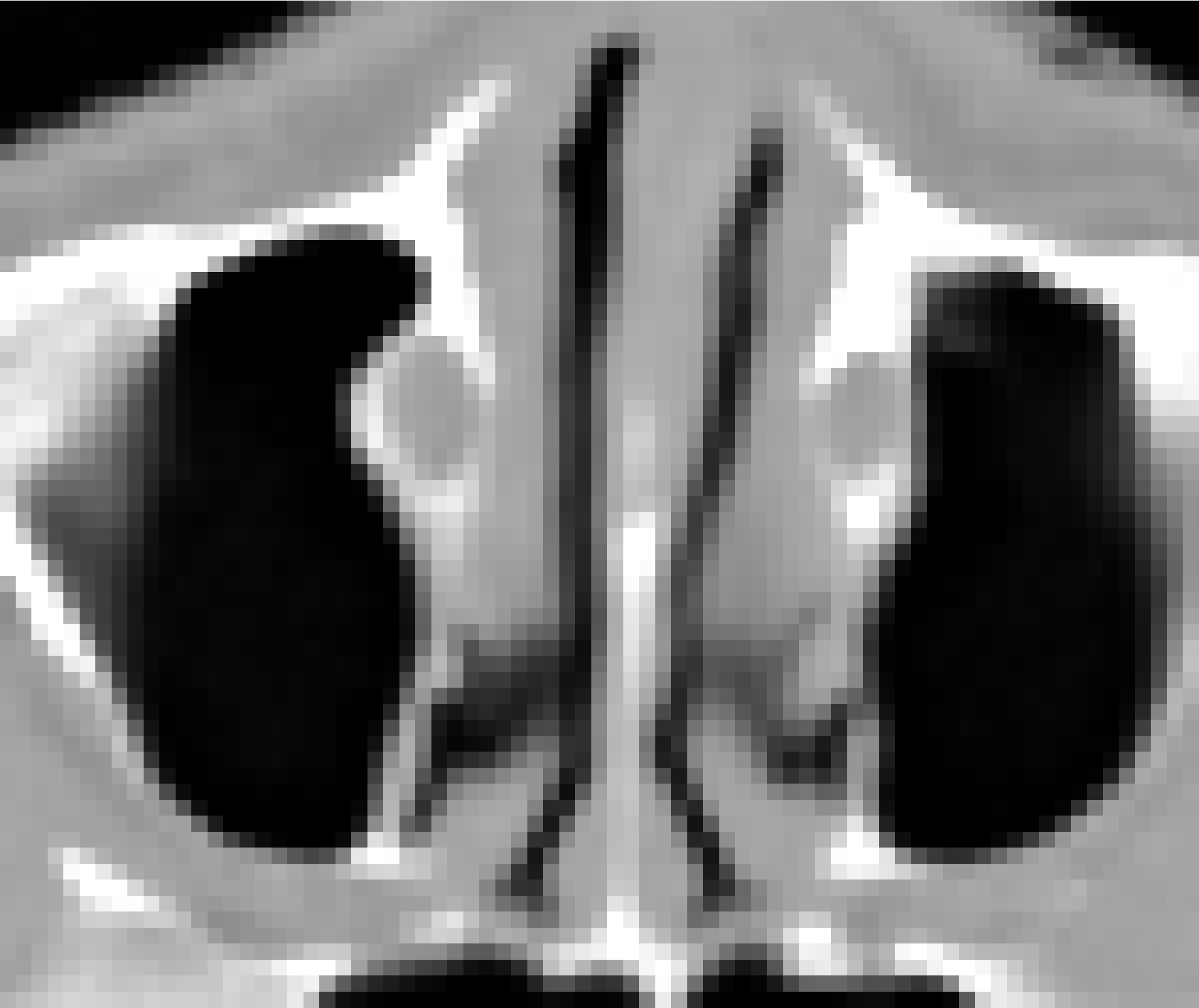}}
				}		
		\end{overpic}}
		\subfloat[Katsevich algorithm]{
			\begin{overpic}[width=\size\columnwidth]{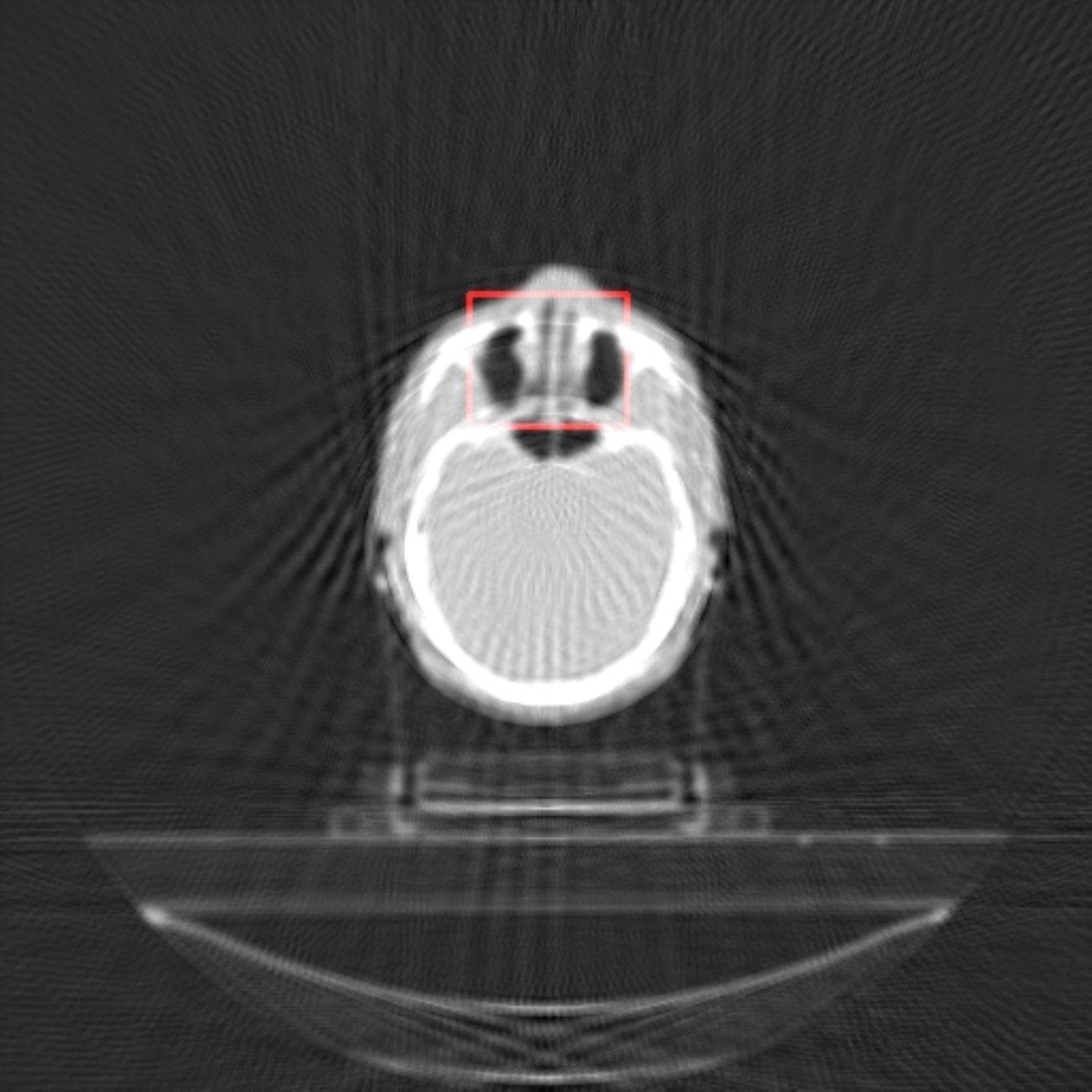}	
				\put(0,0){\color{red}%
					\frame{\includegraphics[scale=0.08]{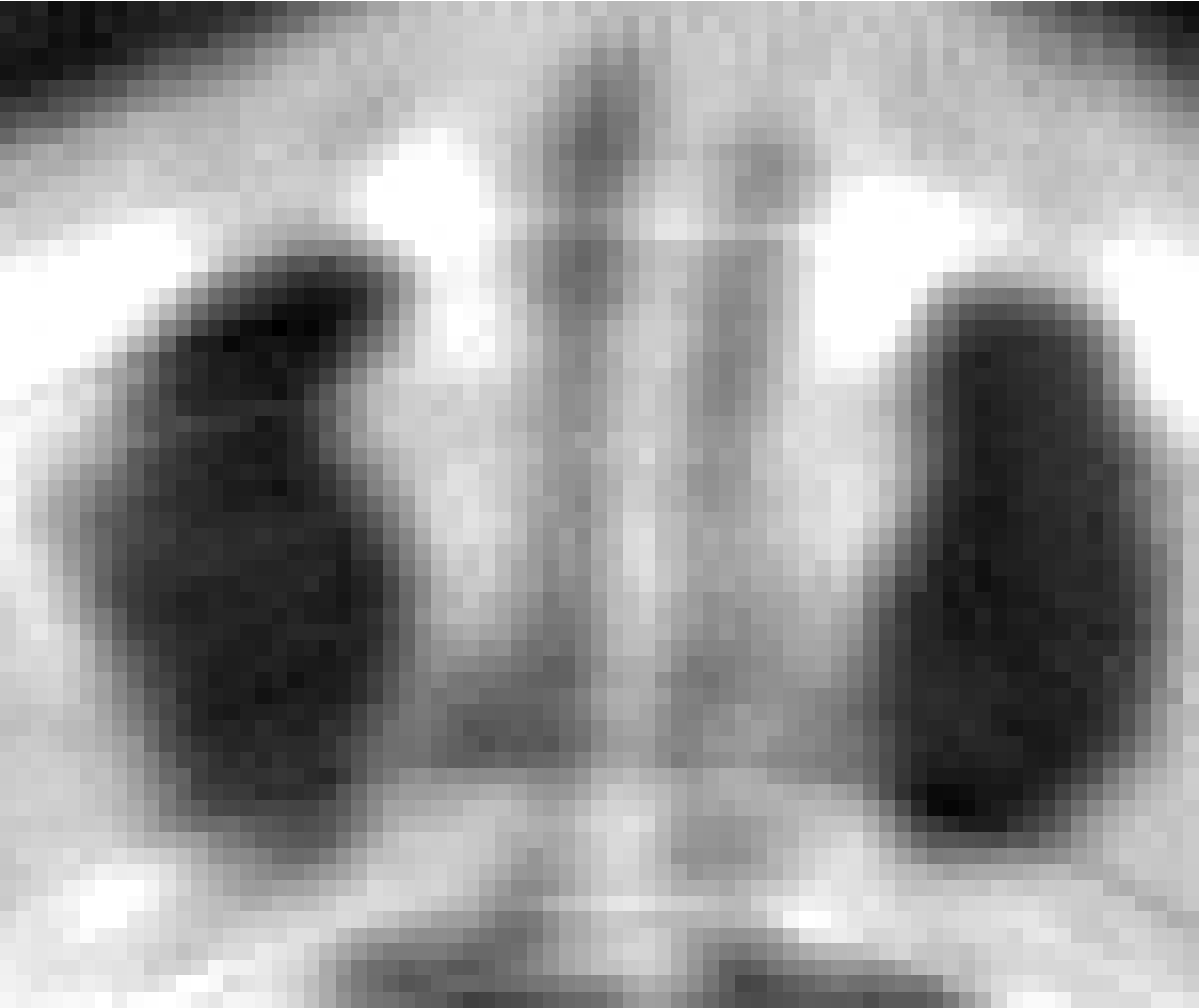}}
				}		
		\end{overpic}}
		\subfloat[FBPConvNet]{
			\begin{overpic}[width=\size\columnwidth]{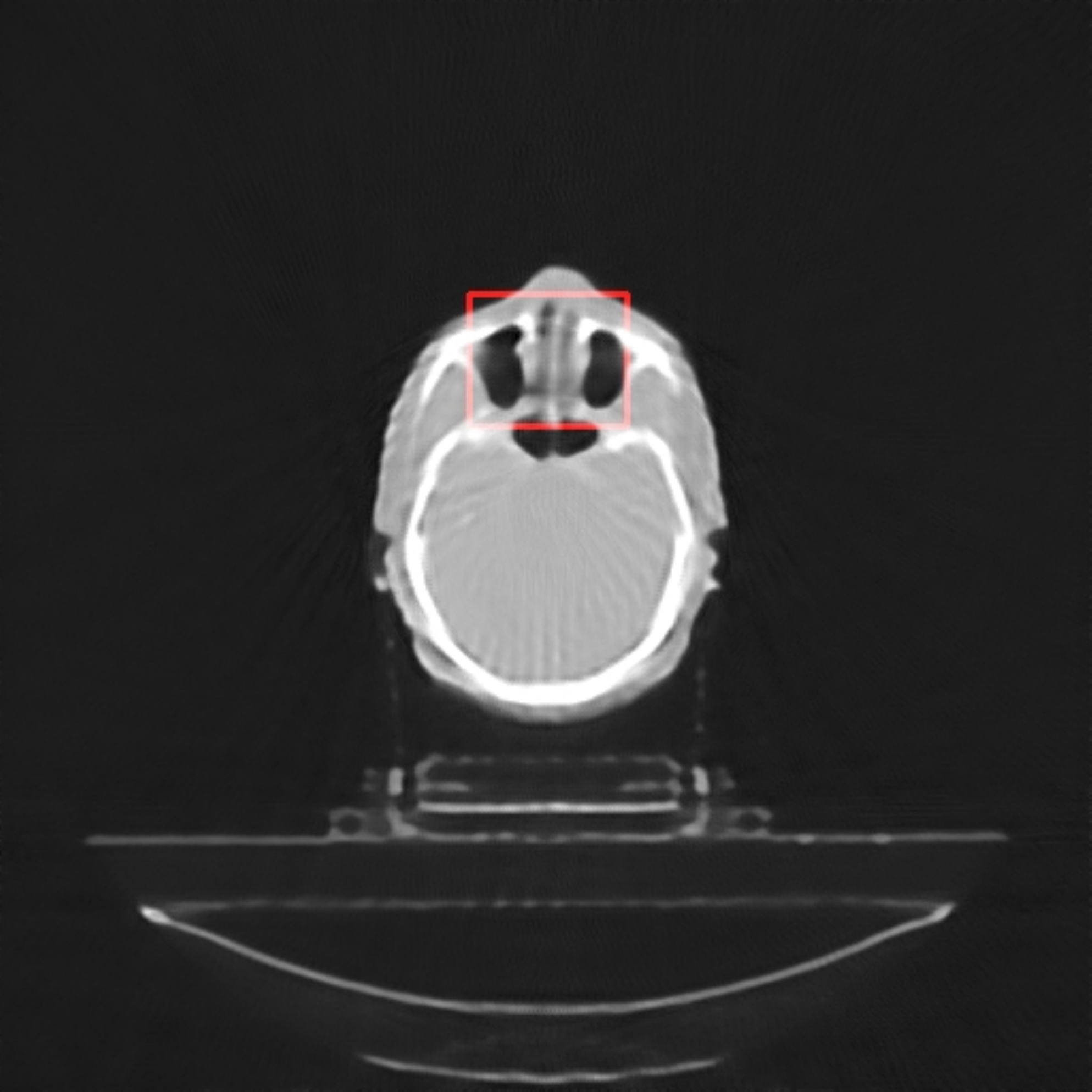}	
				\put(0,0){\color{red}%
					\frame{\includegraphics[scale=0.08]{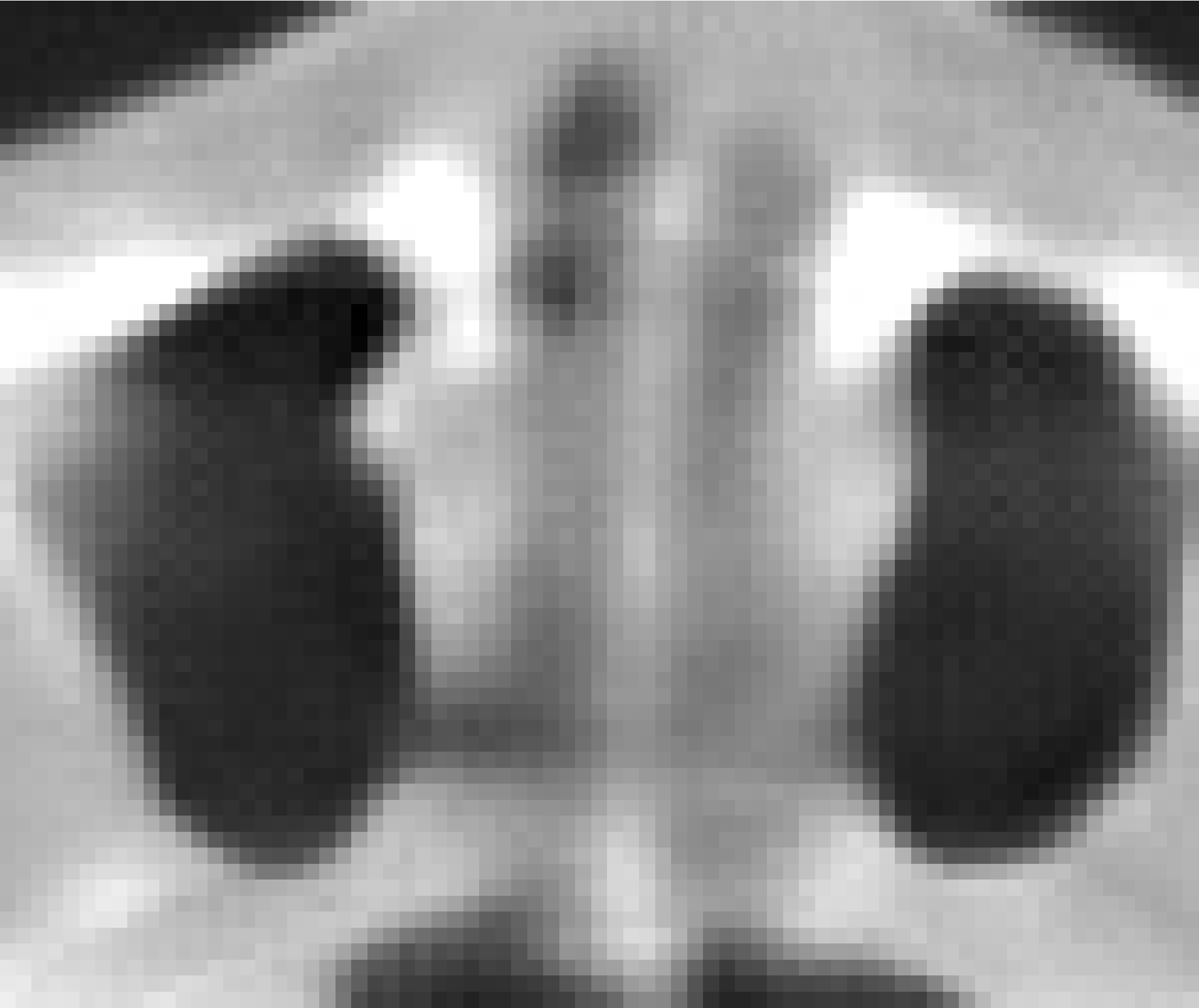}}
				}		
		\end{overpic}}
		\subfloat[Ours]{
			\begin{overpic}[width=\size\columnwidth]{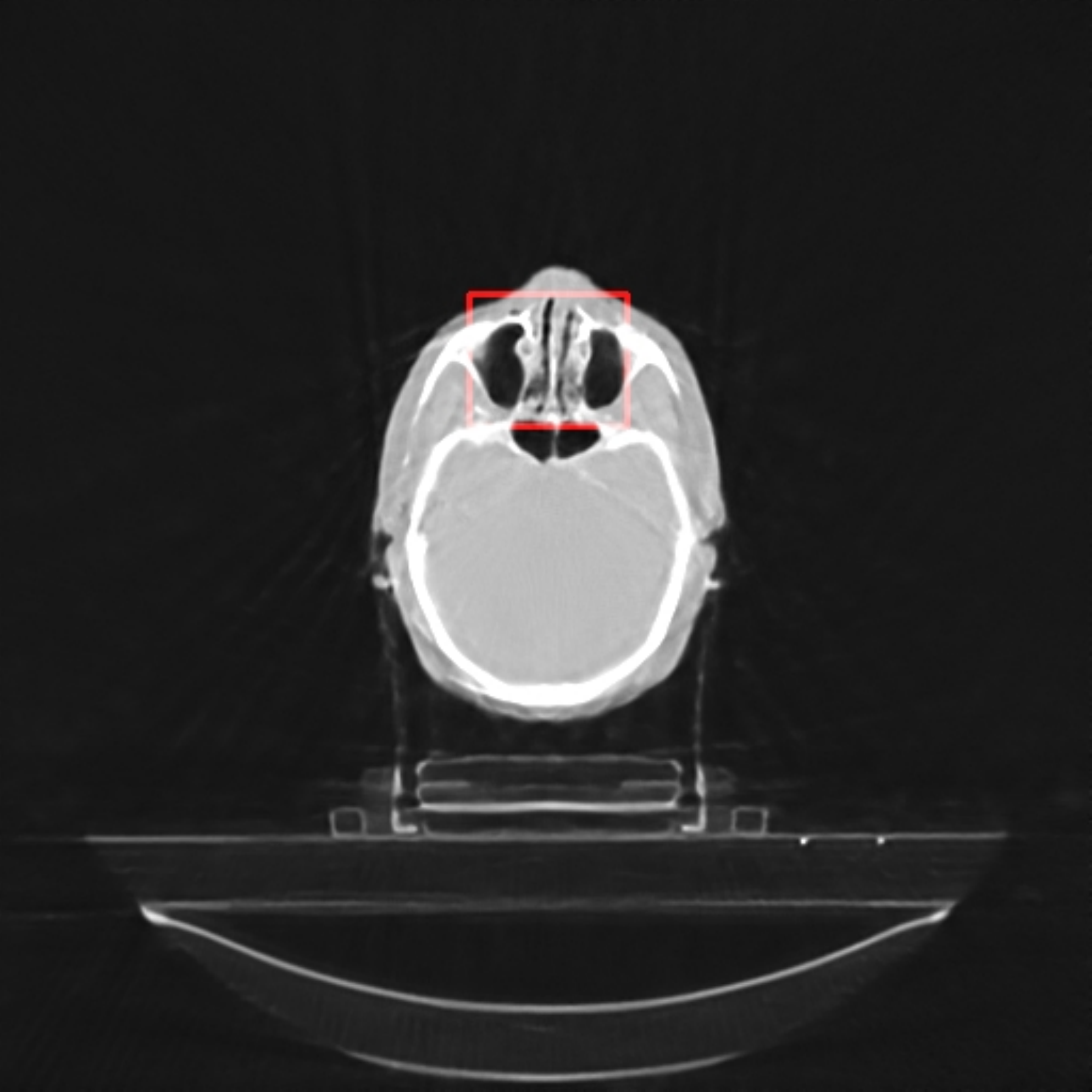}	
				\put(0,0){\color{red}%
					\frame{\includegraphics[scale=0.08]{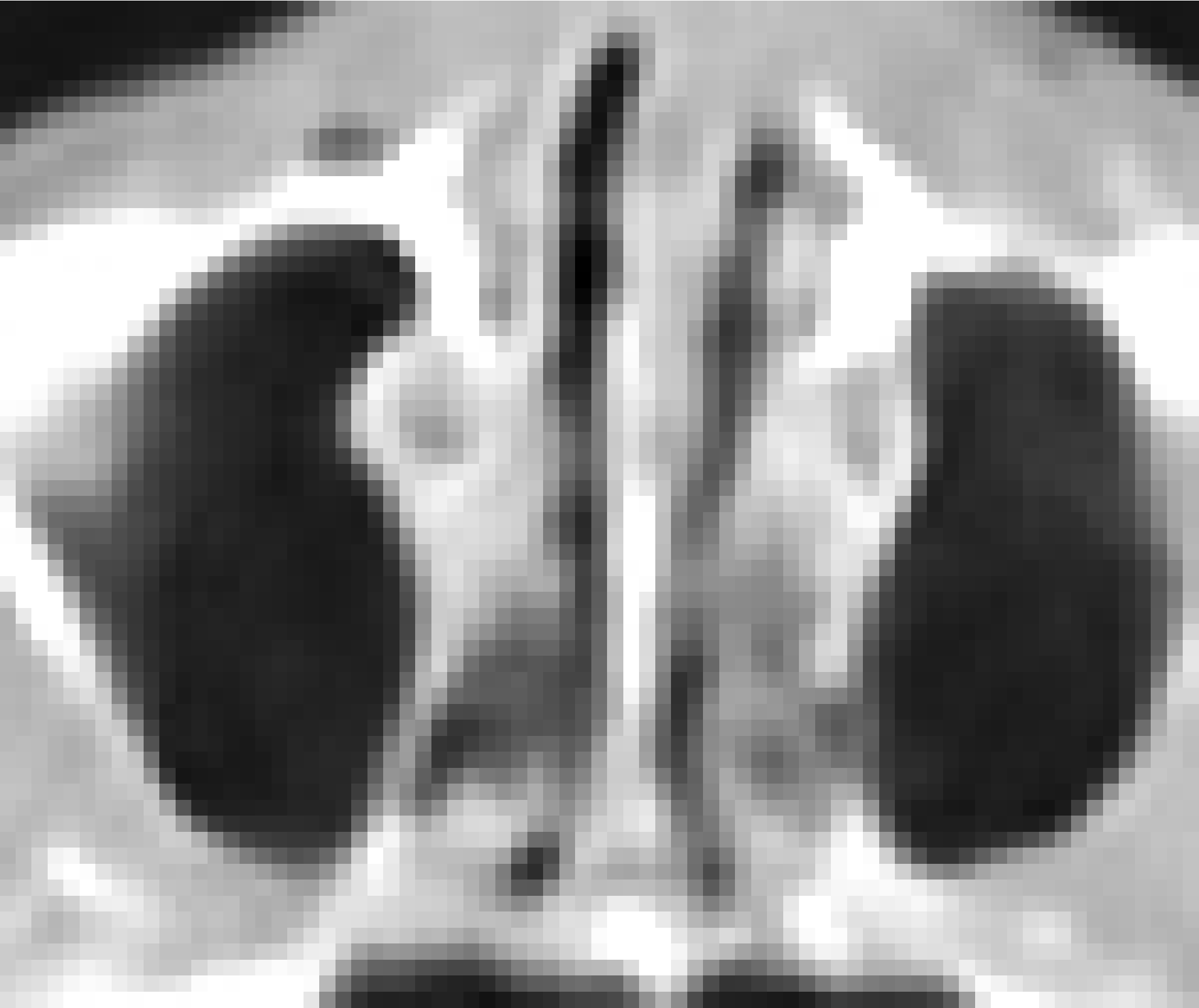}}
				}		
		\end{overpic}}
	}
	\caption{The reconstructed results of head and neck CT data for $p=14\pi$. All images are linearly stretched to [0, 1] and the display window is [0, 0.6].}
	\label{Fig4_neck}
\end{figure*}

To test the generalization of our method, we perform the reconstruction of  our network and FBPConvNet on another type of CT, the head and neck CT, where the head and neck CT dataset \cite{NeckCT} is used to simulate the test data, and  our network and FBPConvNet are  trained by the NIH-AAPM-Mayo  data only. Some reconstructed results of the head and neck CT  are presented in Fig. \ref{Fig2_neck}. We can observe that our method can reconstruct more fine details and boundaries than FBPConvNet, which demonstrates that our method has  good generalization capability. The average RMSE and SSIM of the reconstructed head and neck CT images are listed in Table \ref{T2_neck}. It can be observed that our network has lower RMSE and higher SSIM than  FBPConvNet in average.

\begin{table}[!t]
	\renewcommand{\arraystretch}{1.3}
	\caption{The RMSE and SSIM of the three methods for head and neck CT test data.}
	\label{T2_neck}
	\centering
	\begin{tabular}{cccc}
		\hline
		Pitch & Method & RMSE & SSIM \\
		& Katsevich & 76.657$\pm$12.371 & 0.863$\pm$0.031 \\
		$p=7\pi$& FBPConvNet &  64.633$\pm$12.058& 0.939$\pm$0.009 \\
		& Ours & \textbf{47.807}$\pm$9.870 &\textbf{0.959}$\pm$0.006  \\
		& Katsevich & 77.104$\pm$12.473 & 0.861$\pm$0.031 \\
		$p=14\pi$& FBPConvNet &65.427$\pm$12.627  &0.937$\pm$0.010  \\
		& Ours & \textbf{49.357}$\pm$10.729 & \textbf{0.955}$\pm$0.009 \\
		\hline
	\end{tabular}
\end{table}%

\subsection{Experimental Results for $p=14\pi$}
In this subsection, We present some experimental results to show the performance of our network for pitch $p=14\pi$. 

In Fig. \ref{Fig4_neck}, we present five CT images reconstructed by Katsevich algorithm,  Nonlocal-TV, FBPConvNet and our method from the NIH-AAPM-Mayo test data. We can observe that the images reconstructed by Katsevich algorithm suffer from severe streak artifacts. Nonlocal-TV tends to over-smooth  the reconstructed CT images. Some fine details and boundaries in the CT images reconstructed by FBPConvNet are lost. Compared to these methods, our method can simultaneously suppress the streak artifacts and reconstruct fine details and good boundaries. In Table \ref{T2}, we list the average RMSE and PSNR of the  CT images reconstructed by the four methods, from which we can observe that our method has the lowest RMSE and highest PSNR in average. We can also observe that the average RMSE and PSNR of each method for $p=7\pi$ and $p=14\pi$ vary only a little. Since our network uses Katsevich algorithm to transfer sinograms to CT images and  the inputs to FBPConvNet are also the CT images reconstructed by Katsevich algorithm, it  demonstrates that  Katsevich algorithm is robust to the helical pitch $p$.

We also test the generalization capabilities of FBPConvNet and our network for $p=14\pi$. The CT images reconstructed by  FBPConvNet and our network from the head and neck test data for $p=14\pi$ are shown in Fig. \ref{Fig4_neck}. We can see that the images reconstructed by FBPConvNet are somewhat blurry and lose some fine details while our method can reconstruct CT images with better boundaries and details. The average RMSE and SSIM of the reconstructed CT images are also listed in Table \ref{T2_neck}. We can see that our method gains  lower RMSE and higher SSIM than FBPConvNet in average. 

\section{Discussion and Conclusion}
In this paper, we developed a new GPU implementation of the Katsevich algorithm  \cite{Katsevich2004a} for helical CT, which reconstructs the CT images  pitch by pitch. By utilizing the periodic properties of the parameters of the Katsevich algorithm, our scheme only needs to calculate these parameters once for all  pitches and so  is very suitable for deep learning. By embedding our implementation of the Katsevich algorithm  into the CNN, we proposed an end-to-end deep network for the high pitch helical CT reconstruction with sparse detectors. Experiments show that our end-to-end deep network outperformed the Nonlocal-TV based iterative algorithm and the post-processed deep network FBPConvNet both in subjective and objective evaluations.

One limitation of our method is that it requires the pitch $p$ of the helical scan is constant, which inherits from the  Katsevich algorithm.  In the literature, there exist some exact algorithms for the helical CT reconstruction with variable pitches. However, the  periodic properties of the parameters of these algorithms may not hold. Therefore, we need to compute the parameters for every slice when using these algorithms to reconstruct CT images, which will consume a lot of GPU-memory and isn't fit for deep learning. In our future work, we will research how to solve this issue.

%

%
%
%
%
%

\ifCLASSOPTIONcaptionsoff
  \newpage
\fi

\balance
\bibliographystyle{IEEEtran}
\bibliography{helical}

\begin{thebibliography}{10}
\providecommand{\url}[1]{#1}
\csname url@samestyle\endcsname
\providecommand{\newblock}{\relax}
\providecommand{\bibinfo}[2]{#2}
\providecommand{\BIBentrySTDinterwordspacing}{\spaceskip=0pt\relax}
\providecommand{\BIBentryALTinterwordstretchfactor}{4}
\providecommand{\BIBentryALTinterwordspacing}{\spaceskip=\fontdimen2\font plus
\BIBentryALTinterwordstretchfactor\fontdimen3\font minus
  \fontdimen4\font\relax}
\providecommand{\BIBforeignlanguage}[2]{{%
\expandafter\ifx\csname l@#1\endcsname\relax
\typeout{** WARNING: IEEEtran.bst: No hyphenation pattern has been}%
\typeout{** loaded for the language `#1'. Using the pattern for}%
\typeout{** the default language instead.}%
\else
\language=\csname l@#1\endcsname
\fi
#2}}
\providecommand{\BIBdecl}{\relax}
\BIBdecl

\bibitem{Noo2003a}
F.~Noo, J.~Pack, and D.~Heuscher, ``Exact helical reconstruction using native
  cone-beam geometries,'' \emph{Physics in Medicine \& Biology}, vol.~48,
  no.~23, pp. 3787--3818, nov 2003.

\bibitem{Tam_1998}
K.~C. Tam, S.~Samarasekera, and F.~Sauer, ``Exact cone beam {CT} with a spiral
  scan,'' \emph{Physics in Medicine \& Biology}, vol.~43, no.~4, pp.
  1015--1024, apr 1998.

\bibitem{Danielsson97:F3D}
P.~Danielsson, P.~Edholm, J.~Eriksson, and M.~{Magnusson Seger}, ``Towards
  exact reconstruction for helical cone-beam scanning of long objects. a new
  detector arrangement and a new completeness condition,'' in \emph{Proc.~1997
  Meeting on Fully 3D Image Reconstruction in Radiology and Nuclear Medicine
  (Pittsburgh, PA)}, D.~Townsend and P.~Kinahan, Eds., 1997, pp. 141--144.

\bibitem{Katsevich2004a}
A.~Katsevich, ``An improved exact filtered backprojection algorithm for spiral
  computed tomography,'' \emph{Advances in Applied Mathematics}, vol.~32,
  no.~4, pp. 681--697, May 2004.

\bibitem{Zheng2020}
A.~Zheng, H.~Gao, L.~Zhang, and Y.~Xing, ``A dual-domain deep learning-based
  reconstruction method for fully 3d sparse data helical {CT},'' \emph{Physics
  in Medicine \& Biology}, vol.~65, no.~24, p. 245030, Dec. 2020.

\bibitem{Noo1999}
F.~Noo, M.~Defrise, and R.~Clackdoyle, ``Single-slice rebinning method for
  helical cone-beam {CT},'' \emph{Physics in Medicine \& Biology}, vol.~44,
  no.~2, pp. 561--570, Jan. 1999.

\bibitem{Stierstorfer2004a}
K.~Stierstorfer, A.~Rauscher, J.~Boese, H.~Bruder, S.~Schaller, and T.~Flohr,
  ``Weighted fbp--a simple approximate 3d fbp algorithm for multislice spiral
  {CT} with good dose usage for arbitrary pitch,'' \emph{Physics in Medicine \&
  Biology}, vol.~49, no.~11, pp. 2209--2218, May 2004.

\bibitem{WangSept}
G.~Wang, T.~.~H. Lin, P.~Cheng, and D.~M. Shinozaki, ``A general cone-beam
  reconstruction algorithm,'' \emph{IEEE Transactions on Medical Imaging},
  vol.~12, no.~3, pp. 486--496, 1993.

\bibitem{Guo2011}
J.~Guo, L.~Zeng, and X.~Zou, ``An improved half-covered helical cone-beam {CT}
  reconstruction algorithm based on localized reconstruction filter,''
  \emph{Journal of X-Ray Science and Technology}, vol.~19, no.~3, pp. 293--312,
  2011.

\bibitem{Zhao2007}
J.~Zhao, Y.~Lu, Y.~Jin, E.~Bai, and G.~Wang, ``Feldkamp-type reconstruction
  algorithms for spiral cone-beam {CT} with variable pitch,'' \emph{Journal of
  X-Ray Science and Technology}, vol.~15, no.~4, pp. 177--196, 2007.

\bibitem{Feldkamp1984}
L.~A. Feldkamp, L.~C. Davis, and J.~W. Kress, ``Practical cone-beam
  algorithm,'' \emph{Journal of the Optical Society of America A}, vol.~1,
  no.~6, pp. 612--619, Jun 1984.

\bibitem{Sidky2008}
E.~Y. Sidky and X.~Pan, ``Image reconstruction in circular cone-beam computed
  tomography by constrained, total-variation minimization,'' \emph{Physics in
  Medicine \& Biology}, vol.~53, no.~17, pp. 4777--4807, Aug. 2008.

\bibitem{Liu2012}
Y.~Liu, J.~Ma, Y.~Fan, and Z.~Liang, ``Adaptive-weighted total variation
  minimization for sparse data toward low-dose x-ray computed tomography image
  reconstruction,'' \emph{Physics in Medicine \& Biology}, vol.~57, no.~23, pp.
  7923--7956, Nov. 2012.

\bibitem{Chen2009}
Y.~Chen, D.~Gao, C.~Nie, L.~Luo, W.~Chen, X.~Yin, and Y.~Lin, ``Bayesian
  statistical reconstruction for low-dose x-ray computed tomography using an
  adaptive-weighting nonlocal prior,'' \emph{Computerized Medical Imaging and
  Graphics}, vol.~33, no.~7, pp. 495--500, Oct. 2009.

\bibitem{XuSept}
Q.~Xu, H.~Yu, X.~Mou, L.~Zhang, J.~Hsieh, and G.~Wang, ``Low-dose x-ray {CT}
  reconstruction via dictionary learning,'' \emph{IEEE Transactions on Medical
  Imaging}, vol.~31, no.~9, pp. 1682--1697, 2012.

\bibitem{BaoNov.}
P.~Bao, W.~Xia, K.~Yang, W.~Chen, M.~Chen, Y.~Xi, S.~Niu, J.~Zhou, H.~Zhang,
  H.~Sun, Z.~Wang, and Y.~Zhang, ``Convolutional sparse coding for compressed
  sensing {CT} reconstruction,'' \emph{IEEE Transactions on Medical Imaging},
  vol.~38, no.~11, pp. 2607--2619, 2019.

\bibitem{Sunnegaardh2021}
J.~Sunnegårdh and P.-E. Danielsson, ``Regularized iterative weighted filtered
  backprojection for helical cone-beam {CT},'' \emph{Medical Physics}, vol.~35,
  no.~9, pp. 4173--4185, Nov. 2008.

\bibitem{Yu2021}
W.~Yu and L.~Zeng, ``Iterative image reconstruction for limited-angle inverse
  helical cone-beam computed tomography,'' \emph{Scanning}, vol.~38, no.~1, pp.
  4--13, Nov. 2016.

\bibitem{Katsevich2021}
A.~Katsevich, ``Theoretically exact filtered backprojection-type inversion
  algorithm for spiral {CT},'' \emph{SIAM Journal on Applied Mathematics},
  vol.~62, no.~6, pp. 2012--2026, Nov. 2002.

\bibitem{Zou2004}
Y.~Zou and X.~Pan, ``Image reconstruction on pi-lines by use of filtered
  backprojection in helical cone-beam {CT},'' \emph{Physics in Medicine \&
  Biology}, vol.~49, no.~12, pp. 2717--2731, Jun. 2004.

\bibitem{Zou2004a}
------, ``Exact image reconstruction on pi-lines from minimum data in helical
  cone-beam {CT},'' \emph{Physics in Medicine \& Biology}, vol.~49, no.~6, pp.
  941--959, Feb. 2004.

\bibitem{Ye2021}
Y.~Ye and G.~Wang, ``Filtered backprojection formula for exact image
  reconstruction from cone-beam data along a general scanning curve,''
  \emph{Medical Physics}, vol.~32, no.~1, pp. 42--48, Nov. 2005.

\bibitem{Zou2021}
Y.~Zou, X.~Pan, D.~Xia, and G.~Wang, ``Pi-line-based image reconstruction in
  helical cone-beam computed tomography with a variable pitch,'' \emph{Medical
  Physics}, vol.~32, no.~8, pp. 2639--2648, Nov. 2005.

\bibitem{YanJuly}
G.~Yan, J.~Tian, S.~Zhu, C.~Qin, Y.~Dai, F.~Yang, D.~Dong, and P.~Wu, ``Fast
  katsevich algorithm based on gpu for helical cone-beam computed tomography,''
  \emph{IEEE Transactions on Information Technology in Biomedicine}, vol.~14,
  no.~4, pp. 1053--1061, 2010.

\bibitem{LeeMarc}
H.~Lee, J.~Lee, H.~Kim, B.~Cho, and S.~Cho, ``Deep-neural-network-based
  sinogram synthesis for sparse-view {CT} image reconstruction,'' \emph{IEEE
  Transactions on Radiation and Plasma Medical Sciences}, vol.~3, no.~2, pp.
  109--119, 2019.

\bibitem{Fu2020}
J.~Fu, J.~Dong, and F.~Zhao, ``A deep learning reconstruction framework for
  differential phase-contrast computed tomography with incomplete data,''
  \emph{IEEE Transactions on Image Processing}, vol.~29, pp. 2190--2202, 2020.

\bibitem{ChenDec.}
H.~Chen, Y.~Zhang, M.~K. Kalra, F.~Lin, Y.~Chen, P.~Liao, J.~Zhou, and G.~Wang,
  ``Low-dose {CT} with a residual encoder-decoder convolutional neural
  network,'' \emph{IEEE Transactions on Medical Imaging}, vol.~36, no.~12, pp.
  2524--2535, 2017.

\bibitem{JinSept}
K.~H. Jin, M.~T. McCann, E.~Froustey, and M.~Unser, ``Deep convolutional neural
  network for inverse problems in imaging,'' \emph{IEEE Transactions on Image
  Processing}, vol.~26, no.~9, pp. 4509--4522, 2017.

\bibitem{ZhangJune}
Z.~Zhang, X.~Liang, X.~Dong, Y.~Xie, and G.~Cao, ``A sparse-view {CT}
  reconstruction method based on combination of densenet and deconvolution,''
  \emph{IEEE Transactions on Medical Imaging}, vol.~37, no.~6, pp. 1407--1417,
  2018.

\bibitem{Ge2020}
Y.~Ge, T.~Su, J.~Zhu, X.~Deng, Q.~Zhang, J.~Chen, Z.~Hu, H.~Zheng, and
  D.~Liang, ``Adaptive-net: deep computed tomography reconstruction network
  with analytical domain transformation knowledge,'' \emph{Quantitative Imaging
  in Medicine and Surgery}, vol.~10, no.~2, pp. 415--427, 2020.

\bibitem{Lin15-2}
W.~Lin, H.~Liao, C.~Peng, X.~Sun, J.~Zhang, J.~Luo, R.~Chellappa, and S.~K.
  Zhou, ``Dudonet: Dual domain network for {CT} metal artifact reduction,'' in
  \emph{2019 IEEE/CVF Conference on Computer Vision and Pattern Recognition
  (CVPR)}, 15-2, pp. 10\,504--10\,513.

\bibitem{Zhang2020}
Q.~Zhang, Z.~Hu, C.~Jiang, H.~Zheng, Y.~Ge, and D.~Liang, ``Artifact removal
  using a hybrid-domain convolutional neural network for limited-angle computed
  tomography imaging,'' \emph{Physics in Medicine \& Biology}, vol.~65, no.~15,
  p. 155010, Aug. 2020.

\bibitem{Wang2020}
W.~Wang, X.~G. Xia, C.~He, Z.~Ren, J.~Lu, T.~Wang, and B.~Lei, ``An end-to-end
  deep network for reconstructing {CT} images directly from sparse sinograms,''
  \emph{IEEE Transactions on Computational Imaging}, vol.~6, pp. 1548--1560,
  2020.

\bibitem{ChenJune}
H.~Chen, Y.~Zhang, Y.~Chen, J.~Zhang, W.~Zhang, H.~Sun, Y.~Lv, P.~Liao,
  J.~Zhou, and G.~Wang, ``Learn: Learned experts' assessment-based
  reconstruction network for sparse-data {CT},'' \emph{IEEE Transactions on
  Medical Imaging}, vol.~37, no.~6, pp. 1333--1347, 2018.

\bibitem{HeFeb.}
J.~He, Y.~Yang, Y.~Wang, D.~Zeng, Z.~Bian, H.~Zhang, J.~Sun, Z.~Xu, and J.~Ma,
  ``Optimizing a parameterized plug-and-play admm for iterative low-dose {CT}
  reconstruction,'' \emph{IEEE Transactions on Medical Imaging}, vol.~38,
  no.~2, pp. 371--382, 2019.

\bibitem{PI}
S.~Izen, ``A fast algorithm to compute the $\pi$-line through points inside a
  helix cylinder,'' \emph{Proceedings of The American Mathematical Society},
  vol. 135, pp. 269--276, 07 2006.

\bibitem{7780459}
K.~He, X.~Zhang, S.~Ren, and J.~Sun, ``Deep residual learning for image
  recognition,'' in \emph{2016 IEEE Conference on Computer Vision and Pattern
  Recognition (CVPR)}, 2016, pp. 770--778.

\bibitem{INSPEC:3589863}
R.~Hecht-Nielsen, ``{Theory of the Backpropagation Neural Network},'' in
  \emph{IJCNN: International Joint Conference on Neural Networks}, {1989}, pp.
  {593--605}.

\bibitem{kingma2014adam}
D.~P. Kingma and J.~Ba, ``{Adam: A Method for Stochastic Optimization},''
  \emph{Preprint, arXiv:https://arxiv.org/abs/1412.6980v8}, {2014}.

\bibitem{data}
C.~McCollough, ``{TU-FG-207A-04: Overview of the Low Dose {CT} Grand
  Challenge},'' \emph{Medical Physics}, vol.~{43}, no. {3759-3760}, pp.
  {3759--3760}, {2016}.

\bibitem{ISI:000311351400019}
Y.~Liu, J.~Ma, Y.~Fan, and Z.~Liang, ``{Adaptive-Weighted Total Variation
  Minimization for Sparse Data Toward Low-Dose X-Ray Computed Tomography Image
  Reconstruction},'' \emph{{Physics in Medicine \& Biology}}, vol.~{57},
  no.~{23}, pp. {7923--7956}, {2012}.

\bibitem{zhang2013improved}
J.~Zhang, S.~Liu, R.~Xiong, S.~Ma, and D.~Zhao, ``Improved total variation
  based image compressive sensing recovery by nonlocal regularization,'' in
  \emph{IEEE International Symposium on Circuits and Systems (ISCAS)}, 2013,
  pp. 2836--2839.

\bibitem{ISI:000405701500004}
K.~H. Jin, M.~T. McCann, E.~Froustey, and M.~Unser, ``{Deep Convolutional
  Neural Network for Inverse Problems in Imaging},'' \emph{IEEE Transactions on
  Image Processing}, vol.~{26}, no.~{9}, pp. {4509--4522}, {2017}.

\bibitem{NeckCT}
T.~Bejarano, M.~De~Ornelas-Couto, and I.~B. Mihaylov, ``Longitudinal fan-beam
  computed tomography dataset for head-and-neck squamous cell carcinoma
  patients,'' \emph{Medical Physics}, vol.~46, no.~5, pp. 2526--2537, 2019.

\end{thebibliography}

\end{document}